\documentclass[a4paper,11pt]{article}
\pdfoutput=1 % if your are submitting a pdflatex (i.e. if you have
\usepackage{jcappub} 

\usepackage[T1]{fontenc} % if needed

\usepackage{Extrapackages}
\usepackage{xcolor}
\usepackage{cancel}
\usepackage{arydshln}

\usepackage{wrapfig}
\usepackage{picins}

\newcommand{\Mp}{M_{\mathrm{Pl}}}

\newcommand{\Nf}{{N_{\mathrm{f}}}}
\newcommand{\Lh}{\Lambda_{\mathrm{h}}}
\newcommand{\Lv}{\Lambda_{\mathrm{v}}}

\newcommand{\Lhd}{\Lambda^{\rm DBM}_{\mathrm{h}}}
\newcommand{\Lvd}{\Lambda^{\rm DBM}_{\mathrm{v}}}

\newcommand{\epsilonV}{\epsilon_{\mathrm{V}}}
\newcommand{\etaV}{\eta_{\mathrm{V}}}

\newcommand{\epsiloni}{\epsilon_{\mathrm{i}}}
\newcommand{\etai}{\eta_{\mathrm{i}}}

\newcommand{\Mpl}{\ensuremath{M_{\rm Pl}}}

\newcommand\be{\begin{equation}}
\newcommand\ee{\end{equation}}
\newcommand{\bea}{\begin{eqnarray}}
\newcommand{\eea}{\end{eqnarray}}
\newcommand{\fnl}{\ensuremath{f_{\rm NL}}}

\graphicspath{{/home/hutb2/Documents/Wolfram Mathematica/GRF Inflation/Graphs/}}

\definecolor{mygreen}{rgb}{0.0, 0.5, 0.0}

\definecolor{myred}{rgb}{0.8, 0.0, 0.0}

\title{\boldmath{Manyfield Inflation in Random Potentials}}

\author{Theodor Bjorkmo}
\author{and M.C.~David Marsh}

\affiliation{Department of Applied Mathematics and Theoretical Physics, University of Cambridge\\Wilberforce Road, Cambridge, UK}

\emailAdd{t.bjorkmo@damtp.cam.ac.uk}
\emailAdd{m.c.d.marsh@damtp.cam.ac.uk}

\abstract{We  construct models of inflation with many randomly interacting fields and use these to study the generation of cosmological observables. We model the potentials as multi-dimensional Gaussian random fields (GRFs) and identify powerful algebraic simplifications that, for the first time, make it possible to access the manyfield limit of inflation in GRF potentials. Focussing on small-field, slow-roll, approximate saddle-point inflation in potentials with structure on sub-Planckian scales, we construct explicit examples involving up to 100 fields and generate statistical ensembles comprising of 164,000 models involving 5 to 50 fields. For the subset of these that support at least sixty e-folds of inflation,   we use the `transport method' and $\delta N$ formalism to determine the predictions for cosmological observables at the end of inflation, including the power spectrum and the local non-Gaussianities of the primordial perturbations.
 We find three key results: i) Planck compatibility is not rare, but future experiments may rule out this class of models; ii) In the manyfield limit, the predictions from these models agree well with, but are sharper than,  previous results derived using potentials constructed through  non-equilibrium Random Matrix Theory; iii) Despite substantial multifield effects, non-Gaussianities are typically very small: $\fnl^{\rm loc} \ll 1$. We conclude that many of the `generic predictions' of single-field inflation can be emergent features of complex inflation models.

}

\begin{document}

\maketitle
\flushbottom

\newcommand{\vv}[1]{\mathbf{#1}}
\newcommand{\vh}[1]{\mathbf{\hat{#1}}}
\newcommand{\nb}{\mathbf{\nabla}}
\newcommand{\dv}{\mathbf{\nabla}\cdot}
\newcommand{\cl}{\mathbf{\nabla}\times}
\newcommand{\bs}[1]{\boldsymbol{#1}}
\newcommand{\vvv}{\mathbf{v}}
\newcommand{\vvu}{\mathbf{u}}
\newcommand{\vvx}{\mathbf{x}}
\newcommand{\vvk}{\mathbf{k}} %new
\newcommand{\vvp}{\mathbf{p}} %new
\newcommand{\vvr}{\mathbf{r}}
\newcommand{\vvR}{\mathbf{R}} %new
\newcommand{\vvG}{\mathbf{G}} %new
\newcommand{\mc}[1]{\mathcal{#1}} %new
\newcommand{\vvrd}{\dot{\mathbf{r}}}
\newcommand{\vvrdd}{\ddot{\mathbf{r}}}
\newcommand{\rrd}{\dot{r}}
\newcommand{\rrdd}{\ddot{r}}
\newcommand{\thetad}{\dot{\theta}}
\newcommand{\thetadd}{\ddot{\theta}}
\newcommand{\phid}{\dot{\phi}}
\newcommand{\phidd}{\ddot{\phi}}

\newcommand{\eq}[1]{\begin{equation}{#1}\end{equation}}
\newcommand\nt{\addtocounter{equation}{1}\tag{\theequation}}
\newcommand{\mui}{^{\mu}}
\newcommand{\mli}{_{\mu}}
\newcommand{\nui}{^{\nu}}
\newcommand{\nli}{_{\nu}}
\newcommand{\pd}[2]{\frac{\partial {#1}}{\partial {#2}}}
\newcommand{\td}[2]{\frac{d{#1}}{d{#2}}}
\newcommand{\spd}[2]{\frac{\partial^2 {#1}}{\partial {#2}^2}}
\newcommand{\std}[2]{\frac{d^2 {#1}}{d{#2}^2}}

\newcommand{\dd}{{\rm d}}

\section{Introduction}
\label{sec:intro}
%The inflationary paradigm has so far been very successful. It 
Inflation 
provides a rather simple %and compelling
 explanation of the origin of the primordial density perturbations and 
%it 
successfully 
resolves the flatness and homogeneity problems of the   standard hot big bang cosmology. 
%, and
%the generic predictions for the primordial density perturbations that 
 %the primordial curvature perturbations predicted by inflationary models 
% are  consistent with all cosmological observations. 
 However, little is known about the microscopic origin of inflation and, in particular, what degrees of freedom  it involved.   
  Inflation may have probed energies far above those accessible by terrestrial experiments, and 
  is sensitive to 
  %
  %almost certainly involved {\bf [repeated word]} 
  physics beyond the Standard Model of particle physics. Models of inflation with only a single additional scalar field can be compatible with all current observations, but so may models with multiple fields. Additional scalar fields are common in extensions of the Standard Model that address the gauge hierarchy problem, and ubiquitous in ultraviolet completions realised in string theory.
  %, where large numbers of deformation modes of the higher dimensional compactification manifold appear in the low-energy theory as uncharged scalar fields.
   Determining the field content relevant in the early universe is a fundamental challenge of modern cosmology.

Primordial non-Gaussianity of the local type has been proposed as  a key observable
to observationally distinguish between multifield and single-field  models of inflation.
 In single-field inflation, the levels of local non-Gaussianity can be related to the deviation from scale invariance of the primordial power spectrum \cite{astro-ph/0210603, Creminelli:2004yq},\footnote{This rule applies under some assumptions, which can be violated in special models \cite{Chen:2013aj, Mooij:2015yka}.} which is very small \cite{1502.02114}. Multiple-field effects can significantly enhance the  levels of non-Gaussianity, and amplitudes of the order of $f_{\rm NL} \sim 1$ are realised in some models.\footnote{In this paper, we  focus on non-Gaussianities of the local type, and denote $\fnl^{\rm loc} = \fnl$ without superscript.} Current constraints from Planck observations of the Cosmic Microwave Background give $f_{\rm NL} = 0.8 \pm 5.0$ (68\% c.l.) \cite{Ade:2015ava}, and future surveys of the Large-Scale Structure of the universe  are expected to reach a sensitivity of $\sigma(f_{\rm NL})\sim {\cal O}(1)$ \cite{
Dalal:2007cu,Matarrese:2008nc, Slosar:2008hx, Desjacques:2010jw, 
 Alvarez:2014vva,
 Ferraro:2014jba,
  Baldauf:2016sjb},  probing some subset of models of  multiple-field inflation.
It is now pressing to assess what we realistically can hope to  learn about fundamental physics from these experiments.

 The conditions under which large non-Gaussianity is generated
 during and after inflation have been  studied before by many authors (for a review, see \cite{1002.3110}). However, direct investigations tend to be hampered by the computational complexity of multi-field systems, and most studies have been restricted to models with two or a few fields, or models with greatly restricting  symmetry structures \cite{Vernizzi:2006ve, Battefeld:2006sz, Senatore:2010wk, Peterson:2010mv, Peterson:2011yt, 1312.4035, Price:2015qqb}.  
 %
  %in two-field inflation models (refs) have been studied before, primarily for sum-separable potentials. 
  For a more complete understanding of multifield inflation, it is necessary to go beyond these simplifying assumptions, and allow  both for  more fields and for non-trivial interactions. This is crucial for understanding what  models of inflation can be ruled out if $\fnl$ is constrained to be less than one, or what types of inflationary models are favoured if $\fnl$ of order one is measured. Addressing this question is one of the main aims of this paper.

Multifield inflation models with %non-suppressed 
generic
interactions between the fields  have large numbers of free parameters.
In a low-energy effective theory for $\Nf$ fields valid below the cut-off scale $\Lambda$, these are 
the Wilson coefficients, $c_{a_1 \ldots a_n}$, of all operators that may be
important during inflation, e.g.
\be
V(\phi_1, \ldots, \phi_\Nf) = \Lambda^4\, \sum_{n=0}^{n_{\rm max}} c_{a_1 \ldots a_n} \frac{\phi^{a_1}}{\Lambda} \ldots \frac{\phi^{a_n}}{\Lambda} \, .
\label{eq:V1}
\ee
%renormalisable and non-renormalisable operators. 
Unfortunately, 
the relevant values (or distribution of values) of these parameters are not known from fundamental physics.
One approach, pursued here, is then to search for properties that are rather insensitive to the details of the parameter distribution, and that  depend only on a few effective parameters. The widespread appearance of emergent universality in complex physical and mathematical systems suggests that such robustness may be found as the number of fields, $\Nf$, becomes large \cite{wigner, Kuijlaars, Deift, Erdos}.  
Motivated by this, we  pursue 
a statistical approach: 
we
generate ensembles of multifield scalar potentials $V(\phi_1, \ldots , \phi_\Nf)$ randomly, and determine the distribution of observables as $\Nf\gg 1$.

%\footnote{
%
% We will find that several of the predictions derived from our random potentials agree well 
 %
 % extend more broadly to systems with interacting fields. 
%Inflationary models with many non-interacting or  free fields (e.g.~assisted inflation \cite{astro-ph/9804177})
%
%} 

To 
access the interesting regime of multiple light fields with non-trivial interactions, 
%
%this approach {\color{blue} interesting} and tractable, the randomly generated potentials need to be sufficiently complicated to allow for non-trivial interactions between the fields, but
the potentials need to be
 mathematically simple enough to be computationally tractable. 
One such class of potentials, recently studied in \cite{DBM1, DBM2, DBM3, 1409.5135, Freivogel, 1611.07059, Wang:2016kzp}, can be constructed using non-equilibrium  random matrix theory techniques. 
According to the prescription of \cite{DBM1},  the computational difficulties of multifield inflation can be substantially mitigated by realising
 $V(\phi_1, \ldots, \phi_\Nf)$ only locally along the field trajectory (while being undetermined elsewhere in field space), and by postulating that the Hessian matrix evolves according to Dyson Brownian motion (DBM) along the inflationary path. 
The local Taylor coefficients to quadratic order, defined patch-wise along the path, evolve non-trivially during inflation and implicitly capture the effects of higher-order interaction terms. This  method remains computationally efficient up to very large $\Nf$, making it %for the first time 
possible to determine the  observational predictions\footnote{This method is limited to observables that  can be inferred from information about the potential up to  second order in derivatives as expanded around any point along the field trajectory. As we will review in section \ref{sec:methodperts},  this includes quantities computed from the two-field correlators such as the  primordial power spectrum,  including its spectral index and its running.} in models of inflation with up to a hundred  interacting fields \cite{DBM2, DBM3}. 
In reference \cite{DBM2}, it was shown that the predictions of these `DBM models' become simpler and sharper as the number of fields is increases, and very complicated models with many fields are commonly compatible with Planck constraints on the primordial power spectrum \cite{DBM3}. 
%At large $\Nf$, many of the predictions were shown to simplify
%, and could be  understood from
%due to  eigenvalue repulsion of the eigenvalues of the Hessian matrix. 

However, the random matrix theory method of \cite{DBM1} is not suitable to investigate the generation of primordial non-Gaussianities during inflation: the Brownian motion of the eigenvalues of the Hessian matrix is continuous but not differentiable, and the third derivatives of the potential, required for the computation of the three-point correlation function, are not well-defined in the continuum limit.\footnote{This obstacle may be overcome 
  by regularisation, or by modifying the rules governing the stochastic evolution (cf.~\cite{1409.5135} for one suggestion).}

 \subsubsection*{Manyfield inflation from Gaussian random fields} 
  %this technique is however not suitable for investigating primordial non-Gaussianity from superhorizon evolution, which requires knowledge of the third derivatives of the potential. 
  An alternative  approach 
   is to generate random multifield potentials using Gaussian random fields (GRFs).
   This first was done  in \cite{Frazer:2011tg, 1111.6646}  by expanding $V(\phi_1, \ldots , \phi_\Nf)$ in a set of Fourier modes for potentials with $\Nf \leq 6$ and $\Lambda > \Mp$ (see also \cite{astro-ph/0410281} for $\Nf=1$). However, the interesting regime of multiple-field inflation in potentials with structure on sub-Planckian distances in field space remained intractable.

  % An interesting suggestion for using GRFs for inflation and possibly 
  To 
  access the regime with $\Lambda < M_{\rm Pl}$,  reference   \cite{Bachlechner:2014rqa}
  proposed to generate the potential only locally in field space, e.g.~by gluing together multiple patches  along a path in field space, or by generating the Taylor coefficients of the potential to a sufficiently high order at a single point.  
   %
   %Reference  \cite{Bachlechner} proposed a general method and, in the particular case of  a Gaussian covariance function, studied the scalings of the slow-roll parameters in ensembles of such potentials  \cite{Bachlechner}.
   %This method was further elaborated on in \cite{Masoumi} for more general covariance functions, and studied explicitly for $\Nf=1$ and $\Nf=2$.   
   %
 %   
    These models have well-defined higher derivatives and are arguably simpler than the DBM potentials, but 
    a significant limitation arises 
    %
    %due to the computational limits arising 
    from 
    the need to explicitly specify a very large number of Taylor coefficients, which are not statistically independent.
    % but correlated through  the inverse  of the covariance matrix.
    %
    %
    %the sheer size covariance matrix of the Taylor coefficients, which grows quickly with $\Nf$. 
    For example, 
  a model    with $\Nf=100$ fields 
  and the potential expanded up to fifth order around a single point involves  96,560,546 independent Taylor coefficients. The probability distribution  of these coefficients involves the inverse covariance matrix which has 
      $4.7 \times10^{15}$ independent, and in general non-vanishing, elements. Naively  generating such a matrix numerically 
      is computationally prohibitive, 
      %requires roughly $80$ TB of memory {\color{blue} [old number; accuracy?]}, 
      making explicit studies impractical or impossible. %\footnote{More precisely, 
 %the key obstacle is related to the size of the inverse covariance matrix, which is  a dense matrix even if the covariance matrix is sparse.  
    % } 
  %  Hence, this method has in the past only been applied to the single-field case \cite{Masoumi}, or the case of two fields in which only a single field is sufficiently light to be dynamically important during inflation \cite{Masoumi} (for related earlier studies, see \cite{Tegmark}).

     %and it is more or less impossible to invert it.

    %, in 
   % have i
    %In
    %the past  only been effectively applied to single-field inflation (ref to MVY).

In this paper, we, for the first time, overcome these obstacles and generate multifield GRFs to explicitly study the manyfield limit of inflation in general potentials. We construct models with 
 up to 100 fields by generating the potential locally around an `approximate saddle-point' up to fifth order in the fields,
 and we use an adaption of the `transport method' \cite{1302.3842, 1203.2635, 1502.03125, Dias:2016rjq} to compute cosmological observables from the two-field and three-field correlation functions. 
To make this possible, 
%We circumvent the numerical obstacle of inverting the very large covariance matrices by identifying 
we identify
%Key to this andvance is the 
drastic algebraic simplifications
for GRFs with  a Gaussian covariance function, and we use these to obviate the need for extremely heavy numerics. 
This key advance allows us to study the generation of local non-Gaussianities in random manyfield models of inflation, and assess what levels of \fnl~are generated.

There are three particularly important results in this paper:
\begin{enumerate}
	\item Planck compatible power spectra are not rare for these  models: even for highly complicated manyfield models with millions of non-vanishing interaction terms, the spectral index commonly falls within the observationally allowed range. Interestingly, these models make a sharp statistical prediction for the running of the spectral index, $\alpha_s = \dd  n_s/\dd \ln k$, which can be ruled out by future experiments.
	\item At large $\Nf$, the observational predictions of our GRF models agree well with, but are sharper than, recent predictions derived from DBM potentials. 
%	
%	We show that the GRF potentials we construct are fundamentally different from the DBM potentials
	%, and differ substantially in predictions when $\Nf$ is small. As the number of fields is increased, 
	%
%	the predictions from these  constructions are in good agreement, with our GRF models providing sharper predictions that those derived using RMT methods. 
	As these two constructions are fundamentally different and independent,  this 
	indicates the existence of a `universality class' of  large-$\Nf$ models for which  the  observables are largely insensitive to the details of the underlying potential. 
	\item The amplitude of local non-Gaussianities is typically very small,  $|\fnl|\ll1$. Even when the power spectrum undergoes significant superhorizon evolution, indicative of multifield effects being important, $\fnl$ is 
	typically highly suppressed,
		and even approximately follows the single-field consistency relation: $\fnl = \frac{5}{12}(n_s-1)$. Moreover, in the rare cases where $\fnl \sim {\cal O}(1)$, isocurvature modes remain  unsuppressed at the end of inflation, and a detailed modelling of the reheating dynamics is required to extract reliable predictions. We conclude that 
%
%the ratio of the power in the isocurvature modes to the adiabatic mode was $\gtrsim0.1$, meaning that these values for $\fnl$ have not frozen out, so to make predictions in these cases, one would have to understand the reheating dynamics.
%
constraining $\fnl$ to be smaller than order unity would not rule out manyfield inflation, but a measurement of a large value for $\fnl$ would point to rather special inflationary dynamics. 
\end{enumerate}

We expect that the predictions of this class of models may extend also to other constructions of small-field, slow-roll models of  approximate saddle-point  inflation. However, distinct classes of multifield models (such as large-field models, or models with sharp features in the potential) may well lead to different predictions for some observables.

This paper is organised as follows: in section \ref{sec:method}, we review how GRFs can be used as models for multifield inflationary potentials, and we illustrate the key simplifications that allow us to access the manyfield regime. We furthermore discuss the natural energy scales intrinsic to GRF potentials, their possible interpretation as physical effective field theory potentials, and we critically discuss the tuning required to use these potentials to study multifield inflation. We finally present the ensembles of potentials that we study explicitly, and our method for computing cosmological observables. In sections \ref{sec:result1}--\ref{sec:result3} we discuss the three main results of this paper.  We conclude and discuss further directions in section \ref{sec:conclusions}. A number of additional details, including illustrative case studies,  can be found in the appendices.  

%in section \ref{sec:casestudies} we discuss three particularly interesting inflationary models, and some possible variations in our underlying ensemble of potentials.

Throughout this paper we set the reduced Planck mass to one, $M_\text{Pl} =2.4\times 10^{18}\, {\rm GeV} =1$, but we occasionally reinstate factors of $M_\text{Pl}$ for clarity.  
%
% Moved to conclusions
%
%Consequently, to differentiate between multifield and single field inflation it may become necessary to go beyond the most basic observables, $n_s$ and $\fnl$. It is possible that the full bispectrum would be more useful in differentiating between different models, but there are other things one could look at, such as presence of primordial isocurvature modes or signs of particle production in the power spectrum. The latter is one we intend to pursue in future research.

\section{Gaussian random fields for inflation}
\label{sec:method}

In this section we explain how we use Gaussian random fields (GRFs) to study random multifield inflation.\footnote{%
Previous work on inflation in random potentials include
\cite{Berera:1996nv, astro-ph/0410281, Aazami:2005jf, 
Easther:2005zr,
MarchRussell:2006mj,Tye:2008ef, Tye:2009ff, Frazer:2011tg, 1111.6646, Battefeld:2012qx, Battefeld:2013xwa, Pedro:2013nda, Liu:2015dda, Linde:2016uec, Masoumi:2016eag, He:2017kqc, Liu:2017dzi, Masoumi:2017xbe}. References \cite{Green:2014xqa, Amin:2015ftc, Amin:2017wvc} studied the impact of randomness on particle production during inflation, and references \cite{Agarwal:2011wm,McAllister:2012am, Dias:2012nf} investigated random compactification effects in brane inflation in string theory. 
} The basic idea is to construct the potential locally in field-space as a truncated Taylor series with randomly generated coefficients. 
%The potential, which has some correlation length scale $\Lh$, is then well-approximated in a region with $\Delta\phi<\Lh$. 
By going to sufficiently high order in the Taylor expansion, one can obtain a well-approximated potential in a domain containing the inflaton trajectory.  
%have well-approximated third derivatives within this region, which is needed to compute non-Gaussianity. 
This makes GRFs a powerful tool for studying the observational signatures of generic large-$\Nf$ inflation models.

We begin by briefly reviewing the statistical properties of Gaussian random fields and how the probability distribution function (PDF) for the Taylor coefficients is obtained. This PDF involves the inverse of the covariance matrix and is unfortunately %in general 
of very limited practical use when $\Nf$ is large. However, we find that for a Gaussian covariance function for the random field, there is a drastic simplification which allows us to generate explicit potentials even when $\Nf\gg 1$. %for a very large number of fields. 
 %essentially as many fields as we want.
 
 %In this section we 
We go on to present the class of random inflation models discussed in this paper. Specifically, we discuss the relevant mass-scales of the potentials, and how the GRFs under certain conditions admit an interpretation as proxies for physical effective field theories (EFTs). However, we also point out a challenge for using GRFs to study multifield inflation: the typical mass-splitting of the fields tend to exceed the Hubble parameter, leading to single-field dynamics. We then describe how we choose the initial conditions to generate large ensembles of potentials with multifield dynamics during inflation.  We close this section by briefly explaining the methods used to calculate the background trajectory and the superhorizon evolution of the field perturbations. \\

\subsection{High-dimensional GRFs as random multifield scalar potentials}
\label{sec:overview}
A Gaussian random field has a mean value $\bar V$ and a covariance function,
\begin{equation}
\langle (V(\phi_1)-\bar V)(V( \phi_2)-\bar V)\rangle=C(\phi_1,\phi_2) \, ,
\end{equation}
where the $\phi_i$ are position vectors in field space (with components $\phi_i^a$), which  we take to be flat $\mathbb R^N$. %\footnote{We expect our results to be applicable also to weakly curved spaces, but novel phenomena can appear in inflationary models in singular or highly curved field space geometries \cite{Linde}.}
 Furthermore, we take the GRF to 
%
%have zero mean, and 
be 
%both 
stationary and isotropic with mean zero\footnote{
In the bulk of this paper, we focus exclusively on this simplest class of GRFs. However, in Appendix \ref{sec:casestudies} we briefly discuss a modified GRF that includes a large field-independent cosmological constant, cf.~$\bar V \gg1$.}
 so that, 
\begin{equation}
C(\phi_1,\phi_2)=C(\phi_1-\phi_2)=C(|\phi_1-\phi_2|) \, .
\end{equation}
The covariances for the derivatives are given by the derivatives of the covariance function:
\begin{equation}
\left\langle \frac{\partial^{n_1}V(\phi_1)}{\partial\phi_1^{a_1}\ldots\partial\phi_1^{a_{n_1}}}\frac{\partial^{n_2}V( \phi_2)}{\partial\phi_2^{b_1}\ldots\partial\phi_2^{b_{n_2}}}\right\rangle=\frac{\partial^{n_1+n_2}C(\phi_1,\phi_2)}{\partial\phi_1^{a_1}\ldots \partial\phi_1^{a_{n_1}}\partial\phi_2^{b_1}\ldots\partial\phi_2^{b_{n_2}}}
\, .
\end{equation}
All non-vanishing elements have either $n_1$ and $n_2$ both odd, or both even. To simplify notation, we will from now on write derivatives as,
\eq{
\frac{\partial^{n}V(\phi)}{\partial\phi_1^{a_1}\ldots\partial\phi_1^{a_{n}}}\equiv V_{a_1\ldots a_n}(\phi) \, .
}

\begin{figure}
    \centering
    \includegraphics[width=0.55\textwidth]{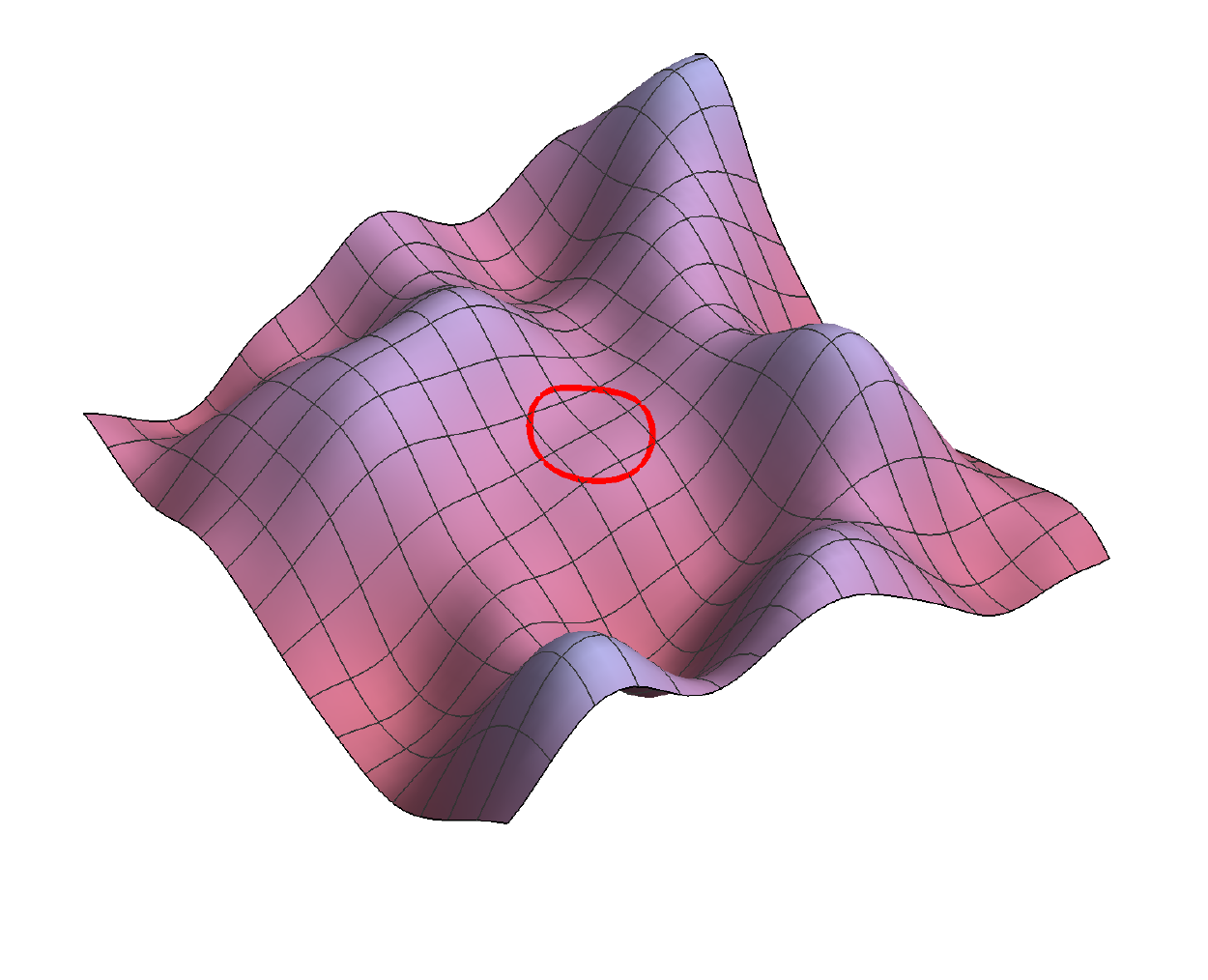}
    %\caption{$\fnl$ values from 1000 random 50-field potentials}
    \caption{An example of a GRF potential with $\Nf=2$ and $n_{\rm max} =175$. Here $\phi \in [-4\Lh,~4\Lh]^2$ and inflation is supported near $\phi=0$. The red circle delineates the region in which a potential truncated at fifth order approximates the full potential to per cent level accuracy.  }
    \label{fig:2dpotential}
\end{figure}

In this paper, we create random multifield potentials by generating Taylor coefficients at a single point in field space. The joint probability distribution of the Taylor coefficients is  a multivariate normal distribution with a covariance matrix given by the derivatives of the covariance function at $\phi_1=\phi_2$. Of course, not all the derivatives are independent, so we only generate the derivatives $V_{abc...}$ with indices ordered such that $a\geq b\geq c$ et cetera. This ensures that all the unique, independent derivatives are included exactly once. If we collectively denote the independent Taylor coefficients of the potential (which includes $V$, $V_a$, $V_{ab}$ et cetera) by  $V_{\alpha}$, where $\alpha$ runs over all the ordered sets of indices for the derivatives we include, the multivariate probability distribution function is given by,
\begin{equation}
P(V_{\alpha})=\frac{\exp\left(-\frac12(V_{\alpha}-\mu_{\alpha})(\Sigma^{-1})_{\alpha\beta}(V_{\beta}-\mu_{\beta})\right)}{\sqrt{\det(2\pi\Sigma)}}\label{eq:pdf}
\, ,
\end{equation}
where $\mu_\alpha=\langle V_\alpha\rangle$ is the expectation value vector and $\Sigma_{\alpha\beta}=\langle V_\alpha V_\beta\rangle$ is the covariance matrix.

Throughout this paper (and just as in \cite{Bachlechner:2014rqa}), we will be working with a Gaussian covariance function,
\begin{equation}
C(\phi_1,\phi_2)=\Lv^8e^{-(\phi_1-\phi_2)^2/2\Lh^2} \, .
\label{eq:Gcovar}
\end{equation}
Here $\Lv$ sets the `vertical energy scale' of the potential and the `horizontal scale', $\Lh$, sets the correlation length of the potential. We are interested in studying models 
in which the potentials have structure on sub-Planckian scales, so we take $\Lh < \Mp$. 
In section \ref{sec:EFT}, we will discuss the physical interpretation of $\Lv$ and $\Lh$, and how  potentials with the covariance function \eqref{eq:Gcovar} may be regarded as proxies for Wilsonian effective field theories.  
%Both $\Lv$ and $\Lh$ have clear physical interpretations which we will discuss shortly. 

Schematically, our procedure for studying manyfield inflation in GRF potentials is as follows: 
we  generate the potential locally in a domain of size $\Lh$ around $\phi=0$,  
%The potential  is  given as the sum,
\begin{equation}
V(\phi)=\sum_{n=0}^{n_\text{max}}V_{a_1...a_n}\frac{1}{n!}\phi^{a_1}...\phi^{a_n} 
=\sum_{n=0}^{n_\text{max}}\Lv^4\tilde{V}_{a_1...a_n}\frac{1}{n!}\frac{\phi^{a_1}}{\Lh}...\frac{\phi^{a_n}}{\Lh} 
%\label{eq:Taylor1}
\, ,
\label{eq:Taylor1}
\end{equation}
up to some order $n_\text{max}$. Throughout this paper we take $n_\text{max} = 5$, unless otherwise specified. This ensures that the third derivatives of the potential, which are required to compute non-Gaussianities, are well-approximated and non-trivial within the domain.\footnote{With the covariance function given in \eqref{eq:Gcovar}, the dimensionless coefficients $\tilde V_{\alpha}$ at order $n$ have rms-values ranging between 1 (all indices different) and $\sqrt{(2n-1)!!}$ (all indices the same). Since these increase slower than $n!$, the Taylor series  convergences as $n_{\rm max}\to \infty$. By going to high orders in the series,
one can therefore construct a large potential landscape, cf.~Figure \ref{fig:2dpotential}. In this paper, we focus  on the inflationary phase in models with small field excursions  (see below), for which an expansion to $n_{\rm max}=5$ suffices.}

If all Taylor coefficients are chosen randomly according to the PDF of equation \eqref{eq:pdf}, the potential is typically much too steep to support inflation. %\footnote{We discuss the tuning required in section \ref{sec:tuning}.}
 However, by  
 choosing a subset of the Taylor coefficients, $V, V_{a_1}$ and $V_{a_1 a_2}$, by hand and generating the remaining coefficients through the corresponding conditional PDF, we can construct multifield scalar potentials that are suitable  for slow-roll inflation around $\phi=0$, but have the random structure of a Gaussian random field away from this point. For example, with $\Nf=100$ and $n_{\rm max} =5$, we specify the $1+100+5,050 = 5,151$ first Taylor coefficients by hand, and generate the remaining 96,555,395 coefficients randomly by using the constrained PDF obtained from equation \eqref{eq:pdf}.
  We will refer to scalar potentials generated by this method as `GRF potentials'.\footnote{By the use of this phrase, we do not suggest that our class of models is unique:  other covariance functions or field space geometries can lead to distinct ensembles of models. For the ease of presentation however, we will in this paper refer to our models as `the' GRF models. } 
  
  As we will discuss in detail in  section \ref{sec:ensemble}, we choose the parameters $V,~V_{a_1}$ and $V_{a_1 a_2}$  so that $\phi=0$ is an approximately saddle-point of the potential with multiple fields with $m^2 \leq H^2$. The `horizontal scale',
%  remaining parameters,  
  %
   $\Lh$, and the number of fields, $\Nf$, both have important effects on the generated model. Finally, $\Lv$ can be  fixed from the normalisation of the primordial perturbations for each model. We will refer collectively to the set $(V, V_{a_1}, V_{a_1 a_2}, \Nf, \Lh)$ as the `hyperparameters' of the GRF potentials. 
 
 %\footnote{Other methods for constructing inflationary potentials using Gaussian random fields include generating the potential in a Fourier expansion \cite{}, and generating the first few Taylor coefficients of the potential along a path in field space \cite{}. For computational reasons, these methods have thus far only been studied in the few-field or single-field cases. }
 
 For each of the potentials that we construct, we study how the fields evolve from the approximate saddle-point, assuming that the 
 field initially `rolls' slowly. Due to the randomness of the potentials, models with the same hyperparameters
but different higher-order Taylor coefficients give rise to different inflationary trajectories, and consequently different numbers of e-folds of inflation. For models supporting at least 60 e-folds of inflation, we compute the evolution of the two-field and three-field correlation functions for the perturbations during inflation using the `transport method' \cite{1302.3842, 1203.2635, 1502.03125, Dias:2016rjq}, and we evaluate the predictions for observables of the models at the end of inflation (for the two-point statistics, our approach is exactly that of \cite{DBM2}).  

By generating large numbers of inflationary models for each fixed set of hyperparameters that we are interested in, we can study the statistical predictions 
for the generation of cosmological observables
in manyfield models of inflation. In particular, we compute the power spectrum of the primordial curvature perturbation, and, upon finding that it is typically well-fitted by a power-law over the scales that are constrained by observations of the Cosmic Microwave Background (CMB), we compute the values of the spectral index $n_s$ and its running $\alpha_s = {\rm d} n_s/{\rm d} \ln k$, and in addition, the tensor-to-scalar ratio, $r$. 
Furthermore, from the two-field correlators, we study the co-evolution of the isocurvature and curvature perturbations during inflation. 
Finally, using the standard $\delta N$ formula \cite{Starobinsky:1986fxa, Sasaki:1995aw, Lyth:2005fi, Vernizzi:2006ve, Battefeld:2006sz} and the three-point function of the fields, we compute the local non-Gaussianity parameter $\fnl = \fnl^{\rm loc}$.
 
% different realisations  
 %the resulting periods of inflation will differ in field trajectories 
 %
 %the inflationary paths differ in extent, field space 

  We emphasise that we only study the generation of observables \emph{during} the inflationary period, and we defer the study of
  % not model 
  the post-inflationary approach to the final vacuum and the reheating process to future studies.

\subsection{A new, efficient, local construction of GRFs}
\label{sec:PDFs}
Given the probability distribution function of equation \eqref{eq:pdf}, it might seem straightforward to just start generating the Taylor coefficients. However, the appearance of the inverse covariance matrix presents a significant complication which has curtailed previous attempts to the single-field or effectively single-field  regimes \cite{Bachlechner:2014rqa, Masoumi:2016eag, Masoumi:2017xbe}.
%, and it is the covariance matrix. 
Even with a sparse covariance matrix, as it is in our case, the inverse covariance matrix  is in general hard to diagonalise, and grows rapidly in size as  the number of fields is increased.

We here identify an algebraic property of the covariance matrix which allows us to 
%What allows us to get around
circumvent this computational hurdle: Gaussian random fields with a Gaussian covariance function have the elegant property that if we know all the derivatives of the same type (even or odd) to some order, then the conditional covariance matrix for the Taylor coefficients at the next order of the same type is diagonal. This result holds to all orders and for any number of fields. This means that all the Taylor coefficients can be generated in a step-by-step fashion  as a set of \emph{independent} Gaussian random variables, without inverting or diagonalising any matrices at all. In practice, the only large matrices that need to be constructed explicitly are those that are used to calculate shifts in the expectation values of higher-order derivatives, caused by fixing the lower-order derivatives. These matrices are sparse and require little memory to be used. All together, this makes it rather easy to construct the GRF potentials even for a very large number of fields, e.g.~$\Nf=100$. In fact, this method shifts the computational bottle-neck for studying manyfield inflation in GRF potentials from generating the potential to solving the equations of motions during inflation.

To provide some practical 
 %but to give the readers some 
intuition for this method, we here illustrate it by looking at the covariance matrices in the case of $\Nf=2$. It is straightforward to check that the covariances vanish between odd and even derivatives  for any stationary, isotropic covariance function. The covariance matrix then becomes block diagonal, and we can treat the odd and even derivatives separately. We will therefore look at the potential, Hessian, and fourth derivatives in this case, which is the simplest non-trivial example.

Suppose we have a collection of non-independently distributed Gaussian random variables, $Z$. If we split them into two parts, they follow the distribution,
\begin{equation}
\begin{bmatrix}
Z_1 \\
Z_2
\end{bmatrix}\sim N\left(
\begin{bmatrix}
\mu_1 \\
\mu_2
\end{bmatrix},
\begin{bmatrix}
\Sigma_{11} & \Sigma_{12}\\
\Sigma_{21} & \Sigma_{22}
\end{bmatrix}\right) \, ,
\label{eq:2Gauss}
\end{equation}
where $(\mu_1,\mu_2)$ is the mean vector and $\Sigma_{ij}$ are block components of the covariance matrix. 
In our construction, $Z_1$ will correspond to lower-order Taylor coefficients, and $Z_2$ to higher-order coefficients in an iterative way which we will make clear below. We may fix the lower-order coefficients by hand (as we  will do for the hyperparameters $\tilde V$, $\tilde V_{a_1}$ and $\tilde V_{a_1 a_2}$ in our construction of inflationary potentials), or by randomly generating them from their marginal probability distribution (as we will do for Taylor coefficients of order three or more). The latter case is greatly simplified by the Gaussianity of the distribution: the marginal probability distribution of a subsystem of Gaussian variables (cf.~the lower-order coefficients)  is simply obtained by truncating the full covariance matrix and mean vector to the variables of the subsystem. For example, %marginalising over $Z_2$ when generating 
the marginal probability distribution of $Z_1$ obtained from equation \eqref{eq:2Gauss} is simply $Z_1 \sim N(\mu_1, \Sigma_{11})$.
If we then fix $Z_1=z_1$, the conditional probability distribution for the remaining variables $Z_2$ is another multivariate Gaussian distribution given by,
\begin{align}
Z_2\sim N\left(\mu_2+\Sigma_{21}\Sigma_{11}^{-1}(z_1-\mu_1),\Sigma_{22}-\Sigma_{21}\Sigma_{11}^{-1}\Sigma_{12}\right) \, .\label{eq:conditional}
\end{align}

We now want to write down the covariance matrices for the potential and its second and fourth derivatives at $\phi=0$. For convenience, we here 
work with the dimensionless fields, $\phi/\Lh$, and the dimensionless potential, $V/\Lv^4$.
%we first rescale the fields and potential as $\phi\to\Lh\phi$ and $V\to\Lv^4V$, making them and the covariance function dimensionless, and then take 
By taking the appropriate derivatives of the covariance function, we find that  the covariance matrix for the potential, second and fourth derivatives is given by,
\be
\Sigma=
\left(
\begin{array}{c;{2pt/2pt} c c c;{2pt/2pt} c c c c c}
 1 & -1 & 0 & -1 & 3 & 0 & 1 & 0 & 3 \\ \hdashline[2pt/2pt]
 -1 & 3 & 0 & 1 & -15 & 0 & -3 & 0 & -3 \\
 0 & 0 & 1 & 0 & 0 & -3 & 0 & -3 & 0 \\ 
  -1 & 1 & 0 & 3 & -3 & 0 & -3 & 0 & -15 \\ \hdashline[2pt/2pt]

 3 & -15 & 0 & -3 & 105 & 0 & 15 & 0 & 9 \\
 0 & 0 & -3 & 0 & 0 & 15 & 0 & 9 & 0 \\
 1 & -3 & 0 & -3 & 15 & 0 & 9 & 0 & 15 \\
 0 & 0 & -3 & 0 & 0 & 9 & 0 & 15 & 0 \\
 3 & -3 & 0 & -15 & 9 & 0 & 15 & 0 & 105 
\end{array} 
\right)
\, ,
\ee
where the first row/column is for the potential, the following three are for the $(1,1)$, $(2,1)$ and $(2,2)$ components of the Hessian, and the final five are for the components of the fourth derivatives in the order $(1,1,1,1)$, $(2,1,1,1)$, et cetera.

 Fixing the zeroth order Taylor coefficient $\tilde V$ and using equation \eqref{eq:conditional}, the covariance matrix for the remaining variables becomes,
\be
\Sigma'=
\left(
\begin{array}{c c c;{2pt/2pt} c c c c c}
 2 & 0 & 0 & -12 & 0 & -2 & 0 & 0 \\
 0 & 1 & 0 & 0 & -3 & 0 & -3 & 0 \\
 0 & 0 & 2 & 0 & 0 & -2 & 0 & -12 \\ \hdashline[2pt/2pt]
 -12 & 0 & 0 & 96 & 0 & 12 & 0 & 0 \\
 0 & -3 & 0 & 0 & 15 & 0 & 9 & 0 \\
 -2 & 0 & -2 & 12 & 0 & 8 & 0 & 12 \\
 0 & -3 & 0 & 0 & 9 & 0 & 15 & 0 \\
 0 & 0 & -12 & 0 & 0 & 12 & 0 & 96 \\
\end{array} 
\right) \, ,
\ee
where we note that the 3-by-3 block matrix in the upper-left corner, corresponding to the three independent components of $\tilde V_{a_1 a_2}$,  has become diagonal. This is the conditional covariance matrix for the second derivatives. Fixing $\tilde V_{ab}$ (either by hand or randomly, by generating three independent Gaussian variables) in addition to $\tilde V$, we find that the covariance matrix for the fourth derivatives is given by,
\eq{
\Sigma''=
\begin{pmatrix}
 24 & 0 & 0 & 0 & 0 \\
 0 & 6 & 0 & 0 & 0 \\
 0 & 0 & 4 & 0 & 0 \\
 0 & 0 & 0 & 6 & 0 \\
 0 & 0 & 0 & 0 & 24 \\
\end{pmatrix} \, , 
}
which again is diagonal. Generating the fourth derivatives randomly now simply involves generating five independent Gaussian random numbers. Note in particular that to construct the Taylor coefficients up to fourth order, we are never required to invert or diagonalise the full covariance matrix. This is the key realisation that allows us to commence the study of manyfield inflation in GRF potentials.

The method illustrated here extends to arbitrary $\Nf$ and to all orders in the Taylor expansion. The general formulae for these covariance matrices and the matrices that shift the expectation values can be found in Appendix \ref{section:formulae}. 
The details and a general proof of this method  will be presented separately in \cite{GRF2}.

\subsection{Physical properties of GRF potentials}
\label{sec:scales}
It is important to note that physical effective field theories supporting manyfield inflation may differ substantially in many details from the mathematically simple GRF models that we study. For example, EFTs with many light fields may reflect the imprints of broken symmetries, such as supersymmetry or axionic shift symmetries for some of the fields. It then appears reasonable to expect that some of the   GRF estimates  (e.g.~of the fine-tuning  of manyfield inflationary models) may differ from that of a physically motivated manyfield theory. However, it is still possible for GRF models of manyfield inflation to be sufficiently complex to capture non-trivial multifield dynamics, and can provide access to  `universal' or robust aspects of manyfield models, if they exist. Motivated by this, our approach here is to engineer manyfield models of slow-roll saddle-point inflation using GRFs, and to search for mechanism that determine the distribution of observables. 

%disagree with computations performed directly in the EFT. 

%In this paper, we use GRF potentials as simple proxies that may capture some aspects of physically motivated 
%effective field theory 
%potentials arising from a particle physics. 
%
%for the multi-field effective field theory potentials that may arise from an ultraviolet completion of particle physics and cosmology. 
To understand the properties of the class of potentials that we study, %and relatedly the limitations of their usefulness, 
it is  important to characterise  the various energy scales that are associated with them. In this subsection, we discuss the distribution of the slow-roll parameters and the typical scale of higher-order terms in the potential. %We review the fine-tuning required for these models to yield sustained periods of inflation. 
We furthermore discuss the conditions under which GRF potentials may be interpreted as proxies for physical effective field theories. Finally, we point out that the mass distribution of GRF potentials is broad compared to the Hubble scale. This raises an additional challenge for using these potentials to study multifield inflation.

\subsubsection{Distributions of the parameters of the potential}

\subsubsection*{The value of the potential:} 
The GRF potentials have mean zero and  typically takes values in the 1$\sigma$ range between $-\Lv^4$ and $\Lv^4$. For the models of inflation that we consider in the bulk of this paper, we take the dimensionless parameter $\tilde V =1$, so that  $V =\Lv^4 \left( 1+ {\cal O}(\phi/\Lh)\right)$. The hyperparameter $\Lv$ then sets the energy scale of inflation. During slow-roll inflation close to the approximate saddle-point at $\phi=0$, the square of the Hubble parameter  is then given by,
\be
H^2 = \frac{1}{3} \frac{\Lv^4}{\Mpl^2} \, .
\ee 

\subsubsection*{The gradient:} 
The typical magnitude of the gradient vector is most easily characterised in terms of the inflationary slow-roll parameter,
\be
\epsilonV=\frac{\Mp^2}{2}\frac{\partial_aV\partial_aV}{V^2} 
=
\frac{1}{2}
\frac{\Mpl^2}{\Lh^2} \tilde V_a \tilde V_a \, .
\ee
For the theory defined by equation \eqref{eq:Gcovar}, the covariance of the dimensionless Taylor coefficients $\tilde V_a$ is given by,
\be
\langle \tilde V_a \tilde V_b \rangle = \delta_{ab} \, ,
\ee
so the typical value of the slow-roll parameter $\epsilonV$  is  given by,
\be
\langle\epsilonV\rangle=2 \Nf \left( \frac{\Mp}{\Lh} 
\right)^2 \gg 1
\, .
\ee
At a typical point in field space, the potential is then too steep to support inflation.  
Since the Taylor coefficients $\tilde V_a$ are $\Nf$ independent Gaussian variables,  the probability  of $\epsilonV$ being no larger than some value $\epsilon_\star$ is given by \cite{Bachlechner:2014rqa},
\begin{equation}
P(\epsilonV\leq\epsilon_{\star})
=
\frac1{(\sqrt{2\pi})^\Nf}
\int_{|\mathbf x|\leq\sqrt{2\epsilon_{\star}}\Lh/\Mp}
d^\Nf x\, e^{-x^2/2}
\simeq
\frac{2}{\Nf\, \Gamma(\frac{\Nf}{2})} 
\left(\sqrt{\epsilon_{\star} }\frac{\Lh}{\Mpl} \right)^\Nf
\, .
\label{eq:epsilontuning}
\end{equation}
Obtaining a small $\epsilonV$ parameter requires tuning of the slope of the potential, and this tuning becomes more severe  as $\Lh$ is decreased from $\Mpl$.  Note however that equation \eqref{eq:epsilontuning} gives the probability of a randomly chosen point having a small $\epsilonV$ parameter, not the probability that a point with a small $\epsilonV$ parameter exists in the field space. 
The latter probability depends on the volume of  field space, which we do not model in this paper. 
%To estimate the latter  probability one would have to
%make further assumptions about the extend of the scalar field space.
% further model the extent of the field space, which we will not be concerned with in this paper. 

\subsubsection*{The Hessian matrix:}
The Hessian matrix, $V_{ab}$, determines the curvature of the potential and its eigenvalues are the squared masses of the fields. From the covariance function \eqref{eq:Gcovar} it is easy to see that the dimensionless Hessian has zero mean and a covariance given by,
\be
\langle \tilde V_{ab} \tilde V_{cd}\rangle= \Big(\delta_{ab}\delta_{cd}+\delta_{ac}\delta_{bd}+\delta_{ad}\delta_{bc} \Big) \, .
\label{eq:Vabcorr}
\ee
The  probability distribution for the Hessian (with all other Taylor coefficients marginalised over), can then be obtained by inverting the covariance matrix $\Sigma_{(ab)(cd)} \equiv \langle \tilde V_{ab} \tilde V_{cd}\rangle$ to find,
\be
\left( \Sigma^{-1} \right)_{(ab)(cd)} = - \frac{1}{2(\Nf+2)} \delta_{ab} \delta_{cd} + \delta_{ac}\delta_{bd} - \frac{1}{2} \delta_{ad} \delta_{bc} \, .
\ee
The marginal probability distribution is then given by \cite{Bray:2007tf},
\eq{
P(\tilde V_{ab})=C_{\mathrm n}\exp\left(-\frac{1}{4}\left(\tilde V_{ab} \tilde V_{ba}-\frac1{\Nf+2}(\tilde V_{aa})^2 \right)\right).
}
Here $C_{\mathrm n}$ is a normalisation factor. 

To elucidate the consequences of this probability distribution, it is useful to consider the large-$\Nf$ limit in which an eigenvalue density  can easily be derived. 
We will denote the physical squared masses by $m^2_a$ and work with the dimensionless eigenvalues $\lambda_a$ of $\tilde V_{ab}$:
\be
m^2_a = \frac{\Lv^4}{\Lh^2}\, \lambda_a \, .
\ee
%where $m^2_a$ is an eigenvalue of $_{ab}$ and $\lambda_a$ is the corresponding eigenvalue of $\tilde V_{ab}$.
To derive the eigenvalue density, we change variables from $\tilde V_{ab}$ to its eigenvalues and eigenvectors, and integrate out the latter. 
Importantly, the probability distribution of the eigenvalues involves the Vandermonde determinant arising from the change of measure, 
\eq{
\prod_{a\leq b}d \tilde V_{ab} \sim \prod_{a<b}|\lambda_a-\lambda_b|\prod_{a=1}^\Nf d\lambda_a \, .\label{eq:measure}
}
The Vandermonde determinant encodes the `eigenvalue repulsion' which is the key driver behind the large-$\Nf$ universality encountered in random matrix theory (see e.g.~\cite{Kuijlaars, Deift, Erdos}). The appearance of the Vandermonde determinant in the probability distribution for the Hessian matrix of GRF potentials is indicative of the close connection between random function theory and random matrix theory. We will return to this connection towards the end of this section, and then again 
 in section \ref{sec:result2}.

By using the eigenvalue density function,
\eq{
\rho(\lambda)=\frac1\Nf\sum_a^\Nf\delta(\lambda-\lambda_a) \, ,
}
the probability distribution for the eigenvalues can be expressed as, 
\begin{align*}
P(\rho)&=C_{\mathrm n}
\exp\Bigg[
%-\frac{\Nf}{4\Lv^8\Lv^{-4}}
-\frac{\Nf}{4} 
\bigg(\int d\lambda \lambda^2\rho(\lambda)
-\left(\int d\lambda \lambda\rho(\lambda)\right)^2\bigg)\\
&\hspace{3.4cm}+\frac{\Nf^2}{2}\int d\lambda d\lambda'\rho(\lambda)\rho(\lambda')\ln(|\lambda-\lambda'|)\Bigg] \, .\nt\label{eq:EVaction}
\end{align*}
The typical distribution of the eigenvalues of the dimensionless Hessian matrix can be found from saddle-point evaluation of equation \eqref{eq:EVaction}. This gives \cite{Bray:2007tf},
\begin{equation}
\rho_{sc}(\lambda)=\frac{1}{2\pi \Nf}\sqrt{4 \Nf -(\lambda-\bar\lambda)^2}\label{eq:deanbray} \, .
\end{equation}
For $\bar \lambda =0$, this spectrum is precisely a Wigner semi-circle, i.e.~the spectrum  of 
the Gaussian Orthogonal Ensemble (GOE)
of 
random symmetric matrices with independent, Gaussianly distributed  entries. For $\bar \lambda \neq 0$, the semi-circle is rigidly shifted to be centred at $\bar \lambda$  \cite{Bray:2007tf}. 

To properly understand the significance of the shift $\bar \lambda$,
it is instructive to
calculate the conditional probability distribution of the Hessian, given that the potential has a certain value, say $V=V_\star$. 
The Hessian matrix and the value of the potential are correlated, and upon using equation \eqref{eq:conditional} for the conditional probability distribution, we find the moments \cite{Bray:2007tf, Bachlechner:2014rqa},
% However, we can only have inflation at points with $V>0$, so we can make this discussion more precise by calculating the conditional distribution for the Hessian with . 
\begin{align}
\langle \tilde V_{ab}\rangle\big|_{V=V_\star}&=-\frac{V_\star}{\Lv^{4}}\, \delta_{ab} \, , \\
%\mathrm{Cov}[V_{ab},V_{cd}|V=V_\star]&=
\langle \tilde V_{ab}\tilde  V_{cd} \rangle\big|_{V=V_\star} - (\langle \tilde V_{ab} \rangle  \langle \tilde V_{cd} \rangle)\big|_{V=V_\star} &=
\Big(\delta_{ac}\delta_{bd}+\delta_{ad}\delta_{bc} \Big) \, .
\label{eq:VabCond}
\end{align}
According to equation \eqref{eq:VabCond}, every unique element of the Hessian is now  statistically independent of the others, and we can write the Hessian as \cite{Bachlechner:2014rqa}, 
\begin{equation}
V_{ab}=\frac{\Lv^4}{\Lh^2 } \left(-\frac{V_\star}{\Lv^{4}}\, \delta_{ab}+ R_{ab}\right) \, ,\label{eq:splithessian}
\end{equation}
where $R_{ab}$ is a random matrix in the Gaussian Orthogonal Ensemble (GOE).
%{\color{blue} with variance \ldots endpoints at $\pm 2\sqrt{\Nf}$}.
The spectrum of the dimensionless Hessian is then given by,
\be
\rho_{sc}(\lambda)=\frac{1}{2\pi \Nf}\sqrt{4 \Nf-\left(\lambda + V_\star/\Lv^4\right)^2}\label{eq:shiftedSemiCircle} \, .
\ee 
Clearly, for points with vanishing vacuum energy, $V_{\star} =0$, the spectrum of the Hessian is precisely that captured by the Wigner semi-circle. For $V_{\star} >0$, which is the case relevant for inflation, the 
typical spectrum
is a semi-circle  rigidly shifted downwards, making comparatively more eigenvalues tachyonic.   However, 
for
  $V_{\star} = \Lv^4$ and $\Nf \gg 1$, this shift is small: 
 the endpoints of the semi-circle spectrum of $R_{ab}$ are located at $\pm 2 \sqrt{\Nf}$, and the  spectrum of $V_{ab}$ is a shifted semi-circle with endpoints at $(\pm 2 \sqrt{\Nf} -1)\Lv^4/\Lh^2$. 

We  define the slow-roll $\etaV$ parameter as,
\be
 \etaV =\Mp^2\frac{m_\text{min}^2}{V} \, ,
 \label{eq:etaV}
\ee
with $m^2_{\rm min}$ denoting the smallest eigenvalue of the Hessian matrix.
An immediate consequence of equation \eqref{eq:splithessian} is that $\etaV$
tends to be very large and negative for the typical, slightly shifted semi-circle spectrum \cite{Bachlechner:2014rqa}: 
\be
\etaV =   -(2 \sqrt{\Nf}+1) \left( \frac{\Mpl}{\Lh} \right)^2 \, .
\ee
Smaller magnitudes of $\etaV$ can be obtained 
%{ \color{blue} either from an upward fluctuation of the value of the potential, or  }
if the spectrum of the Hessian is in a rare configuration in which no eigenvalue is very tachyonic. 
%Furthermore, to get any inflation at all we also need all the mass-matrix eigenvalues to be at or above $\etai V_0\Mp^{-2}$, or inflation would end very quickly (and the spectral index would be incompatible with observations). From
Given
 equation \eqref{eq:splithessian} for $V = V_{\star}$, the probability of $|\etaV|$ being no larger than $|\eta_{\star}|$ is given by, 
\eq{
P\left(
|m^2_{\rm min}| < |\eta_\star| \frac{V_{\star}}{\Mpl^2} 
\right)
%
%\lambda_i\geq\etai V_0\Mp^{-2})=
=P_\text{GOE}
\left(\lambda_{\rm min} > 1+\eta_{\star} \frac{\Lh^2}{\Mpl^2} 
\right) \, ,
}
where $\lambda_{\rm min}$ denotes the smallest eigenvalue of the GOE matrix $R$, and $P_\text{GOE}$ denotes its probability distribution. 
%where the RHS is the probability of a matrix of the Gaussian Orthogonal Ensemble having all eigenvalues above $\lambda_\text{lim}=V_0\Lv^{-4}(1+\etai\Mp^{-2}\Lh^2)\simeq V_0\Lv^{-4}$. This can be calculated accurately, see for example Dean and Majumdar \cite{DeanMajumdar}, but we are primarily interested in the rough scaling with $\Nf$. 
%The radius of the semi-circle is $2\sqrt{\Nf}$ which is significantly larger than $\lambda_\text{min}$ when $\Nf$ is large,  we can approximate the probability as
For $\Nf \gg 1$, the radius of the typical  semi-circle configuration is $2\sqrt{\Nf} \gg 1$, so that,
\be 
P_\text{GOE}
\left(\lambda_{\rm min} > 1+\eta_{\star} \frac{\Lh^2}{\Mpl^2} 
\right) \approx 
P_\text{GOE}
\left(\lambda_{\rm min} > 1
\right) =
\exp\left(-c\, \Nf^2\right) \, ,
\label{eq:etatuning}
\ee
where $c= \tfrac{1}{108} \left(35+16 \sqrt{7}+27 \ln (18)-54 \ln \left(\sqrt{7}-1\right)\right) \approx 1.19$. 
In the last step we have used the fluctuation probability computed for the subset of  `fluctuated spectra' of the GOE with no negative eigenvalue \cite{cond-mat/0609651}. 
Hence, small slow-roll parameters are very infrequent in GRF potentials with many fields.

 We close this section by 
 noting that
 the tight connection between our GRF models and random matrix ensembles also has strong implications for the distribution of vacua \cite{Bachlechner:2014rqa, Bray:2007tf}.
% 
% the severe fine-tuning of equation \eqref{eq:etatuning} also has direct implications for the distribution of vacua in GRF potentials . 
Metastability of  Minkowski and de Sitter critical points requires $m^2_{\rm min} > 0$.
According to the RMT analysis, such points are exceedingly rare: 
\be
P(m^2_{\rm min} >0 | V\geq 0) \leq P_{\rm GOE}(\lambda_{\rm min} \geq 0)  \, .
\label{eq:metastab}
\ee 
%
%
%: since $P_\text{GOE}\left(\lambda_{\rm min} > 0
%\right) \ll 1$ for large $\Nf$, metastability is very rare and the vast majority of all positive vacuum energy critical points are unstable. 
The rarity of metastable de Sitter vacua is a common feature also of other classes of random potentials, such as random supergravity theories \cite{1112.3034, Bachlechner:2012at, Bachlechner:2014rqa}.

Equation \eqref{eq:metastab} implies that the fraction of metastable de Sitter vacua in our GRF potentials is bounded from above by the probability of large fluctuations of one of the simplest random matrix ensembles. As a consequence, the frequency of metastable de Sitter vacua  scales with $\Nf$ like $\ln(P_{\rm GOE}) \sim  - \Nf^2$.
%
%, cf. equation \eqref{eq:etatuning}.
%It may appear surprising that 
%
% the  recent argument of reference \cite{Easther:2016ire} 
%
%
%
%may appear surprising: 
Recently however, 
the authors of \cite{Easther:2016ire} (see also \cite{Bachlechner:2014rqa}) found that vacua in GRF potentials comprise a fraction of $\sim {\rm exp}\left(-\alpha \Nf\right)$ of all critical points of GRFs (for some constant $\alpha$), which  far exceeds the metastability estimate of \eqref{eq:metastab}. 
 This apparent discrepancy is resolved by noting that the vast majority of the metastable vacua found in \cite{Easther:2016ire} are located deep down in the potential, at $V \lesssim -2\sqrt{\Nf} \Lv^4$, in our mean-zero GRF models.\footnote{
The analogous result in random supergravities is that most metastable de Sitter vacua realised in the `approximately supersymmetric' regime \cite{1112.3034, Bachlechner:2014rqa}. Note however, that known constructions of de Sitter  string compactifications tend to rely on non-random `structures'  to enhance the probability of metastability (see e.g.~\cite{Kachru:2003aw, Ferrara:2014kva, Westphal:2006tn, Achucarro:2006zf, Dudas:2006vc, Rummel:2014raa, Blaback:2015zra, Cicoli:2015ylx, Marsh:2014nla, Gallego:2017dvd}). Thus it is certainly possible that the simple GRF models may capture some rather robust aspects of manyfield inflation in fundamental theory, but  fail to accurately describe their vacuum structure.}
%
%
%One may be tempted to simplistically `fix' this issue by adding the potential a large field-independent cosmological constant that will lift the negative energy vacua (required for boundedness of the potential) to vanishing or positive net cosmological constant. We will briefly discuss the challenges of studying manyfield inflation in such `uplifted' GRF potentials in section \ref{sec:uplifted}. 
 %A better approach however may be to seek guidance from fundamental theory: the past decade's progress in constructing de Sitter vacua in string theory (see e.g.~\cite{})
 %indicates that metastability commonly relies on particular string theory or supergravity `structures' of the theory, which are not realised by the GRF models. \footnote{Some evidence for such failures have been established  in the Large Complex Structure region of the moduli space of type IIB and F-theory compactifications, where both GRF and RMT models have been shown unable to correctly describe the moduli spectra \cite{}. }   It would be interesting to further investigate these questions in explicit classes of string compactifications. 
%
For such large and negative values of the potential, the semi-circle spectrum is rigidly shifted upwards so that metastability is common. 
Due to the simple relation of equation \eqref{eq:splithessian} (and its generalisation for other covariance functions), any carefully phrased question about the vacuum statistics in GRF models map into precise questions about the eigenvalue statistics of random matrices. 

\subsubsection*{Cubic and higher-order terms:} 
GRF potentials have non-trivial, randomly generated interaction terms at cubic and higher orders. At order $n$, these are of the order of,
\be
V_{a_1 \ldots a_n} \sim \frac{\Lv^4}{\Lh^n} \, .
\ee
In particular, each of the  cubic order terms are then  of the order of,
\be
V_{a b c} \sim  \frac{\Lv^4}{\Lh^3} \sim  \frac{H^2 \Mpl^2}{\Lh^3}  \sim 
\left( \frac{\Lv}{\Lh}\right)^2 \left(\frac{\Mpl}{\Lh} \right)\, H
\, . 
\label{eq:cubic}
\ee
The vertical scale $\Lv$ factors out of the evolution equations for both the background and the perturbations and only serves as a normalisation factor for the scale of the  scalar perturbations. We find in all cases that $(\Lv/\Lh)^2 < \Lh/\Mpl$, and the  cubic terms of equation \eqref{eq:cubic} tend to be smaller than $H$ in the models that we consider. However,  since $\Lv$ does not affect the field equations, it also does not affect our predictions for the spectral index, its running, or the local non-Gaussianity parameter, $\fnl$. Consequently,
the  predictions of our models also apply to models in which $V_{abc} \approx H$, but for which the amplitude of the scalar perturbations is larger than the observationally inferred value.\footnote{Another way to achieve a relative enhancement of the cubic terms with respect to the Hubble parameter is to set the zeroth order Taylor coefficient much below its rms value: $\tilde V \ll 1$. This however makes the fine-tuning required to achieve small slow-roll parameters more severe.}

  %\vspace*{20pt}

\subsubsection{GRF potentials and physical effective field theories} 

\label{sec:EFT}

%While 
Gaussian random fields provide a mathematically convenient construct, but 
are not directly derived as effective field theories (EFTs)   from particle physics or string theory.\footnote{Simple GRFs can under certain assumptions be related to
the potentials of 
 multi-axion theories \cite{Bachlechner:2017hsj}.}
%to properly understand  
In this section, we discuss how the GRF potentials exhibit some `EFT-like' properties 
with important consequences for the cosmology.
% and how these properties have some important consequences. 
%
%
%under certain conditions 
%be understood as proxies for physical effective field theories (EFTs).
%
%under certain conditions be related to physical EFTs.
%
%
%they  
%We here clarify how GRF potentials
 %and {\color{blue} we point out that these naturally capture several well-known features of non-supersymmetric effective theories}. As a direct consequence of this, we show 
%Moreover, 
Relatedly, 
we  note that 
these potential naturally have  spread-out mass spectra, and additional tuning is required to construct models with non-trivial multifield dynamics.

In quantum field theory, unprotected dimensionful operators are naturally large. For the UV cutoff $\Lambda$, the potential is typically of the order of $V \sim H^2 \Mpl^2 \sim \Lambda^4$, and the scalar masses are of order $m^2 \sim \Lambda^2 \sim H \Mpl \gg H^2$ (cf.~equation \eqref{eq:V1}). 
In
the GRF potentials considered in this paper, cf~equation \eqref{eq:Taylor1},  the scale $\Lambda$  is replaced by the two parameters $\Lv$ and $\Lh$. The horizontal scale $\Lh$ sets the coherence length of the potentials, and can be interpreted as the UV-cutoff of the theory. 
Since all operators are suppressed by the same cut-off scale,  sharp features over distances $\ll \Lh$ are very rare. We expect this to be a general feature of models with generic interactions suppressed by a single, common cut-off scale.

The parameter  $\Lv$ sets the natural energy scale of the GRF models.
\begin{figure}
\centering
    \includegraphics[width=0.75\textwidth]{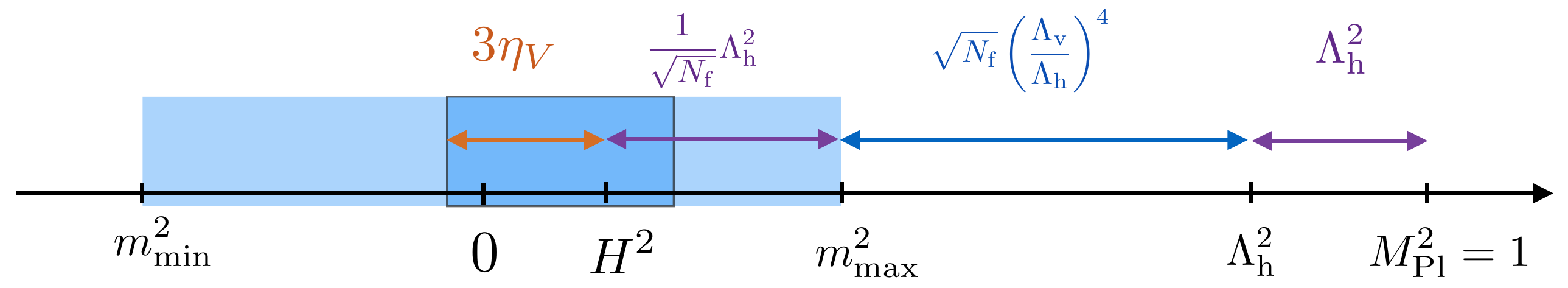}
%  \end{center}
%  \vspace{0pt}
  \caption{Schematic illustration of the relevant energy scales of the GRF potentials. The light blue shaded region indicates the equilibrium spectrum of the Hessian; the darker region corresponds to our chosen `flat spectrum' initial condition. 
  }
  %  \vspace{-200pt}
  \label{fig:Escales}
\end{figure} 
%
%As discussed in section 
%For a GRF potential of the form \eqref{eq:Taylor1} with dimensionless  Taylor coefficients %($\tilde V$, $\tilde V_{a_1}$, etcetera) 
%of order unity, the coherence scale $\Lh$ has a natural interpretation as the ultraviolet (UV) cut-off of the EFT: $\Lh \equiv \Lambda_{\rm UV}$. 
During inflation, 
%the dynamical energy scale is $H$, and 
a consistent EFT interpretation requires $H^2 \ll \Lh^2$. If reheating proceeds rapidly after the end of inflation, 
the stronger condition of $V  \sim  \Lv^4 \sim T_{\rm rh}^4 \ll \Lh^4 $ applies.
%The effective field theory interpretation 
%of the GRF models can then be valid during inflation if
%
%breaks down if during inflation
 % However, for a rapid reheating process at the end of inflation, the EFT description can be violated 
 %already for. 
By taking $\Lv < \Lh$, even the latter condition is generically satisfied. 
 A Wilsonian EFT is obtained by integrating out states more massive than the UV cutoff, 
leaving only states with $m^2 \leq \Lambda^2$. 
Consequently, 
in the context of GRF potentials, % that admit a physical interpretation, 
we expect the eigenvalues of the Hessian matrix to be no larger  than $\sim\Lh^2$. For $\Lv < \Lh$, this condition is  satisfied unless $\Nf$ is very large: 
\begin{equation}
\frac{m_\text{max}^2}{\Lh^2} \approx %\frac{2\sqrt{\Nf}\Lv^4/\Lh^2}{\Lh^2}=
\left(2\sqrt{\Nf} -1\right) \frac{\Lv^4}{\Lh^4} \, .
\end{equation}
Figure \ref{fig:Escales} illustrates the relevant energy scales of GRF potentials discussed in this paper.

%\subsection*{GRFs for multifield inflation:  a critical assessment}
We now note  a serious obstacle for using GRF potentials to study multifield inflation:  the mass spectrum is generically spread out over energy scales $\gg H$. This is immediately evident from the width of the (shifted) Wigner semi-circle distribution, which predicts a typical eigenvalue spacing for the Hessian of, 
\be
\frac{\frac{1}{\Nf}\left(m^2_{\rm max} - m^2_{\rm min}\right)}{H^2}=
%\frac{2\sqrt{4\Nf}\Lv^4/\Lh^2}{\Lv^4/3\Mp^2}=
12 \frac{1}{\sqrt{\Nf} } \left( \frac{\Mpl}{\Lh} \right)^2 \, .
\label{eq:massspread}
\ee
For $\Lh < \Mpl$, this then implies that only systems with a very large number of fields, $\Nf \gtrsim (\Mpl/\Lh)^4$, can be expected to exhibit non-trivial multiple field effects. This large spread in the masses explains why recent attempts at using GRFs to study multiple-field inflation \cite{Masoumi:2017xbe} have only captured single-field dynamics. %with only one light field, the system is effectively  single-field, and cannot be expected to capture non-trivial multiple-field dynamics. 
To use these models to study non-trivial multifield dynamics, one has to further tune the initial conditions.
%\footnote{We emphasise that we 
%
%
%regard the GRF models as mathematically simple proxies of physically motivated models of multifield inflation. As such, the GRF models may be useful in capturing non-trivial multifield effects, but do not necessarily provide accurately indicate 
%}   

The particular $\Nf$ dependence of 
equation \eqref{eq:massspread} follows from the form of the covariance function, cf.~equation \eqref{eq:Gcovar}. % in particular from its explicit $\Nf$ independence. %However, 
The width of the eigenvalue distribution may be changed by modifying the   covariance function, e.g.~to $C(\phi_1-\phi_2)=\Nf\Lv^8e^{-(\phi_1-\phi_2)^2/2\Nf\Lh^2}$,  
which gives  an $\Nf$ independent eigenvalue distribution of the Hessian, and an additional suppression by $1/\sqrt{\Nf}$ in equation \eqref{eq:massspread}. However, such a modified covariance function enhances the effective coherence length of the potential to  $\sqrt{\Nf}\Lh$, which becomes super-Planckian for large $\Nf$.\footnote{This lesson applies somewhat more generally: 
 the width of the mass spectrum relative $H^2$ is controlled by $C^{(4)}(0)/C(0)$. Compressing the spectrum requires decreasing this ratio, but for many covariance functions $C^{(p)}(0)/C(0) \sim (C^{(4)}(0)/C(0))^{p/4}$, 
and an overall suppression of these ratios translates directly into an increased coherence length. Compressing the mass scale while keeping the coherence scale associated with the interaction terms fixed then requires covariance functions with multiple scales. We will not consider such modifications further in this paper. }

 % This way, theories with partially protected dimensionful operators provide the key physical motivation for cosmological models with multiple interacting  fields. 

The broad spread in the distribution of masses of GRF potentials 
%limits the applicability of these models as  proxies for physical multifield systems, at least for generic spectra. This feature of GRF models
 is not surprising
 as these  models do not incorporate protective approximate  symmetries %Inflation in models with many axions has previously been considered in \cite{}. 
 %
 %Approximate symmetries 
 (e.g.~broken supersymmetry, or approximate shift-symmetries), which  can lower the natural scale of dimensionful operators, and make inflation less fine-tuned.
 It would be interesting to extend our method  to construct manyfield models of inflation in random supergravity models with spontaneously  broken supersymmetry, following the ideas proposed in \cite{1112.3034, Bachlechner:2012at,Bachlechner:2014rqa}. 

\subsection{The statistical ensembles of models}
\label{sec:ensemble}

As discussed in section \ref{sec:overview}, the hyperparameters of our GRF potentials are:
\be
(\tilde V, \tilde V_{a}, \tilde V_{ab}, \Nf, \Lh, \Lv) \, .
\ee 
An understanding of pre-inflationary physics in fundamental theory could potentially provide us with prior probability distributions on these parameters. Lacking such priors, we compute the observational predictions for certain ranges of the hyperparameters, and investigate how these predictions change as hyperparameters are varied.  

Specifically, we construct ensembles of manyfield models of inflation as follows:
First, we set 
$\tilde V$ to its rms value, $\tilde V =1$, so that $\Lv$ sets the scale of the potential  at $\phi=0$. 
In slow-roll inflation, the parameter $\Lv$  has no effect on the equations of motion for either the background field or the perturbations around it.  
We exploit this by evolving each model with a fiducial value of $\Lv$ and,
for each model yielding a sufficiently long period of inflation,  rescale $\Lv$ at the end of inflation so that the amplitude of the primordial curvature perturbation at the `pivot scale'  is consistent with the value determined by the Planck experiment \cite{1502.02114}.\footnote{To identify the precise e-fold at which the pivot scale crossed the horizon during inflation requires a detailed modelling of the reheating phase (cf.~\cite{Lyth:1998xn} for a review). For the baseline parameters, we find that the pivot scaled exited the horizon $58-N_{\rm rh}$ e-folds before the end of inflation, where $N_{\rm rh}$ parametrises the expansion between the end of inflation and the onset of the hot big bang. For concreteness, we will assume throughout this paper that the pivot scale crossed the horizon 55 e-folds before the end of inflation. } This fixes $\Lv$ separately for each model.

To obtain sufficiently flat potentials that  support inflation, the gradient and Hessian have to be tuned. We set $\tilde V_{a}$ so that $\epsilonV$ is sufficiently small that models with at least 60 e-folds of inflation are not too infrequent. This leads us  explore values of $\epsilonV$ in the range of $2\times10^{-8}$ to $10^{-11}$. We henceforth take, 
\be
\epsiloni \equiv \epsilonV\big|_{\phi=0} \, ,
\ee 
to parameterise the magnitude of the gradient vector. This vector obviously also has a direction, and we explore the effect of its alignment relative to the eigenvectors of $\tilde V_{ab}$, as we now discuss.

We have seen in section \ref{sec:scales} that the 
%eigenvalue distribution of the Hessian matrix 
curvature of the potential needs to be tuned to give rise to sustained inflation and multifield dynamics. In particular, the smallest eigenvalue of the Hessian, which we  parametrise  by,
\be
\etai \equiv \etaV\big|_{\phi=0} \, ,
\ee 
 must be close to zero. We numerically investigate values of $\etai$ in the range $-10^{-1}$ to $-10^{-4}$. 
 To explore the non-generic spectra relevant for multifield inflation, we consider two (non-random) initial conditions for the spectrum of $V_{ab}$: 
 \bea
 \begin{array}{l l l}
 {\rm Flat~spectrum:}&~&m^2_{a} \big|_{\phi=0} :~~ {\rm Uniformly~distributed~in~}\left(3 \etai H^2 ,~ \frac{9}{4} H^2 \right) \, , \\
  {\rm Compressed~spectrum:}&~&m^2_{a} \big|_{\phi=0} :~~ {\rm Uniformly~distributed~in~}\left(3 \etai H^2 ,~ -3\etai 
  H^2 \right)\, .
  \end{array}
  \label{eq:spectra}
 \eea
Field perturbations with an effective squared mass  greater than $9/4 H^2$ are exponentially suppressed already at horizon exit.\footnote{More precisely, the effective squared masses of the perturbations are the eigenvalues of the matrix $
M_{ab} =V_{ab} - \frac{1}{a^3} \partial_t \left(
a^3 \frac{\dot \phi_a \dot \phi_b}{H} 
\right)
$ \cite{Mulryne:2009ci, Chen:2009zp}
. 
}
This motivates the upper bound of the `flat spectrum'. The (extremely) compressed, nearly degenerate spectrum is specifically chosen to maximise the chances of non-trivial multifield effects, and is included for reasons that will become apparent in section \ref{sec:result3}. We note that these initial spectra will `relax'  to the generic spectrum for a GRF over distances of $\sim{\cal O}(\Lh)$. During most of inflation, the field is slowly rolling and this relaxation is very slow in e-fold `time', but we will see that towards the end of inflation,  multiple fields develop tachyonic masses, $m_{a}^2 < 0$. 

We now return to the question of the relative alignment of $V_{a}$ with the eigenvectors of $V_{a b}$. In a Gaussian random field
the gradient and Hessian are uncorrelated, $\langle V_{a} V_{ab}\rangle =0$, and generically, $V_{a}$ has support along all eigenvectors of $V_{ab}$. 
However, slow-roll inflation makes the field follow the gradient descent along the potential, which  tends to quickly align $V_{a}$ with the smallest eigenvalue direction of the Hessian matrix, which we will denote by `1' (for a more detailed discussion on this, see e.g.~\cite{DBM3}). Motivated by this, we consider two classes of orientations of $V_{a}$:
 \bea
 \begin{array}{l l l}
 {\rm Random:}&~~~~&V_{a} \big|_{\phi=0}  \sim \sqrt{2\epsiloni }\frac{\Lv^4}{\Mpl}\, {\rm Uniform}\left(S^{\Nf-1} \right) \, , \\
  {\rm Aligned:}&~~~&V_{a} \big|_{\phi=0} 
= \sqrt{2\epsiloni }\frac{\Lv^4}{\Mpl}\, \delta_{a\,1}
 \, .
  \end{array}
 \eea
 With the initial conditions $(\tilde V, \tilde V_{a}, \tilde V_{ab})$ fixed, we generate the higher-order Taylor coefficients randomly using the conditional PDFs derived as discussed in section \ref{sec:PDFs}.
 
Finally, we explore the numerically accessible a range of values for the remaining hyperparameters $\Nf$ and $\Lh$: for $\Nf$ very large or $\Lh$  small, inflation is only supported if the slow-roll parameters are highly tuned, which can cause numerical accuracy problems.\footnote{The computations reported in this paper did not require supercomputer capabilities, but 
potentials with  $\Nf\gg1$ places some restrictions on memory access.
Our largest simulations ran on a computing system with 144 CPUs and 516 GB RAM.  }    
%
%We find that we can easily create large ensembles of roughly 1000 inflationary models for $\Nf = 5, \ldots , 50$, $\Lh =0.4$, a randomly oriented gradient and the spectrum 1 for the Hessian. We also explore particularly examples  with more fields, up to $\Nf=100$, but due to the computational cost of computing the perturbations in theories with many millions of interaction terms, we do not construct large ensembles of models with $\Nf>50$.
%
%
A summary of the hyperparameter choices that we explore in this paper can be found in Appendix \ref{app:params}. Some of our results are best illustrated for a fixed choice of parameter. We take as our baseline model,
\be
\mbox{\bf Baseline:~} \Nf=10 \, ,~~
\Lh = 0.4 \Mp \, , ~~
\epsiloni =2\times 10^{-9} \, ,
 ~~\etai = - 10^{-4}\, , 
 \label{eq:baseline}
\ee
with the flat spectrum of the Hessian, cf.~equation \eqref{eq:spectra}, and a randomly directed gradient vector. For this choice of hyperparameters, obtaining at least 60 e-folds of inflation is not uncommon, and $\Nf$ is sufficiently large for multifield effects to be clearly manifest.

%{\color{blue} pasted:}
%The parameter $\Lv$ then sets the energy scale of inflation, as $V|_{\phi =0}=\Lv^4$. Our analysis of the perturbations is substantially simplified by noting that,  in slow-roll, 

%We may then 

%\section{Inflation in random potentials}
%In this section we \ldots

%, which at large $\Nf$ becomes numerically heavy.

%\begin{itemize}
%	\item Mass scales, GRF usually give single field models, compressed spectra
%	\item $N_e(\epsilon_v,d)$ and $N_e(r_\perp)$
%	\item Comparison with DBM, $N(\eta)$ and eigenvalues
%\end{itemize}

%Include these somewhere in this section
%\begin{itemize}
%	\item Dependence of the number of e-folds on $\Lh$, $\Nf$ \checkmark
%	\item Does the number of e-folds depend on the mass spectrum or starting direction? \checkmark
%	\item Evolution of the masses. When does the second element of the Hessian turn negative, and what fraction of eigenvalues become negative? \checkmark
%	\item Basin of attraction?\checkmark
%\end{itemize}

%\begin{itemize}
%	\item Energy scales
%	\item The mass spectrum
%	\item $\epsilon$ distribution
%	\item Isotropy assumptions
%	\item How GRF is not a good model for, say, string vacua, but how it is a very useful tool.
%	\item In this section and others: large $N_f$ universality: $\eta$ independence, distribution of $n_s$, suppression of isocurvature
%\end{itemize}

%\subsection{Aspects of GRFs as scalar potentials}

\subsection{Method: background}
Given a randomly generated multifield potential, we evolve the fields numerically using the coupled Klein-Gordon and Friedmann-Robertson-Walker equations,
\be
\ddot\phi_a+3H\dot\phi_a=-\partial_aV~ ,~ ~~ H^2=\frac{\frac12\dot\phi_a \dot\phi_a+V}{3\Mp^2} \, .
\ee
%Inflation occurs and is sustained as long as we have both
%\begin{align}
%\epsilonH&=-\frac{\dot{H}}{H^2}<1 & \etaH&=\frac{\dot\epsilonH}{H\epsilonH}\ll1.
%\end{align}
%It is convenient to define two slow-roll parameters given by
%\begin{align}
%\epsilonV&=\frac{\Mp^2}{2}\frac{\partial_aV\partial_aV}{V^2} & \etaV&=\Mp^2\frac{m_\text{min}^2}{V}.
%\end{align}
%where $m_\text{min}^2=\text{Min}(\text{Eig}(\partial_a\partial_bV))$. When these are small, we can work in a 
In the 
slow-roll approximation that we use throughout this paper, these equations become,
\be
3H\dot\phi_a=-\partial_aV~ ,~ ~~ H^2=\frac{V}{3\Mp^2} \, .
\ee
Expressed with respect to the number of e-folds, $N$, the slow-roll Klein-Gordon equation is simply given by, 
\eq{
\frac{d\phi_a}{dN}=-\partial_a\ln V\label{eq:SR} \, .
}
Equation \eqref{eq:SR} makes it clear that the vertical scale, $\Lv$, has no impact on the background field evolution in slow-roll. 

% This approximation is very much justified, as the initial value for $\epsilon$ is usually of order $10^{-9}$ and the its value remains very small until very close to the end of inflation. Computationally it is also advantageous; the potentials we work with contain millions of terms, making them slow to work with, and the slow-roll approximation simplifies and speeds up the calculations of the background trajectory and the evolution of the perturbations, enabling us to do the thousands of simulations required for a proper analysis.

\subsection{Method: perturbations}
\label{sec:methodperts}

To calculate 
the observational predictions of the manyfield models of inflation, we use the `transport method' \cite{1302.3842, 1203.2635, 1502.03125, Dias:2016rjq} (see also \cite{Dias:2011xy, Anderson:2012em, Seery:2016lko, Mulryne:2016mzv, Ronayne:2017qzn}). This formalism allows us to evolve the two-field and three-field correlators on superhorizon scales from horizon crossing to the end of inflation.\footnote{The transport method can be applied to both slow-roll and non-slow-roll systems, and also on sub-horizon scales \cite{1302.3842}. For our purposes, it suffices to consider the superhorizon evolution of the field perturbations during slow-roll inflation.} %As discussed in \cite{DBM3}, this method is highly efficient as $\Nf$ becomes large because, in very small patches, the solution for the evolution of the field perturbations in flat gauge can be found analytically,  and involves only background quantities. This makes it possible to evolve the perturbations without having to solve the coupled differential equations for the perturbations numerically. 
%allows us to track the evolution of the power spectrum and the non-Gaussianities on superhorizon scales.
Analytic solutions for this method exist for certain potentials, just like in the $\delta N$ formalism, but the main advantage of it is that it 
allows for accurate and efficient numerical solutions, regardless of the form of the potential. In this subsection, we briefly review the key elements of the transport method. We furthermore recall how multifield dynamics  can cause  the curvature perturbation to evolve on superhorizon scales, and we define the isocurvature and curvature correlators. We close this section by briefly reviewing the $\delta N$ formula for the non-Gaussianity amplitude $\fnl$. 

\subsubsection*{The transport method}
%

 %The transport formalism has been developed elsewhere, but for the readers who are unfamiliar with it, we briefly outline how it works here. 
 %In this formalism we work i
 In the spatially flat gauge, we can write the perturbations at the end of inflation as an expansions in the perturbations at horizon exit:
\begin{equation}
\delta\phi_a=\Gamma_{ab}\delta\phi^\star_b+\frac12\Gamma_{abc}\delta\phi^\star_b\delta\phi^\star_c+...
\end{equation}
where the horizon exit perturbations have been marked with a $^\star$. Using the separate-universe approach \cite{Lyth:1984gv, Wands:2000dp}, we expand the slow-roll equations of motion, equation \eqref{eq:SR}, around the background trajectory to obtain evolution equations for $\delta\phi_a$. It is then easy to see that the `propagators' $\Gamma_{ab}$ and $\Gamma_{abc}$ must obey the differential equations, 
\begin{align}
\frac{d\Gamma_{ab}}{dN}=&\hspace{4pt}u_{ac}\Gamma_{cb} \, , \\
\frac{d\Gamma_{abc}}{dN}=&\hspace{4pt}u_{ad}\Gamma_{dbc}+u_{ade}\Gamma_{db}\Gamma_{ec} \, ,
\end{align}
where,
\bea
u_{ab}&=&-\partial_a\partial_b\ln V \, , 
\label{eq:uab}
\\ 
 u_{abc}&=&-\partial_a\partial_b\partial_c\ln V 
 \label{eq:uabc}
 \, ,
\eea
with the derivatives evaluated on the background trajectory. These differential equations have the following formal solutions:
\begin{align}
\Gamma_{ab}(N) = \Gamma_{ab}(N, N^\star)&={\cal P}\, \exp\left(\int^N_{N^\star} dN'u_{ab}(N')\right) \, ,
\label{eq:GammaabSol}
 \\
\Gamma_{abc}(N)= \Gamma_{abc}(N, N^\star)&=\int^N_{N^\star}dN'\Gamma_{a\mu}u_{\mu\nu\rho}(N')\Gamma_{\nu b}\Gamma_{\rho c} \, , 
\label{eq:GammaabcSol}
\end{align}
where ${\cal P}$ is a path ordering operator. In equation \eqref{eq:GammaabcSol}, we have used Greek indices as a short-hand for propagators evolving to or from $N'$, e.g.~$\Gamma_{\nu b} = \Gamma_{\nu b}(N', N^\star)$.

Once the field perturbations at the end of inflation, $\delta \phi^{\rm end}_a$,  are known, the curvature perturbation is given by a gauge transformation:
\begin{equation}
\zeta=N_a\delta\phi^{\rm end}_a+\frac12N_{ab}\delta\phi^{\rm end}_a\delta\phi^{\rm end}_b+...,\label{eq:zeta}
\end{equation}
where the coefficients  $N_a$ and $N_{ab}$ are given by \cite{1302.3842},
\begin{align}
N_a&=\frac{1}{\sqrt{2\epsilon}}\frac{V_a}{\sqrt{V_bV_b}} \, , 
\label{eq:Na}
\\
N_{ab}&=\frac{VV_{ab}}{V_cV_c}
+ \frac{V_a V_b}{V_c V_c}\left(1+2 \frac{V}{(V_e V_e)^2} V_f V_{fd} V_d \right) - \frac{V}{(V_c V_c)^2} \left( V_a V_{bc} V_c + V_b V_{ac} V_c \right) \, .
%-\left(\frac{1}{V_cV_c}-\frac{2VV_cV_{cd}V_d}{(V_eV_e)^3}\right)V_aV_b+V_{\{a}
%\left(\frac{V_{b\}}}{V_cV_c}-\frac{VV_{b\}d}V_d}{(V_cV_c)^2} \right) \, ,
%+V_b \left(\frac{V_a}{V_cV_c}-\frac{VV_{ab}V_b}{(V_cV_c)^2} \right) \, .
\label{eq:Nab}
\end{align}
%{\color{blue} simplify}
%where in the last term we have symmetrised over $a$ and $b$: $A_{\{a} B_{b\}} \equiv =A_a B_b + A_b B_a$. 
%Here,
%\eq{
%A_a= \left(\frac{V_a}{V_bV_b}-\frac{VV_{ab}V_b}{(V_cV_c)^2} \right) \, .
%}

\subsubsection*{The curvature perturbation}
 We write the field two-point correlator as,
\begin{equation}
\langle\delta\phi_a^\star(\mathbf k_1)\delta\phi_b^\star(\mathbf k_2)\rangle=(2\pi)^3\delta^{(3)}(\mathbf k_1 + \mathbf k_2)\frac{2\pi^2}{k^3} \Sigma_{ab} \, .
\end{equation}
In slow-roll and with a slowly turning field trajectory at horizon crossing, we can take the initial condition of the field correlators to be given by,
\be
\Sigma_{ab}^\star=\frac{H^2(N^\star)}{4\pi^2} \, \delta_{ab} \, .
\label{eq:Sigma}
\ee 
In \cite{DBM3}, it was show that this approximation works well for manyfield models of approximate saddle-point inflation. 
The curvature power spectrum at some later time, $N$, is then given by,
\begin{equation}
P_\zeta(N, k)=N_a(N)N_c(N)\, \Gamma_{ab}(N, N_{\star})\, \Gamma_{cd}(N, N_{\star})\, \Sigma_{bd}^\star \, .
\end{equation}

\subsubsection*{Isocurvature perturbations}
\label{sec:isopert}
To linear order in the field perturbations, the curvature perturbation of equation \eqref{eq:zeta}
is given by field fluctuations along the instantaneous background trajectory,
\be
\zeta = \frac{1}{\sqrt{2 \epsilonV}} \delta \phi_{\parallel} \, ,
\ee
where $\delta \phi_{\parallel} =  n_a \delta \phi^a$ for $n_a = V_a/ |V_b|$.
Field perturbations along the $\Nf-1$ perpendicular directions give rise to
`entropic' or
 `isocurvature' perturbations.
We can decompose the field fluctuations as,
\be
{\delta\phi}^a \equiv \delta \phi_{\parallel}\,  n^{a}+ \delta \phi_{\perp}^j  \,v^a_j 
\, ,
\label{eq:decomp}
\ee
where $v^a_j(N)$ 
denotes a 
 generic orthonormal frame of basis vectors in  directions 
perpendicular to $V_a$. Here
$a$ is a vector  index $a$ and with $j=1, \ldots, \Nf-1$.
In analogy to $\zeta$ (and just as in \cite{DBM3}), we define the isocurvature ${\cal S}^i$ as,
\be
\label{eq:iso}
{\cal S}^i \equiv \frac{1}{\sqrt{2 \epsilon_V}}  \delta \phi_{\perp}^i \, .
\ee
In slow-roll and on superhorizon scales, the curvature and isocurvature evolve as \cite{GrootNibbelink:2001qt, DBM3},
\bea
\zeta' &=&  2 \left(n^a \frac{V_{ab}}{V} v^b_i\right) {\cal S}^i \, , 
\label{eq:zetaprime}
\\
%\left( \ln {\cal S} \right)' &=& \left(n^a \frac{V_{ab}}{V} n^b\right) - \left(v^a \frac{V_{ab}}{V} v^b\right) - 2 \epsilon_V \, .
({\cal S}^i)' &=& (n^a \frac{V_{ab}}{V} n^b - 2 \epsilon_V) {\cal S}^i - v^a_i\, \frac{V_{ab}}{V}\, v^b_k\; {\cal S}^k \, . 
\label{eq:Sprime}
\eea
Equations \eqref{eq:zetaprime} and \eqref{eq:Sprime} reflect the well-known fact that isocurvature can source superhorizon evolution of the curvature perturbation $\zeta$,  but the curvature perturbation does not source isocurvature \cite{GarciaBellido:1995qq, Gordon:2000hv}. 

 The isocurvature correlations %(suppressing the momentum-conserving delta function)
 are then given by,
%\be
%P_{\rm iso}^{ij}(N)= 
%\frac{k^3}{(2\pi)^3}\, 
%\langle {\cal S}^i {\cal S}^j \rangle
%=
%\frac{1}{ 2\epsilon_V} %\sum_{j=1}^{N_f-1}  
%v^{a}_i \,  \Sigma^{ab}\, v^{b}_j \, .
%\ee
\be
\langle {\cal S}^i(\mathbf k_1) {\cal S}^j(\mathbf k_2) \rangle=
(2\pi)^3\delta^{(3)}(\mathbf k_1+\mathbf k_2)\frac{k^3}{2\pi^2}P_{\rm iso}^{ij}(N)\,,
\ee
with,
\eq{
P_{\rm iso}^{ij}(N)= 
\frac{1}{ 2\epsilon_V} %\sum_{j=1}^{N_f-1}  
v^{a}_i \,  \Sigma^{ab}\, v^{b}_j \, .
}
We refer to the isocurvature power spectrum (without indices) as,
\be
P_{\rm iso} = \delta_{ij} P^{ij}_{\rm iso} = 
\frac{1}{ 2\epsilon_V} %\sum_{j=1}^{N_f-1}  
v^{a}_i \,  \Sigma^{ab}\, v^{b}_i \, .
\label{eq:Piso}
\ee

%We can apply the transport approach to the curvature perturbations to obtain differential equations governing their evolution. We define
%\begin{equation}
%Q_a\equiv\frac{\delta\phi_a}{\sqrt{2\epsilon}},
%\end{equation}
%in terms of which the curvature and isocurvature perturbations may be written
%\begin{align}
%\mathcal{R} &=n_aQ_a & \mathcal S_a &=Q_a-(n_b Q_b)n_a,
%\end{align}
%where $n_a=V_a/(V_bV_b)^{1/2}$. Using the transport equations it is then straightforward to find the evolution equation for $\mathcal R$, which is
%\begin{equation}
%\frac{d\mathcal R}{dN}=-2\frac{n_aV_{ab}}{V}\mathcal S_b.\label{eq:Rsourcing}
%\end{equation}

\subsubsection*{Non-Gaussianity}
Equation \eqref{eq:zeta} is related to  the commonly used `$\delta N$' formulas, which involve the field perturbation at horizon crossing, by,
%
%The expression for the curvature perturbation in equation \eqref{eq:zeta} is not the same formula as that of the $\delta N$ formalism. Those quantities, however, are related by
\begin{align}
N_a^{\delta N}&=N_b\Gamma_{ba} \, ,\\
N_{ab}^{\delta N}&=N_c\Gamma_{cab}+N_{cd}\Gamma_{ca}\Gamma_{db} \, .
\end{align}
To compute the parameter $\fnl$, we use the $\delta N$-formalism expression, which is given by \cite{Vernizzi:2006ve},\footnote{We adopt the sign convention of \cite{Vernizzi:2006ve}  for $\fnl$.}
\begin{equation}
-\frac65\fnl=\frac{r}{16}(1+f)+\frac{N_a^{\delta N}N_b^{\delta N}N_{ab}^{\delta N}}{(N_c^{\delta N}N_c^{\delta N})^2},\label{eq:fnl}
\end{equation}
where $0\leq f\leq 5/6$ is momentum dependent,  and $r$ is the tensor-scalar ratio. In the small-field inflation models studied in this paper, $r \ll 10^{-3}$, and the first term is negligible.

While the calculation of the power spectrum typically is insensitive to small numerical errors, $\fnl$ is not. The dominant, second term of equation \eqref{eq:fnl} can itself  be expressed as the sum of two terms,
\eq{
-\frac65\fnl=\frac{N_a^{\delta N}N_b^{\delta N}N_{ab}^{\delta N}}{(N_c^{\delta N}N_c^{\delta N})^2}=\frac{N_c\Gamma_{ca}N_d\Gamma_{db}(N_{ef}\Gamma_{ea}\Gamma_{fb}+N_{e}\Gamma_{eab})}{(N_{g}\Gamma_{gi}N_{h}\Gamma_{hi})^2}
\, 
.
}
Quite commonly, both these terms can be large (say $\mathcal O(10)$), but cancel each other to a very high degree (say  
down to $\mathcal O(10^{-2})$). This delicate cancellation calls for high precision of the numerical evaluation of   the background and  the $\Gamma$ coefficients. We briefly discuss our numerical implementation of the evolution of the perturbations in Appendix \ref{sec:numpert}.

\section{Result I: Planck compatibility is not rare, but future experiments may rule out this class of
models}
\label{sec:result1}
We are now ready to discuss the results of our simulations of manyfield models of inflation in random potentials. %We begin by studying %the statistical predictions of ensembles of inflationary models, focussing 
In this section we focus on 
observables related to the two-point correlation function, such as the primordial power spectra of curvature and isocurvature perturbations. 
Our first key result is that despite multifield effects typically being non-negligible, power spectra tend to be very smooth, and observational compatibility is not rare in these  models. 

This section is organised as follows: we first discuss the evolution of the classical background, and we highlight and explain the particularly strong `eigenvalue repulsion' effect on the smallest eigenvalue of the Hessian  in GRF models. %We furthermore discuss the e-fold distribution as a function of the hyperparameters. 
We then discuss the primordial  perturbations of these models: we validate that the power spectra are well-approximated by approximately scale-invariant power laws over the scales relevant for the CMB. We then note that essentially all models predict small deviations from the strict power-law form, and we compute the predictions for the spectral tilt, $n_s$ and its running, $\alpha_s$, as functions of the hyperparameters. 
This leads us to establish a surprisingly robust prediction of these models, which makes it possible to rule them out with future experiments. 
We furthermore find that multifield effects are typically important, but that, importantly, isocurvature tends to decay during inflation. 
%Finally, we illustrate  how  a simple single-field model can capture several aspects of 
 %the background evolution of   manyfield model, but fails to generate consistent predictions for observables. 

Several of the results found in this section are directly analogous to results recently observed in models of  manyfield inflation in DBM potentials \cite{DBM3}, while others differ substantially. In section \ref{sec:result2} we compare these setups in detail.

 \subsection{Background evolution in GRF inflation}
 We first briefly discuss some key elements of the evolution of the inflationary background in the GRF potentials. 
To get an intuition for these models, it is instructive to first consider an example. We here take a  randomly generated 100-field model as our case-study. 
This model was generated from the hyperparameters $\Lh =0.4$,  $\epsilon=5\times10^{-10}$,   $\etai=-10^{-4}$, a flat spectrum, cf.~equation \eqref{eq:spectra}, and a randomly directed gradient vector at $\phi=0$. This particular model gives a total number of e-folds $N^{\rm end} =80.6 $ over a total field displacement of $\Delta \phi = \sqrt{\phi^a \phi^a}=0.187\Lh $.

\begin{figure}
    \centering
    \begin{subfigure}{0.48\textwidth}
    \centering
    \includegraphics[width=0.96\textwidth]{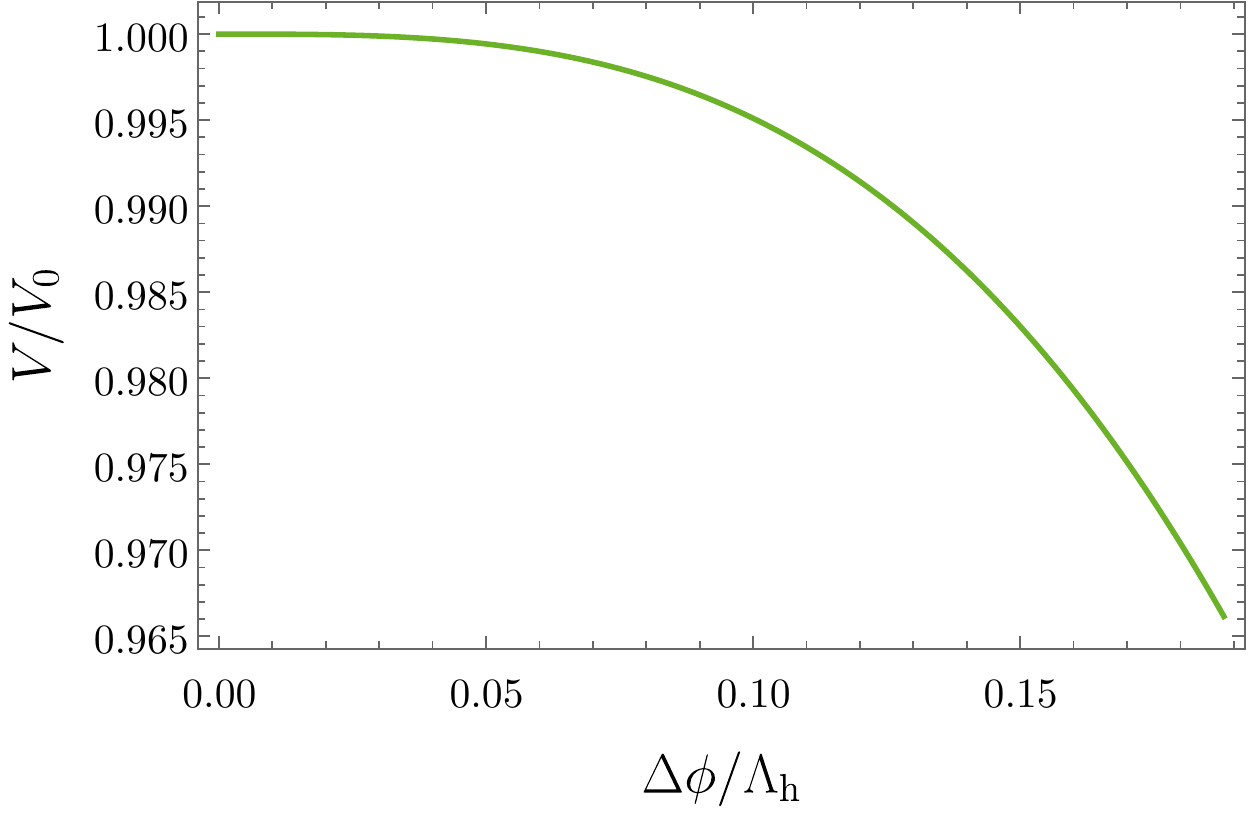}
    %\caption{$\fnl$ values from 1000 random 50-field potentials}
    \end{subfigure}
    ~ %add desired spacing between images, e. g. ~, \quad, \qquad, \hfill etc. 
    \begin{subfigure}{0.48\textwidth}
         \includegraphics[width=1\textwidth]{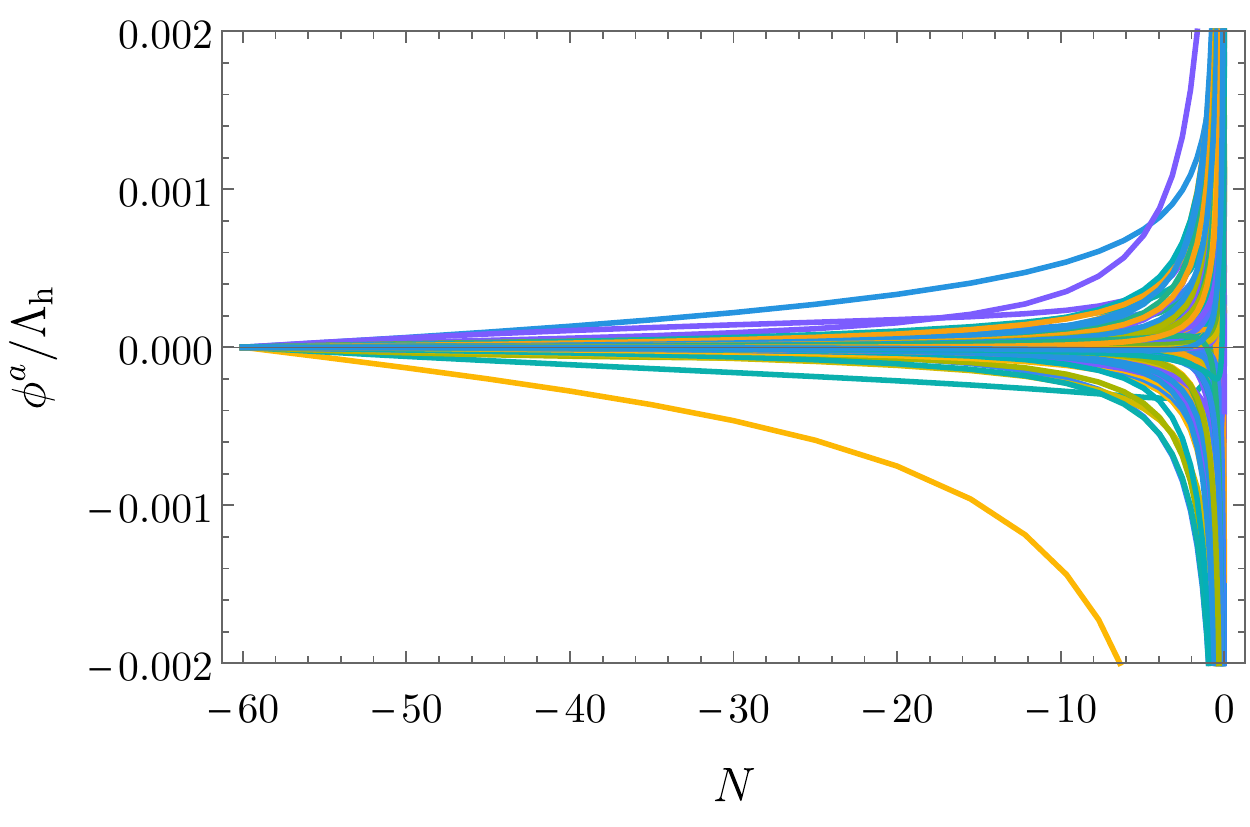}
   % \caption{Fraction of negative eigenvalues}

    \end{subfigure}
    \caption{The value of the potential as a function of the field displacement (left), and the evolution of multiple components of the fields (right) in a random 100-field GRF model.}
    \label{fig:100fieldVphiN}
\end{figure}

The left plot of Figure \ref{fig:100fieldVphiN} shows the normalised value of the potential energy as a function of $\Delta\phi $. %While the potential along a randomly sampled direction would exhibit bumps and other features, t
The potential along the descending inflationary trajectory  is very smooth and featureless. Since $\epsilonV$ is initially very small and $|\dot \phi| = \sqrt{2\epsilonV}$, the field rolls very slowly initially, but accelerates super-exponentially towards the end of inflation.
%
%
 %the inset of the right plot of 
 Figure \ref{fig:100fieldVphiN} shows %the total field displacement as a function of $N$. The right plot of Figure \ref{fig:100fieldVphiN} also illustrates 
 how multiple components of the field evolve during inflation, and indicates that the inflationary trajectory turns as the field descends the potential. 

The eigenvalues of the Hessian matrix are not constant in a general inflationary model, and we expect the eigenvalues of the GRF models to relax from the fine-tuned initial configuration to the (slightly off-centred) semi-circle spectrum. Figure  \ref{fig:100fieldEVs} shows the evolution of the squared masses as a function of the field displacement during inflation (left plot), and as a function of the number of e-folds (right plot) for our 100-field example. Indeed, as the fields evolve from $\phi=0$, the spectrum spreads out. 
Half of the fields, initially heavier than the others, tend to become even more massive during inflation, and are not very  important  for either the background evolution or the spectrum of the perturbations.  By contrast, the lighter half of the fields  become even lighter, and many even go tachyonic: with small variations over all the models we have considered, almost precisely half the fields have $m^2 < 0$ at the end of inflation.\footnote{On approach to the final vacuum configuration after inflation, these eigenvalues will again become positive.}  When plotted as a function of $\Delta \phi$, the bundle of eigenvalues is conical, which is indicative of the dominance of the cubic terms in the potential.

%A curious property of Figure \ref{fig:100fieldEVs} is that the smallest eigenvalue of the Hessian strays from the bundle, and evolves to tachyonic values at a noticeably quicker rate than the other eigenvalues. While this `straying' is not a general property observed in all models we have constructed, it is very common for $\Nf\gtrsim 10$. Only  the smallest eigenvalue appears susceptible to stray from the bundle.
%
%This behaviour is surprising, and apparently related to the dynamics of multi-field slow-roll inflation. When evolving the fields along a randomly chosen path in field space, the eigenvalues of the Hessian matrix spread out in a tight bundle, with no outliers. It is well known that as the inflationary trajectory descends the potential along the negative gradient, $\dot \phi$ tends to align with direction of the most tachyonic eigenvalue of the Hessian (see e.g.~\cite{DBM3} for a recent discussion). 
%However, that gradient descent should also lead to `straying' of the smallest eigenvalue in systems with $\Nf\gg 1$ is to us unexpected. 
%In sections \ref{sec:iso} and \ref{sec:result3}, we will see that this phenomenon contributes in part to some of the observational predictions of manyfield inflation in GRF potentials. 
\subsubsection{The `straying' smallest mass-squared}
\label{sec:stray}

\begin{figure}
    \centering
    \begin{subfigure}{0.48\textwidth}
         \includegraphics[width=1\textwidth]{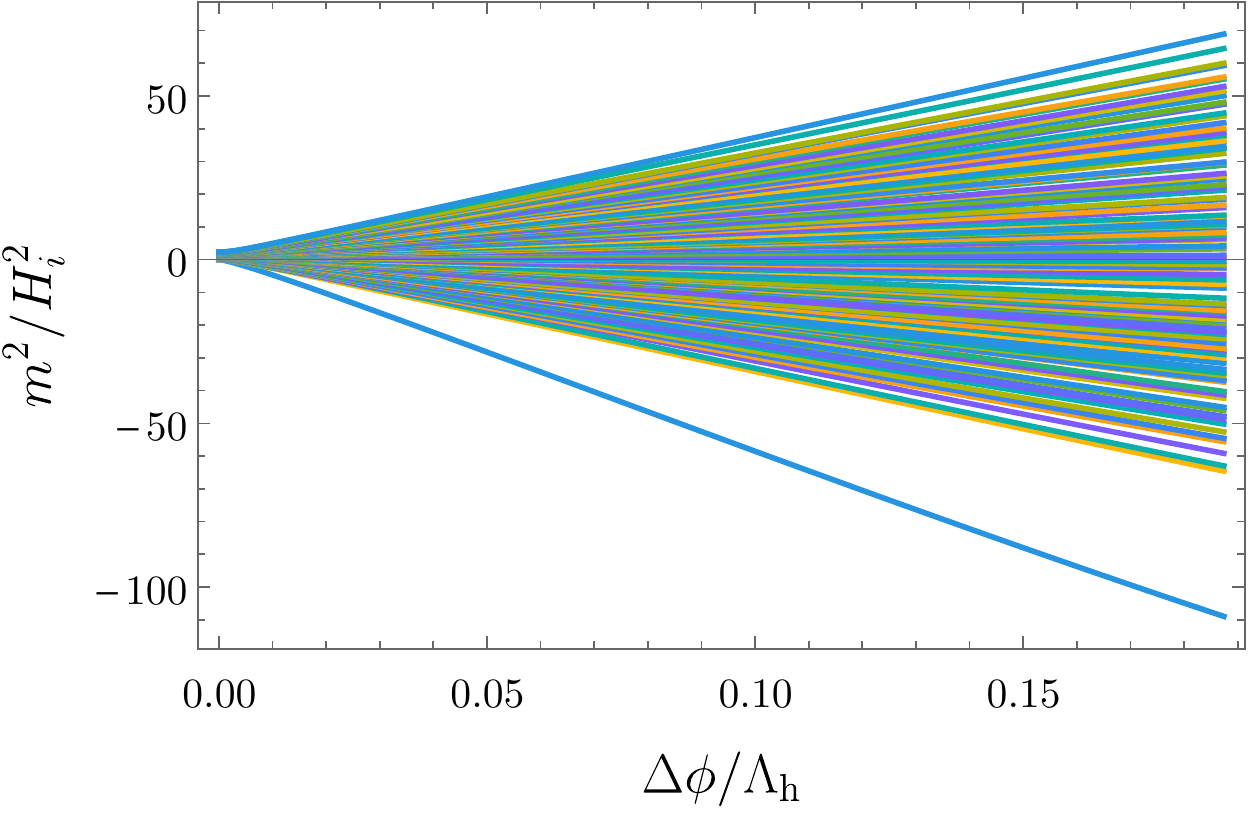}
   % \caption{e-folds before end when a second direction becoms tachyonic}
    \end{subfigure}
    ~ %add desired spacing between images, e. g. ~, \quad, \qquad, \hfill etc. 
    \begin{subfigure}{0.48\textwidth}
         \includegraphics[width=1\textwidth]{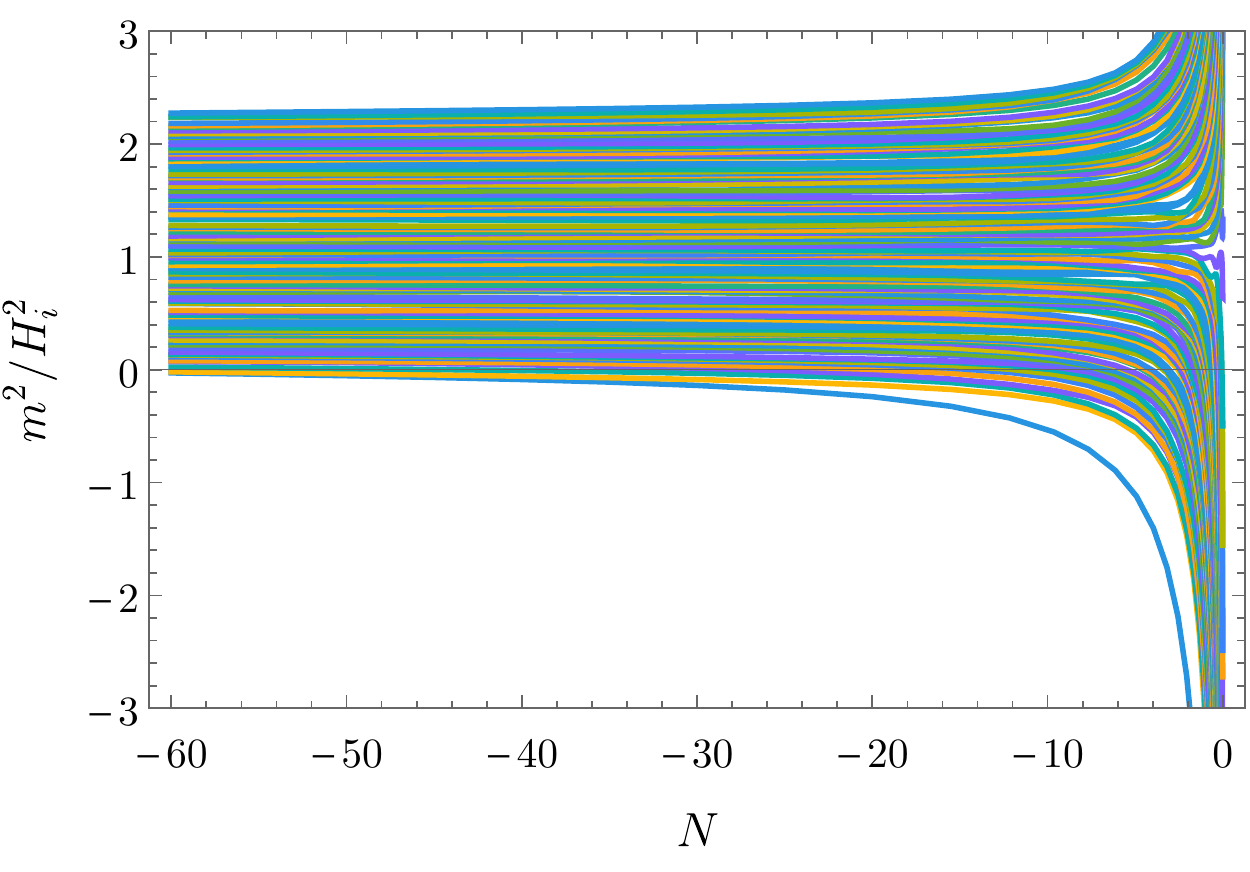}
   % \caption{Fraction of negative eigenvalues}
    \end{subfigure}

    \caption{Eigenvalue evolution of the Hessian in a 100-field example, starting from a flat spectrum.
     }\label{fig:100fieldEVs}
\end{figure}

Figure \ref{fig:100fieldEVs} also illustrates a curious and important feature of these models: the smallest %mass-squared 
eigenvalue decreases more rapidly than the others, and `strays' from the conical bundle towards more tachyonic values. This `straying' behaviour of the smallest eigenvalue 
has to our knowledge not been discussed previously in the literature, but 
appears  for large $\Nf$  in essentially all inflationary models that we have studied. In sections \ref{sec:iso} and \ref{sec:result3}, we will see that it contributes to some of the most interesting predictions of the inflationary GRF models.

While the rapid evolution of the smallest eigenvalue of the Hessian may appear surprising, it has a simple explanation in terms of the properties of the GRF potentials, and the dynamics of multifield slow-roll inflation. 
 In our inflationary models, the initial values of both the gradient and the Hessian matrix are very small. This means that after a short field excursion, which typically involves some turn, the 
 %background evolution 
gradient and Hessian  
 become dominated by the third-order terms. Without loss of generality, we may take the `1'-axis to be aligned with the  field excursion at this point, with $\phi_1>0$. We then have,
\begin{align}
V_a(\phi)&\simeq\frac12V_{a11}\phi_1^2\, , & V_{ab}(\phi)&\simeq\frac12V_{ab1}\phi_1\, .
\label{eq:Vabexpl}
\end{align}
With the initial gradient set to be small, the third derivatives are drawn from a distribution with a mean that is very close to zero and variances given by,
\eq{
\mathrm{Var}(\tilde V_{abc})=
\begin{cases}
    6       & \quad \text{if all indices are equal}\\
    2       & \quad \text{if only two are equal}\\
    1       & \quad \text{if none are equal.}
  \end{cases}
}
We then see that the magnitude of $V_1(\phi)$ is expected to be larger than the other components of the gradient. Furthermore, since $\phi_1>0$, we expect that $\dot \phi_1\propto -V_1(\phi)>0$, in which case $V_{111}$ must be be negative. We can therefore expect $V_{11}(\phi)=V_{111}\phi_1$ to be larger than the other elements of the Hessian matrix, and negative. Moreover, since $\mathrm{Var}(V_{a11})>\mathrm{Var}(V_{ab1})$ for $b\neq1$ (and $a\neq b$) the off-diagonal row-vector $V_{1a}$ is expected to be larger in magnitude than the other row vectors.
%
%
%, and since $V_1(\phi)$ must have the opposite sign of $\phi_1$, we also see that $V_{11}(\phi)$ must be negative. 
This will typically lead to a large negative mass-squared eigenvalue with an eigenvector approximately aligned with the gradient direction. This is precisely what we observe through the `straying' smallest eigenvalue of the Hessian. 

%The mass-squared in the gradient direction is therefore expected to be large and negative, and this would give the large, negative mass-squared eigenvalue that we always see. 

  \begin{figure}
    \centering
    \begin{subfigure}{0.44\textwidth}
         \includegraphics[width=1\textwidth]{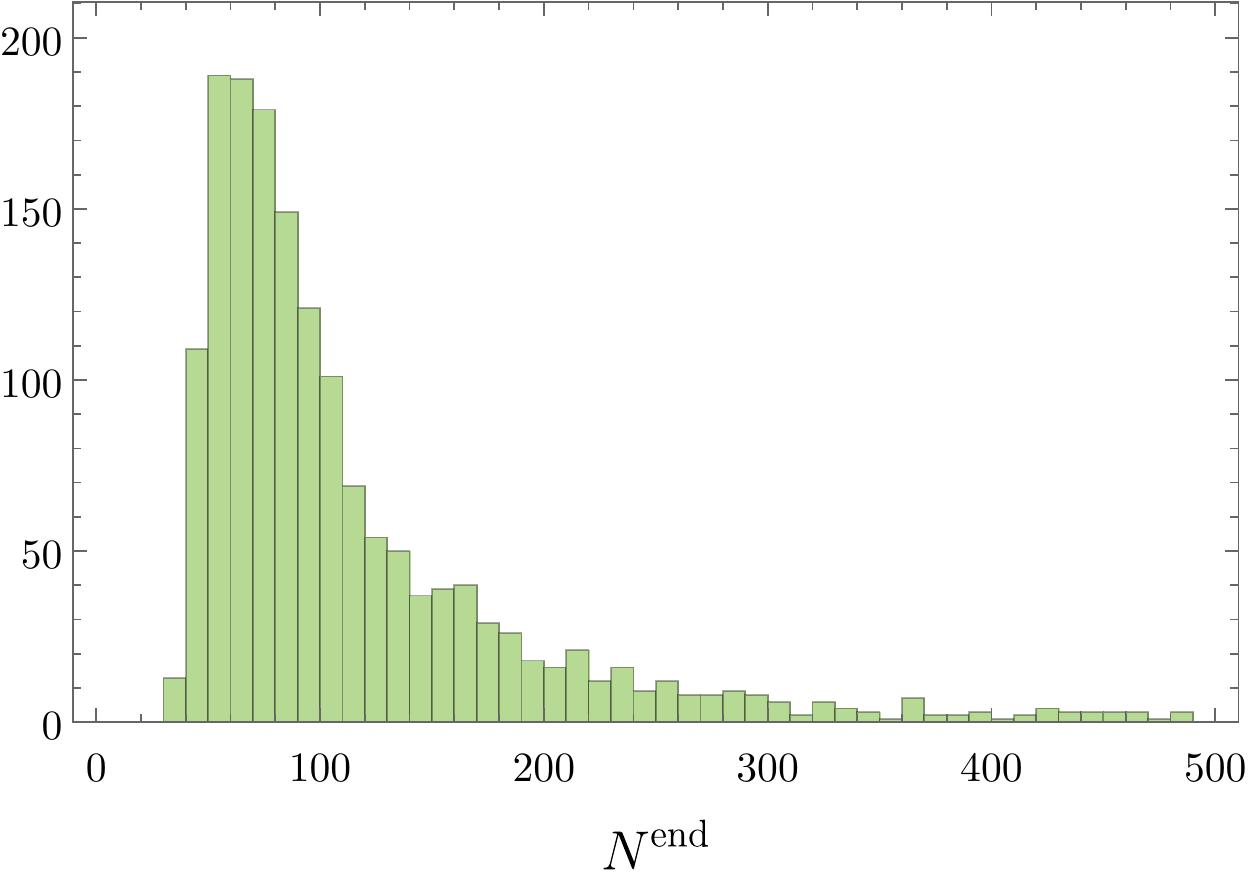}
    %\caption{Calculated average values for $N^\text{end}$ as a function of $\Lh$, shown together with the with the function calculated in the toy model.}
    \label{fig:NendLh}
    \end{subfigure}
    \begin{subfigure}{0.47\textwidth}
         \includegraphics[width=1\textwidth]{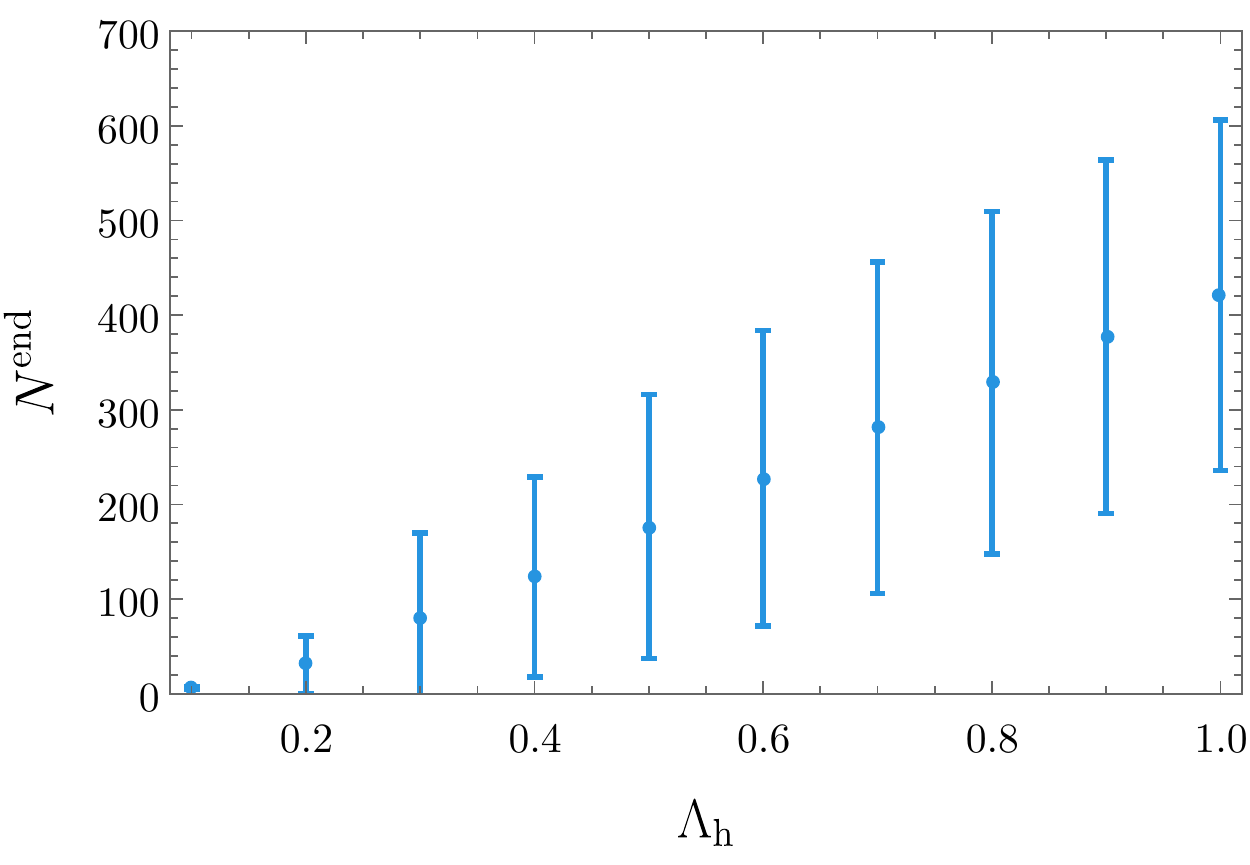}
    %\caption{Calculated average values for $N^\text{end}$ as a function of $\Lh$, shown together with the with the function calculated in the toy model.}
    \label{fig:NendLh}
    \end{subfigure}
    ~ %add desired spacing between images, e. g. ~, \quad, \qquad, \hfill etc. 
    \begin{subfigure}{0.48\textwidth}
         \includegraphics[width=1\textwidth]{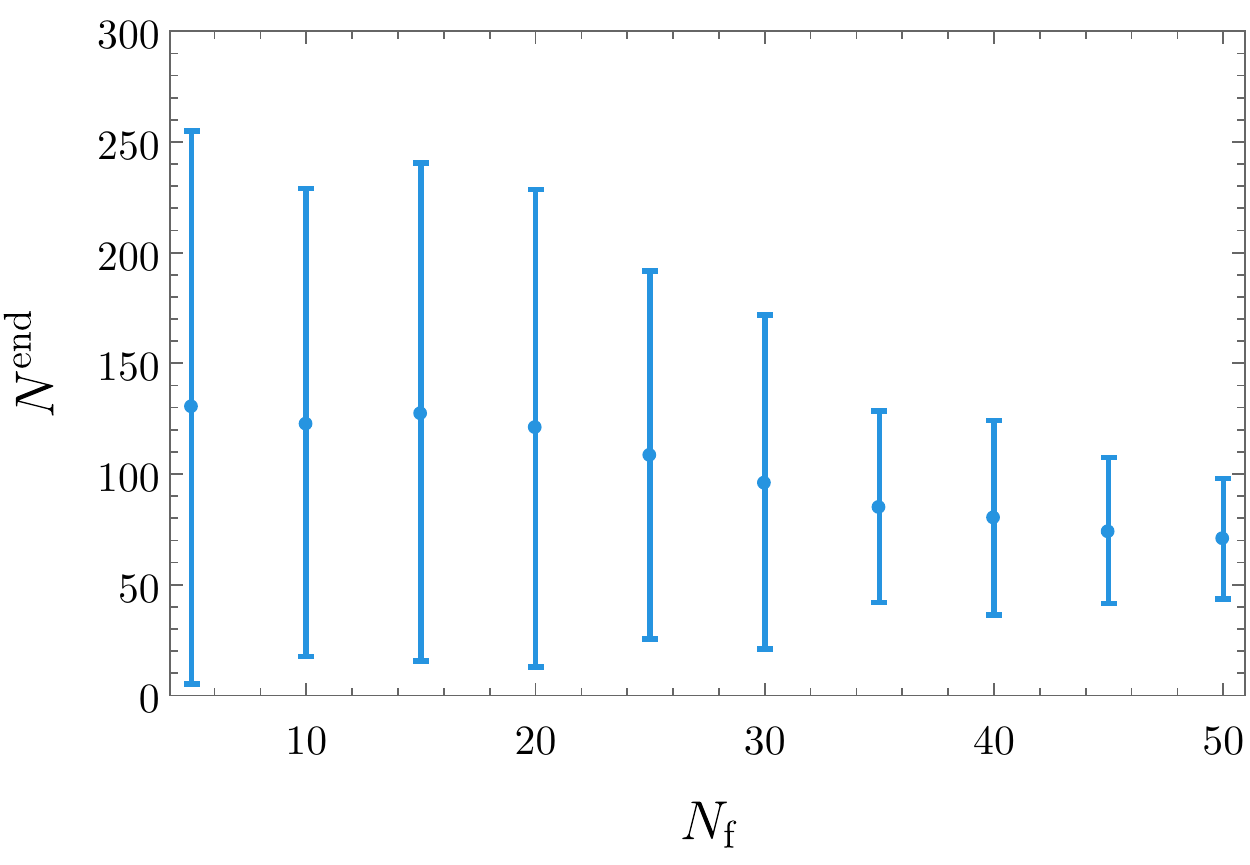}
    %\caption{Calculated average values for $N^\text{end}$ as a function of $\Lh$, shown together with the with the function calculated in the toy model.}
    \label{fig:Nendd}
    \end{subfigure}
        ~
        \begin{subfigure}{0.49\textwidth}
         \includegraphics[width=1\textwidth]{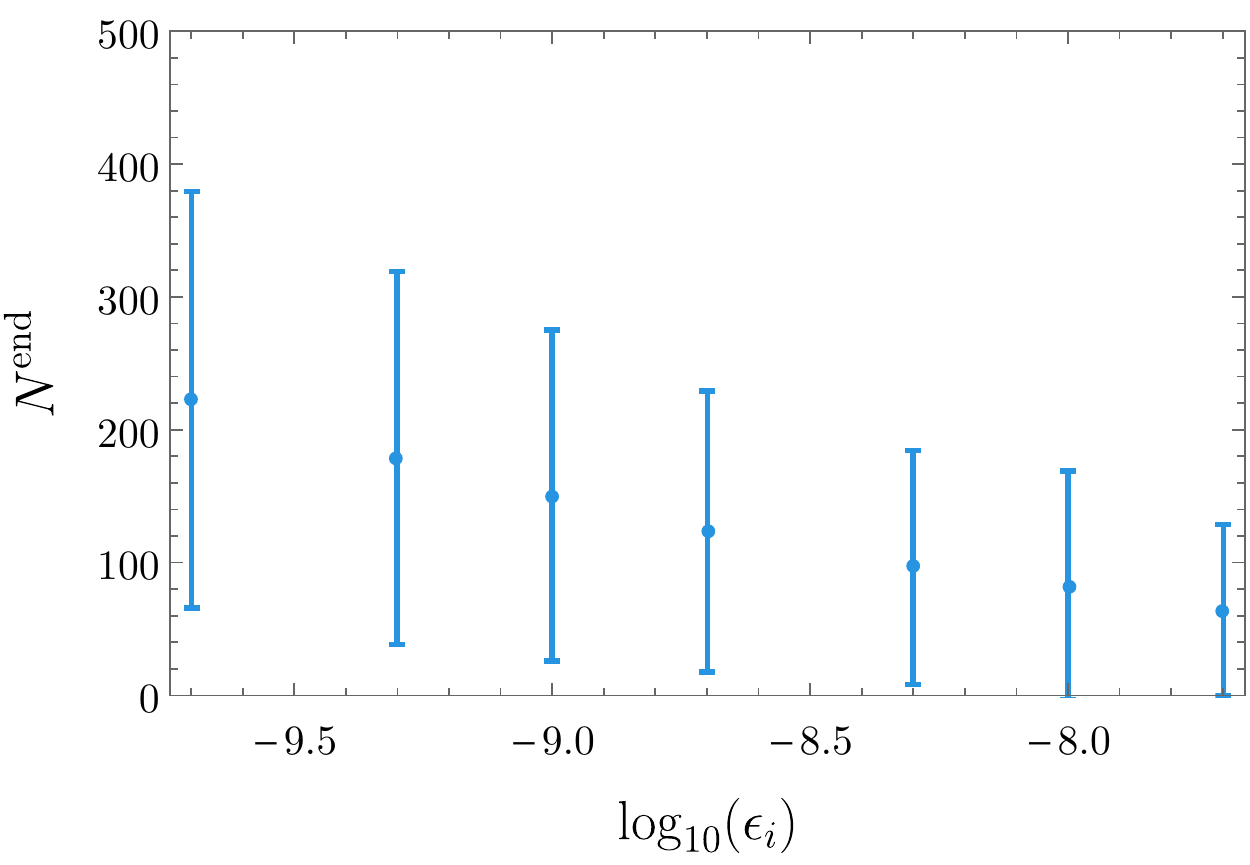}
   % \caption{Fraction of negative eigenvalues}

    \end{subfigure}
    \caption{Histogram of the e-fold distribution of the baseline model  (cf.~equation \eqref{eq:baseline}), and the dependence of the mean and standard deviation on the hyperparameters in one-parameter variations from the baseline. Each data point is generated from an ensemble of 2000 inflationary models.
    }\label{fig:Nend}
\end{figure}

This evolution of the squared masses importantly affect the inflationary evolution of the 
field perturbations (as we will discuss in section \ref{sec:iso}), and also the background dynamics. 
In slow-roll inflation, the acceleration of the field is given by,
\eq{
%\frac{d n_a}{dN}=\frac{n_bV_{ba}}{V}-\frac{n_bV_{bc}n_c}{V}n_a
\phi''_a = \sqrt{2 \epsilonV} \left(
\frac{V_{ab} n_b}{V} - 2 \epsilonV n_a
\right) \, ,
\label{eq:acc}
}
where $n_a = V_a/|V_a|$. We see that if $n_a \approx - \delta_a^1$,  the acceleration tends to be large and positive in the `1'-direction.  This will make the trajectory
`straighten' during this phase, and $n_a$ will become closer and closer aligned with the direction of the smallest eigenvalue of the Hessian matrix.

%  trajectory will therefore be rather straight during this phase, and there will be little sourcing of the curvature pertubation by the isocurvature modes, as can be seen from equation \ref{eq:zetaprime}.

\subsubsection*{The e-fold distribution of GRF models}
%In the next section we will discuss the dependence of observables on the hyperparameters. 
An important factor influencing the observables is the distribution of the number of e-folds of inflation. Figure \ref{fig:Nend} shows the mean values and standard deviations of this distribution for various one-parameter variations from the baseline parameter choice of equation \eqref{eq:baseline}.

Unsurprisingly, flatter spectra lead to more e-folds of inflation. As the number of fields is increased, the e-fold distribution slowly shift to lower values, but the dependence is not very strong.  For a given choice of hyperparameters, the distribution of the number of e-folds typically exhibits a broad peak and a `heavy', polynomially decreasing tail corresponding to models with a large number of e-folds.  
  
%  The mass matrix typically does not change much over the majority of the e-folds, since this small-field inflation and most of the field-space movement happens near the end of inflation. At the end, however, we typically see that the mass-spectra we set initially spread out rapidly. This is not unexpected, since the typical mass-separation is much larger than the ones we set.
  
  %However, as we increase the number of fields, and the initial mass matrix becomes more densely populated, we see that field directions become tachyonic earlier. Figure \ref{fig:neg} illustrates this between 10 and 50 fields. The 5-field data was not included, as sometimes no other field direction became tachyonic. The second graph in this figure illustrates another interesting behaviour: regardless of the number of fields present, one would expect at least half the field directions to become tachyonic before inflation ends, even if all but one start out as massive. This behaviour is not surprising; as we typically set the initial eigenvalues to be small, we expect the eigenvalue repulsion to cause them to spread out and approach the Wigner semi-circle distribution. This is indeed the case, and we usually see the masses spreading out linearly with $\Delta\phi$ in a cone. In terms of $N$, however, the eigenvalues stay more or less constant until the final few e-folds where they spread out very quickly.

 \subsection{Smooth and simple power spectra from  complex inflationary models}
%[Full power spectra over ~60 e-folds] 
We now turn to observables generated by these models, focussing in this section on the power spectrum of the curvature perturbation, $P_{\zeta}(k)$. While many of the simplest models of single-field or few-field inflation naturally generate very simple, almost scale-invariant power spectra, there is no guarantee that highly complicated and random manyfield models should also do so. Turns of the field trajectory or bumps in the potential could generate strong deviations from scale-invariance, and highly featured power spectra.  Quite remarkably however, we here find that even random models involving several dozens of fields and millions of interaction terms typically produce extremely smooth and simple power spectra.

\begin{figure}
    \centering
    \begin{subfigure}{0.48\textwidth}
         \includegraphics[width=1\textwidth]{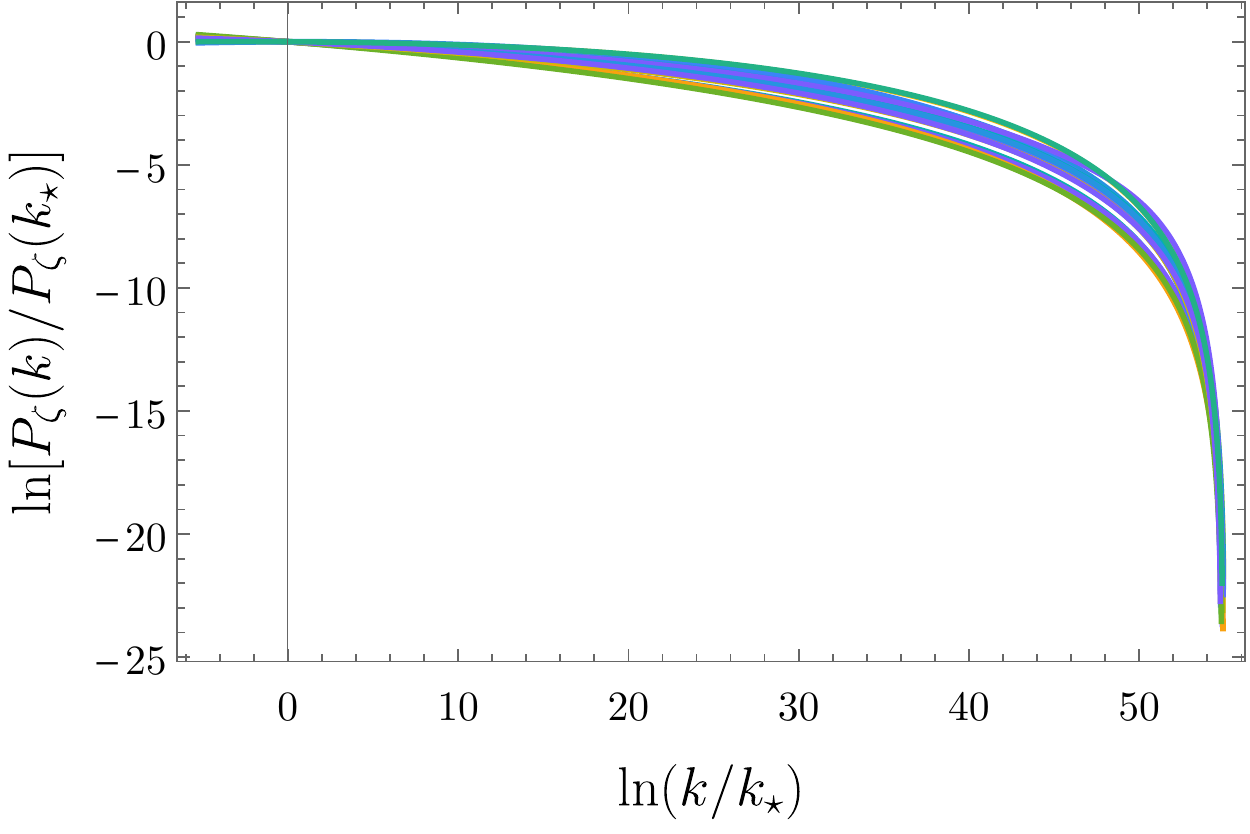}
   % \caption{e-folds before end when a second direction becoms tachyonic}
    \label{fig:fullps10}
    \end{subfigure}
    ~ %add desired spacing between images, e. g. ~, \quad, \qquad, \hfill etc. 
    \begin{subfigure}{0.48\textwidth}
         \includegraphics[width=1\textwidth]{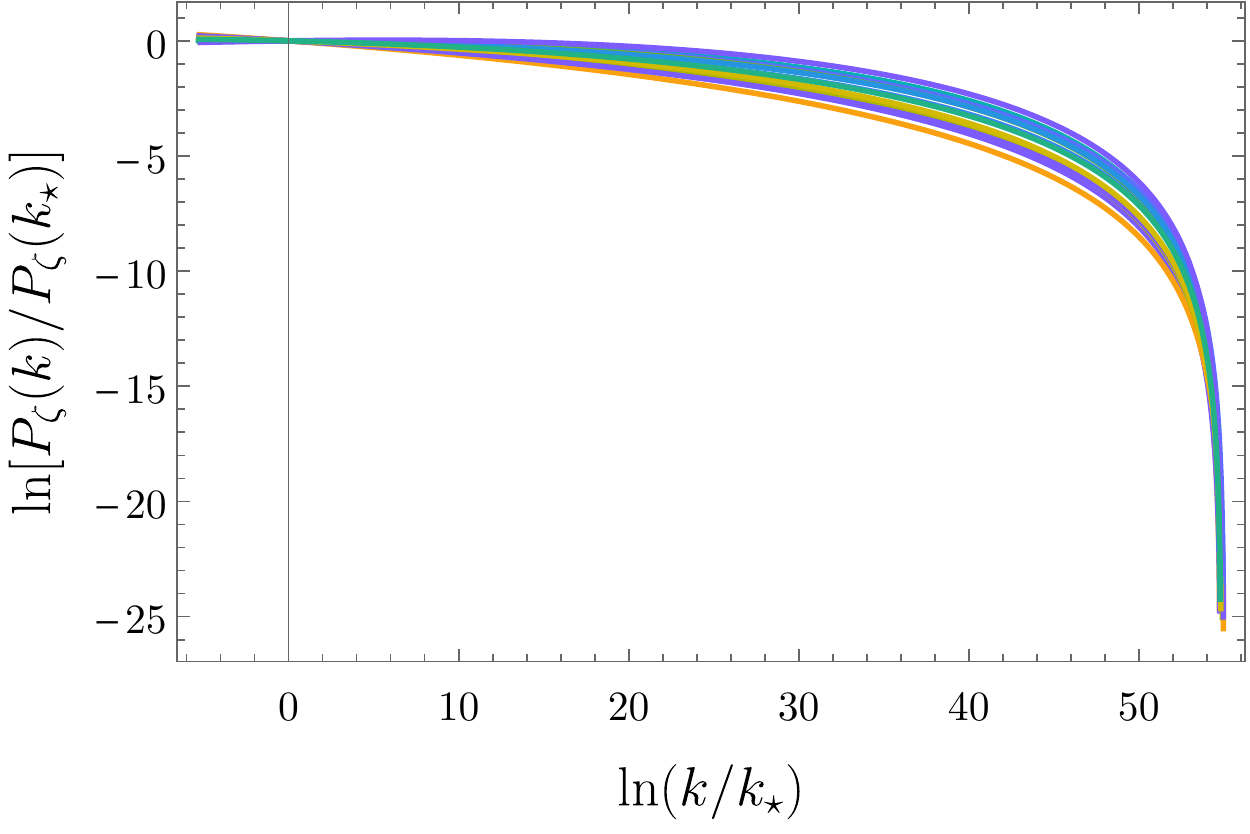}
   % \caption{Fraction of negative eigenvalues}
    \label{fig:fullps50}
    \end{subfigure} \\
     \begin{subfigure}{0.48\textwidth}
         \includegraphics[width=1\textwidth]{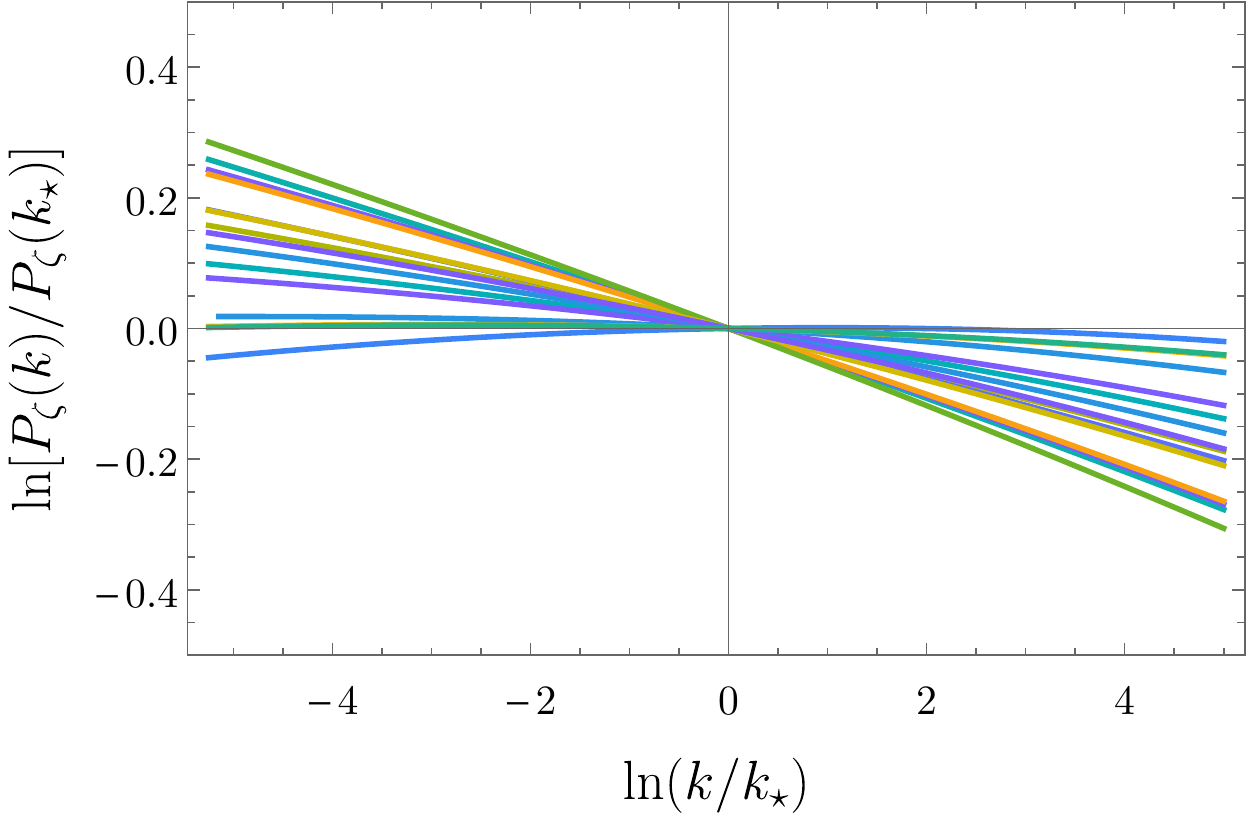}
   % \caption{e-folds before end when a second direction becoms tachyonic}
    \label{fig:ps10}
    \end{subfigure}
    ~ %add desired spacing between images, e. g. ~, \quad, \qquad, \hfill etc. 
    \begin{subfigure}{0.48\textwidth}
         \includegraphics[width=1\textwidth]{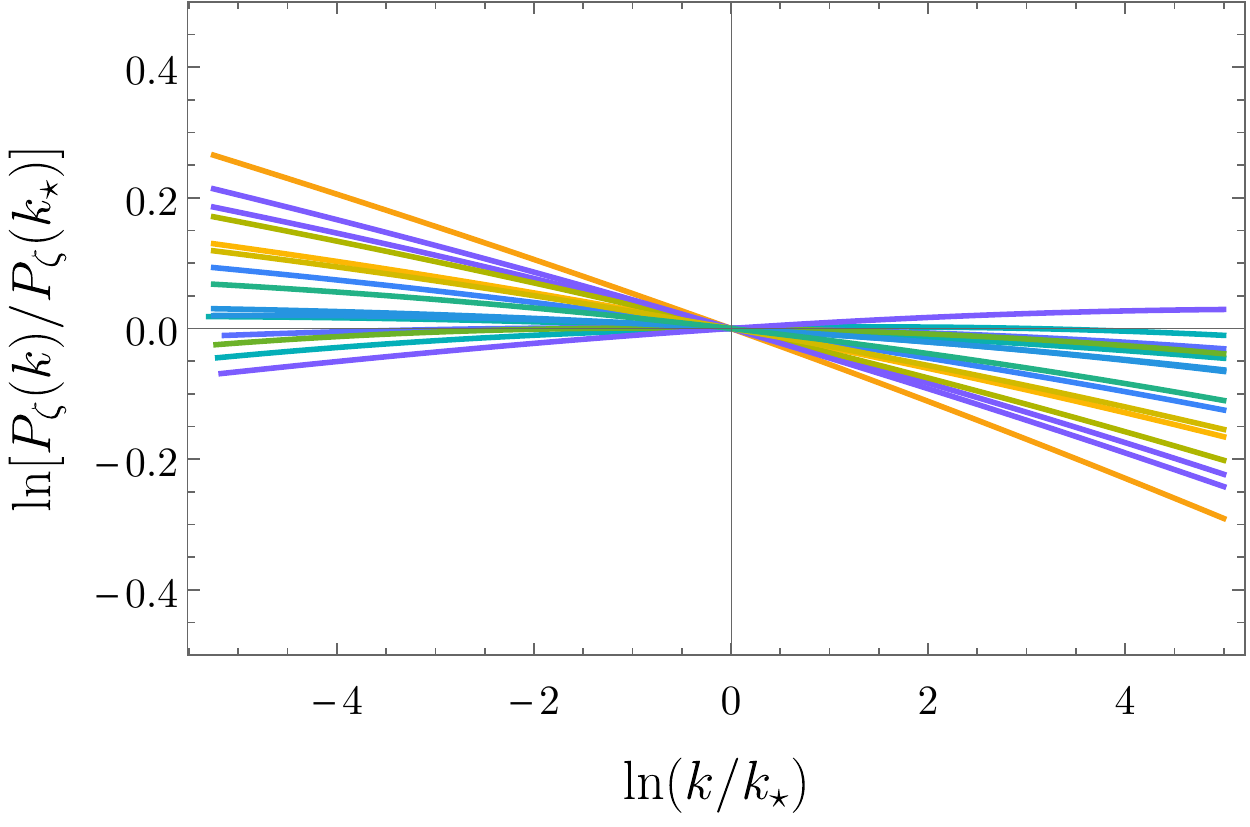}
   % \caption{Fraction of negative eigenvalues}
    \label{fig:ps50}
    \end{subfigure}
    \caption{Examples of power spectra from 15 randomly generated models of GRF inflation for $\Nf=10$  (left) and $\Nf=50$ (right). Hyperparameters other than $\Nf$ are as in the baseline case, cf.~equation \eqref{eq:baseline}.    }\label{fig:fullps}
\end{figure}

A sense of the typical properties of the generated power spectra can be inferred from 
%To provide a sense of t
%The typical properties of the generated power spectra are illustrated by
  Figure \ref{fig:fullps},
which shows the 
 power spectra for 15 randomly generated models with 10 (left) and 50 (right) fields. The top row shows the power spectra evaluated over the full range of scales 
  exiting the horizon within the last sixty e-folds of inflation, while the plots of  the bottom row zooms in on the 10 e-fold range centred at the `pivot scale' corresponding to modes exiting the horizon 55 e-folds before the end of inflation.
  All plotted power spectra are  evaluated at the end of inflation, $P_{\zeta}(N_{\rm end}, k)=P_{\zeta}(k)$, and normalised at the pivot scale $k_{\star}$.%\footnote{In single-field inflation, the curvature perturbation freezes out and stops evolving after crossing the horizon. As we reviewed in section \ref{sec:isopert}, multifield effects can induce superhorizon evolution of the curvature perturbation, and in section \ref{sec:iso}, we will see that the superhorizon evolution of the power spectrum during inflation is often substantial in manyfield GRF models, i.e.~$P_{\zeta}(N_{\star}, k) \neq P_{\zeta}(N_{\rm end}, k)$.
%  In general, the power spectra may continue to evolve after the end of inflation and through the reheating phase, which we do not model in this work. We will in section \ref{sec:iso} however, that there are reasons to believe that reheating may have a very small effect on the generated power spectra.}

  Over the full range of scales spanning 60 e-folds, the power spectra show strong deviations from scale-invariance, with rapidly decreasing power spectra for very small-scale modes. This drop in power is related to the rapid growth of the slow-roll parameter $\epsilonV$ towards the end of inflation, which causes a speed-up of the field and a suppression of the power of the curvature perturbation. On zooming in on the scales most relevant for CMB observations, the generated power-spectra are very simple, and of an approximate power-law form. This simple form of the power spectra  is common to all GRF models we have studied, independent of the precise choice of hyperparameters. We will now discuss the dependence of the detailed predictions of the models on the hyperparameter choices.

 \subsection{Distributions of $n_s$ and $\alpha_s$}
 \label{sec:Planck}

The simple form of the power spectra around the pivot scales  justifies fitting them by an approximate power-law,
\be
P_{\zeta}(k) = A_s \left(\frac{k}{k_{\star}} \right)^{n_s-1} \, ,
\ee
where we allow for a non-vanishing running of the spectral index, $\alpha_s ={\rm d}n_s/{\rm d} \ln k|_{k=k_{\star}}$.

\begin{figure}
    \centering
    \includegraphics[width=0.5\textwidth]{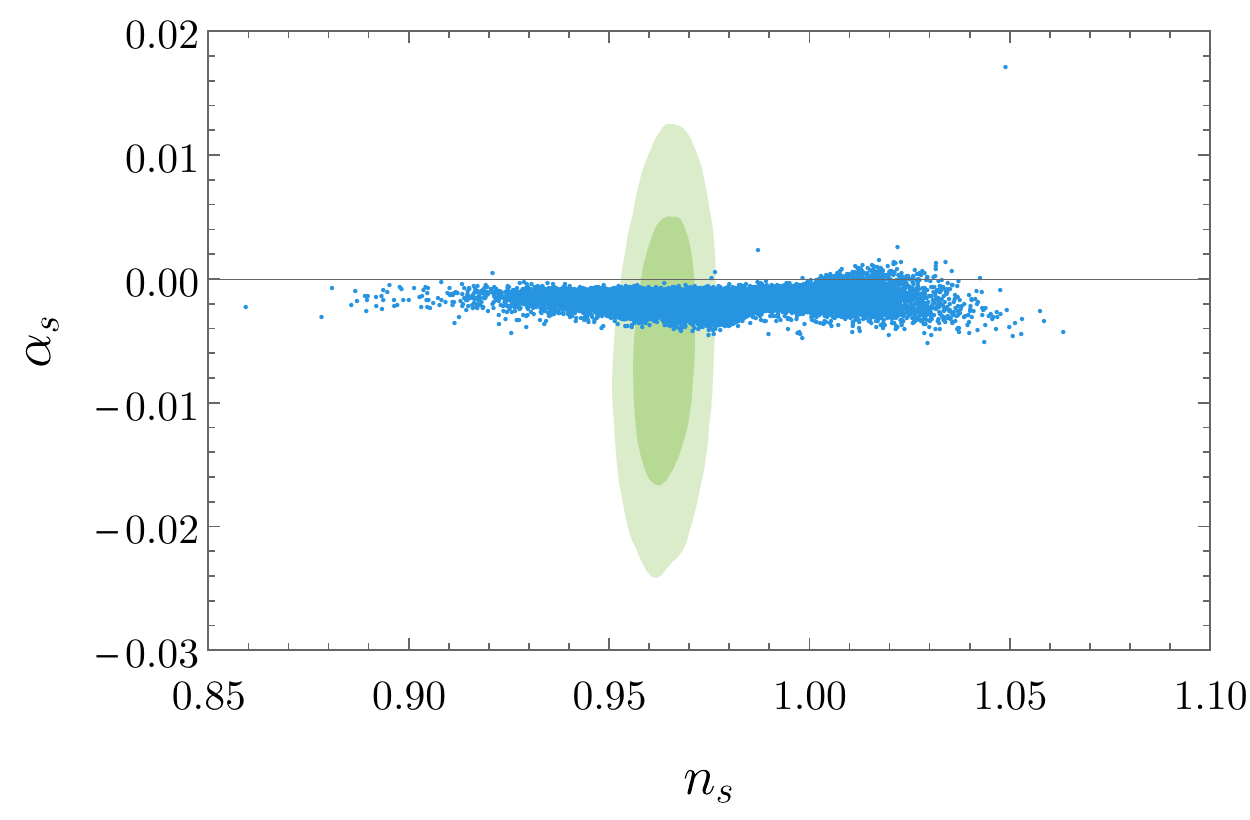}
    \caption{The distribution of $n_s$ and $\alpha_s$ for 25,000 inflation models spanning values of $\Nf$ between 5 and 50, with the 68\% and 95\% confidence contours from Planck (TT+TE+lowP+lensing) \cite{1502.02114}.}\label{fig:nsalphas}
\end{figure}

 %It is instructive to first consider the distribution of $n_s$ and $\alpha_s$ for randomly generated models with a given set of hyperparameters. In
  Figure \ref{fig:nsalphas} shows the aggregated values of  $(n_s, \alpha_s)$ for 25,000 models of GRF inflation with $\Nf$ ranging between 5 and 50, and for varying values of the other hyperparameters.\footnote{This aggregate consists of all models in the tables `varying $\Nf$' and  `$\Nf=50$' in Appendix \ref{app:params}.
} The distribution for $n_s$ indicates that the power spectra are approximately scale-invariant, and that the spectra are more commonly red than blue (around 85\% were red). For these values of the hyperparameters,  the distribution for $n_s$ is broader than current Planck constraints, but Planck-compatible values are not rare. 
 
 The statistical prediction for the running of the spectral index, $\alpha_s$, is remarkably sharp. A small and negative running is vastly favoured (especially among the models with Planck-compatible spectral indices), and these models could be ruled out should future experiments infer a positive or substantially negative running of the spectral index. Indeed, over 99\% of these models, and all of those in the Planck 68\% c.l.~for $n_s$, fall in the range $-0.04<\alpha_s<0$.
 For the baseline hyperparameters, cf.~equation \eqref{eq:baseline}, we find $n_s =0.970\pm0.018$ and $\alpha_s = -0.00143\pm0.00034$. Normalising the amplitude of scalar perturbations
 fixes $\Lv$, for the baseline models we find $\Lv=9.6\, (\pm1.7) \times10^{-5}$.

  The tensor-to-scalar ratio is very small in all models we have constructed. For the baseline parameters, we find $r =3.07\, (\pm0.28)\times10^{-8}$. Since the total field displacement during inflation is $\Delta \phi =   0.36\, (\pm 0.04)\, \Lh$ for these parameters, we see that the `Lyth bound'  \cite{Lyth:1996im}  is far from saturated: 
in single-field models of inflation,
\begin{equation}\label{eq:lythold}
r=16\epsilonV<8\left(\frac{1}{N_{\rm exit}}\right)^{2}\left(\frac{\Delta\phi}{M_{\rm{Pl}}}\right)^{2} \, ,
\end{equation}
if $\epsilonV$ is constant or monotonically increasing, $N_{\rm exit}$ denotes the e-fold when the pivot scale crossed the horizon (in our case $N_{\rm exit} =55$), and $\Delta \phi$ denotes the total field displacement during inflation. 
Thus, for the mean-value base-line parameters, we find the bound $r < 5.5\times 10^{-5}$. %, which is comfortably saturated.
%
%
%. For our model this bound is $r < 3 \times 10^{-5}$, which is clearly far from being saturated.
 There are two reasons for the non-saturation of the Lyth bound. First, the field initially evolves very slowly, but speeds up super-exponentially towards the end of inflation.   Second, we will see in section \ref{sec:iso} that isocurvature modes tend to enhance the amplitude of the scalar perturbation, but leave the tensor perturbations untouched. 
 This further suppresses the tensor-to-scale ratio $r$.
 
 %This further decreases the  the value of $r$ compared to a single-field estimate. 

 Figure \ref{fig:nsalphas} shows that 
manyfield GRF models can be compatible with current observational constraints on the power spectrum, but provide a sharp prediction for its running, and can be ruled out by future experiments. We now investigate how these predictions depend on the hyperparameters.
 %
% 
 % The distributions of $n_s$ and $\alpha_s$ depend on some of the hyperparameters, and we will see that certain regions of hyperparameter space can be ruled out already by current observational constraints.

\begin{figure}
    \centering
    \begin{subfigure}{0.48\textwidth}
         \includegraphics[width=1\textwidth]{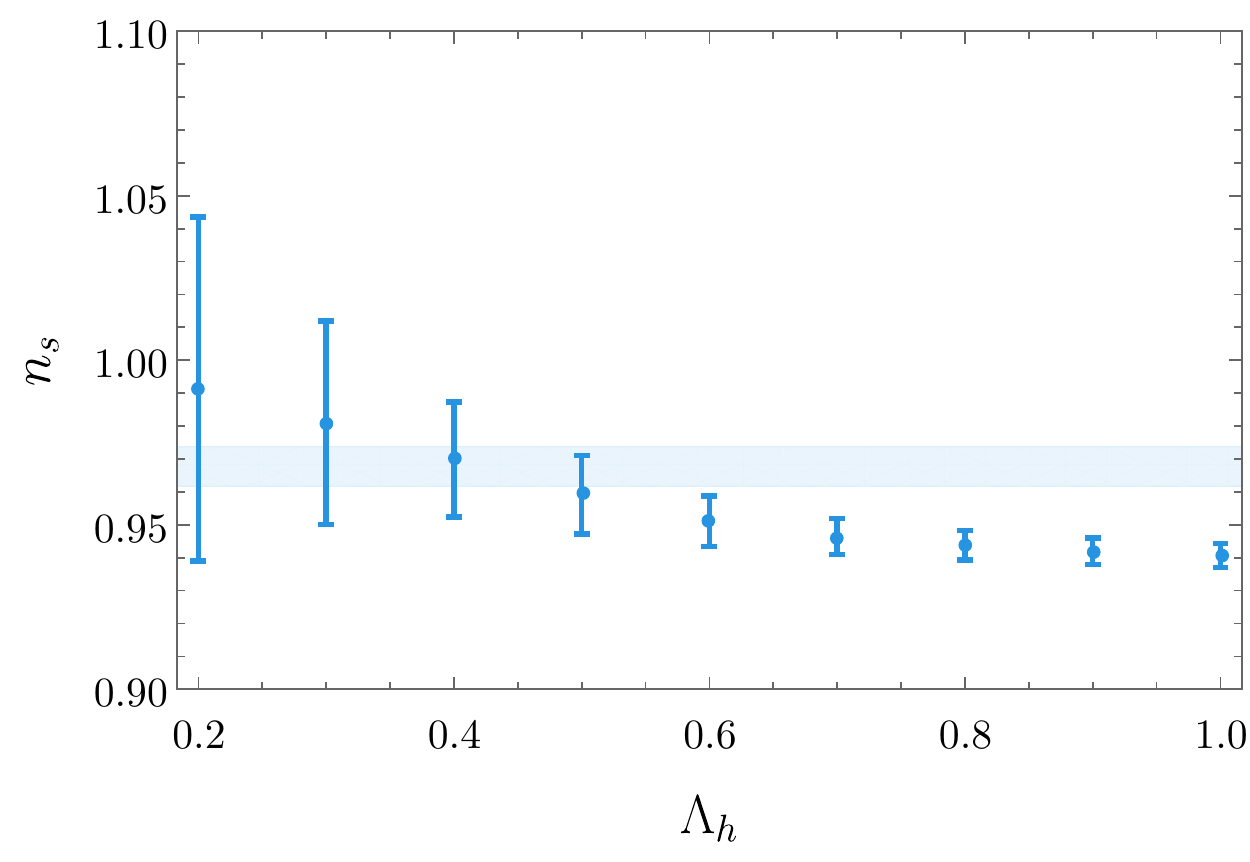}
   % \caption{Fraction of negative eigenvalues}
%    \label{fig:diso}
    \end{subfigure}
%    \caption{These graphs show the effect of varying the }\label{fig:dSHiso}
%\end{figure}
%
%\begin{figure}
%    \centering
~
    \begin{subfigure}{0.48\textwidth}
         \includegraphics[width=1\textwidth]{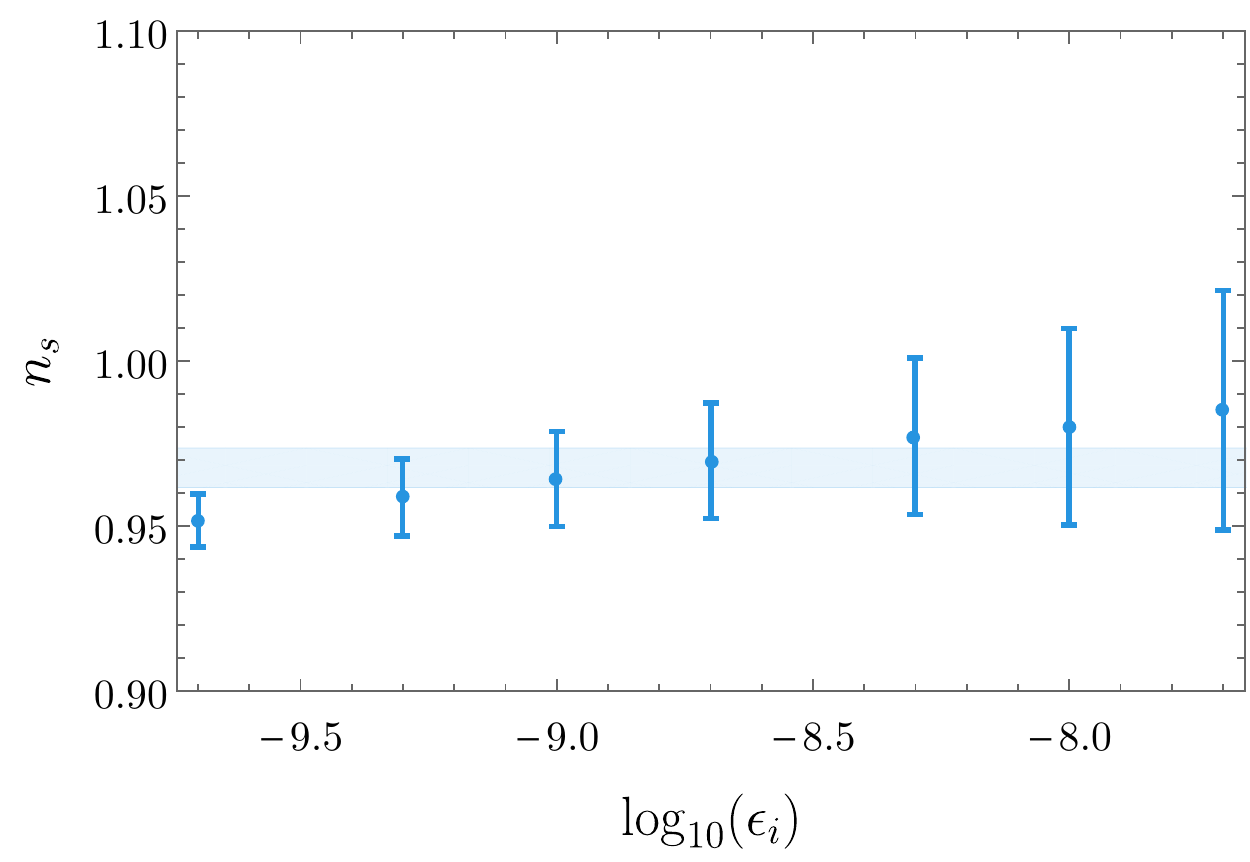}
   % \caption{e-folds before end when a second direction becoms tachyonic}
%    \label{fig:dns}
    \end{subfigure}
      %(or a blank line to force the subfigure onto a new line) \\
      \\
          \begin{subfigure}{0.48\textwidth}
         \includegraphics[width=1\textwidth]{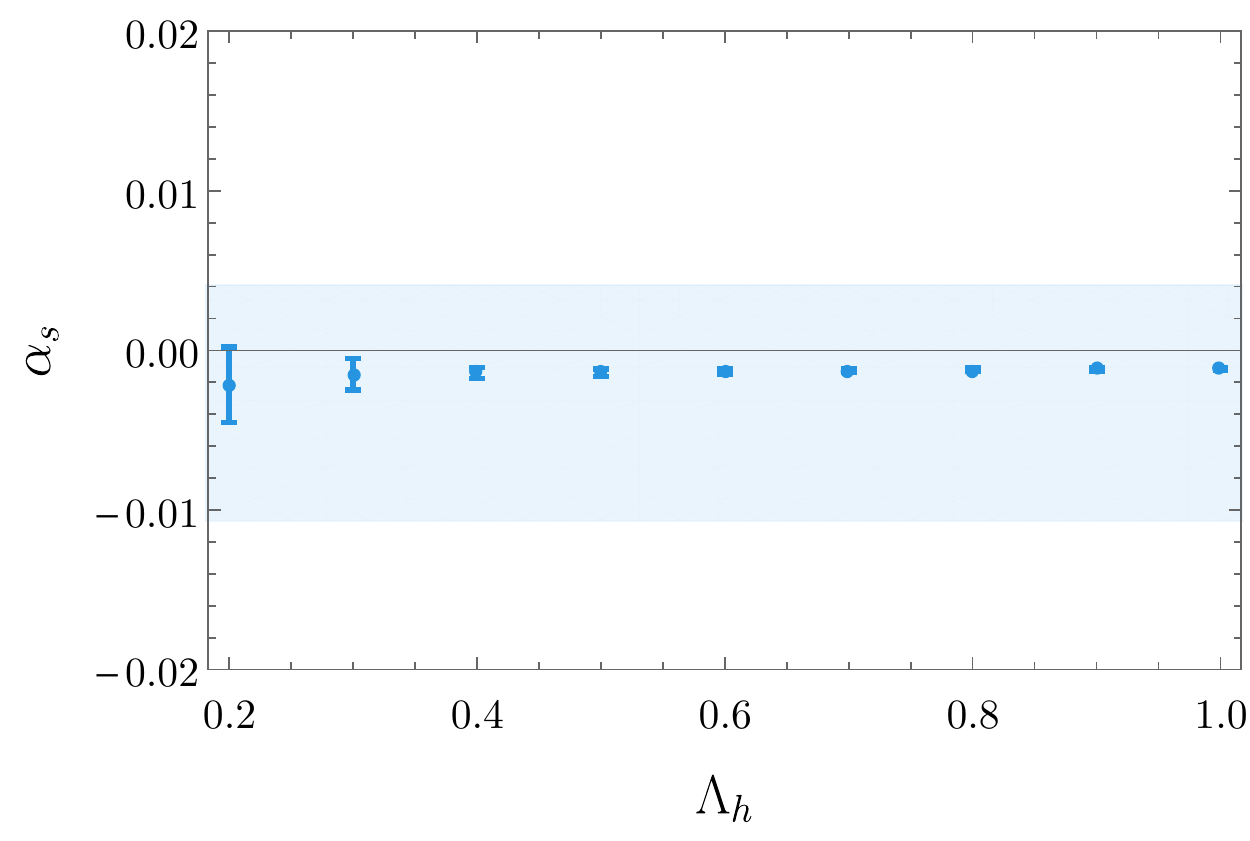}
   % \caption{Fraction of negative eigenvalues}
%    \label{fig:diso}
    \end{subfigure}
%    \caption{These graphs show the effect of varying the }\label{fig:dSHiso}
%\end{figure}
%
%\begin{figure}
%    \centering
~
    \begin{subfigure}{0.48\textwidth}
         \includegraphics[width=1\textwidth]{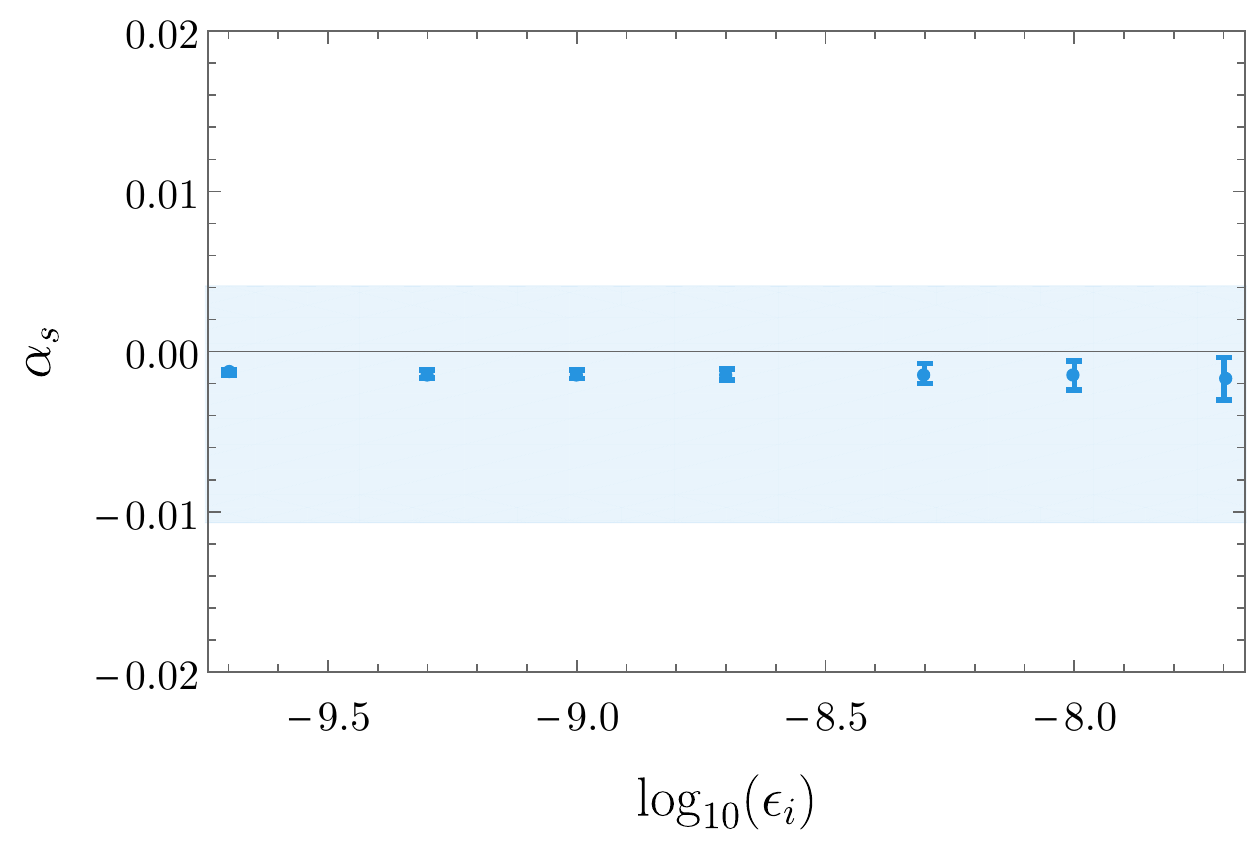}
   % \caption{e-folds before end when a second direction becoms tachyonic}
%    \label{fig:dns}
    \end{subfigure}
      %(or a blank line to force the subfigure onto a new line) \\
    \caption{The spectral index and its running as functions of the smoothness of the potential in one-parameter variations from the baseline hyperparameters. The blue regions indicate  the 68\% c.l.~from Planck (TT+lowP+lensing) \cite{1502.02114}. } \label{fig:nsUniform}
\end{figure}

   Figure \ref{fig:nsUniform} shows the dependence of $n_s$ and $\alpha_s$ of $\epsiloni$ and $\Lh$. All data points are based on at least 1000 models except those with $\Lh\leq0.3$ or $\epsilon_i\geq10^{-8}$ where fewer models gave sufficient number of e-folds, and the data points are determined from several  hundred realisations. 
 We first note that taking $\Lh$ large or $\epsiloni$ small both have the effect of `flattening' the potential, either globally or locally around $\phi=0$. Figure \ref{fig:nsUniform} indicates that such a flattening  makes the spectrum more red, and the statistical predictions for $n_s$ and $\alpha_s$ become sharper. For large $\Lh$, this reddening of the spectra make the models significantly discrepant with current observational bounds on $n_s$, making it possible to rule out this particular region of hyperparameter space with current observations (however, by increasing $\epsiloni$, the spectral indices at large $\Lh$ can be made compatible with Planck again).  The distribution of the running of the spectral index is, by comparison, remarkably robust under changes to the hyperparameters. All sampled models are compatible with current constraints on $\alpha_s$, and the prediction of  a small negative running  remains  sharp  as either $\epsiloni$ is decreased or $\Lh$ is increased.

The observed relation between a flatter potential and a redder spectral index may appear surprising at first, as  flatter potentials commonly give rise to  more scale-invariant spectra. There is however a rather simple explanation of this empirical relation.  The spectral index of the perturbations depends not only on the flatness of the potential, but also on its curvature. In a general multifield model,\footnote{Assuming  that at horizon crossing $\Sigma_{ab} = \delta_{ab}\, H^2_{\star}/2$, as we do throughout this paper. } the spectral index at the end of inflation is given by \cite{Dias:2011xy},
\be
n_s-1 = - 2 \left(
\epsilon^\star + e_a\, u_{ab}^\star\, e_b
\right) =2\frac{e_a V_{ab}^\star e_b}{V^\star}-2\left(\epsilon^\star +\frac{e_aV^\star_aV^\star_be_b}{V^\star}\right)\, ,
\ee
   where the $\star$ subscript denotes quantities evaluated at horizon crossing, the symmetric tensor $u_{ab}$ is defined as in equation \eqref{eq:uab}, and the unit vector $e_a$, which encodes the possible superhorizon evolution of the spectral index, is defined as,
\be
 e_a =  \frac{N_b \Gamma_{ba} }{|N_c \Gamma_{cd}|}\, .
 \label{eq:nstransport}
\ee
Physically, it 
relates the adiabatic perturbation at the end of inflation to field perturbations at horizon crossing. 
%
%tells us from what directions in field space the curvature perturbation at the end of inflation is primarily sourced. 
If there is no superhorizon evolution, $e^a$ is aligned with the tangent vector of the field trajectory at $N^\star$ (i.e.~$e_a \propto \partial_a V(\phi_\star)$). However, if the spectral index evolves on superhorizon scales, $e^a$ becomes misaligned, typically towards the most tachyonic directions. % that will source  $\zeta$ the most, which would be the most tachyonic directions.

In the approximate saddle-point models that we consider, the potentials need to be very flat around $\phi=0$ in order to support  sufficiently long periods of inflation, cf.~Figure \eqref{fig:Nend}. The spectral index will therefore be dominated by the term involving the Hessian.

\begin{figure}
    \centering
    \begin{subfigure}{0.48\textwidth}
         \includegraphics[width=1\textwidth]{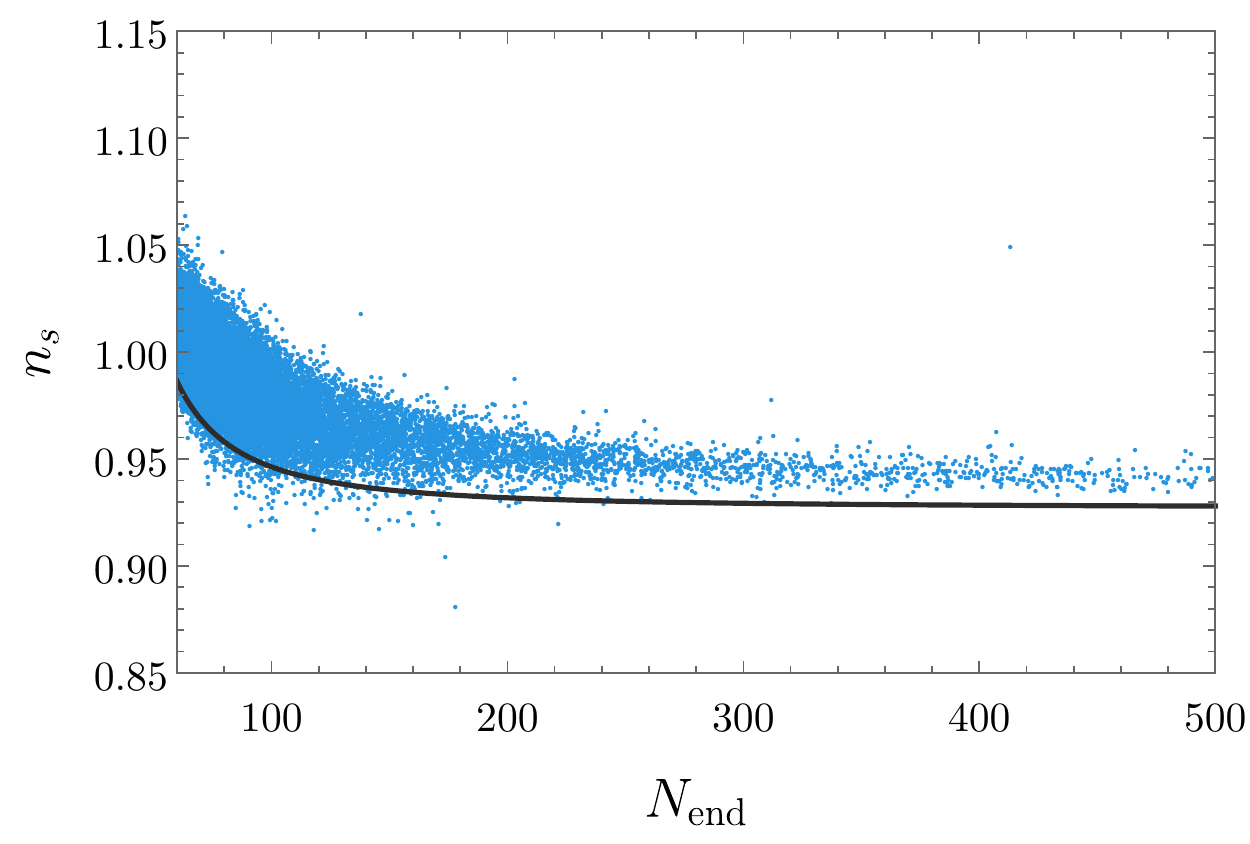}
   % \caption{e-folds before end when a second direction becoms tachyonic}
%    \label{fig:dLog10SHUniform}
    \end{subfigure}
    ~ %add desired spacing between images, e. g. ~, \quad, \qquad, \hfill etc. 
  \begin{subfigure}{0.48\textwidth}
         \includegraphics[width=1\textwidth]{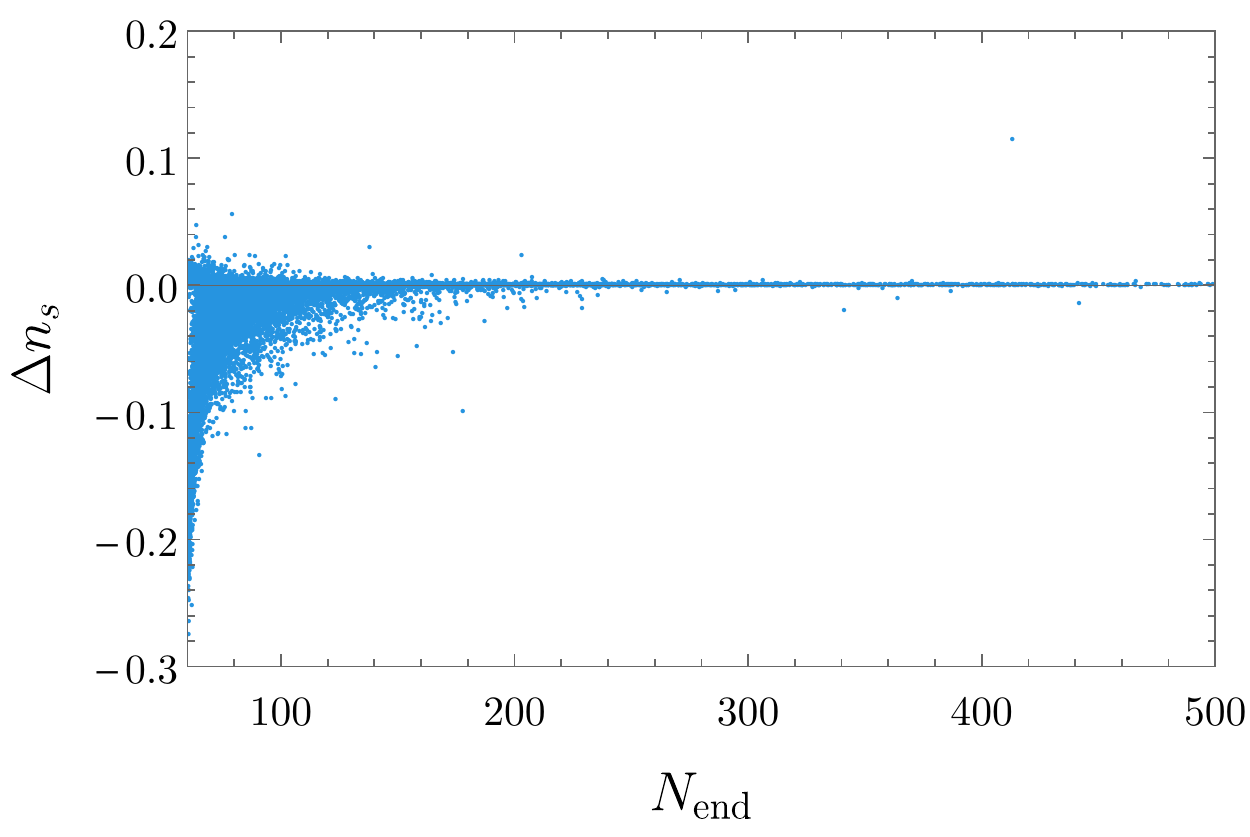}
   % \caption{e-folds before end when a second direction becoms tachyonic}
%    \label{fig:dLog10SHUniform}
    \end{subfigure}
    \caption{%The graph on the left shows t
    The distribution of $(N^\text{end},\, n_s)$ at the end of inflation (left) and the change of the spectral index on superhorizon scales, $\Delta n_s = n_s^\text{end} - n_s^\star$, (right), for about 19,000 inflation models with uniform mass spectra, again spread over values of $\Nf$ varying between 5 and 50 (the same models as in Figure \ref{fig:nsalphas}, but without the compressed spectra models). The black curve in the left graph shows the prediction of the single-field model discussed in Appendix \ref{subsection:toymodel}.}\label{fig:Nendns19000}
\end{figure}

%But after the field leaves $\phi=0$, the Hessian matrix relaxes from the non-generic initial spectrum to a (slightly off-centred) semi-circle distribution. The smallest eigenvalues of $V_{ab}/V$, which tend to provide the dominant contribution of $e_a\, u_{ab}^\star\, e_b$, then has a longer time to relax, and evolves further, to more tachyonic values. This explains the reddening of the spectra as the hyperparameter $\epsiloni$ shrinks or $\Lh$ grows.
In inflationary realisations giving not much more than 60 e-folds of inflation,  the pivot scale exits the horizon relatively early during inflation. At this point,  the masses will not have had time to spread out much, and the gradient will in general not be aligned with the most tachyonic direction. 
The curvature perturbation will then typically undergo some evolution on superhorizon scales, and the vector $e_a$ will develop components 
in both the adiabatic direction (at $N^\star$) and the more tachyonic directions.\footnote{This also explains why the spectral index becomes redder due to superhorizon evolution, cf.~Figure \ref{fig:Nendns19000}.} 
A wide distribution of the spectral index is therefore expected in this case. 

By contrast, in models supporting $\gg60$ e-folds of inflation,  the pivot scale typically exits the horizon when the gradient is dominated by the third-order coefficient in the Taylor series. In this case, the masses will have spread out more and both  the gradient vector at $N^\star$ and $e_a$ will, to good approximations, be aligned with the direction of the smallest eigenvalue of the Hessian.
%
%
%, , and . Isocurvature modes only source the curvature perturbation if the trajectory turns, i.e. if the gradient is not aligned with the smallest eigenvalue of the Hessian, so the sourcing will now be minimal and t
%The vector $e_a$ will now point in the direction of the smallest eigenvalue. 
%
This results in a redder power spectrum, since the smallest mass eigenvalue has had time to decrease further. Moreover, the variance of the spectral index is smaller in these models, since the direction of $e_a$ is much less random. These effects are visible in Figure \ref{fig:Nendns19000}.

The relation between flatter potentials and redder power spectra is now easy to understand. With all else the same,  a flatter potential generates more e-folds of inflation so that the horizon crossing of the pivot scale, 55 e-folds before the end of inflation,  happens correspondingly later (in e-fold time) after the field has left $\phi=0$. As per the discussion above, we then expect to see redder power spectra with smaller variances, which is precisely what we see when $\epsiloni$ is decreased or $\Lh$ is increased.

% \begin{figure}
%    \centering
%    \includegraphics[width=0.5\textwidth]{Nendns19000.pdf}
%    \caption{The distribution of $N^\text{end}$ and $n_s$ at the end of inflation (blue) and horizon exit (green) for about 19,000 inflation models with uniform mass spectra, again spread over values of $\Nf$ varying between 5 and 50.}\label{fig:Nendns19000}
%\end{figure}

%Figure \ref{fig:nsUniform} also shows another  important property of multifield inflation in GRF potentials: as $\epsiloni$ grows or $\Lh$ shrinks, the variance of the predictions increases. This effect is dramatic for the spectral index, but also discernible for its running. %We note that each data point in Figure  \ref{fig:nsUniform} correspond to at least {\color{blue} XXX} inflationary models
%{\color{blue} This is related to the number of e-folds decreasing, but how and why? This is important to understand.}

\begin{figure}
    \centering
    \begin{subfigure}{0.48\textwidth}
         \includegraphics[width=1\textwidth]{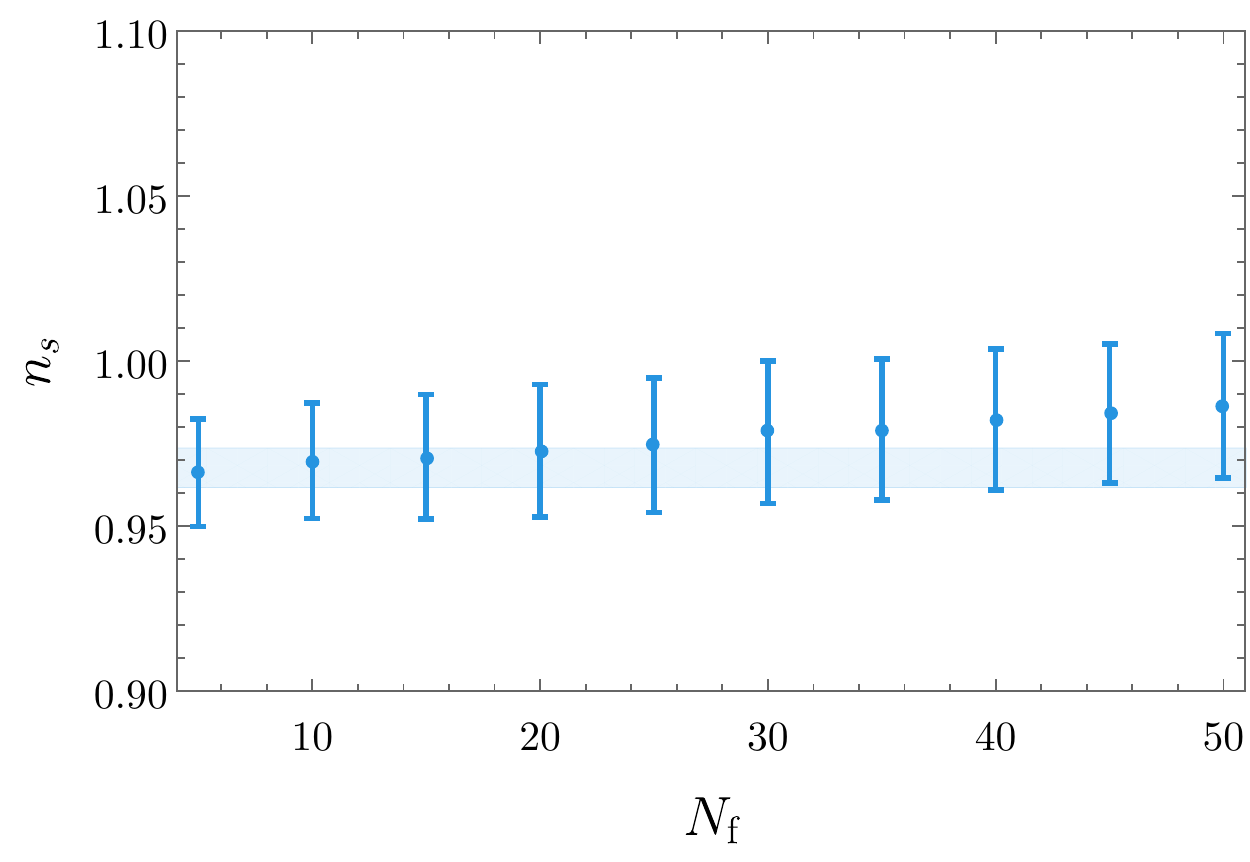}
   % \caption{e-folds before end when a second direction becoms tachyonic}
%    \label{fig:dLog10SHUniform}
    \end{subfigure}
    ~ %add desired spacing between images, e. g. ~, \quad, \qquad, \hfill etc. 
  \begin{subfigure}{0.48\textwidth}
         \includegraphics[width=1\textwidth]{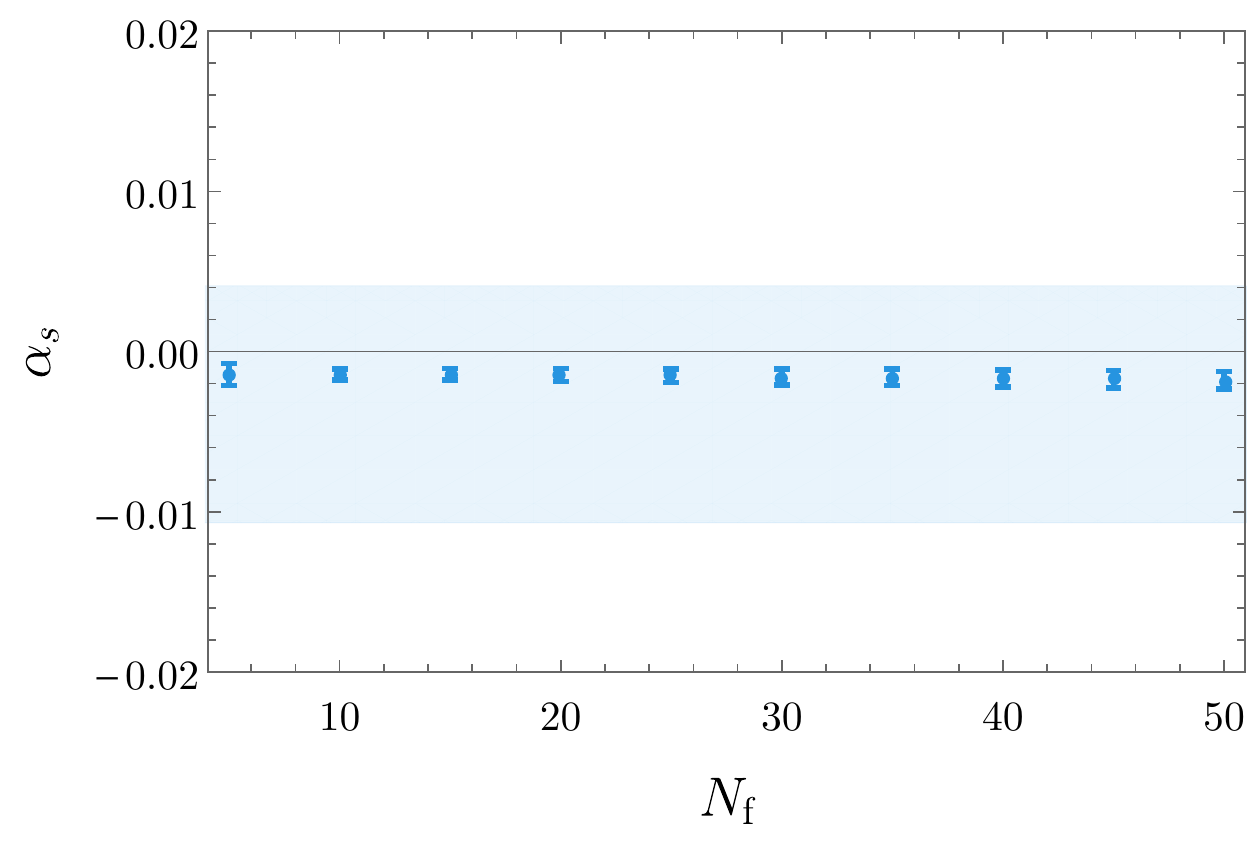}
   % \caption{e-folds before end when a second direction becoms tachyonic}
%    \label{fig:dLog10SHUniform}
    \end{subfigure}
    \caption{The spectral index and its running as a function of $\Nf$
    in one-parameter variations from the baseline  together with the 68\% c.l.~from Planck \cite{1502.02114}. }\label{fig:nsalphaNf}
\end{figure}

 We now turn to the effects of the number of fields, $\Nf$, on the power spectrum. Figure \ref{fig:nsalphaNf} shows the how the mean values and standard deviations of $n_s$ and  $\alpha_s$ are weakly dependent on $\Nf$. First, we note that  as $\Nf$ is increased, the spectra become less red %Interestingly however, the upward shift of $n_s$ is in this case not accompanied by a large increase in the variance of the distribution, in contrast to Figure \ref{fig:nsUniform} when $\epsiloni$ is increased or $\Lh$ decreased.{\color{blue} Explained by two competing effects? Small $N$ push for increased variance and large $\Nf$ increased eigenvalue repulsion, which in itself should lead to smaller error bars.
and the variance of the spectral index also increases, albeit slowly. This may again be explained by the correlation between $n_s$ and the total number of e-folds of inflation: as the number of fields is increased, the models tend to give fewer e-folds of inflation, cf.~Figure \ref{fig:Nend}, which leads to $e_a$ developing non-vanishing components along multiple directions in field space.

 %This is however somewhat dependent on the initial mass spectrum. E.g, with a very compressed initial mass spectra there is a much higher sensitivity to $\Nf$ in the spectral index, but the running still remains small and negative at high $\Nf$. 

 In sum, we have found that the distribution for the spectral index in our GRF models tends to favour red, approximately scale-invariant spectra. Some regions of the parameter space lead to sharp predictions of excessively red spectra, and can be ruled out already with current observations. However, large regions of hyperparameter space are compatible with current constraints from the Planck experiment. %The spectral index distribution can in principle always be made Planck-compatible by adjusting one of the hyperparameters. 
 More importantly, we have found that these models predict, sharply and robustly, a small negative running of the spectral index. The Planck experiment has  constrained the running of the spectral index to $\alpha_s = -0.0033\pm 0.0074$, but future experiments may reach a sensitivity of $\sigma(\alpha_s)=10^{-3}$ \cite{1612.05138}. A future observation of $\alpha_s \gtrsim 0$ or $\alpha_s \lesssim-0.004$ would rule out all Planck-compatible  models that we have constructed.

 \subsection{Substantial superhorizon evolution, but also decaying isocurvature}
 \label{sec:iso}
 
 We have seen in section \ref{sec:Planck} that the predictions from the random GRF models are remarkably simple, despite the underlying potentials being highly non-trivial functions of many fields. Indeed, the prediction of an approximate scale-invariant power spectrum with a small running of the spectral index agree with two of the `generic' predictions of \emph{single-field} models of slow-roll inflation.  In this section, we investigate to what extend multifield dynamics is important for the predictions of the manyfield GRF models.

%It is tempting to read Figures \ref{fig:fullps} and \ref{fig:ps} as indications a simple underlying inflationary dynamics, possibly admitting a single-field description. 
%However, our GRF models involve many dynamically important fields that interact through large numbers of non-suppressed interactions terms, and multifield effects are important. 

We begin by considering the superhorizon evolution of the pivot-scale modes exiting the horizon 55 e-folds before the end of inflation. 
We first recall that
to linear order in the field perturbations (and upon suppressing the  $k$ dependence), the modes at the end of inflation are related to the modes at horizon exit by the transfer equation \cite{Amendola:2001ni, Wands:2002bn, Byrnes:2006fr},
\be
\left(
\begin{array}{c}
\zeta \\
{\cal S}^i
\end{array}
\right)_{\rm end}
=
\left(
\begin{array}{c c}
1 & T_{\zeta {\cal S}^j} \\
0 & T_{{\cal S}^i {\cal S}^j}
\end{array}
\right)
\left(
\begin{array}{c}
\zeta \\
{\cal S}^j
\end{array}
\right)_{\star} \, .
\label{eq:transfer}
\ee
For $\Sigma^*_{ab} \propto \delta_{ab}$ (as we assume in this paper, cf.~equation \eqref{eq:Sigma}), the superhorizon evolution of the curvature perturbation is given by,
\be
\frac{P_{\zeta}(N_{\rm end}, k)}{ P_{\zeta}(N_{\star}, k)} = 1 + \sum_{i=1}^{\Nf-1}  T_{\zeta {\cal S}^i}^2 \, ,
\label{eq:superhor}
\ee
so that, under these assumptions, superhorizon evolution can only lead to a net increase in the power of the curvature perturbation. 
If $P_{\zeta}(N_{\rm end}, k)/ P_{\zeta}(N_{\star}, k)-1  \lesssim 10^{-3}$ for a range of $k$ modes, the observational predictions of the model can be regarded as safely independent of multifield effects, and the horizon crossing power spectrum determines the observational predictions for e.g.~$n_s$ and $\alpha_s$. This rarely happens in manyfield models of inflation in GRF potentials.

 \begin{figure}
  \begin{subfigure}{0.48\textwidth}
  
    \includegraphics[width=1\textwidth]{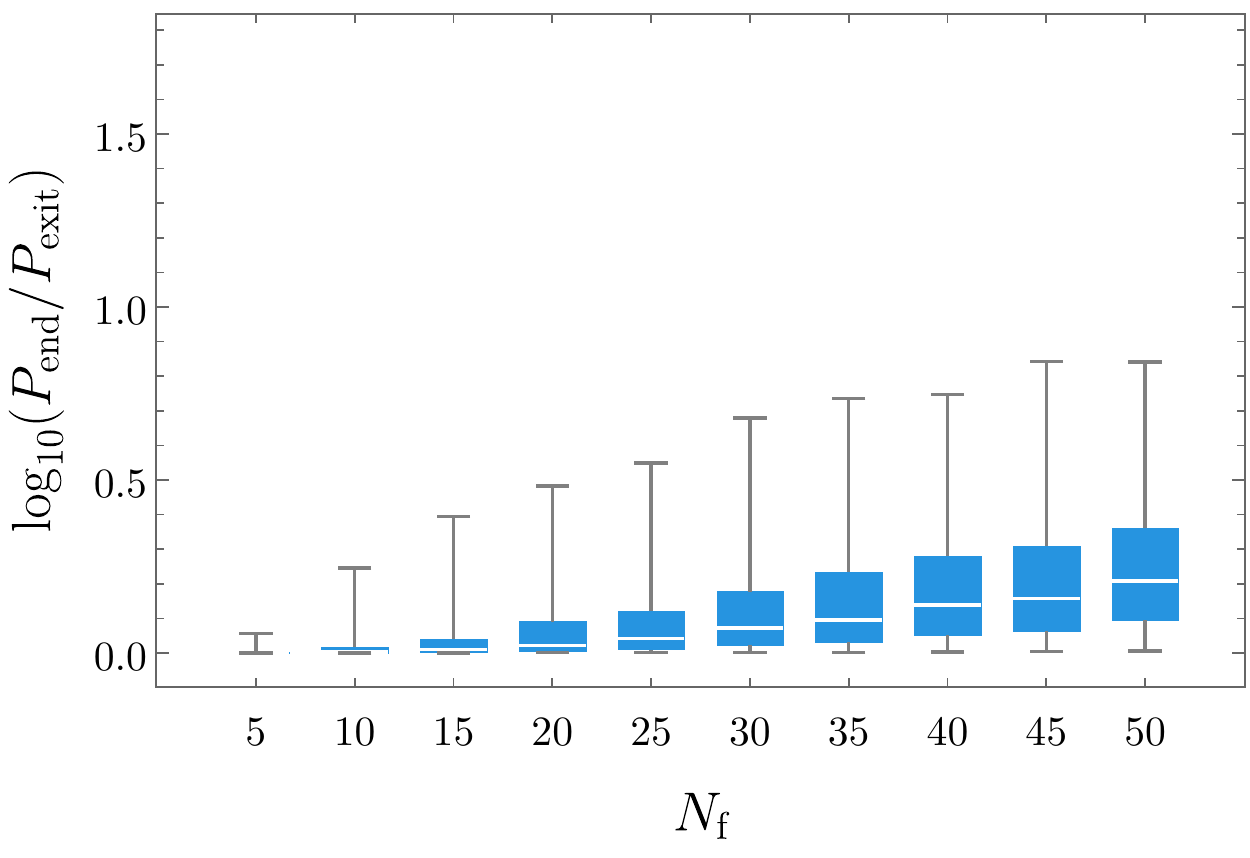}
   % \caption{e-folds before end when a second direction becoms tachyonic}
%    \label{fig:dns}
\end{subfigure}
~
    \begin{subfigure}{0.48\textwidth}
         \includegraphics[width=1\textwidth]{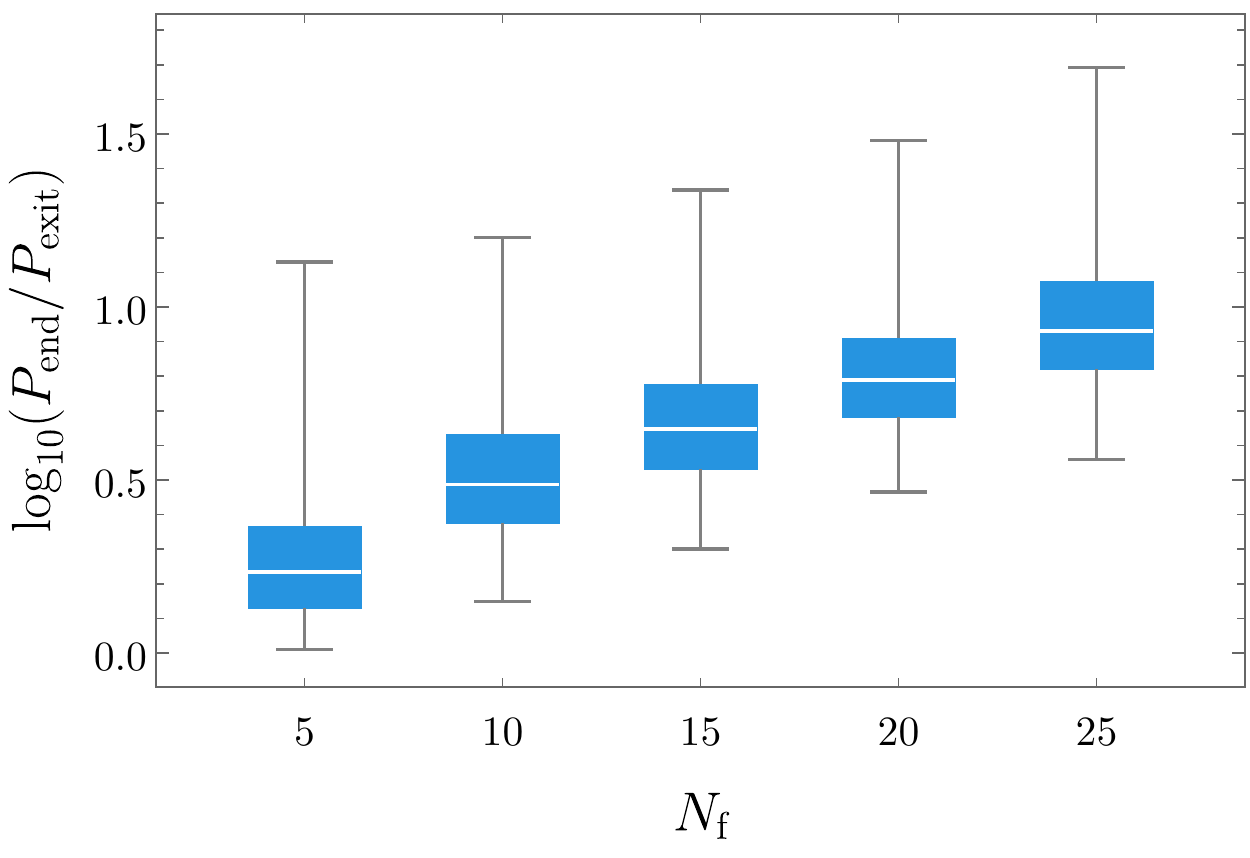}
   % \caption{Fraction of negative eigenvalues}
%    \label{fig:diso}
    \end{subfigure}
\caption{
Superhorizon evolution of the mode exiting the horizon 55 e-folds before the end of inflation for the flat initial spectrum (left) and the compressed spectrum (right), cf.~equation \eqref{eq:spectra}.
Boxes indicate first and third quartile, together with the median; `whiskers' indicate 1st and 99th percentile. 
The left graph shows a one-parameter variation from the baseline; the right shows models with $\Lh=0.4$, $\epsiloni=10^{-10}$ and $\etai=-10^{-4}$.
}
\label{fig:SH}
      \end{figure}

 The box plots in Figure \ref{fig:SH} show the effects of varying $\Nf$ on the distributions of $\log_{10}(P_{\zeta}(N_{\rm end}, k_\star)/ P_{\zeta}(N_{\star}, k_\star))$ for both the flat  (left) and compressed (right) spectra.
 Each box is generated from  over 1000 inflationary models, except for $\Nf=20$ and $25$ for the compressed spectrum, which were generated from 600 and 200 models, respectively. Unsurprisingly, models with more fields and more compressed initial spectra exhibit larger superhorizon evolution. However, even for the flat spectrum, models with more than 5 fields tend to evolve substantially on superhorizon scales, so that the predictions at horizon crossing do not automatically give the predictions for observables at the end of inflation. Evidently, multi-field effects are important in manyfield inflation in GRF potentials.
 
% {\color{blue} Possibly paragraph on how $n_s$ evolves on superhorizon scales, if there is evidence for such evolution for both flat and compressed spectra.}\textcolor{purple}{ This has now been answered earlier in the text}

In multifield models of inflation, the curvature perturbation may evolve well past the end of inflation, through the reheating phase. 
In many models of multifield inflation in the literature, this problem is dealt with by ensuring that the fields enter an approximately single-field `adiabatic limit' in which all but a single mode are very massive (i.e.~$m^2> H^2$) and the isocurvature modes decay exponentially. Once the isocurvature perturbations have decayed, the curvature perturbation ceases to evolve and the predictions 
are expected to 
become 
insensitive to
the details of the reheating phase. As illustrated by Figure \ref{fig:100fieldEVs}, the spectrum of the Hessian matrix of the GRF models we consider typically contains multiple tachyonic eigenvalues at the end of inflation so that, clearly, no standard adiabatic limit is reached. However, we will now see that isocurvature still becomes exponentially suppressed during inflation. 

\begin{figure}
   \begin{subfigure}{0.48\textwidth}
         \includegraphics[width=1\textwidth]{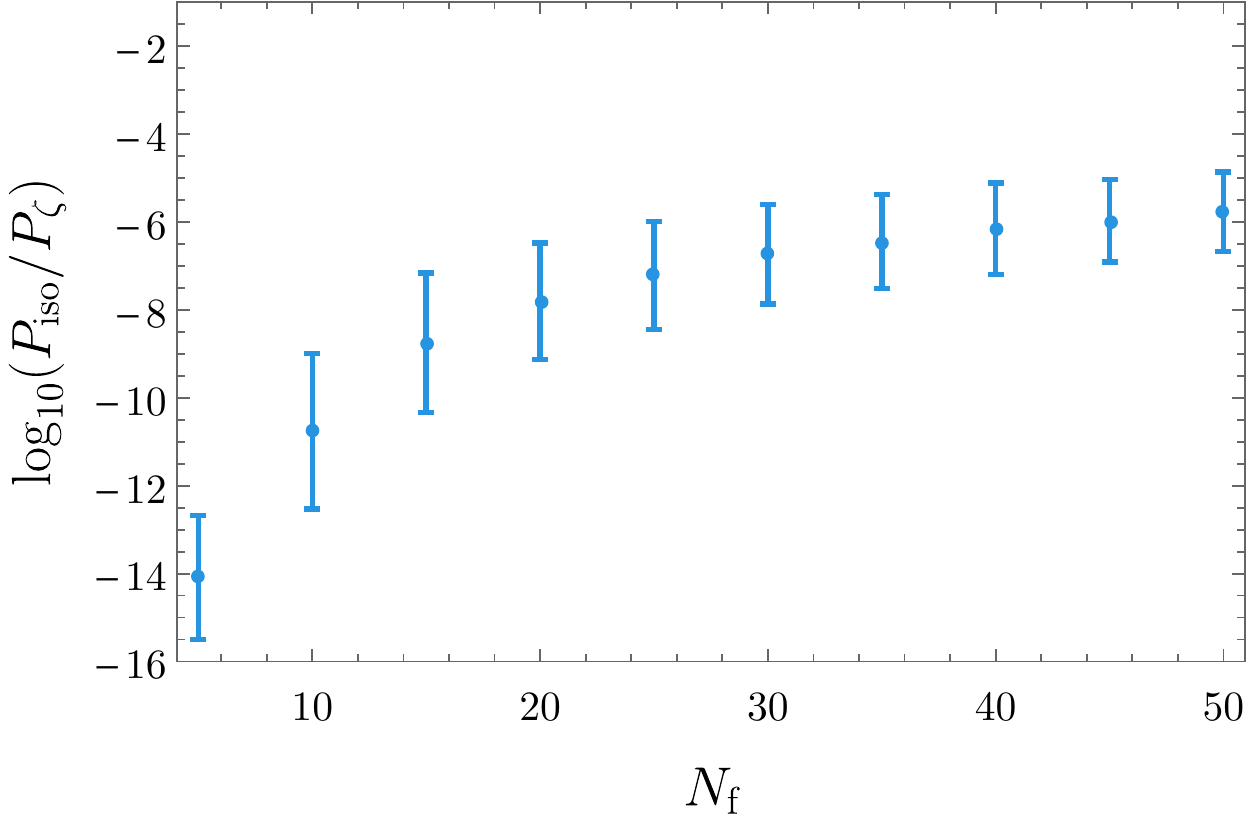}
   % \caption{e-folds before end when a second direction becoms tachyonic}
%    \label{fig:dns}
    \end{subfigure}
      ~
  \begin{subfigure}{0.48\textwidth}
         \includegraphics[width=1\textwidth]{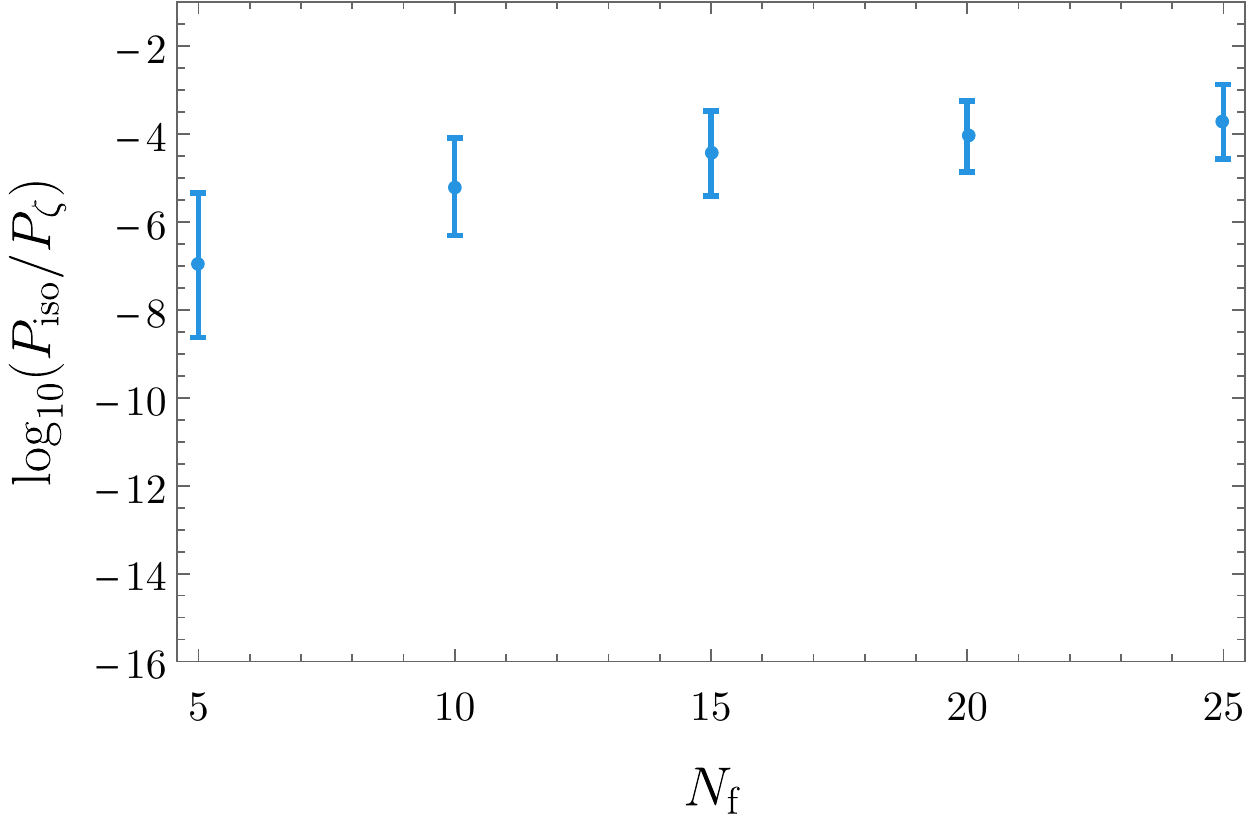}
   % \caption{Fraction of negative eigenvalues}
    \end{subfigure}
    \caption{Isocurvature-to-curvature ratio at the end of inflation for flat (left) and compressed (right) initial spectra. Other hyperparameters  as in Figure \ref{fig:SH}.}
    \label{fig:iso}
\end{figure}

At horizon crossing, $\Sigma_{ab} \propto \delta_{ab}$ and $P_{\rm iso}/P_{\zeta} = (\Nf-1)$, cf.~equation \eqref{eq:Piso}. Figure \ref{fig:iso} shows the mean values and standard deviations of the ratio $P_{\rm iso}/P_{\zeta}$ evaluated at the end of inflation in our ensembles of models. Strikingly, the power in the isocurvature mode evolves during inflation from dominating over the curvature perturbation to becoming exponentially suppressed. For the flat initial spectrum and $\Nf=5$, the ratio $P_{\rm iso}/P_{\zeta}$ falls below the numerical accuracy of our simulations. For larger $\Nf$, this ratio typically remains exponentially suppressed. Models with the highly compressed initial spectrum feature  larger levels of isocurvature at the end of inflation, but even in this extreme case, the superhorizon evolution suppresses the isocurvature perturbations by several orders of magnitude.  

The suppression of the isocurvature despite multiple tachyonic directions  can be understood as a consequence of the inflationary slow-roll dynamics, as discussed in  \cite{DBM3}. To see this, we may re-express the components of the transfer matrix \eqref{eq:transfer} in terms of the transport coefficients $\Gamma_{ab}(N, N_{\star})$, using the decomposition of the fluctuations into instantaneous adiabatic and entropic fluctuations, cf.~equation \eqref{eq:decomp}:
\bea
1 &=& \left(\frac{\epsilon_\star}{\epsilon_N} \right)^{1/2}\, n_a(N)\, \Gamma_{ab}\, n_b(N_\star) \, , 
\label{eq:transfer1}
\\
T_{\zeta {\cal S}^i} &=& \left(\frac{\epsilon_\star}{\epsilon_N} \right)^{1/2}\, n_a(N)\, \Gamma_{ab}\, v^i_b(N_\star) \, , 
\label{eq:transfer2}
\\
T_{{\cal S}^i {\cal S}^j} &=& \left(\frac{\epsilon_\star}{\epsilon_N} \right)^{1/2}\, v^i_a(N)\, \Gamma_{ab}\, v^j_b(N_\star) \, .
\label{eq:transfer3}
\eea
Here $\epsilon_N = \epsilon(N)$ and equation \eqref{eq:transfer1} follows from the conservation of $\zeta$ in the absence of entropic perturbations. 
The appearance of multiple negative eigenvalues of the Hessian matrix leads to multiple growing field perturbations (and multiple eigenvalues of $\Gamma^{\rm T} \Gamma$ that are greater than 1). % in spatially flat gauge. To linear order, the adiabatic perturbation corresponds to fluctuations along the background trajectory, cf.~equations \eqref{eq:zeta} and \eqref{eq:Na}. 
In slow-roll inflation, the field velocity tends to align with the smallest eigenvalue of the Hessian matrix, cf.~equation \eqref{eq:acc}. This makes the adiabatic field perturbation grow faster than the each of the less tachyonic entropic perturbations. However, from equation \eqref{eq:transfer1} we see that the growth of the adiabatic field perturbation (in the absence of additional sourcing from entropic modes) is directly related to the growth of $\epsilon(N)$. 

%
%the background trajectory follows the negative of the  gradient direction, and in these models the gradient direction tends to dynamically align with the eigenvector of the most tachyonic eigenvalue. Hence, the inflationary background evolution tends to make the adiabatic direction correspond to the fastest growing field perturbation. 
%
%
%
%
%
%
%The %uniform density gauge   
%curvature perturbation $\zeta$ of equation \eqref{eq:zeta} and the isocurvature perturbations ${\cal S}^i$ of equation \eqref{eq:iso} are both related to the flat gauge field perturbations by a gauge transformation, which involves  a factor of $1/\sqrt{2 \epsilonV}$. Clearly, as $\epsilonV$ grows during inflation, this factor suppresses the magnitude of $\zeta$ and ${\cal S}^i$. Indeed, it follows from equation \eqref{eq:zetaprime} that if the gradient direction is perfectly aligned with the eigenvector of the smallest eigenvalue of the Hessian matrix, the growth of the flat gauge field perturbation is precisely cancelled by the competing suppression from the gauge transformation, and $\zeta'=0$. 
For entropic modes that grow slower than the adiabatic perturbation, the decaying prefactor $\sqrt{\epsilon_\star/\epsilon_N}$
cause a net suppression of isocurvature during inflation.  This explains why the isocurvature ${\cal S}$ can decay during inflation, despite the presence of multiple tachyonic fields.  

This discussion also makes it clear that the `straying' behaviour of the smallest eigenvalue of the Hessian, discussed in section \ref{sec:stray},  leads to a further suppression of isocurvature modes during inflation. Furthermore, the associated `straightening' of the field trajectory leads to fewer opportunities for the isocurvature to source the curvature perturbation through turns in field space, cf.~equation \eqref{eq:zetaprime}. 

 In sum, in this subsection we have seen that multifield effects are typically important in manyfield inflation, but  that entropic perturbations tend to decay. While no single-field `adiabatic limit' is reached during inflation in theses models, the large suppression of isocurvature 
 %suggests that 
 %observables evaluated at the end of inflation
% may 
%remain 
%provide a good approximation of the final values of these observables in many classes of reheating models. 
%observables are somewhat shielded 
may shield observables from subsequent superhorizon evolution during the post-inflationary reheating phase. It would be interesting to apply our construction of GRF potentials to investigate the %post-inflationary
 evolution of the adiabatic and entropic perturbations in reheating models with many interacting fields in more detail.

%{\color{blue} Concluding paragraph for entire section.} 

\section{Result II: At large $\Nf$, GRF and RMT  models largely agree}
\label{sec:result2}
A key motivation behind this work is to use mathematically simple constructions of manyfield systems to search for mechanisms that may drive observables to simple and robust predictions. Identifying such mechanisms could prove very helpful in analysing  more complicated manyfield models motivated by fundamental theory. However, even for a given class of mathematically simple  models, it can be hard to separate its particularities   from the  properties that may be more broadly applicable. It is therefore important to test the predictions of any class of models by comparing to the predictions of independent constructions. To make such comparisons useful, the different classes of models should  share some rough similarities, but be fundamentally different in their details. For our purposes, we are interested in models of small-field, slow-roll, saddle-point inflation with many interacting fields.  Fortunately, our construction of GRF models in this paper and the construction of random DBM potentials of \cite{DBM1, DBM3} now provide two such classes of theories, thus allowing the first critical tests of possible `universal' predictions of random manyfield models of inflation.

%While the GRF and DBM approaches to random multifield inflation are rather different, there are a number of very similar features in two models. In this section, we discuss some of these similarities, in particular the simplicity at large $\Nf$, the importance of multifield effects and how eigenvalue repulsion drives predictions. Similarities that emerge at large $\Nf$ are very intersting, as they could hint at some large $\Nf$ universality which is independent of the detailed form of the potentials.

In this section, we first briefly review  the differences and similarities  between the DBM models and our GRF models in section \ref{sec:differences}. 
Our discussion highlights the differences in the evolution of the Hessian matrix, but  also clarifies the context of 
 %of recent claims of discrepancies based on 
 recent discussions on the equilibrium spectra of  single-field and effectively single-field GRF models \cite{Masoumi:2016eag, Masoumi:2017xbe}. 
%
%
%, and explain 
%why these proposed discrepancies are not realised in the multifield models of inflation  that we study. 
In section \ref{sec:tests},
we then provide a first detailed comparison between the observational predictions of the DBM and GRF models. 
%In particular, we test which of the `seven lessons' from manyfield in DBM potentials also hold for manyfield inflation in GRF potentials.
%Strikingly, we will see that 
 %these constructions, despite being very different, agree well on the predictions for key properties and observables.  
 A brief review of the DBM construction can be found in Appendix \ref{app:DBM}. We refer the reader to references \cite{DBM1, DBM3} for a more detailed discussion on the properties of DBM potentials.
 
\subsection{Comparison of DBM and GRF potentials}
\label{sec:differences}
The DBM construction of \cite{DBM1} led to the first  explicit studies of models of inflation with many interacting fields in random potentials.
%
%
%As we have shown in this paper,  Gaussian random fields can be used  to study of inflation in models with many interacting, dynamically important fields, and thus 
%provide an alternative prescription 
%through which the predictions of the  DBM models can be compared. 
%In this subsection, we briefly discuss the similarities and differences between the DBM and GRF models. %Our discussion may help to clarify the context and range of validity  of some  rather drastic claims on the relation of RMT techniques and GRF models  in the recent literature \cite{Masoumi}. 
%, both allowing the study of inflation in models with multiple interacting, dynamically important fields.  
%
As we mentioned in section \ref{sec:intro} however, this prescription differs from GRF models in that the cubic and higher-order terms of the potential are regulator-dependent. In this section, we will discuss how some properties the Hessian matrices differ between the GRF and DBM potentials. 
For the purpose of  clarity, we first compare the \emph{`equilibrium spectra'} of the models, i.e.~the eigenvalue distribution of the Hessian matrix at some randomly chosen point in field space, and we then turn to the \emph{evolution} of the Hessian, e.g.~from a fine-tuned initial configuration to the equilibrium.

%An obvious difference  between the DBM and GRF potentials  concerns cubic and higher-order derivatives of the potential. As reviewed in Appendix \ref{app:DBM}, the DBM potentials are not defined locally to more than quadratic order. Moreover, due to the intrinsically Brownian nature of DBM, the evolution of the eigenvalues of the Hessian is (almost surely) everywhere continuous, but also (almost surely) nowhere differentiable. Thus, cubic and higher-order derivatives are not well-defined in the continuum limit, and regulator-dependent in any discretisation. By contrast, the GRF potentials are well-defined to any order. 
%As we discussed in section \ref{sec:intro}, this curtails studies of non-Gaussianities in DBM models.  
%
%
%. For example, 
%
%as discussed in more detail in \cite{DBM3}. 
%
%: those involving the potential, locally only up to quadratic order may be addressed (e.g.~to compute the power spectrum of the perturbations, as in \cite{DBM2, DBM3}), but those requiring higher derivatives may not (including the computation of non-Gaussianities).  
%prescription may only be useful to address questions that involve the potential locally in field space
%As discussed in section \ref{sec:intro}, this has enabled us to 
%
%the third order terms are not we
%The properties of the Hessian matrix are of particular importance  for manyfield models of inflation. 

\subsubsection*{The equilibrium spectra of GRF and DBM models}
In equation \eqref{eq:splithessian}, we saw that the Hessian matrix in the GRF models consists of a GOE matrix and shift, proportional to the unit matrix times the negative of the value of the potential. %This will be the case for any Gaussian covariance function. 
We also noted that for $\Nf\gg1$ and  typical values of the potential, this shift had a very small effect on the spectrum of the GRF. This way, the spectrum of the simplest GRF models is very similar to that of the simplest  DBM model, which omits the shift entirely. 
In the light of this, it may appear surprising that a recent study claimed that the equilibrium spectrum constitutes a fundamental difference between the GRF and DBM models \cite{Masoumi:2017xbe}. We here provide the context for these claims. 

First, the variance of the GOE matrix in the simplest GRF model is fixed by the choice of covariance function. For equation \eqref{eq:Gcovar}, this leads to a width of the Wigner semi-circle spectrum of $4\sqrt{\Nf}$, cf.~equation \eqref{eq:deanbray}. In the DBM model, the variance is a free parameter which is typically chosen so that the spectrum has an $\Nf$-independent width. This choice makes it convenient to compare systems with different number of fields within a fixed mass-range, but other choices are possible, and clearly,
the width of the equilibrium eigenvalue spectrum can hardly be described as  a fundamental difference between the DBM and GRF constructions. 

Second, one can try to make the shift of the spectrum more important even during inflation. To do so, one may attempt to inflate near the bottom of the potential, where the semi-circle spectrum is significantly `up-shifted' from the centred Wigner semi-circle law. However, for the mean zero GRF models, any upward shift of spectrum  only occurs for negative values of the potential, making inflation impossible. To construct models in which the shift is important, one may  add to the mean-zero GRF a large, field-independent cosmological constant of size ${\cal O}(2 \sqrt{\Nf} \Lv^4)$.\footnote{ The addition of  a large field-independent cosmological constant may appear ad-hoc, and to our knowledge,  lacks a clear physical motivation. For example, sources of energy density in string compactifications tend to be moduli-dependent in the Einstein frame.}
%$V_{\rm tot} = V_{\rm GRF} + $, where we have denoted the mean zero GRF by $V_{\rm GRF}$. 
The uplifted potential will then have a typical, $1 \sigma$  range of $(2\sqrt{\Nf} \pm1) \Lv^4$, as opposed to $\pm \Lv^4$ for the mean-zero GRF.  By construction, the equilibrium spectrum for small values of the potential is now a Wigner semi-circle with the left edge shifted to zero, and no tachyonic eigenvalues. We briefly discuss manyfield inflation in this class of potentials in Appendix \ref{sec:casestudies}.

The substantially shifted spectrum of the Hessian of the modified GRF potential is (by construction) discrepant with the centred Wigner semi-circle, and thereby the equilibrium spectrum of the standard DBM model. This was  key to the argument  of reference \cite{Masoumi:2017xbe}, which proposed this discrepancy as a fundamental difference between DBM and GRF models. However, due to the simplicity of equation \eqref{eq:splithessian}, it is straightforward to modify the DBM model to capture the spectrum of any such modified GRF model.\footnote{For example,  in direct analogy with equation \eqref{eq:splithessian}, one may take  $v_{ab}^{\rm tot} = v_{ab}^{\rm DBM} + \delta_{ab} f_{\rm shift}(v_0)$, where  only $v_{ab}^{\rm DBM}$ undergoes Dyson Brownian Motion, and the new  term  encodes the desired shift of the spectrum.} Thus, it appears challenging to use simplistic arguments based on the equilibrium spectra of the DBM and GRF models to identify  fundamental differences between these constructions.

\subsubsection*{The evolution of the Hessian matrix}
%
%We note that the DBM and GRF potentials can also differ in their predictions for the typical spectra of the Hessian matrix \cite{Bray:2007tf, Bachlechner:2014rqa}. However, as we discuss in Appendix \ref{app:DBM}, it is hard to construe this difference to be particularly important.
%
 %
%
 %\end{enumerate}
%We close this section by emphasising  that  GRF and DBM models 
%differ fundamentally in t
The \emph{evolution} of the Hessian matrix as the field traverses some path in field space constitutes a fundamental difference between the DBM and GRF constructions, even if the equilibrium spectra coincide. This difference is evident in the relaxation of the eigenvalues of the Hessian from an atypical initial configuration to the equilibrium configuration, as can bee seen by comparing  the 100-field DBM model of Figure 4 of reference \cite{DBM3} to our 100-field GRF model of Figure \ref{fig:100fieldEVs}.
The spectra of the GRF models  relax in a much more linear, regular fashion. 
Moreover,  in section \ref{sec:stray} we showed that the statistical properties of the cubic terms in GRF potentials lead to `straying' smallest eigenvalues in slow-roll inflation. This phenomenon 
has no counterpart in DBM models.
%can to our knowledge not be reproduced by DBM models, even in principle. 

%However, they do comprise fundamental differences between the DBM and GRF models, and indicate that these constructions are independent.

In sum, while both DBM and GRF models can be used to describe manyfield inflation, the two constructions are independent and differ substantially in several ways. 
Thus, by comparing the predictions of these two classes of models, we may  search for  robust and model-independent signatures of many-field dynamics during inflation.

\subsection{Comparison of DBM and GRF predictions}
\label{sec:tests}
In this section, we assess the robustness of the predictions from manyfield models of inflation by comparing our results derived in this paper to those derived from DBM models in  \cite{DBM3}. The results of reference \cite{DBM3} were organised into `seven lessons'. We here test each of them. 

\begin{enumerate}
\item {\bf  \color{mygreen} Manyfield inflation is not single-field inflation.} One immediate aspect of  multifield models of inflation is that they typically contain several fields with masses not much larger than the Hubble parameter. Such `light' fields cannot be integrated out, and commonly contribute to multifield effects that impact observables. In this sense, manyfield models of inflation are clearly not identical to single-field models.

In \cite{DBM1} however, it was shown that some aspects of the DBM models at large $\Nf$ (such as the distribution of e-folds), could be modelled by a single-field model. Reference \cite{Freivogel} elaborated on this single-field model to estimate the spectral index of the large-$\Nf$ DBM models, however this single-field estimate was discrepant with the actual distribution computed from the DBM multifield models \cite{DBM2, DBM3}. Thus, single-field models have had a limited success in describing the properties of manyfield DBM models. Moreover, intrinsically multifield effects such as superhorizon evolution of the curvature perturbation are common  in DBM models, which indicates that  manyfield inflation   differ from single-field inflation.

We have seen that in manyfield GRF models, multifield effects are also common: the field explores multiple directions in field space, isocurvature modes can be important, and the curvature perturbation typically evolves on superhorizon scales. These effects cannot be captured by a single-field model so that, evidently, manyfield inflation is different from single-field inflation.\footnote{
Recently, reference \cite{Masoumi:2017xbe} studied a GRF-motivated `multifield' system with one light and many heavy fields, finding that this model gives rise to single-field dynamics. As we are interested in inflation with many dynamically important fields, our assumptions for the initial configuration differ from that of \cite{Masoumi:2017xbe}, and the question that we explore here -- whether manyfield systems can be effectively described as single-field models -- also differ significantly from whether an effectively single-field system is well-described by a single-field system.
} It can still be interesting to explore how well a simple single-field model can capture the results of the GRF manyfield models. In Appendix \ref{sec:singlefieldmodel}, we construct such a single-field model and show  that its predictions qualitatively (but not quantitatively) agree with the more complicated GRF models.  

%derive its predictions. We find that while some aspects of the background dynamics can be modelled, the observational predictions from this single-field model  are incompatible with the manyfield GRF models.  

In conclusion, for both GRF and DBM models, manyfield inflation is different from single-field inflation.

\item {\bf \color{myred} The larger the number of fields, the simpler and sharper the predictions.} For small $\Nf$, the power spectra  generated by the DBM models are heavily featured, and deviate strongly from scale-invariance. However, as $\Nf$ is increased, the predictions of these models  become simpler, and the power spectra much more regular \cite{DBM2}. For $\Nf \gtrsim 10$, the power spectra tend to be well described by an approximate power law, with a spectral index close to unity, and with a small negative running \cite{DBM2, DBM3}.

By contrast, the GRF models studied here give simple predictions already for small $\Nf$. As $\Nf$ is increased, the generated power spectra remain simple, but many of the predictions are only weakly dependent on $\Nf$. 

% the `straying' behaviour of the smallest mass-squared becomes more prominent

Thus,   the DBM and GRF constructions differ in that the predictions  of the former become simple as $\Nf$ is increased, while those of the latter are simple already for small $\Nf$. At large $\Nf$, both constructions predict simple power spectra.
%
%At small $\Nf$, we know of no reason why the prediction of DBM and GRF models should agree, and indeed, the power spectra generated from small-$\Nf$ GRF models are 
%

\item {\bf \color{mygreen} Planck compatibility is not rare, but future experiments may rule out this class of models.} For $\Nf \gtrsim 10$ in DBM models, the power spectra tend to be well described by an approximate power law, with a spectral index close to unity, and with a small negative running. These models can easily be compatible with current observational constraints on the power spectrum, and make a rather sharp prediction for a small negative running of the spectral index: $-0.004 \lesssim \alpha_s \lesssim 0$ at $1\sigma$ \cite{DBM2, DBM3}.

 In section \ref{sec:Planck}, we saw that the GRF models  predict a spectral index close to unity, and make a sharp prediction of a small negative running. Comparing Figure \ref{fig:nsalphas} to Figure 10 of \cite{DBM3}, we see that the predictions of the two constructions also qualitatively agree: red spectra tend to be favoured, but (for a wide range of hyperparameters) not so red as to be incompatible with Planck observations. The prediction for the running is in both cases sharper than that for $n_s$,   and the prediction from the GRF model agrees quantitatively with, but is sharper than, the prediction from the DBM models. %Thus,  the prediction of a small negative running appears to be a very robust predictions of manyfield inflation in random potentials. 
 
 We conclude that observational compatibility and the prediction of a small negative running appear to be quite robust predictions of manyfield inflation in random potentials. 

\item {\bf \color{mygreen} The smoother the potentials, the sharper the predictions.} Reference \cite{DBM3} found that 
flattening the DBM potential by increasing  $\Lh^{\rm DBM}$ or decreasing $\epsiloni$ led to sharper statistical predictions. In Figure \ref{fig:nsUniform}, we have seen that the same sharpening occurs also for GRF models. 

\item {\bf \color{mygreen} Hyperparameters can transition from stiff to sloppy} A key finding of  reference \cite{DBM1} was that `eigenvalue repulsion' sharply reduces the duration of inflation near a critical point of the DBM potentials: even if the curvature of the potential is fine-tuned to be small at the critical point, small cross-couplings in the Hessian cause the curvature to grow in the neighbourhood of the critical point.
As the field evolves from $\phi=0$ in DBM models, the eigenvalues of the Hessian matrix relax towards the equilibrium configuration, and quickly spoil any initial fine-tuning of the $\etaV$ parameter.  As a consequence, it was shown in \cite{DBM1} that 
the number of e-folds becomes independent of $\etai$ for $|\etai| \lesssim 0.01$. Reference \cite{DBM3} furthermore showed that also the spectral index and its running are independent of $|\etai|$, if similarly  small, and interpreted this behaviour  as a `stiff-to-sloppy' transition of the hyperparameter $\etai$.

\begin{figure}
    \centering
   \begin{subfigure}{0.48\textwidth}
         \includegraphics[width=1\textwidth]{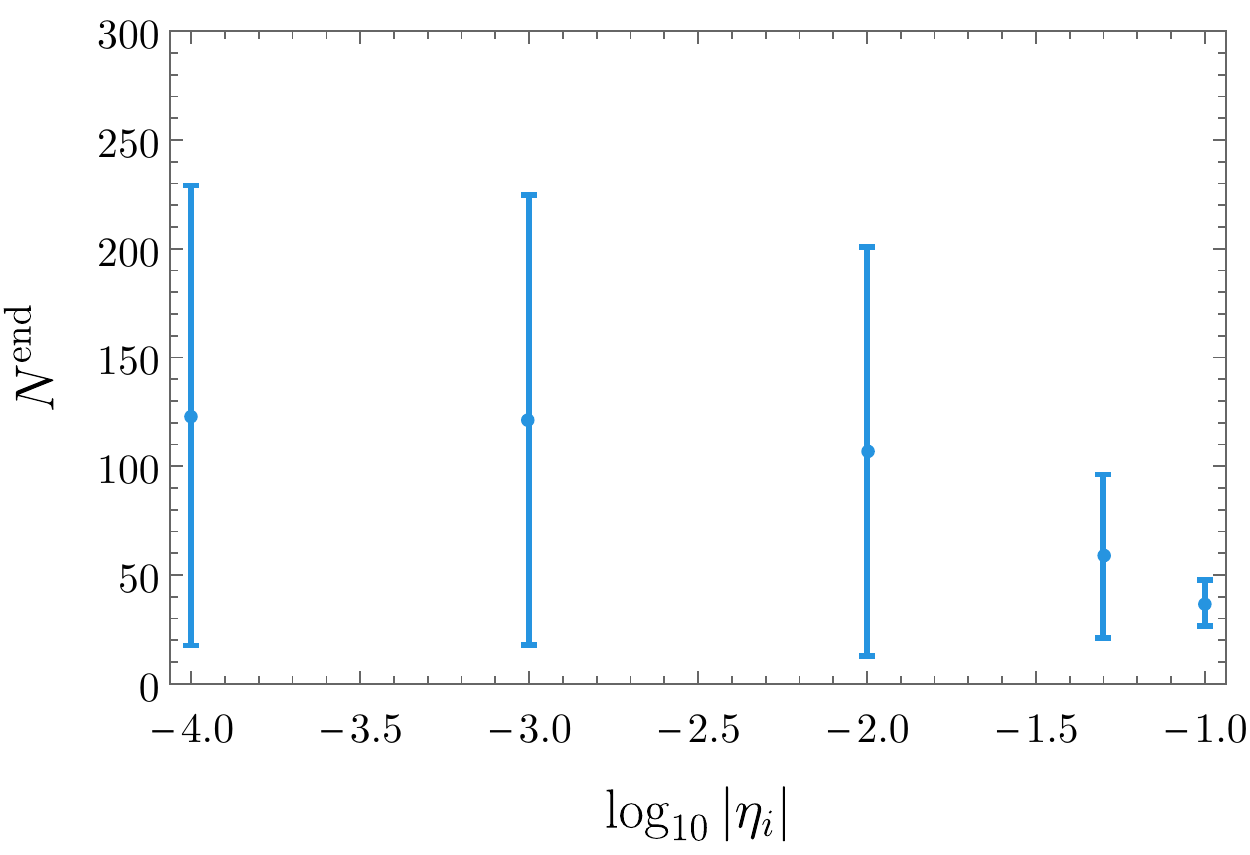}
   % \caption{Fraction of negative eigenvalues}
    \end{subfigure}
~  
   \begin{subfigure}{0.48\textwidth}
         \includegraphics[width=1\textwidth]{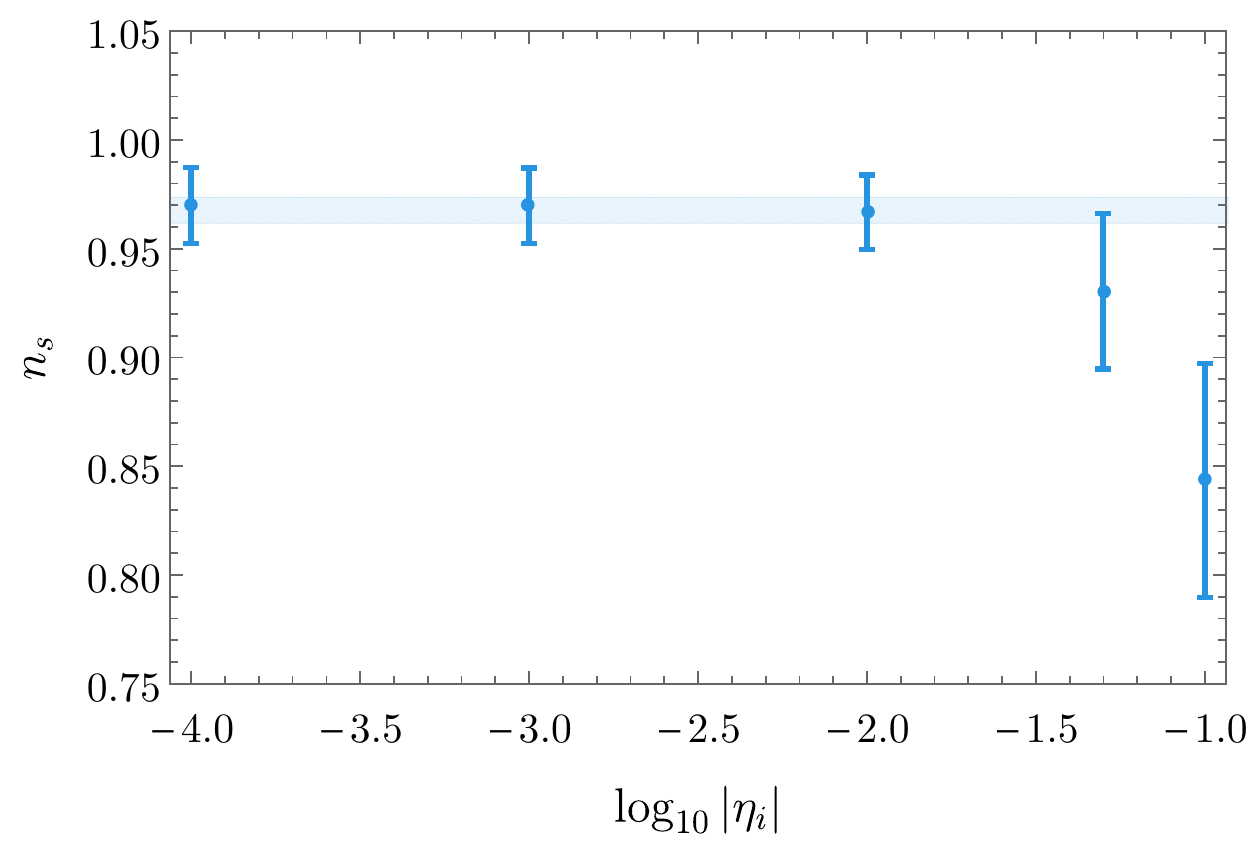}
   % \caption{Fraction of negative eigenvalues}
    \end{subfigure}
    \caption{The total number of e-folds and $n_s$ as functions of $\etai$, with other parameters as in the baseline  %\ref{eq:baseline} 
    and  together with the 68\% c.l.~from Planck \cite{1502.02114}.
     }\label{fig:nseta}
\end{figure}

As discussed in section \ref{sec:differences}, the evolution of the Hessian matrix in the DBM and GRF models differ significantly, and the $\etai$-dependence of the predictions of the GRF models provides a non-trivial test of the robustness of the DBM results. This is particularly interesting as  reference \cite{Masoumi:2017xbe} recently used an effectively single-field system motivated by uplifted GRF potentials to propose that the `steepening' responsible for the $\etai$ independence is absent in GRF models.\footnote{The authors of references \cite{Masoumi:2016eag, Masoumi:2017xbe} also state that this `steepening' leads to strong deviations from scale-invariance in DBM models of inflation, but the power spectra of these models are commonly consistent with small deviations from scale invariance 
 \cite{DBM2, DBM3}.  } 

For the GRF models studied in this paper,  Figure \ref{fig:nseta} settles this question.  %shows the $\etai$ dependence of the number of e-folds of inflation and the spectral index. 
For $-\etai \gtrsim 0.01$, the predictions depend strongly on $\etai$, as  the initial curvature of the potential curbs the duration of inflation. 
 By contrast, for $-\etai \lesssim 0.01$ the predictions become  independent of the precise value of $\etai$, as the initial fine-tuning of the Hessian matrix is quickly spoiled when the field moves away from the saddle-point. The distribution of the number of e-folds and the predictions for $n_s$ then stabilise, and become independent of $\etai$. The running, $\alpha_s$, while not plotted, is  independent of $\etai$ whenever $n_s$ is.  
 Thus, while the evolution of the Hessian matrix differ between GRF and DBM models,  the  prediction 
of $\etai$ independence holds for both constructions.  
 This strongly suggests that in models in which the eigenvalue spectrum relaxes from an initial, fine-tuned spectrum to a more generic spectrum that includes some  tachyonic eigenvalues, the predictions become 
independent of the initial curvature of the potential for small $|\etai|$.

\item {\bf \color{mygreen} Despite tachyons, isocurvature can decay.} 
In section \ref{sec:iso}, we found that despite the presence of multiple tachyons, isocurvature tends to decay during inflation in GRF models. 
This suppression of isocurvature was previously observed in DBM models in \cite{DBM3}, and was there similarly  explained as a dynamical consequence of multi-field slow-roll inflation. Comparing our Figure \ref{fig:iso} to Figure 17 of \cite{DBM3}, we see that dependence of the end-of-inflation values of $P_{\rm iso}(k_{\star})/P_{\zeta}(k_{\star})$ on $\Nf$ qualitatively agree between DBM and GRF models: for small $\Nf$, the suppression is most severe, but it remains exponential   for large $\Nf$.  

Our work provides suggestive evidence for a rather model-independent  suppression of isocurvature perturbations  in  small-field slow-roll inflation. This is non-trivial, as no single-field `adiabatic limit' is reached in these models, which typically contain many tachyons. %This suppression has interesting consequences for  reheating, and affects how signals of multiple fields may be manifest in the CMB. We hope to return to these questions in future work. 

\item {\bf \color{mygreen} Eigenvalue repulsion drives the predictions.} % In section \ref{sec:scales}, we noted the strong link between the Hessian of the GRF potentials and the simplest RMT models. In particular, we emphasised how the change of variables from the independent elements of the Hessian to its eigenvalues and eigenvectors gives the Vandermonde determinant as a Jacobian. The presence of the Vandermonde determinant ensures that the probability of two eigenvalues becoming degenerate vanishes, 
%and is key to the mutual repulsion 
%of eigenvalues. 
 %Moreover, we reviewed how the Hessian matrix, and any fixed value of the potential, is simply given by a GOE matrix plus a shift, proportional to the unit matrix  (cf.~equation \eqref{eq:splithessian}) \cite{Bray:2007tf, Bachlechner:2014rqa}.
% Clearly, any carefully phrased question about the spectrum of GRF models can then be translated into precise questions in random matrix theory. {\color{blue} repetitive}
%
 %
 In DBM models, several of the predictions at large $\Nf$ can be explained by eigenvalue repulsion \cite{DBM1, DBM2, DBM3}. In particular, the non-generic spectrum in the initial patch  quickly relaxes towards the Wigner semi-circle distribution as a consequence of   eigenvalue repulsion. This relaxation explains the independence on $\etai$, the tendency towards red spectral indices, the negative running, and the observed regularity of the power spectra  for large $\Nf$.
 
 In GRF models, %the distribution for the Hessian matrix clearly involves eigenvalue repulsion through the Vandermonde determinant, but the variation of the Hessian over the field space differs substantially from the DBM models. 
 the eigenvalues of the Hessian repel in a linear fashion over small field-space distances, leading to the cone of eigenvalue trajectories observed in Figure \ref{fig:100fieldEVs}. 
 In section \ref{sec:stray}, we showed that  the statistical properties of the cubic terms of GRF potentials  lead to a `straying' behaviour of the smallest eigenvalue in slow-roll inflation, which is then repelled to  tachyonic values at a faster rate than other eigenvalues. Also for the GRF models, we have been able to relate the predictions of the model to properties of the relaxation of the spectrum from a fine-tuned initial configuration to the (slightly shifted) semi-circle. Thus, also for the GRF models, eigenvalue repulsion drives the predictions. 

\end{enumerate}

In sum, six out of the seven `lessons' from manyfield inflation in DBM potentials derived in \cite{DBM3} apply also to manyfield inflation in GRF potentials. The single lesson for which the predictions  differ involves the properties of the models for small $\Nf$, in which the details of the constructions evidently are very important. For $\Nf\gg1$, the predictions of these very different constructions agree, which may be indicative  of an emergent limit of inflation in which disparate classes of potentials make the same `universal' predictions.

\section{Result III: $\fnl\sim {\cal O}(1)$ is very rare in manyfield  inflation}
\label{sec:result3}

\begin{figure}
    \centering
    \begin{subfigure}{0.48\textwidth}
    \centering
    \includegraphics[width=0.96\textwidth]{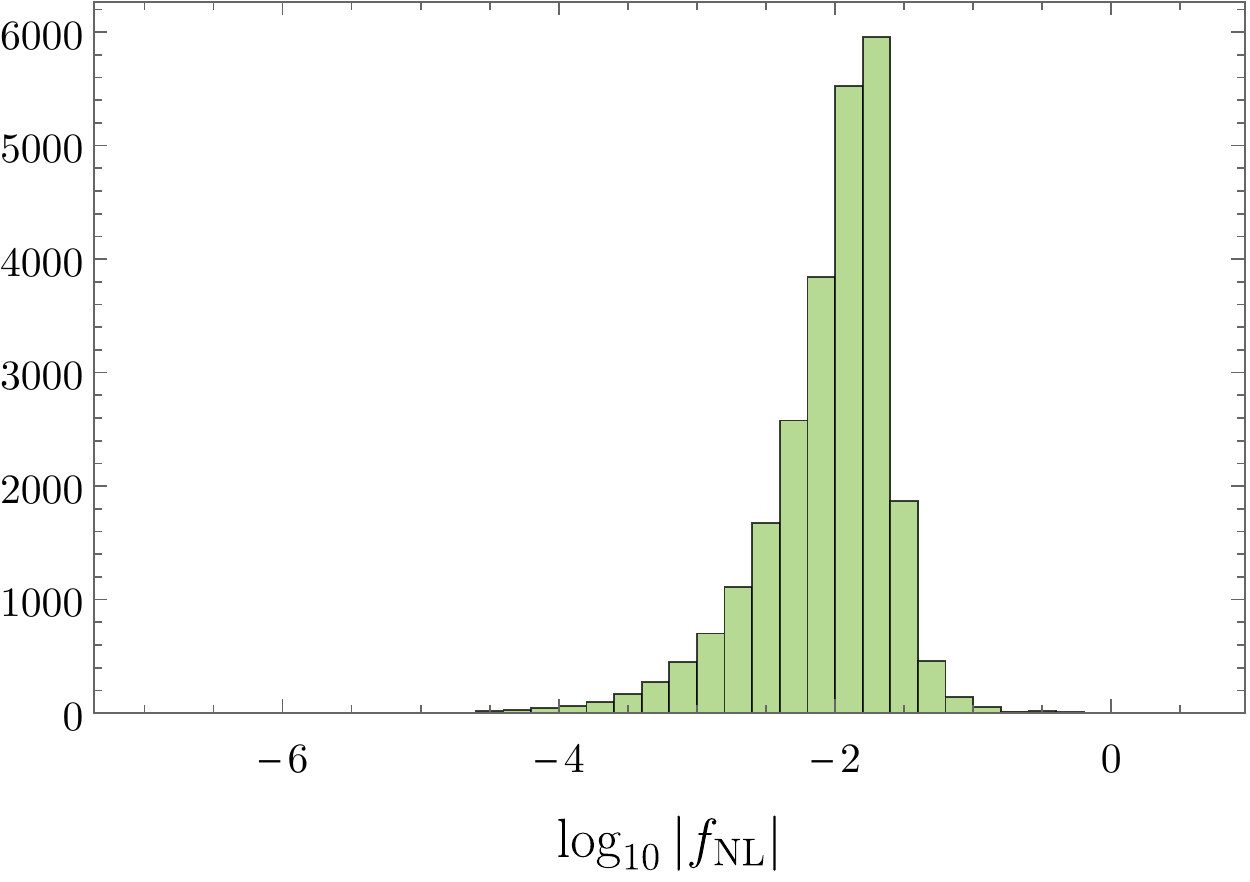}
    %\caption{$\fnl$ values from 1000 random 50-field potentials}
    \end{subfigure}
    ~ %add desired spacing between images, e. g. ~, \quad, \qquad, \hfill etc. 
    \begin{subfigure}{0.48\textwidth}
         \includegraphics[width=1\textwidth]{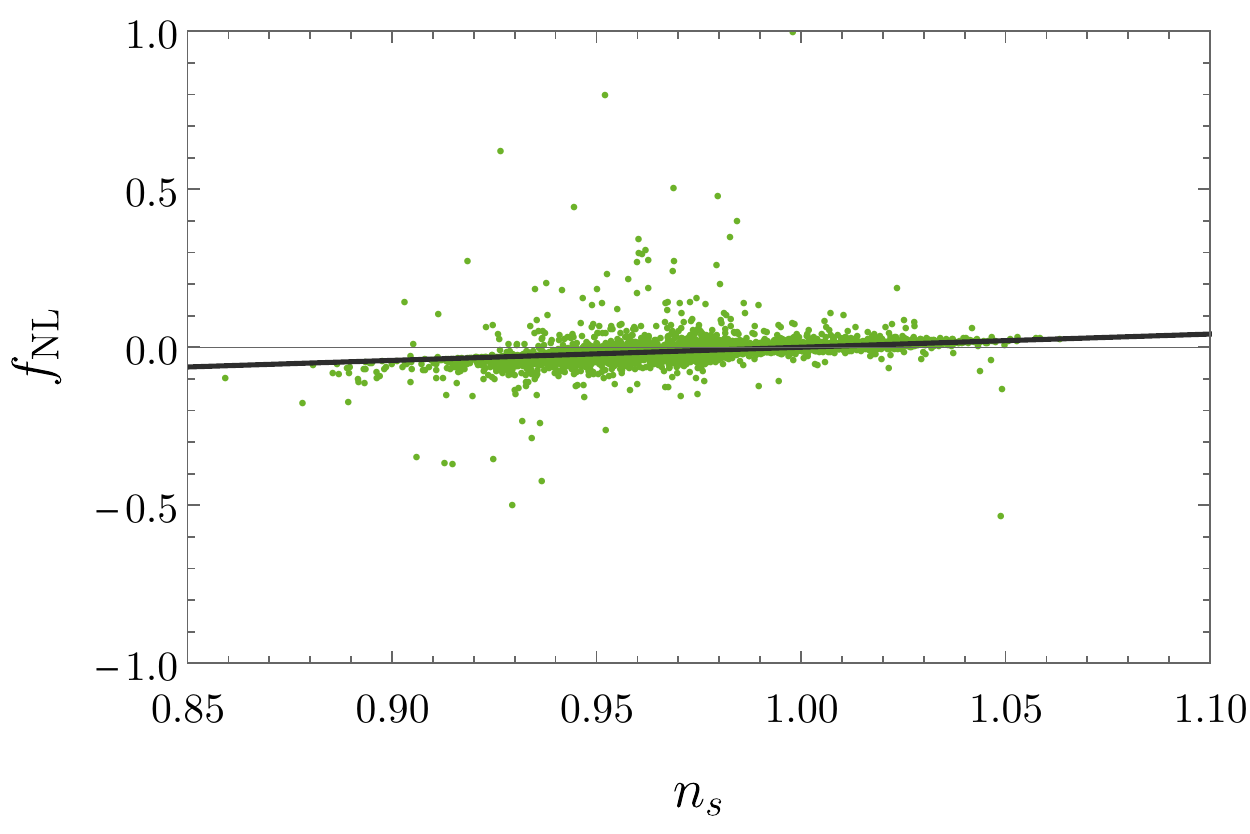}
   % \caption{Fraction of negative eigenvalues}

    \end{subfigure}
    \caption{%Values for $\fnl$ with 50 fields and compressed initial mass spectrum.%%%
    Values for $\fnl$ and $n_s$ for 25,000 random inflation models, spanning values of $\Nf$ between $5$ and $50$ (the same as in Figure \ref{fig:nsalphas}). 
    The black line in the right plot indicates the single-field consistency condition: $\fnl = \tfrac{5}{12}(n_s-1)$.
    NB: the graph on the right excludes six points with $|\fnl|>1$.}\label{fig:50fieldfNL1}
\end{figure}

We are now ready to discuss  the main result of this paper: the levels of primordial non-Gaussianities (NGs) generated in models of manyfield inflation with random potentials. 
Upcoming cosmological experiments are set to
target $\fnl \equiv \fnl^{\rm local}$, and are expected to reach a sensitivity of $\sigma(\fnl)\sim {\cal O}(1)$  over the next few years \cite{
Dalal:2007cu,Matarrese:2008nc, Slosar:2008hx, Desjacques:2010jw, 
 Alvarez:2014vva,
 Ferraro:2014jba,
  Baldauf:2016sjb}. 
The  results presented in this section provide important insights into what we can realistically hope to learn from these experiments. 

In inflationary models with multiple canonically normalised fields, the level of non-Gaussianity at horizon exit is commonly very small  \cite{Vernizzi:2006ve, Battefeld:2006sz}. Substantial amplitudes of local NG, i.e.~$\fnl\sim {\cal O}(1)$, can be generated through superhorizon evolution of the curvature perturbation, either during inflation or after the end of inflation (cf.~e.g.~\cite{Bernardeau:2002jy, Bernardeau:2002jf, Rigopoulos:2005us, Rigopoulos:2005ae,  Bartolo:2001cw, Wang:2010si, Vernizzi:2006ve, Battefeld:2006sz} and \cite{1002.3110} for a review). 
 %
 %The generation of $\fnl$ has been studied by many authors and in a large number of models. Still, both analytical and numerical 
% However, existing computations of $\fnl$
 %tend to apply only to particular classes of few-field models
 %\cite{Peterson:2010mv}, or manyfield models with very simple potentials \cite{Battefeld:2006sz}.  
 %Little is known about  $\fnl$ in models with many dynamically important fields subject to general potentials. 
In this section, we %address this issue 
%by using 
use the  transport method and $\delta N$ formalism to compute 
$\fnl$ in our ensembles of randomly generated models of manyfield inflation.

%The main result of this section is illustrated by Figure \ref{fig:50fieldfNL1}: here $n_s$ and $\fnl$ are plotted for 1000 random GRF models with $\Nf=50$, evolved from the highly compressed initial spectrum of equation \eqref{eq:spectra}. The level of NG is small for all models in this ensemble, $|\fnl| \ll 1$. Among these models, ${\rm max}(|\fnl|) = 0.05$ {\color{blue} check}, and most realisation even approximately follow the single-field consistency condition between $\fnl$ and $n_s$. As we will now discuss, this statistical prediction is quite robust, and  largely independent of the hyperparameters. 

The main result of this section is illustrated by Figure \ref{fig:50fieldfNL1}: here $n_s$ and $\fnl$ are plotted for an aggregate of 25,000 random inflation models, spanning values of $\Nf$ between $5$ and $50$, with both flat and compressed initial mass spectra, cf.~equation \eqref{eq:spectra}. The levels of non-Gaussianity is generally very small for these models, with the vast majority having $\fnl\sim\mathcal O(0.01)$. Out of the 25,000 models, only six had values of $|\fnl| >1$ (these fall outside the boundaries of  the right plot of Figure \ref{fig:50fieldfNL1}). Moreover, most realisations even approximately follow the single-field consistency condition between $\fnl$ and $n_s$. %As we will now discuss, this statistical prediction is quite robust, and  largely independent of the hyperparameters. 
For the baseline ensemble of 1000 models (with parameters as in equation \eqref{eq:baseline}), we find $\fnl = -0.012 \pm 0.008$ (at $68\%$ confidence level).

\begin{figure}
    \centering
    \begin{subfigure}{0.48\textwidth}
    \centering
    \includegraphics[width=1\textwidth]{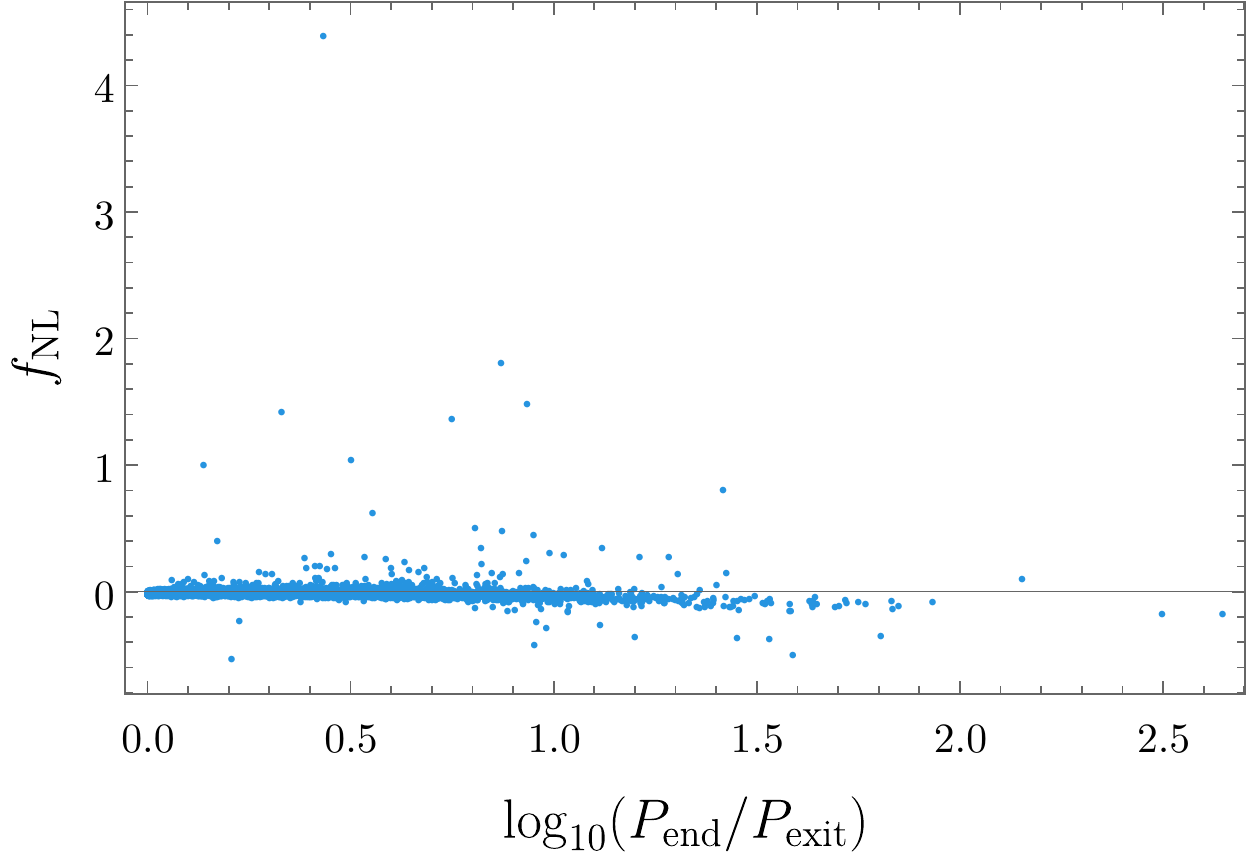}
    %\caption{$\fnl$ values from 1000 random 50-field potentials}
    \end{subfigure}
    ~ %add desired spacing between images, e. g. ~, \quad, \qquad, \hfill etc. 
    \begin{subfigure}{0.48\textwidth}
         \includegraphics[width=1\textwidth]{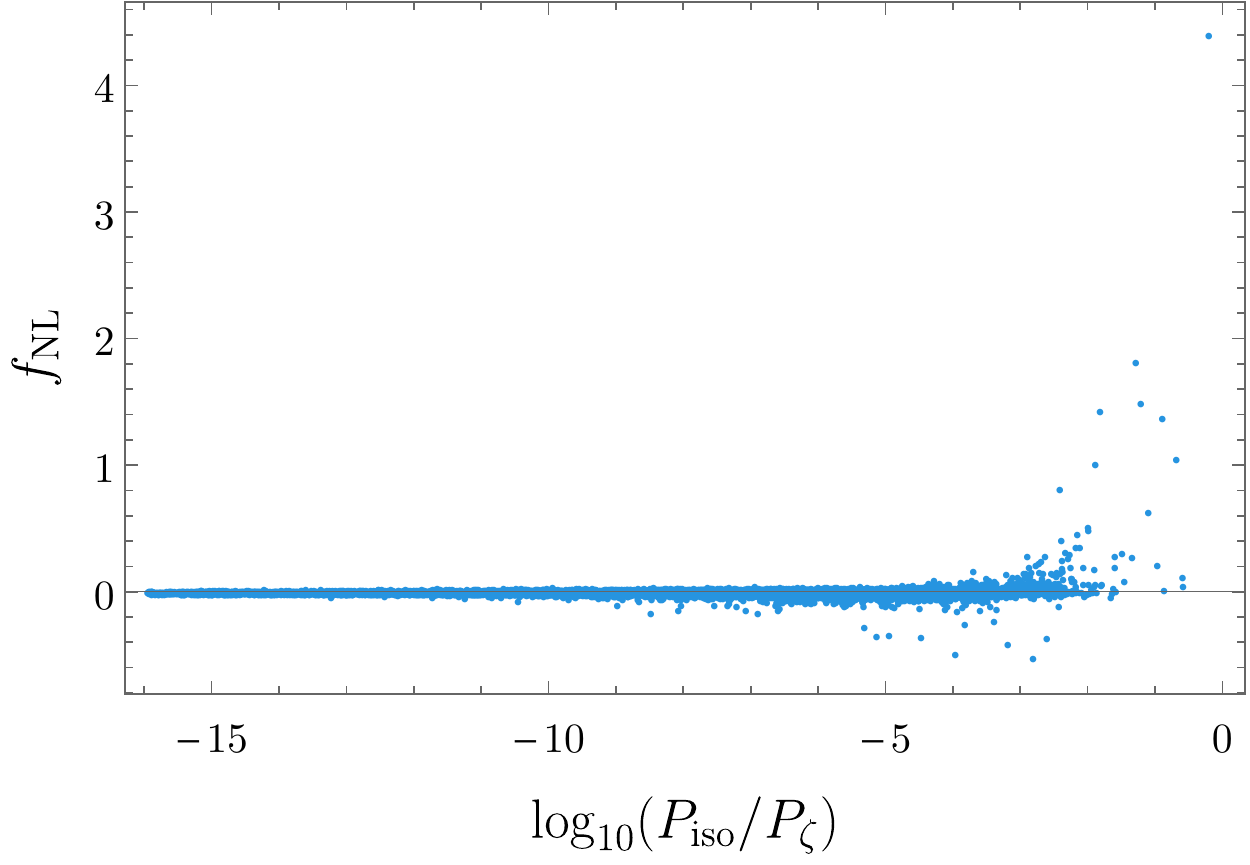}
   % \caption{Fraction of negative eigenvalues}

    \end{subfigure}
    \caption
    {Superhorizon evolution and surviving isocurvature for the same  models as in Figure \ref{fig:50fieldfNL1}.}\label{fig:SHisofnl}
\end{figure}

Single-field inflation generates only small levels of NG, and multifield effects are necessary for large $\fnl$. However, multifield effects do not suffice to ensure $|\fnl| \sim {\cal O}(1)$. The left plot of Figure \ref{fig:SHisofnl} shows the relation between $\fnl$ and the superhorizon evolution, as given by  $\log_{10}\left(P_{\zeta}(N_{\rm end})/P_{\zeta}(N_{\star})\right)$, for these 25,000 models of inflation. Large values of $\fnl$ are only observed in models with some level of superhorizon evolution, but many models with a large ratio of $P_{\zeta}(N_{\rm end})/P_{\zeta}(N_{\star})$ produce low levels of non-Gaussianity.

There is however a stronger relation between large $\fnl$ and the amount of surviving power in the isocurvature modes at the end of inflation, as the right plot of  Figure \ref{fig:SHisofnl} shows. All the cases with large $\fnl$ have a ratio of isocurvature modes to adiabatic modes (at $k= k_\star$) of at least $\mathcal O(0.01)$.
In these models, the curvature perturbation may continue to evolve after the end of inflation, and it is necessary to  model the reheating phase  to determine  the final value of $\fnl$ relevant for CMB and Large Scale Structure (LSS) experiments.
%
%
%By studying the evolution of $\fnl$ during inflation in specific examples, we find that it is typically sourced monotonically and towards the end of inflation, if at all.  
%In models with large non-Gaussianities,  $\fnl$ grows  in roughly linear proportion to the field displacement during inflation.
%
%It is not the case here that $\fnl$ typically becomes large during inflation and is then suppressed. 
Only in a handful instances with the highly compressed initial mass spectrum did $\fnl$ increase to ${\cal O}(1)$ during inflation, but decrease again by the end of it. 
 In Appendix \ref{sec:casestudies}, we provide case studies of a typical 100-field model (with small $\fnl$) and one of the rare cases of a 25-field model yielding  $\fnl \sim {\cal O}(1)$.

The statistical prediction of small $\fnl$ is robust under changes to the hyperparameters. 
The number of fields, $\Nf$, has no noticeable effect on $\fnl$: large NGs are rare for all values we have considered.
%\footnote{This is very similar to the generation of $\fnl$ in model with many non-interacting fields \cite{Battefeld:2006sz}.}  
We find a weak dependence on the flatness of the potential: when the potential becomes very flat and the superhorizon evolution decreases (cf.~our discussion in section \ref{sec:Planck}), the values of $\fnl$ follow the single-field consistency relation very closely, and large values of $\fnl$ become more rare.
We find that $\fnl$ is independent of $|\etai|$, except the largest values of $|\etai|$ we investigate, for which large values of $\fnl$ become slightly more common.

The initial mass spectrum at $\phi =0$ does however have a clear impact on the levels of $\fnl$ generated. For the flat spectrum with the eigenvalues of the Hessian uniformly distributed between $3 \eta H^2$ and $9 H^2/4$, large values of $\fnl$ are exceedingly rare: in 19,000 examples with values of $\Nf$ varying between 5 and 50 we found only one model with large $\fnl$ (see Appendix \ref{sec:casestudies}). By contrast, with the (rather extremely) compressed initial spectrum where the masses are spread between $3 \eta H^2$ and $-3 \eta H^2$, we saw 5 in a sample of around 6,000. Thus, while large values of $\fnl$ are still rare, near-degenerate initial spectra appear to make large NGs more frequent.

\subsection{Why so small?}

We have found that in randomly generated models of inflation involving many coupled fields, %and commonly exhibiting non-trivial multifield dynamics, 
large values of $\fnl$ are very rare. In this section, we discuss the main reasons for this suppression of non-Gaussianities.

The smallness of $\fnl$ in our class of models resonates with previous results on the generation of non-Gaussianity through non-derivative interactions during multifield inflation. Reference \cite{Battefeld:2006sz} derived an analytic formula for $\fnl$ in multifield models of inflation with sum-separable potentials (thereby generalising the two-field result of  \cite{Vernizzi:2006ve}), and found that large non-Gaussianities are very rare in slow-roll models with $\Nf$ free fields. In particular, $\fnl$ was found not to be enhanced as $\Nf$ was increased, consistent with our findings. Moreover, two-field models generating large non-Gaussianities during inflation tend to require substantial levels of fine-tuning, cf.~\cite{Byrnes:2008wi, Byrnes:2008zy, Peterson:2010mv,  1002.3110, Tanaka:2010km}.

To understand the smallness of $\fnl$ in these random manyfield models, it is illuminating to 
consider the 
re-expression of 
 the $\delta N$ formula for $\fnl$ derived in  \cite{Peterson:2010mv, Peterson:2011yt}. %the $\delta N$ formula for $\fnl$ was re-written in a particularly illuminating way. 
In our notation, generating large $\fnl$ in slow-roll, slow-turn models of multifield inflation requires a large contribution from the term, 
\be
\fnl \supset \frac{5}{6}
\sqrt{2 \epsilon_\star}\,
\frac{ T_{\zeta {\cal S}}^2}{\left(1+ T_{\zeta {\cal S}}^2\right)^2}\, 
h^j \partial^*_j  T_{\zeta {\cal S}}  
%+ \ldots
\,  .
\label{eq:PT}
\ee
Here $T_{\zeta {\cal S} }\equiv \left(\sum_i T^2_{\zeta {\cal S}_i}\right)^{1/2}$, and $h^i = e_b\, v_b^i(N^*)/ |e_c\, v_c^j(N^*)|$, cf.~equations \eqref{eq:decomp}, \eqref{eq:nstransport} and \eqref{eq:transfer2}. Here also,  $\partial^*_j$ denotes a derivative with respect to the field-space position of the trajectory at horizon crossing in an entropic direction. 
Equation \eqref{eq:PT} has two important consequences: first, to generate large NG, some superhorizon evolution of the curvature perturbation is necessary ($T_{\zeta {\cal S}} \neq0$). However, too much superhorizon evolution suppresses $\fnl$. Second, large $\fnl$ is only possible if the level of superhorizon evolution of $P_{\zeta}$ is a very sensitive function of the initial conditions, so that $|h^j \partial^*_j  T_{\zeta {\cal S}}  |\gtrsim {\cal O}(1/\sqrt{\epsilon_\star})$.

In our GRF models, $T_{\zeta{\cal S}}$ is commonly ${\cal O}(1)$, and  the factor $ T_{\zeta {\cal S}}^2/(1+ T_{\zeta {\cal S}}^2)^2$ does not strongly suppress $\fnl$. However, the amount of superhorizon evolution is rarely a highly sensitive function of the initial conditions: trajectories separated by some small, initial perturbations $|\delta \phi_{\perp}^*| \ll \Lh$ tend to follow very similar paths in field space, and do not generate drastically different $T_{\zeta {\cal S}}$. Consequently, $|h^j \partial^*_j  T_{\zeta {\cal S}}  |$ is typically not large enough to generate appreciable levels of NGs.

The decay of isocurvature in multifield, saddle-point models of inflation (cf.~our discussion in section \ref{sec:iso}) contributes to the typical smallness of $\fnl$. When the entropic perturbations ${\cal S}^i$ have decayed sufficiently, $T_{\zeta {\cal S}}$ ceases to be sourced, and no additional initial condition dependence is induced.  This way, the decay of ${\cal S}^i$ limits the period during which %features of the potential could possibly induce 
large NGs could be generated. We also note that $T_{\zeta {\cal S}}$ tends to be sensitive to the initial conditions precisely when entropic, off-trajectory perturbations are important. This provides a heuristic explanation for why the isocurvature tends to be large in the few examples we found with  $\fnl \sim {\cal O}(1)$.

%of the primordial perturbations. 
%Reversely, we note that models in which the background trajectory is very sensitive to small fluctuations can yield large $\fnl$, but typically also induces large levels of isocurvature. 
%To generate substantial amplitudes of NGs, isocurvature modes are necessary. However, we showed in section \ref{sec:result1} that isocurvature typically decays during inflation, and we explained this decay as dynamical consequence of slow-roll inflation in our approximate saddle-point models (this phenomenon was first discussed in \cite{DBM3}). In our GRF models, the smallest eigenvalue of the Hessian `strays' from the other eigenvalues, effectively enhancing the suppression of isocurvature, and the potential `straightens', which dampens the transfer of power from isocurvature to curvature modes.\footnote{We note however, that turns are common in the initial and final stages of inflation in our models.} We expect that in most of the models that we have generated, the suppression of isocurvature during inflation is the main reason for the smallness of the generated $\fnl$. 

We close this section by noting how large $\fnl$ may be more frequent in modifications  of our construction. 
%expect that some modifications to our manyfield models could lead to enhanced levels of NGs. 
%
Models with very sharp turns or in which nearby classical background trajectories rapidly diverge  can lead to  substantial non-Gaussianities (see e.g.~\cite{Chen:2011zf, Achucarro:2010jv}). In our construction of the potentials, cf.~equation \eqref{eq:Taylor1}, all terms in the potential are suppressed by the same `UV cutoff scale' $\Lh$, which makes features or sharp turns on scales $\ll \Lh$ very rare, even if the interaction terms are random. Large $\fnl$ may be more common in multi-scale potentials with features on small scales, at least if these appear along the trajectory before the isocurvature has decayed. 

Finally, our results do not preclude large values of $\fnl$ being generated after inflation, during the reheating phase, as is the case in many `spectator' models, cf.~\cite{Linde:1996gt, Lyth:2001nq, Moroi:2002rd, Lyth:2003, 1612.05248}. However, in models with general interactions and exponentially suppressed isocurvature at the end of inflation, generating large $\fnl$ through reheating dynamics may remain challenging. 

To get a rough sense of the reheating dynamics required in order for isocurvature to become relevant during reheating, we consider a modified version of the curvaton scenario \cite{Lyth:2001nq} with two fields: the inflaton, $\phi$, and the curvaton, $\sigma$. For simplicity, we assume that the fields are free, without mutual interactions, and decay instantaneously at $H(t_{\phi})=\Gamma_{\phi}$ and $H(t_{\sigma})=\Gamma_{\sigma} < \Gamma_{\phi}$, respectively. The inflaton is assumed to decay into radiation, which initially dominates the energy density, while $\sigma$ oscillates around the vacuum, with an energy density that red-shifts like matter. At the time $t_{\sigma}$, the curvaton is assumed to have come to dominate the energy density. The total energy density during the period $t_{\phi} < t< t_{\sigma}$ is given by,
\be
 \rho= \rho_{r}  + \rho_{\sigma}=  \rho_{r}(t_\phi) \left(\frac{a(t_{\phi})}{a(t)}\right)^4 + \rho_{\sigma}(t_\phi)\left(\frac{a(t_{\phi})}{a(t)}\right)^3 \, .
 \label{eq:curvatonrho}
 \ee
 With these assumptions, the total curvature perturbation is given by  \cite{Lyth:2001nq},
 \be
 \zeta = \frac{4\zeta_r + 3 \frac{\rho_{\sigma}}{\rho_r} \zeta_\sigma}{4+ 3  \frac{\rho_{\sigma}}{\rho_r} } \, ,
 \ee
where $\zeta_r$ and $\zeta_\sigma$ are separately conserved and respectively correspond to the curvature perturbations induced by $\phi$ and $\sigma$. 
Highly suppressed  isocurvature at the end of inflation corresponds to $\zeta_r \gg \zeta_\sigma$ (this assumption differs from those of the curvaton scenario in which $\zeta_r$ is taken to be negligibly small).  The curvaton will become important during reheating if,
\be
\frac{\rho_{\sigma}(t_{\sigma})}{\rho_{r}(t_{\sigma})} \gtrsim \frac{\zeta_r}{\zeta_\sigma} \, .
\ee 
Using equation \eqref{eq:curvatonrho}, we may write this condition as,
\be
\left(\frac{\rho_{\sigma}(t_\phi)}{\rho_{r}(t_\phi)}\right)\left( \frac{\rho_\sigma(t_\phi)}{3 \Gamma_{\sigma}^2 \Mpl^2}\right)^{1/3} \gtrsim \frac{\zeta_r}{\zeta_\sigma} \, .
\label{eq:decayratebound}
\ee
%For $\rho_\sigma(t_\phi) \gg \rho_r(t_\phi)$, this inequality is easily satisfied, but this scenario can be hard to realise in practice. Decay rates and the energy contained in a field generally increase with the mass, and since $\Gamma_\sigma\ll\Gamma_\phi$ we would  expect it to have a much smaller mass and hence be much less energetic.
%More natural is to consider the case 
Writing 
$ \rho_\sigma(t_\phi) = \beta \,\rho_r(t_\phi)$ for $\beta\ll 1$, this equation %\eqref{eq:decayratebound}
 can  be written as a simple condition on the decay rates,
\be
 \frac{\Gamma_\phi}{\Gamma_\sigma} \gtrsim \frac{1}{\beta^2} \left(\frac{\zeta_r}{\zeta_s}\right)^{3/2} \, .
\ee
This indicates that an exponential hierarchy of decay rates is necessary for initially suppressed isocurvature to become important during reheating. We note in closing that this argument is simplified and relies on assumptions that are not expected to hold for GRF potentials (e.g.~the absence of interactions). We expect to return to the question of manyfield reheating in more detail in future work (see also \cite{Leung:2012ve, Meyers:2013gua, Hardwick:2016whe, Hotinli:2017vhx} for some recent studies of this and related questions).

\section{Conclusions}
\label{sec:conclusions}
In this paper, we have studied inflation in models with multiple fields subject to randomly generated interaction terms. We have used Gaussian random fields (GRFs) with a Gaussian covariance function to model the scalar potentials locally around an approximate saddle-point in field space. 
These potentials admit an interpretation as proxies for physical effective field theories,  
and exhibit structure 
%with 
%potentials with structure over 
 %valid below the UV-scale 
 over field space distances of 
 $\Lh < \Mpl$.
  By identifying and systematically applying algebraic simplifications to the covariance matrix of the Taylor coefficients, we have for the first time been able to use this method to construct explicit scalar potentials with many interacting fields. Our examples include 100-field models involving 97 million independent, randomly generated couplings. %, encoded in potentials with structure on sub-Planckian scales.

We  used these potentials to construct models of slow-roll inflation with many dynamically important fields, i.e.~models of manyfield inflation.  By using the transport method and the $\delta N$ formalism, we computed the primordial perturbations generated during inflation, including the curvature and isocurvature modes, and the amplitude of local non-Gaussianity, $\fnl$. These studies led to several novel results. 

\subsubsection*{Summary of findings}

As the fields evolve from the approximate saddle-point where the mass spectrum is fine-tuned, the eigenvalues of the Hessian  `relax' towards a shifted semi-circle distribution. This relaxation is, over short distances, driven by the cubic terms in the potential. We have shown that a combination of the statistical properties of Gaussian random fields and the dynamics of multifield slow-roll inflation leads to a particularly fast relaxation of the smallest eigenvalue of the Hessian, making it  `stray' from the other eigenvalues. This `straying' has important consequences for the observational predictions of the models.  

The generated power spectra of the GRF models are very simple, close to scale-invariant and well-described by an approximate power law. We have shown that large regions of the (hyper-)parameter space are compatible with current observational constraints on the spectral index. However, some regions, in particular those describing very flat potentials, are already ruled out by the Planck experiment. 

%A  robust prediction of these models regards
 These models make a robust prediction for 
 the running of the spectral index. A small negative running is vastly preferred, and a future observation of $\alpha_s$ outside the range $-0.004 \leq \alpha_s \leq 0$ would, together with existing Planck constraints, rule out all models that we have constructed. For our baseline 10-field model (cf.~equation \eqref{eq:baseline}), we find $\alpha_s = -0.00143\pm0.00034$ (at $68\%$ confidence level).

A striking aspect of these models is that while several fields go tachyonic during inflation and the curvature perturbation commonly evolves significantly on superhorizon scales, the power in the isocurvature modes decays during inflation. We have explained this phenomenon, also observed in \cite{DBM3}, as a consequence of multifield slow-roll dynamics, in which the adiabatic mode tends to align with the most rapidly growing field perturbation. The suppression is further enhanced by the `straying' of the smallest eigenvalue of the Hessian matrix. The exponential suppression of isocurvature during multifield slow-roll inflation makes the predictions  less sensitive to the details of the reheating phase, and could make models involving large numbers of dynamically important fields comfortably compatible with CMB constraints on isocurvature.  

We have furthermore critically assessed the similarities and differences between our models and those recently constructed using the random matrix theory `DBM' technique \cite{DBM1, DBM2, DBM3}. We have argued that these constructions provide fundamentally different descriptions for the evolution of the eigenvalues of the Hessian matrix,\footnote{By contrast,  the equilibrium spectra of these models only differ superficially, and  are easily rendered compatible. Our discussion in sections \ref{sec:scales} and \ref{sec:result2} may be useful for readers interested in the context of some  recent results comparing GRF and RMT models \cite{Masoumi:2016eag, Easther:2016ire, Masoumi:2017xbe}.
} yet strikingly, six out of the seven `lessons' from manyfield inflation in DBM potentials found in reference \cite{DBM3} also hold for our GRF models, with the only difference being related to properties of the models at small $\Nf$. This suggests that, at large $\Nf$, these very different constructions may fall in the same `universality class' of inflationary models.

Finally, we computed the level of local  non-Gaussianity (NG) generated by multifield effects on superhorizon scales and found that, typically,  it is very small,  with $|\fnl|\ll1$. For the baseline parameters we found  $\fnl = -0.012 \pm 0.008$ (at $68\%$ confidence level). The
 smallness of $\fnl$ is largely independent of the choice of hyperparameters, and constitutes  a robust statistical prediction of these models. We found that $\fnl$ is typically small even in models in which multifield effects are important and in which the superhorizon evolution of the perturbations is substantial. In a handful of models, we found $\fnl \sim {\cal O}(1)$. However, all  models with large $\fnl$ that we found also have comparatively large levels of isocurvature remaining at the end of inflation, which may affect the predictions of the models through the reheating phase. 
%We explained the suppression of $\fnl$ as a consequence of the decay in isocurvature,  
We have argued that the suppression of $\fnl$ 
%is related to a robustness of the predictions for the power spectrum under
is a consequence of the robustness  of the multifield evolution to
 small modifications of the initial conditions, and the decay of isocurvature during inflation. 
%is primarily affected by the decay of isocurvature, 
%but that it is also enhanced by the `straying' of the smallest eigenvalue and the corresponding `straightening' of the inflationary trajectory when the potential is dominated by the cubic terms. %As isocurvature decays during inflation also in the very different DBM models of manyfield inflation, we expect that the generic prediction of $|\fnl|\ll1$ holds for a much broader  set of models than the ones we have considered here.

Our results indicate that multifield models of inflation do not generically predict $\fnl \sim {\cal O}(1)$, and that large classes of models including the slow-roll, saddle-point GRF models we have constructed, typically yield $|\fnl| \ll 1$.
% future experiments that 
%are sensitive to  will not be able to probe
% in potentials with structure on . 
This suggests that   a future detection of $\fnl$ of order unity would point to  rather special, non-random inflationary dynamics.

%and we argued that it was also related to the `straying' of the smallest eigenvalue and the corresponding `straightening' of the inflationary trajectory when the potential is dominated by the cubic terms. 
\subsubsection*{Future directions}

There are a number of possible extensions to this work. We anticipate that our method can be varied and applied to other scenarios involving GRF potentials. Our focus in this paper has been on inflation for typical values of the potential in the simplest mean-zero GRFs, and we have only briefly discussed some variations involving  a field-independent cosmological constant in Appendix \ref{sec:casestudies}.  There, we found a remarkable agreement of the predictions between saddle-point models of inflation for the mean-zero and the `uplifted' GRF models. It would be interesting to obtain a broader understanding of the possible range of multifield inflationary scenarios that can be investigated through GRF potentials. Moreover, our choices for the initial conditions at the approximate saddle point have been motivated by simplicity, and could be generalised. 
%, and we expect generalisations to be straightforward. 

In section \ref{sec:EFT}, we noted that our class of potentials share many properties with physical effective field theories valid below the cutoff scale  $\Lh$.  Features on scales $\ll \Lh$ are very rare in such models. It would be interesting to investigate multiple-scale extensions of this construction, in which such sharp features would be more common.
Relatedly,
it would  be interesting to apply our methods to theories in which the mass-scale of the fields is naturally of the order of the Hubble parameter, for example by constructing supergravity theories with randomly generated superpotentials  \cite{Bachlechner:2014rqa}. Such a description could also make it possible to tighten the connection between mathematically simple models of manyfield inflation and fundamental physics. 

%GRF potentials constructed from other covariance functions, and, in particular, that 
%, and if their predictions can be made to differ from the GRF models that we have studied.
%\footnote{By contrast, our method only applies to Gaussian covariance functions.}

%Other initial conditions ? off-set trajectories. Fine-tuning (A shift of size $0.01\Lh$ in a random direction orthogonal to the potential gradient will typically result in no inflation.)

%Non-slow-roll.

%Our construction can be applied to study manyfield dynamics in the post-inflationary reheating phase, which may lead to further insights into the 
%evolution and fate of isocurvature perturbations and  non-Gaussianities. 
%Moreover, by truncating the potential at a very high order, $n_{\rm max} \gg 1$, it is possible to use our method to describe large potential `landscapes', at least for models with a few fields (cf.~Figure \ref{fig:2dpotential}). These potentials can be large enough to include false vacua, inflationary regions, and `final' vacua with small  cosmological constants, and may serve as testing grounds for theoretical ideas.   

By truncating the potential at a very high order, $n_{\rm max} \gg 1$, it is also possible to use our method to describe large potential `landscapes' for models with a few fields (cf.~Figure \ref{fig:2dpotential}). This can be applied to study manyfield dynamics in the post-inflationary reheating phase, which may lead to further insights into the evolution and fate of isocurvature perturbations and non-Gaussianities. With a larger landscape it becomes possible to follow both the fields and their perturbations through the inflationary phase and then down to the vacuum. This has however has been left for future work. These high-order, low-dimensional potentials can also be large enough to include false vacua, inflationary regions, and `final' vacua with small  cosmological constants, and may serve as testing grounds for various other theoretical ideas.

Finally, we have limited our studies to geometrically flat field spaces ($\phi \in \mathbb{R}^{\Nf}$), and it would be very interesting to  include derivative interactions and non-trivial field space geometries. 
Highly curved field spaces can have strong impact on the background dynamics (see e.g.~\cite{Silverstein:2003hf, Achucarro:2010jv,
Renaux-Petel:2015mga, Renaux-Petel:2017dia,
1612.04505, Brown:2017osf}), but already  weak interactions
can lead to bispectrum signals beyond the local shape. For example, interactions between a single, light, self-interacting isocurvature field, $\Sigma$, and the inflaton, $\Phi$,
of the form, 
\be
{\cal L}_{\rm mix} = - \frac{1}{2} \frac{\Sigma}{\Lambda} (\partial \Phi)^2 \, ,
\ee
can lead to large amplitude non-Gaussianities, even if %beyond the local shape of the bispectrum. 
$\Lambda\gg H$  \cite{Green:2013rd, Assassi:2013gxa}.
 %It would be very interesting to 
%It would be very interesting to understand the corresponding constraint in more general models of interacting manyfield inflation. 
%
% to a broader set of bispectrum shapes that may be constrained by CMB observations.
Finally, these models could be extended to include non-minimal couplings of the fields to gravity, cf. e.g.~\cite{Kaiser:2013sna, Schutz:2013fua, DeCross:2015uza, DeCross:2016fdz, DeCross:2016cbs}.

%In sum, our construction provides a new window to manyfield models of inflation subject to general non-derivative interactions. 

In closing, we note that the 
method developed in this paper opens a new window towards general models of inflation with many fields subject to non-trivial interactions. Our 
findings speak for the robustness  of the inflationary paradigm:  adiabatic, approximate scale-invariant and nearly Gaussian perturbations are commonly regarded as the `generic predictions' of single-field slow-roll inflation. In this paper, we have shown that even highly complex models of manyfield inflation
%and random interactions
produce very similar predictions. 
%These results may inform the interpretations of future cosmological observations. 
We anticipate that these results can be useful in interpreting the outcomes of future cosmological observations. 
%We have furthermore showed that drastically different constructions of manyfield inflationary models produce equivalent predictions. 
%This suggests that the simple predictions of the inflationary paradigm may be emergent consequences of complex inflationary systems. 

 %Constraining $\fnl$ to be smaller than order unity would not rule out multifield inflation, but a measurement of a large value for $\fnl$ would on the other hand
 
%  Consequently, to differentiate between multifield and single field inflation it may become necessary to go beyond the most basic observables, $n_s$ and $\fnl$. It is possible that the full bispectrum would be more useful in differentiating between different models, but there are other things one could look at, such as presence of primordial isocurvature modes or signs of particle production in the power spectrum. The latter is one we intend to pursue in future research.

\section*{Acknowledgements}
We are very grateful to Mustafa Amin, Tobias Baldauf, Daniel Baumann,  James Fergusson,  Bogdan Ganchev, Alan Guth, Liam McAllister, Christopher Moore, Sonia Paban, Enrico Pajer and Paul Shellard for stimulating discussions. 
We would in particular like to thank Thomas Bachlechner, Mafalda Dias and Jonathan Frazer for interesting discussions and valuable comments on a draft of this paper. 
TB is funded  by an STFC studentship at DAMTP, University of Cambridge. 
DM is funded by  Stephen Hawking Advanced Fellowship from the Centre for Theoretical Cosmology, DAMTP, University of Cambridge.

\appendix

\section{Formulae for the covariance matrices  \label{section:formulae}}
Once all the lower-order Taylor coefficients of the same type have been fixed, the covariance matrix is a diagonal matrix where the entries are given by a simple combinatorial factor. This factor is the product of the factorials of the number of times each number appears in the set of indices. Equivalently, this is just the number of ways one can pair up numbers of the same values from two identical sets of indices.

Every time we fix some Taylor coefficients, we need to shift the expectation values of the higher-order Taylor coefficients, as shown in equation \ref{eq:conditional}. Let us for convenience denote the matrix corresponding to $\Sigma_{21}\Sigma_{11}^{-1}$ by $E_{\alpha\beta}$, where $\beta$ corresponds to some set of indices at the order that was just fixed and $\alpha$ corresponds to some higher-order indices. For $E_{\alpha\beta}$ to be non-zero it must be possible to simultaneously pair up every index in $\beta$  with an identical index in $\alpha$. For a given sets of indices $\alpha$ and $\beta$, the value of the component $E_{\alpha\beta}$ is determined as follows:
%will be given by:
\begin{enumerate}
	\item Multiply the number of ways the indices in $\alpha$ can be paired with identical indices in $\beta$ by the number of ways the remaning indices in $\alpha$ can be paired up with each other (again, indices can only be paired up with others of the same value).
	\item There is an overall minus sign if the orders differ by an odd multiple of two.
	\item Finally divide by the above-mentioned combinatorial factor for $\beta$. 
\end{enumerate}
These matrices are generally sparse, and remain easy to use and store even as the number of fields becomes large.

\section{Numerical method}
\label{sec:numpert}

\subsubsection*{Background evolution}

Equation 
\eqref{eq:SR} 
comprises a set of $\Nf$ coupled, non-linear first-order ordinary differential equations, and its general solution is not known. For large $\Nf$, solving it numerically can also be challenging: the right hand side may involve many of millions of terms encoding the various interactions between the fields. We now discuss our method for evolving the fields.

To solve equation \eqref{eq:SR}, we approximate the full potential in very small regions around the trajectory to quadratic order, and solve  the evolution of the background in a step-by-step manner with a multiderivative method.  
%
%is straightforward to solve numerically; it can be solved step by step with series expansion. 
More precisely, for some small e-fold step $\Delta N$, we write the solution as,
\eq{
\Delta\phi_a(\Delta N)=\Delta\phi_a^{(1)}\Delta N+\frac12\Delta\phi_a^{(2)}\Delta N^2 + {\cal O}(\Delta N^3) \, .
}
Substituting this Ansatz into the slow-roll equations and matching order-by-order in $\Delta N$, we find,
\bea
\Delta\phi_a^{(1)}&=&-\frac{V_a}{V} \, , \\
 \Delta\phi_a^{(2)}&=&\frac{V_{ab}V_b}{V^2}-\frac{V_aV_bV_b}{V^3} \, .
\eea
When implementing this solution, it is of course important to make sure that the second order term in the solution is much smaller than the first-order term; otherwise the series is not a good approximation to the solution. 
The number of small patches needed depends on the individual realisation, and on hyperparameters such as the initial spectrum, and the number of fields. 
For 50 fields, we typically find that breaking up the inflationary trajectory to around 2000 small patches suffices to keep the step-size small enough for this method to be numerically accurate. 

When working with potentials with many millions of interaction terms, finding the local values of the Taylor coefficients in each patch can be come computationally intensive. To ameliorate this problem, we approximate the fifth order Taylor expanded potential in moderately small regions by  lower order Taylor series involving fewer terms. For 50 fields, we may 
approximate the potential to fourth order in  around 60 such moderately small regions for each inflationary realisation. 
We then use this lower order potential to compute the second order Taylor coefficients in the very small local patches used in the solution of equation \eqref{eq:SR}. We note however that the calculations of non-Gaussianity are very sensitive to numerical errors, so care is needed to ensure that the computation is sufficiently accurate.

%
%This approach works very well with the way the superhorizon evolution of the perturbations are computed, which will be explained in the next sub-section. 
%The only issue is that as the number of fields is increased, the potential becomes increasingly hard to work with, in due to the number of fifth-order terms. With 100 fields, there are over 92 million independent fifth-order terms, which drastically slows down the evaluation speed. To get around this, it is convenient to 

\subsubsection*{Perturbations}

As first pointed out in \cite{DBM2, DBM3}, the transport method is easily implemented in a patch-by-patch manner. 
The propagator $\Gamma_{ab}$ transports the perturbation between spatially flat hypersurfaces
and satisfies the chain rule,
\begin{equation}
\Gamma_{ab}(N_3, N_1)=\Gamma_{ac}(N_3, N_2)\Gamma_{cb}(N_2, N_1) \, .
\end{equation}
In a sufficiently  small patch, say $N_2-N_1= \Delta N \ll 1$ the propagator \eqref{eq:GammaabSol} simplifies to,
\be
\Gamma_{ab}(N_2, N_1) = {\rm exp} \left( \Delta N\, u_{ab} \right) \, ,
\ee
where $u_{ab} = u_{ab}(N_1+ \Delta N/2) \approx u_{ab}(N_1) \approx u_{ab}(N_2)$.
The full propagator from horizon crossing to the end of inflation is then obtained by left-multiplication of all subsequent propagators \cite{DBM2, DBM3}, i.e.
\be
\Gamma^{\rm tot}_{ab} (N_{\rm end}, N^\star) = \Gamma_{a c_p}(N_{\rm end}, N_p)\,  \Gamma_{c_p c_{p-1}}( N_p, N_{p-1}) \ldots
\Gamma_{c_2 c_{1}}( N_2, N_{1})\, 
 \Gamma_{c_1 b}(N_{1}, N^\star) \, .
\ee 

%This allows us to split the trajectory up into many parts, within each the potential and its derivatives are more or less constant, and
%\begin{equation}
%\Gamma^{i\to i+1}_{ab}\simeq\exp\left(u^i_2\delta N^i\right)_{ab}
%\end{equation}
%to leading order, where we assume $u_{ab}$ ($u_2$) to be constant in the step. Note that this imposes a further constraint on the step-size on the ODE solver.
%If $u_{ab}$ is assumed to increase linearly in the step, one can use the Baker-Campbell-Haussdorff formula to calculate the leading order correction to this. The result is
%\begin{equation}
%\Gamma^{i\to i+1}_{ab}\simeq\exp\left(u^i_2\delta N^i+\frac{1}{12}[\delta u_2,u_2](\delta N^i)^2\right)_{ab},
%\end{equation}
%where $u_2$ is evaluated in the middle of the step, and $\delta u_2$ is the change in $u_2$ over the step.
%The total $\Gamma_{ab}$ coefficent is then given by combining all the $\Gamma^{i\to i+1}_{ab}$.

Similarly,  $\Gamma_{abc}$ can be simplified by splitting  up the integral \eqref{eq:GammaabcSol} into many parts,
\bea
\Gamma_{def}(N_{i+1}, N_i) &=& 
\int^{N^{i+1}}_{N^{i}}dN'\Gamma_{d\mu}(N_{i+1}, N') u_{\mu\nu\rho}(N')\Gamma_{\nu e}(N', N_i)\Gamma_{\rho f}(N', N_i) \, , \\
\Gamma_{abc}(N, N^\star)&=&
\sum_{i=0}^{p}
\Gamma_{ad}(N, N_{i+1}) \Gamma_{def}(N_{i+1}, N_i) \Gamma_{eb}(N_i, N^\star) \Gamma_{fc}(N_i, N^\star)
\, ,
\eea
where $N_0 = N^\star$ and $N_{p+1} = N$. 
%
%$n$ is the number of steps, and
%\begin{equation}
%\Gamma^{i\to i+1}_{def}=\int^{N^{i+1}}_{N^{i}}dN'\Gamma_{d\mu}u_{\mu\nu\rho}(N')\Gamma_{\nu e}\Gamma_{\rho f}.
%\end{equation}
Assuming the step size is sufficiently small, we can with good accuracy evaluate $\Gamma_{def}(N_{i+1}, N_i)$ as,
%\begin{align*}
%\Gamma^{i\to i+1}_{def}&\simeq\int^{N^{i+1}}_{N^{i}}dN'\exp(u^i_2(N^{i+1}-N'))_{d\mu}u_{\mu\nu\rho}^i\exp(u^i_2(N'-N^i))_{\nu e}\exp(u^i_2(N'-N^i))_{\rho f}\\
%&\simeq\sum_{k=1}^n\left( I+u^i_2\frac{\delta N^i}{n}\right)_{d\mu}^{n-k}I^i_{\mu\nu\rho}\left( I+u^i_2\frac{\delta N^i}{n}\right)^{k-1}_{\nu e}\left( I+u^i_2\frac{\delta N^i}{n}\right)^{k-1}_{\rho f},\numberthis
%\end{align*}
%as long as $\Nf$ is taken to be sufficently large, with
%\begin{align*}
%I_{\mu\nu\rho}^i&=u_{\mu\nu\rho}^i\delta N^i/n+\frac12\left(u_{\mu\sigma}u_{\sigma\nu\rho}+u_{\mu\sigma\rho}u_{\sigma\nu} +u_{\mu\nu\sigma}u_{\sigma\rho}\right)(\delta N^i/n)^2+\frac16\big(u_{ae}u_{ed}u_{dbc}+\\
%&~~~~+u_{adc}u_{de}u_{eb}+u_{abd}u_{de}u_{ec}+u_{ad}u_{dec}u_{eb}+u_{ad}u_{dbe}u_{ec}+u_{ade}u_{db}u_{ec}\big)(\delta N^i/n)^3\nt
%\end{align*}
%\begin{equation}
%\Gamma^{i\to i+1}_{def}\simeq\int^{N^{i+1}}_{N^{i}}dN'\exp(u^i_2(N^{i+1}-N'))_{d\mu}u_{\mu\nu\rho}^i\exp(u^i_2(N'-N^i))_{\nu e}\exp(u^i_2(N'-N^i))_{\rho f},
%\end{equation}
\be
\Gamma_{def}(N_{i+1}, N_i)
\simeq
\int^{N^{i+1}}_{N^{i}}dN'\, e^{
\left[
(N^{i+1}-N') u_{d\mu}
\right]
}\, u_{\mu\nu\rho}(N') \, 
e^{\left[(N'-N^i)u_{\nu e} \right]}\, 
e^{\left[(N'-N^i)u_{\rho f}\right]}
\, ,
\ee
which is easily evaluated numerically. Once the $\Gamma_{ab}(N_{i+1}, N_i)$ and $\Gamma_{def}(N_{i+1}, N_i)$ have been calculated for all steps, they can be used to calculate power spectrum and $\fnl$  modes crossing the horizon at any point during inflation. Moreover, by doing the gauge transformation to a constant energy density surface (cf.~equations \eqref{eq:Na}, and \eqref{eq:Nab}) at any point during inflation, we can follow the evolution of the power spectrum and $\fnl$ on superhorizon scales. 

%, as well as the complete evolution of the power spectrum or $\fnl$ for any mode we want.

\section{Ensembles of models}
\label{app:params}
For each of the initial conditions below we ran 2,000 simulations, giving us more than 1,000 successful inflation models ($N\geq60$)  for all ICs except near certain fringes of the parameter space (e.g. very small $\Lh$ or large $\epsiloni$). The analysis of the perturbations was only done for the models which gave at least 60 e-folds of inflation.

\subsubsection*{Varying $\Nf$}

\begin{center}
    \begin{tabular}{ | l | l | l | l | l | p{3.5cm} |}
    \hline
   $\Lh$ & Mass spectrum & Gradient direction & $ \epsiloni$ & $\etai$& $\Nf$  \\ \hline
   $0.4$ & Uniform & Random & $ 2\cdot10^{-9}$ & $-10^{-4}$ &5, 10, 15, 20, 25, 30,\newline 35, 40, 45, 50\\ \hline
   $0.4$ & Compressed & Random & $1\cdot 10^{-10}$ & $-10^{-4}$ &5, 10, 15, 20, 25\\ \hline
   $0.4$ & Uniform & Aligned& $5\cdot 10^{-10}$ & $-10^{-4}$ &5, 10, 15, 20, 25, 30\\ \hline
   $0.4$ & Uniform, uplifted & Random & $ 2\cdot10^{-9}$ & $-10^{-4}$ &5, 10, 15, 20, 25\\ \hline
   $0.4$ & Compressed, uplifted & Random & $1\cdot 10^{-10}$ & $-10^{-4}$ &5, 10, 15, 20\\ \hline
    \end{tabular}
\end{center}

\subsubsection*{Varying $\Lh$}

\begin{center}
    \begin{tabular}{ | l | l | l | l | l | p{3.5cm} |}
    \hline
   $\Nf$ & Mass spectrum & Gradient direction & $ \epsiloni$ & $\etai$& $\Lh$  \\ \hline
   $10$ & Uniform & Random & $ 2\cdot10^{-9}$ & $-10^{-4}$ & 0.1, 0.2, 0.3, 0.4, 0.5,\newline 0.6, 0.7, 0.8, 0.9, 1.0\\ \hline
   $10$ & Compressed & Random & $1\cdot 10^{-10}$ & $-10^{-4}$ & 0.1, 0.2, 0.3, 0.4, 0.5,\newline 0.6, 0.7, 0.8, 0.9, 1.0\\ \hline
   $10$ & Uniform & Aligned& $5\cdot 10^{-10}$ & $-10^{-4}$ & 0.1, 0.2, 0.3, 0.4, 0.5,\newline 0.6, 0.7, 0.8, 0.9, 1.0\\ \hline
    \end{tabular}
\end{center}

\subsubsection*{50-field runs}

\begin{center}
    \begin{tabular}{ | l | l | l | l | l| l |}
    \hline
   $\Nf$ & $\Lh$ & Mass spectrum & Gradient direction & $ \epsiloni$ & $\etai$\\ \hline
   $50$ & $0.4$ & Uniform & Random & $ 2\cdot10^{-9}$ & $-10^{-4}$\\ \hline
   $50$ & $0.4$ & Compressed & Random & $ 2\cdot 10^{-11}$ & $-10^{-4}$ \\ \hline
   $50$ & $0.4$ & Uniform & Aligned& $1\cdot10^{-10}$ & $-10^{-4}$ \\ \hline
    \end{tabular}
\end{center}

\subsubsection*{Varying $\epsiloni$}

\begin{center}
    \begin{tabular}{ | l | l | l | l | l| p{4.5cm}|}
    \hline
   $\Nf$ & $\Lh$ & Mass spectrum & Gradient direction & $\etai$ & $ \epsiloni$\\ \hline
   $10$ & $0.4$ & Uniform & Random &  $-10^{-4}$ & $2\cdot10^{-10}$, $5\cdot10^{-10}$, $1\cdot10^{-9}$, $2\cdot10^{-9}$, $5\cdot10^{-9}$, $1\cdot10^{-8}$, $2\cdot10^{-8}$ \\ \hline
   $10$ & $0.4$ & Compressed & Random &  $-10^{-4}$ & $1\cdot10^{-11}$, $2\cdot10^{-11}$, $5\cdot10^{-11}$, $1\cdot10^{-10}$, $2\cdot10^{-10}$, $5\cdot10^{-10}$, $1\cdot10^{-9}$ \\ \hline
   %$10$ & $0.4$ & Uniform & Aligned&  $-10^{-4}$ & 2 \\ \hline
    \end{tabular}
\end{center}

\subsubsection*{Varying $\etai$}

\begin{center}
    \begin{tabular}{ | l | l | l | l | l| p{4.5cm} |}
    \hline
   $\Nf$ & $\Lh$ & Mass spectrum & Gradient direction & $ \epsiloni$ & $\etai$\\ \hline
   $10$ & $0.4$ & Uniform & Random & $ 2\cdot10^{-9}$ & $-10^{-4}$, $-10^{-3}$, $-10^{-2}$,$-5\cdot10^{-2}$, $-10^{-1}$\\ \hline
   $10$ & $0.4$ & Compressed & Random & $ 2\cdot 10^{-11}$ & $-10^{-4}$, $-10^{-3}$, $-10^{-2}$,$-5\cdot10^{-2}$, $-10^{-1}$\\ \hline
  % $10$ & $0.4$ & Uniform & Aligned& $1\cdot10^{-10}$ & $-10^{-4}$, $-10^{-3}$, $-10^{-2}$,$-5\cdot10^{-2}$, $-10^{-1}$\\ \hline
    \end{tabular}
\end{center}

\section{A single-field toy model \label{subsection:toymodel}}
\label{sec:singlefieldmodel}
We have seen in section \ref{sec:Planck} that the observational predictions of manyfield models of inflation coincide with some of the `generic predictions' of single-field slow-roll inflation: an approximately scale-invariant power spectrum over observable scales, with a small running of the spectral index. While we have also seen in section \ref{sec:iso} that multifield are typically important in the full manyfield models, it is interesting to investigate the extent to which our results can be understood through simpler single-field models. Such models may capture the most important aspects of the more complicated manyfield models, but are simple enough  to admit an analytic treatment. %{\color{blue} Possibly mention DBM case.}
In this section, we construct such simple class of single-field models, and discuss how its predictions compare against our numerical simulations of the full manyfield models.

%To gain some insight into the dynamics of these random inflation models, we look at a single-field inflation toy model where some analytic results can be obtained.

 We expand the single-field potential to cubic order around the approximate critical point at $\phi =0$,
\begin{equation}
V(\phi)=V_0\left(1-c_1\phi-\frac{c_3}{3!}\phi^3\right) \, .
\label{eq:Vsingle}
\end{equation}
where the $c_i$ all are positive. We have here set the second order term at $\phi =0$ to zero, since, as we will justify below, this term is overwhelmed by the third order term already for small field displacements. 
%
%We will see that neglecting the second-order terms in the potential is well-motivated already for very small field discplacements
%We truncate the Taylor expansion at third order since we expect it to be this term which will cause inflation to end, as will be verified shortly. 
%
%Furthermore, the reason we take these constant to be positive, making the derivatives negative, is that in the multifield case there will almost always be some direction in which the second and third-order Taylor coefficients are negative.
%
% It is a good approximation where either the gradient of the potential is aligned with the most tachyonic direction, or all the masses of the fields are of order $\etai V/\Mp^2$. As we will see, this model will give a good approximation of the background dynamics, but not the perturbations. 
%
%
%and the single-field model that we will consider in the rest of this section is then given by,
 %\begin{equation}
%\label{eq:Vsingle2}
%\end{equation}
We furthermore assume  that the potential remain approximately constant during inflation, $V \approx V_0$, which simplifies the analytic expressions for the slow-roll parameters in this model:
\begin{align}
\epsilonV&\simeq \Mp^2 \frac{(c_1+c_3\phi^2/2)^2}{2} \, ,& \etaV&\simeq-\Mp^2c_3\phi \, .
\end{align}
%and we see that i

We will now use this model to compute the expected total number of e-folds generated during inflation and the predictions for the spectral index and its running. 
We first note that inflation ends when 
$\epsilonV  \approx -\dot H/H^2 = 1$, which happens when the value of the field is, 
\be
\phi^\text{end}\equiv\frac{2^{3/4}}{\sqrt{(c_3\Mp)}} \, .
\ee
The number of e-folds generated as the field travel from $\phi=0$ to some value $\phi$ is given by,
\begin{equation}
N(\phi)=\int_0^{\phi} \frac{d\phi'}{\Mp\sqrt{2\epsilonV}}=\frac{\tan^{-1}(\sqrt{c_3/2c_1}\phi)}{\Mp^2\sqrt{c_1c_3/2}}\label{eq:Nofphi} \, .
\end{equation}
%
%If we substitute in $\phi^\text{end}$ into
Upon evaluating equation 
 \eqref{eq:Nofphi} for $\phi = \phi^{\rm end}$, we see that the argument of the inverse tangent function becomes very large so that $\tan^{-1}(\sqrt{c_3/2c_1}\phi^{\rm end}) \approx \pi/2$, and the total number of e-folds is approximately given by,
\begin{equation}
N^\text{end}\approx N^\text{max}\equiv\frac{\pi}{\Mp^2\sqrt{2c_1c_3}} \, .
\label{eq:Nend}
\end{equation}
Inflation ends before $N$ becomes exactly $N^\text{max}$, but for the initial conditions we are interested in, it is a good approximation.

%Equation \eqref{eq:Nofphi} can of course also be integrated backwards, and one finds that the whole trajectory would give twice that number of e-folds. There is a caveat to the last statement, however. In the multifield case, the inflaton field may have reached the starting point by rolling down in a direction with a heavy mass. If this is the case, then we expect fewer e-folds before than after.

We can of course also invert equation \eqref{eq:Nofphi}  to give $\phi$ as a function of $N$:
\begin{equation}
\phi(N)=\sqrt{\frac{2c_1}{c_3}}\tan\left(\frac{\pi N}{2N^\text{max}}\right) \, ,\label{eq:phiofN}
\end{equation}
which is valid for $N \leq N^{\rm end} < N^{\rm max}$. 
%
%There appears to be a divergence here, but recall that $N^\text{end}< N^\text{max}$, so inflation ends and our approximation becomes invalid before this happens.
%
%We can also see that the contribution from the third derivative will come to dominate the gradient after half that number of e-folds, and using the same numbers as before, we find that the second derivative will become dominated by the contribution from the third derivative within a fraction of an e-fold. This means that our approximation appears to be consistent, and can expect $\epsilonV$ to remain very small throughout the majority of the trajectory. Indeed, one can check that inflation will actually already have ended before $\epsilonV=\etaV$. We can consequently safely treat $\epsilonV$ small compared to $\etaV$.
%
Using equation \eqref{eq:phiofN}, it is straightforward to compute the spectral index and its running analytically for this toy model. 
For the models we are interested in, 
 $|\etaV|\gg\epsilonV$, and the spectral index is given by,
\begin{equation}
n_s-1=2\etaV-6\epsilonV\simeq-2\Mp^2c_3\phi_\star=-\frac{2\pi}{N^\text{max}}\cot\left(\frac{\pi\Delta N}{2N^\text{max}}\right),
\label{eq:nssingle}
\end{equation}
where we defined $\Delta N=N^\text{max}-N_\star$ (in our multifield simulations, we take $\Delta N = 55$). For $N^\text{max} \geq \Delta N$, the spectrum is red and $n_s$ has the limit $1-4/\Delta N$ as $N^\text{max}\to\infty$, and it is easy to see that this is a lower bound. Using ${\rm d}\ln k\simeq {\rm d} N_\star$, we find that the running is given by,
\begin{equation}
\alpha_s=\frac{dn_s}{dN_\star}=-\left(\frac{\pi}{N^\text{max}}\right)^2\csc^2\left(\frac{\pi\Delta N}{2N^\text{max}}\right),
\end{equation}
which is manifestly negative and has the limit $-4/\Delta N^2$ as $N^\text{max}\to\infty$, which is an upper limit.

We are now interested in comparing this class of single-field models to  the full  multifield models with  potential \eqref{eq:Taylor1}. To do so, we identify $V_0=\Lv^4$ and $c_1=\sqrt{2\epsiloni}\Mp^{-1}$.  A non-vanshing second-order term could be identified with $c_2=|\etai|\Mp^{-2}$. 
The coefficient $c_3$ then corresponds to a randomly generated third-order  derivative, which, as we will detail below, we take to be of  ${\cal O}(\Lh^{-3})$. Already for small field displacements, $\Delta \phi/\Lh \gtrsim |\etai| (\Lh/\Mpl)^2$, the third derivative comes to dominate over the second order term. This justifies dropping the second order term from the potential. To see roughly how $N^\text{end}$ scales with the various parameters, we fix $c_3$ to,
\begin{equation}
 c_3=\frac{ \sqrt{\sum_a \langle V_{a11}^2 \rangle }}{\Lv^4 }
 \approx
 \sqrt{2 \Nf} \frac{1}{\Lh^3}
 \, ,
\end{equation}
where we have denoted the initial gradient direction by `1'.\footnote{ The approximation comes from taking the contribution from $a=1$, $\langle(V_{111}^i)^2\rangle=6\Lv^8\Lv^{-6}$,  to be the same as for $a\neq 1$, $\langle(V_{a11}^i)^2\rangle=2\Lv^8\Lv^{-6}$. Note that since $V_a$ is already fixed to be very small, the (conditional) mean of $V_{abc}$ is zero to a very good approximation.} Since we are using the rms value of the third order coefficients in the multifield model to fix $c_3$, we expect that predictions made from the single-field model may capture the mean values of $N^{\rm end}$ (up to some $\mathcal O(1)$ coefficient), which in turn tells us how the mean values of $n_s$ and $\alpha_s$ will scale.\footnote{Of course, to find the mean values of all these quantities, one should write them as funtions of the $V_{a11}$ and integrate over the PDF. Since our single-field model makes several approximations, however, there is no need to work with such precision (but we did check that for $N^\text{end}$ the answer is very close).}

Plugging  our expressions for $c_1$ and $c_3$ into equation \eqref{eq:Nofphi}, we see that the  total number of e-folds of the single field model is is given by,
\begin{equation}
 N^\text{end} \approx \frac{\pi}{2}\frac{1}{( \Nf \epsiloni)^{1/4}}\left(\frac{\Lh}{\Mp} \right)^{3/2}\, .\label{eq:NEV1}
\end{equation}

By comparing equation \eqref{eq:NEV1} to the results of the numerical simulations plotted in Figure \ref{fig:Nend}, 
we see that the scaling of $N^{\rm end}$ with $\Lh$, $\Nf$ and $\epsiloni$ are not followed very closely. 
%
%these scalings are not followed very closely at all by the models with flat spectra and random initial gradient. 
However, in the more special cases when we start with the gradient aligned with the smallest eigenvalue or when we use the compressed spectrum, these scalings are reasonably accurate (but the $\mathcal O(1)$ coefficient is incorrect). %where 
%In these cases, 
%the initial mass matrix has little effect on the background evolution

Figure \ref{fig:Nendns19000} shows the single-field prediction of equation \eqref{eq:nssingle} together with the numerical simulations from the full GRF models. 
%
%The predicted form of $n_s(N^\text{end})$ 
%
Qualitatively, the single-field model is in good agreement, and captures both the decrease of $n_s$ for $N^{\rm end}$ not too large, and its asymptotical constancy for $N^{\rm end} \gg 60$. However, the precise predictions for $n_s$ are inaccurate. The single-field  limit for $\alpha_s$ is quite close to value we observe for the multi-field models. For $\Delta N=55$, the value is $\alpha_s=-0.00132$, which  agrees with the baseline model prediction, $\alpha_s=-0.00143\pm0.00034$.

Altogether, we see that the single-field toy model captures several of the qualitative features of the multifield models, but does not produce quantitatively accurate  predictions. This is not surprising, since the single-field model neither takes into account turns of the trajectory nor the superhorizon evolution of the power spectra. To make accurate predictions, the full multifield treatment is needed.

\section{The DBM construction of random manyfield potentials}
\label{sec:DMBreview}
\label{app:DBM}
In this subsection, we briefly review the construction of random scalar field potentials using non-equilibrium random matrix theory, and we discuss the most relevant properties and predictions of these models.

A  key motivation for the construction of \cite{DBM1} is that
 inflation is  only sensitive to the scalar potential in the vicinity of the field trajectory, while being independent of its properties elsewhere in field space.  
One may take advantage of this fact by generating the  scalar potential only along the dynamically determined field trajectory by gluing together nearby patches in which the potential is locally defined up to some fixed, low order. This method avoids the steep computational cost that limited early studies of multifield inflation in GRF potentials to only involving a few fields, with structure only over super-Planckian field-space distances \cite{1111.6646}.

The starting point of the `DBM construction' is  the scalar potential defined up to quadratic order around the point $p_0$,
\be
V = (\Lvd)^4 \sqrt{\Nf} \left(
v_0 + v_a \frac{\phi^a}{\Lhd} + \frac{1}{2} v_{ab} \frac{\phi^a}{\Lhd} \frac{\phi^b}{\Lhd}
\right)
\, .
\label{eq:Vdbm}
\ee
Here, $\Lvd$ sets the vertical scale of the potential, the convention for the prefactor $\sqrt{\Nf}$ is explained in \cite{DBM1, DBM3}, and $\Lhd$ sets the horizontal scale of the potential (we will shortly return to the interpretation of this parameter). At a nearby point in field space, say $p_1$ separated from $p_0$ by $\delta \phi^a$, the potential admits a local Taylor expansion in which the coefficients $v_0$, $v_a$, and $v_{ab}$ only differ from those at $p_0$ by a small amount:
\bea
&v_0\big|_{p_1} =& v_0\big|_{p_0} + v_a \big|_{p_0} \frac{\delta \phi^a}{\Lhd} \, , ~~~~v_a\big|_{p_1} = v_a\big|_{p_0} + v_{ab} \big|_{p_0} \frac{\delta \phi^b}{\Lhd} \, ,
\nonumber
 \\
&
v_{ab} \big|_{p_1}  =& v_{ab} \big|_{p_0} + \delta v_{ab} \big|_{p_0\to p_1} \, . 
\label{eq:DBMstep}
\eea
Here $\delta v_{ab}$ captures the effects of cubic (and higher-order) terms on the second derivatives of the potential. Clearly, by stipulating the rules for how $\delta v_{ab}$ is generated, any potential may be locally generated in this fashion.  In a given small patch, the slow-roll equations for the background and the evolution equations for the perturbations are easily solved, making it  possible to follow the evolution of the system along a string of points, $p_0, p_1, p_2,$ etcetera, on the dynamically determined inflationary trajectory. By repeating the procedure of \eqref{eq:DBMstep}, the potential is  `charted' as the field evolves. 

The prescription for constructing $\delta v_{ab}$ determines the generated potential. To study multifield inflation with randomly interacting fields, reference  \cite{DBM1} considered a stochastic evolution law for $\delta v_{ab}$, leading to an ensemble of random scalar potentials  for each initial choice of parameters. In \cite{DBM1}, the law governing the generation of $\delta v_{ab}$ was then chosen so that, over large distances, $v_{ab}$ samples the Gaussian Orthogonal Ensemble of random symmetric matrices. A simple example of such a law is to take the independent matrix elements of $v_{ab}$ evolve  with the Brownian motion of independent harmonic oscillators. More precisely,  the independent elements of $\delta v_{ab}$ are generated as Gaussian random numbers with the first two moments given by,
\bea
\langle \delta v_{ab} \big|_{p_i \to p_{i+1}} \rangle &=& - v_{ab}\big|_{p_i} \frac{|\delta \phi^a|}{\Lhd} \, , \nonumber \\
\langle \delta v_{ab}^2 \big|_{p_i \to p_{i+1}} \rangle &=& \sigma^2 \left( 1+\delta_{ab} \right) \frac{|\delta \phi^a|}{\Lhd} \, .
\label{eq:DBM}
\eea
This is `Dyson Brownian motion' (DBM), originally proposed as an out-of-equilibrium extension of the `Coulomb gas' statistical picture of random matrix theory. 
%{\color{blue} $\delta v_{ab}^2$ or Var($\delta v_{ab}$)} 
Given any initial configuration of $v_{ab}(0)$ at $p_0$, the DBM evolution continuously relaxes the Hessian matrix to a random sample of the GOE. The probability distribution of $v_{ab}$ then becomes a function of the path length, $s$, in units of $\Lhd$ \cite{Dyson, Uhlenbeck},
\be
P(v_{ab}(s)) \sim {\rm exp}\left[ -\frac{{\rm tr}\left((v_{ab}(s) - q v_{ab}(0))^2\right)}{2 \sigma^2(1-q^2)}\right] \xrightarrow[s \gg 1~~]{}
 {\rm exp}\left[ -\frac{{\rm tr}(v_{ab}(s)^2 )}{2 \sigma^2}\right] 
% = P_{\rm Wigner}({\cal H}(s))
 \, ,
 \label{eq:P(vab)}
\ee
where $q = {\rm exp}(-s)$. Thus, $\Lhd$ has the interpretation of the coherence length over which the Hessian randomises, the corresponding eigenvectors `delocalise', and the potential exhibit significant random structure. 

In \cite{DBM1}, the DBM construction was used to gain access to inflation in scalar potentials with multiple interacting fields, and in \cite{DBM2, DBM3}, these were used to, for the first time, study the observational predictions of manyfield models of inflation. Some modifications and extensions of this prescription were discussed in \cite{1409.5135,  1611.07059, Wang:2016kzp}, and in \cite{Freivogel} the predictions of a single-field approximation to the large-$\Nf$ DBM models was elaborated on. (Note however that the observational predictions of this single-field model
%, which include a very red spectral index, 
were already falsified in \cite{DBM2}, as subsequently discussed in \cite{DBM3}.)  

\section{Case studies and a modified GRF potential}
\label{sec:casestudies}
In the main body of this paper, we focussed on the statistical predictions of ensembles of manyfield models. In this appendix, we discuss two particular examples of randomly generated inflation models that highlight the general results discussed in this paper. We furthermore discuss the case of `uplifted' potentials mentioned in section \ref{sec:differences}.

The first inflation model we look at is a 100-field model which, despite significant superhorizon evolution of the power spectrum, gives little non-Gaussianity. 
%, and we will use it to show the general features of these models. 
The second case is a 25-field model which is one of the rare examples with significant non-Gaussianity at the end of inflation. 
%, but where the final isocurvature is non-negligible. 
It is in fact the only model with the uniform mass spectrum that we found to  produce large non-Gaussianity, and we will highlight what distinguishes this model from the others.

\subsection{A 100-field model}

\begin{figure}
    \centering
    \begin{subfigure}{0.48\textwidth}
    \centering
    \includegraphics[width=1\textwidth]{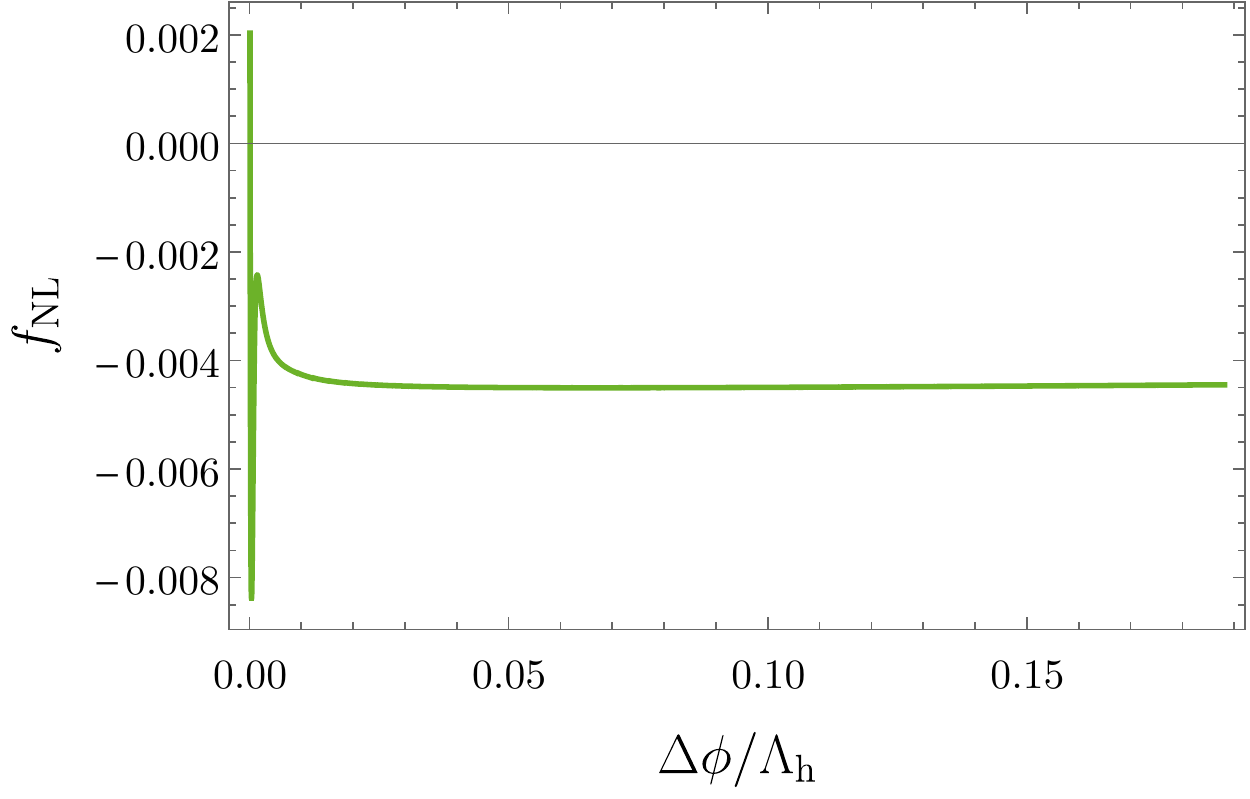}
    %\caption{$\fnl$ values from 1000 random 50-field potentials}
    \end{subfigure}
    ~ %add desired spacing between images, e. g. ~, \quad, \qquad, \hfill etc. 
    \begin{subfigure}{0.48\textwidth}
         \includegraphics[width=0.96\textwidth]{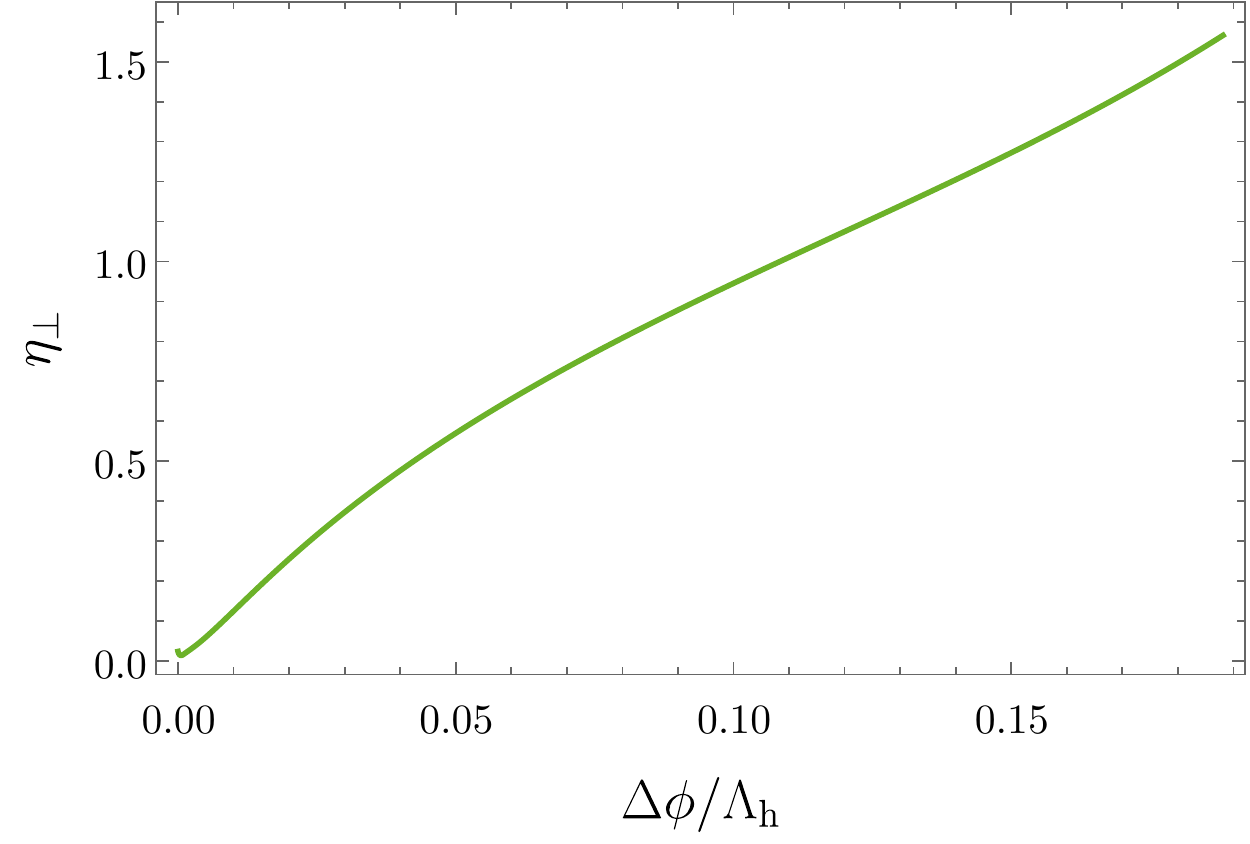}
   % \caption{Fraction of negative eigenvalues}

    \end{subfigure}
    
    \centering
    \begin{subfigure}{0.48\textwidth}
    \centering
    \includegraphics[width=1\textwidth]{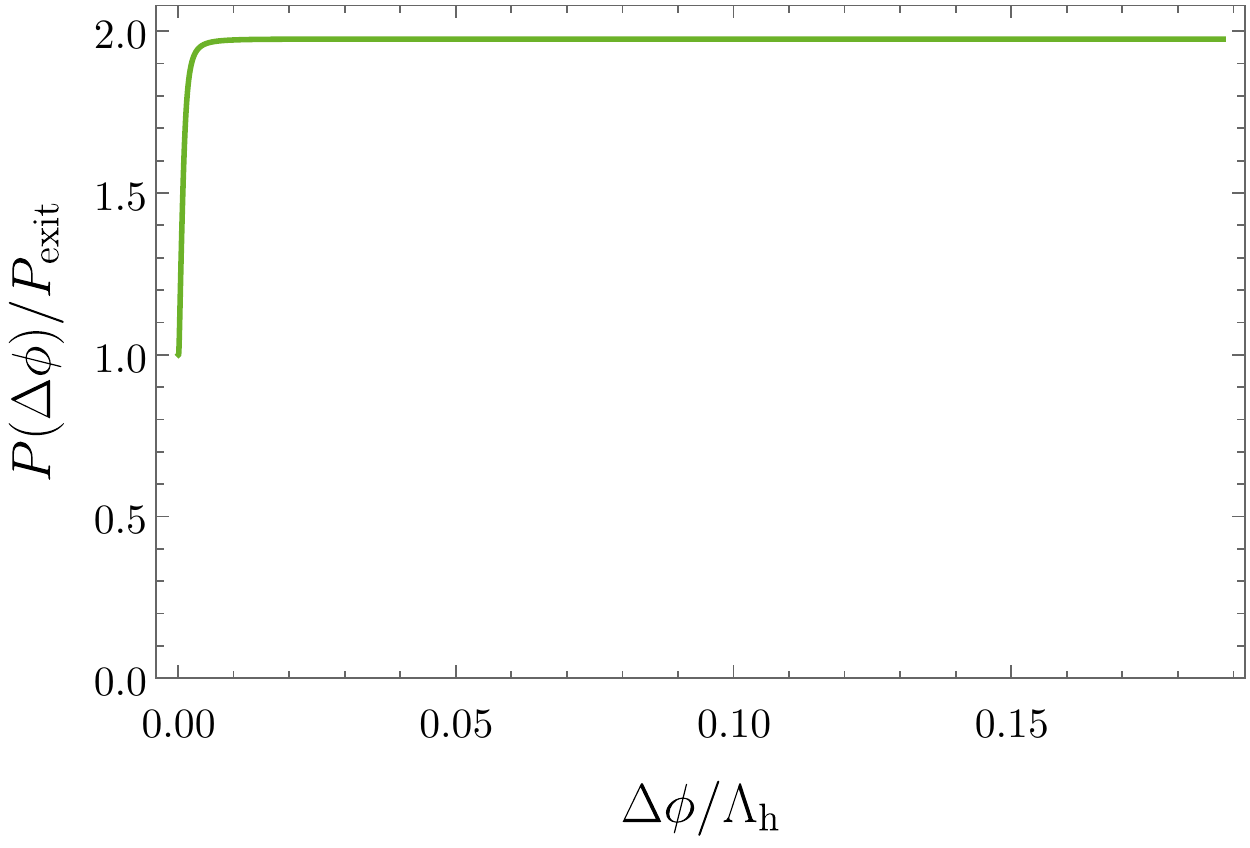}
    %\caption{$\fnl$ values from 1000 random 50-field potentials}
    \end{subfigure}
    ~ %add desired spacing between images, e. g. ~, \quad, \qquad, \hfill etc. 
    \begin{subfigure}{0.48\textwidth}
         \includegraphics[width=1\textwidth]{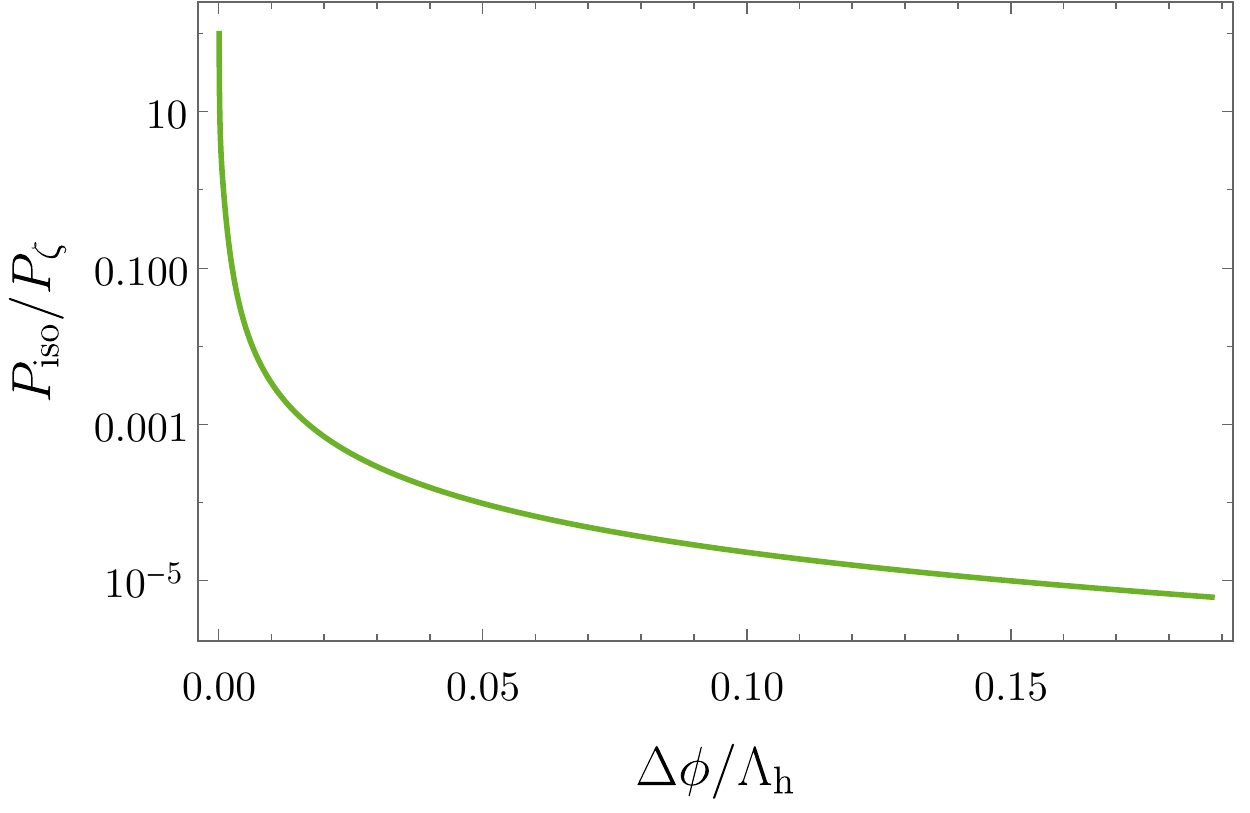}
   % \caption{Fraction of negative eigenvalues}

    \end{subfigure}
    \caption{Multifield aspects of the 100-field example.}\label{fig:100fieldgraphs}
\end{figure}
The spectrum of the random 100-field models that we discuss in this section is  shown in Figure \ref{fig:100fieldEVs}, and its background evolution if further illustrated by 
 Figure \ref{fig:100fieldVphiN}. Recall that the initial conditions for this model are $\Lh=0.4$, $\epsiloni=5\times10^{-10}$, mass-squareds evenly distributed between $\etai V_0\Mp^{-2}$ and $(3H/2)^2$ with $\etai=-10^{-4}$. 
 
 The  spectral index of this model is $n_s=0.978$, and its running is given by $\alpha_s=-0.0018$. The amplitude of  local non-Gaussianity at the end of inflation is given by $\fnl=-0.004$. % essentially consistent with zero. %As a whole, this patricular inflation model is indistinguishable from single-field slow-roll inflation.

%As Figure \ref{fig:100fieldVphiN} shows, the majority of the field movement is at the very end of inflation, and the potential only decreases a little bit during inflation. %This justifies some of the approximations made in the single-field toy model.

While the observables produced by this model are simple, the superhorizon dynamics of the perturbations is not. Figure \ref{fig:100fieldgraphs} shows that the scalar power at the pivot scale doubles after horizon crossing, and that for several e-folds, the isocurvature-to-curvature ratio is greater than one (recall that almost all the field-space movement happens towards  the end of inflation). Nevertheless, by the end of inflation the isocurvature becomes heavily suppressed, the power spectrum freezes out, and $\fnl$ remains small.

To further understand the multifield aspects of this model, we define the vector,
\eq{
\eta_{\perp\, i}=\frac{n_aV_{ab}v_b^i}{V}\, ,
}
so that, according to equation \eqref{eq:zetaprime}, $\zeta' = 2 \eta_{\perp\, i} {\cal S}^i$. Thus, the  norm  $\eta_{\perp} \equiv |\eta_{\perp i}|$ determines the strength of the isocurvature-to-curvature sourcing. Figure \ref{fig:100fieldgraphs} shows that $\eta_{\perp}$ increases during inflation in the 100 field model, and becomes ${\cal O}(1)$ towards the end of inflation. However, at this point the isocurvature has decayed exponentially, so that $\zeta$ remains essentially constant.

\subsection{A 25-field model giving large non-Gaussianity}

\begin{figure}
  \centering
    \begin{subfigure}{0.48\textwidth}
    \centering
    \includegraphics[width=1\textwidth]{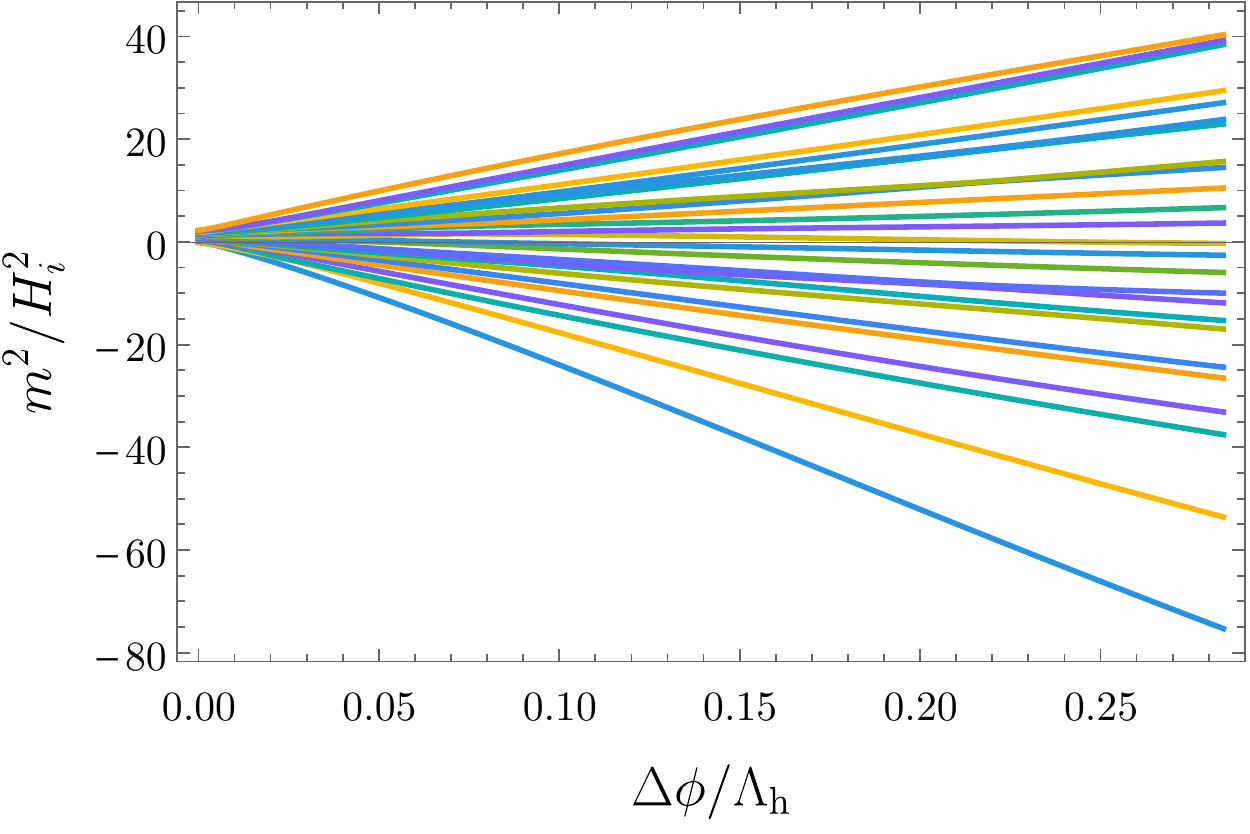}
    %\caption{$\fnl$ values from 1000 random 50-field potentials}
    \end{subfigure}
    ~ %add desired spacing between images, e. g. ~, \quad, \qquad, \hfill etc. 
    \begin{subfigure}{0.48\textwidth}
         \includegraphics[width=1\textwidth]{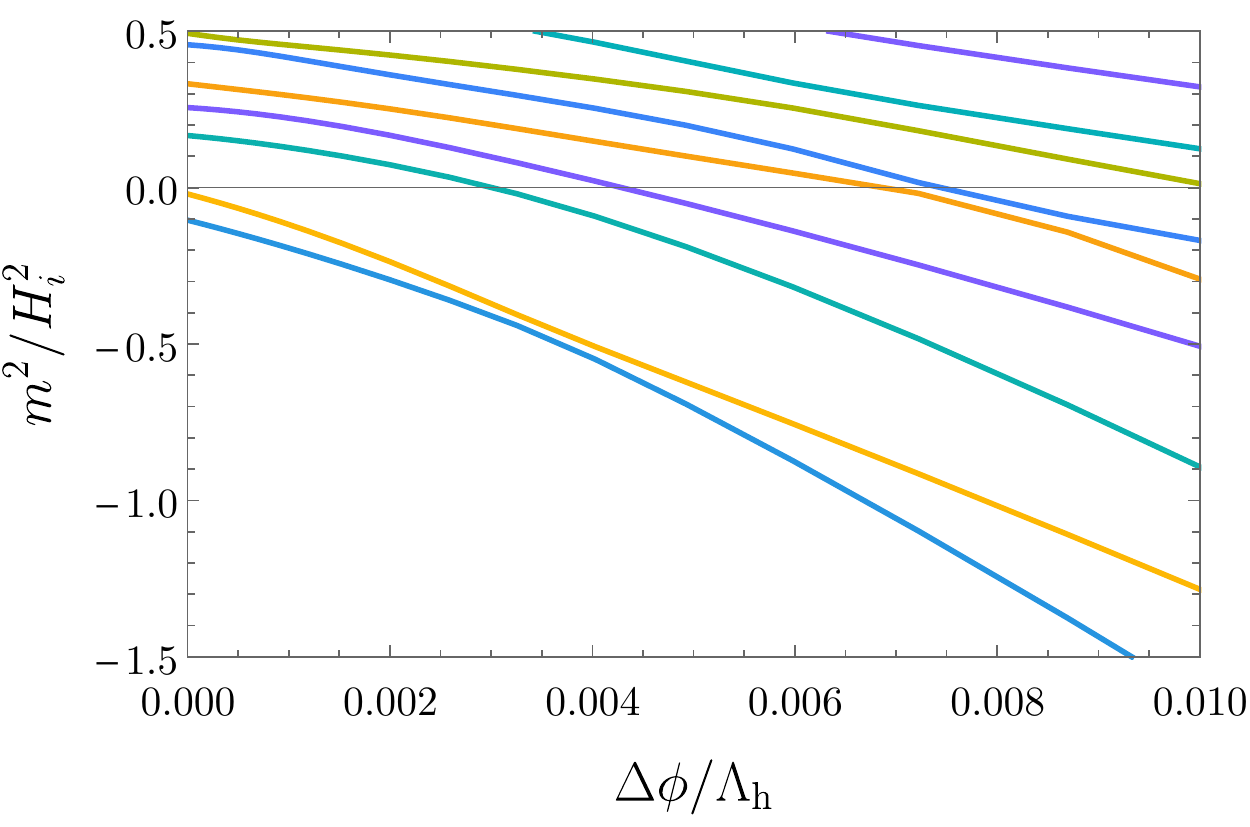}
   % \caption{Fraction of negative eigenvalues}

    \end{subfigure}
    \centering
    \begin{subfigure}{0.48\textwidth}
    \centering
    \includegraphics[width=1\textwidth]{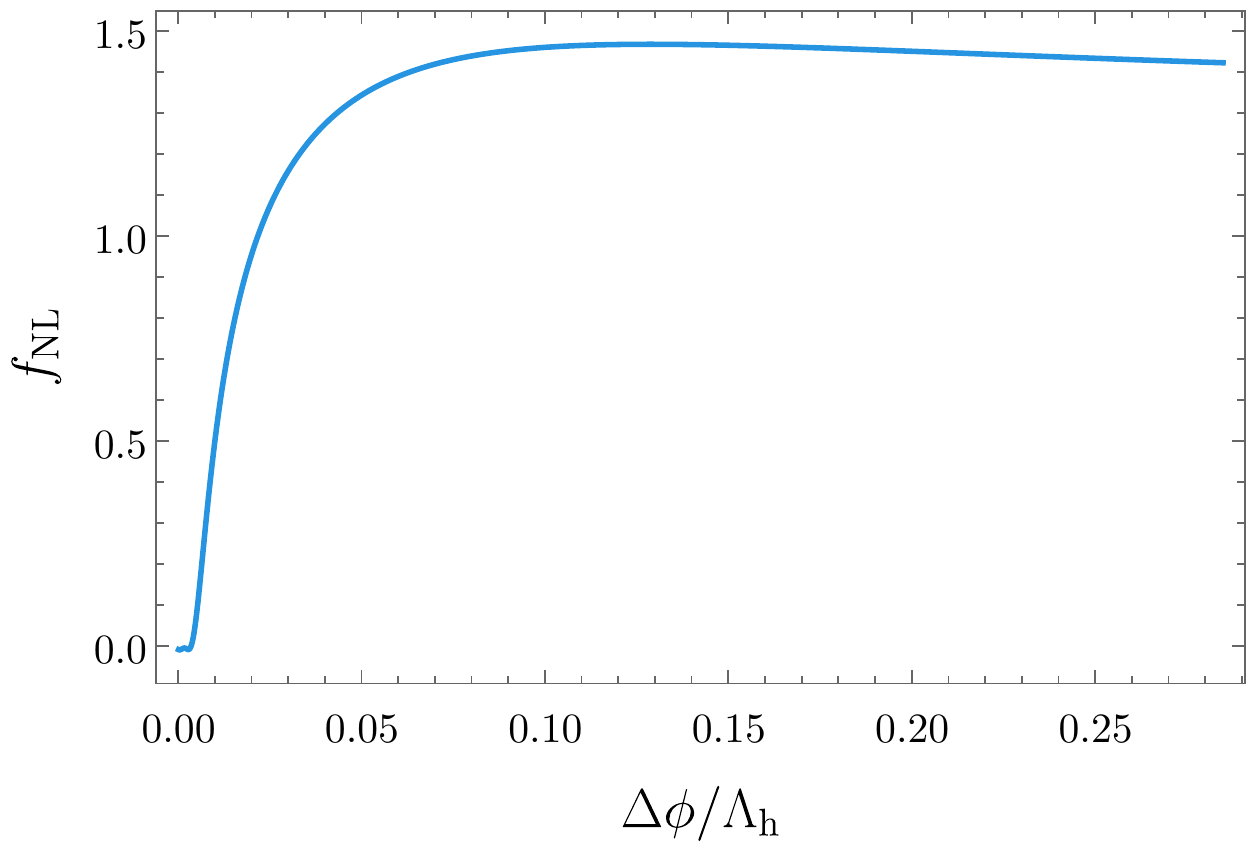}
    %\caption{$\fnl$ values from 1000 random 50-field potentials}
    \end{subfigure}
    ~ %add desired spacing between images, e. g. ~, \quad, \qquad, \hfill etc. 
    \begin{subfigure}{0.48\textwidth}
         \includegraphics[width=0.96\textwidth]{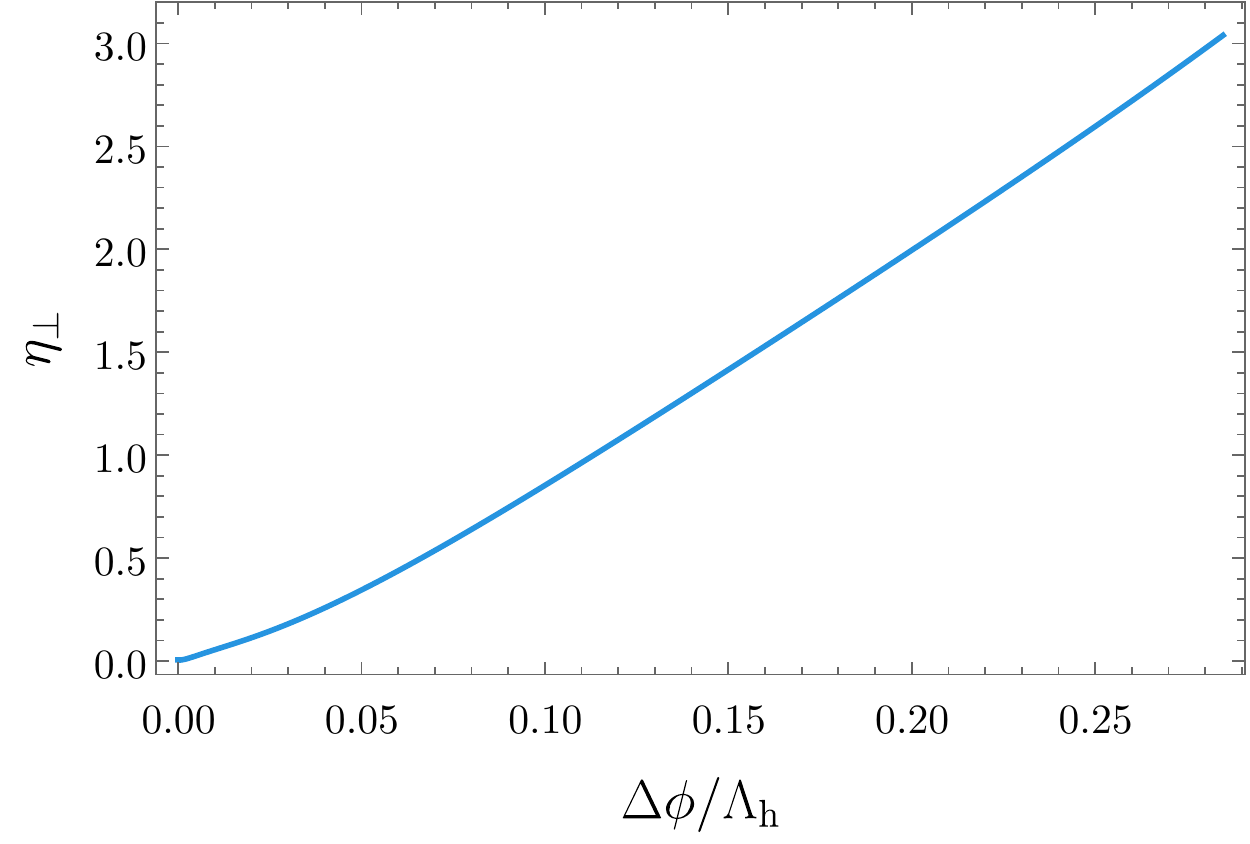}
   % \caption{Fraction of negative eigenvalues}

    \end{subfigure}
    
    \centering
    \begin{subfigure}{0.48\textwidth}
    \centering
    \includegraphics[width=1\textwidth]{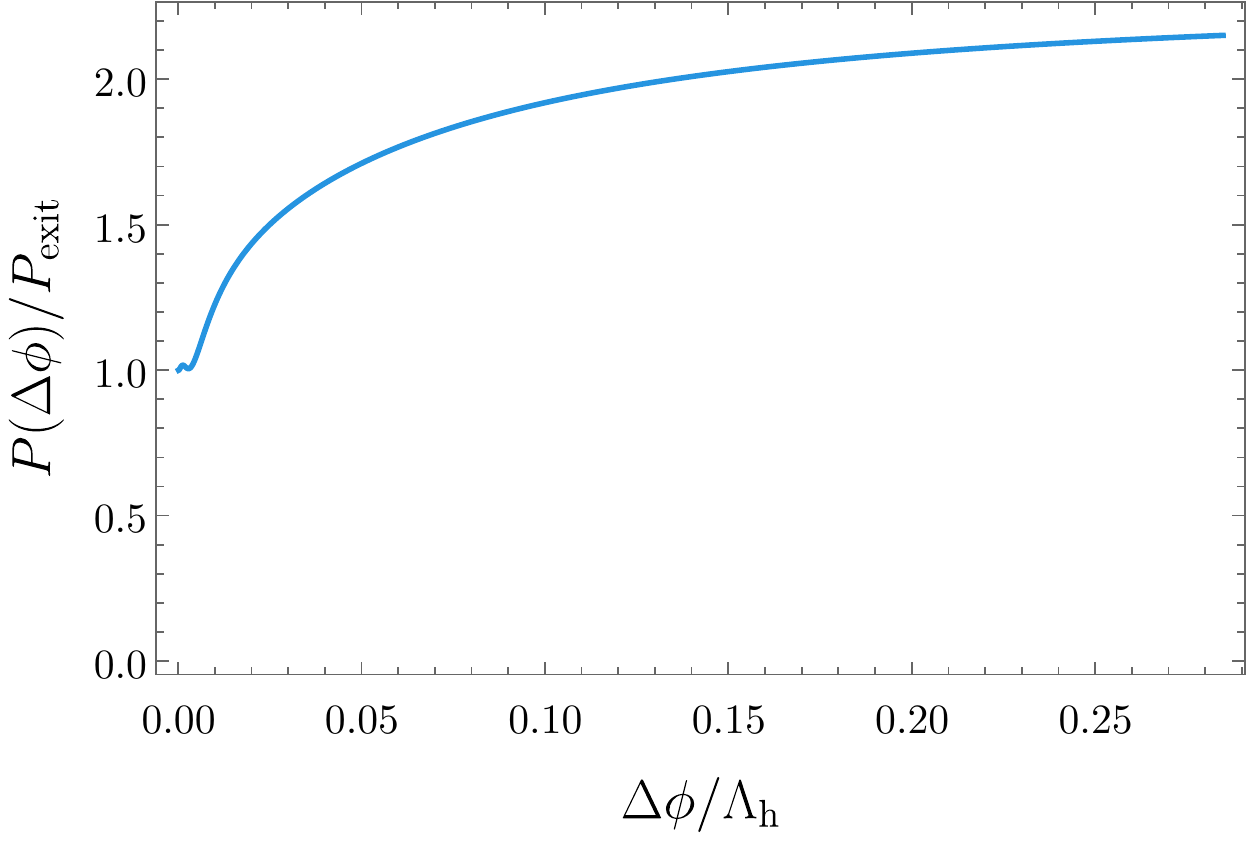}
    %\caption{$\fnl$ values from 1000 random 50-field potentials}
    \end{subfigure}
    ~ %add desired spacing between images, e. g. ~, \quad, \qquad, \hfill etc. 
    \begin{subfigure}{0.48\textwidth}
         \includegraphics[width=1\textwidth]{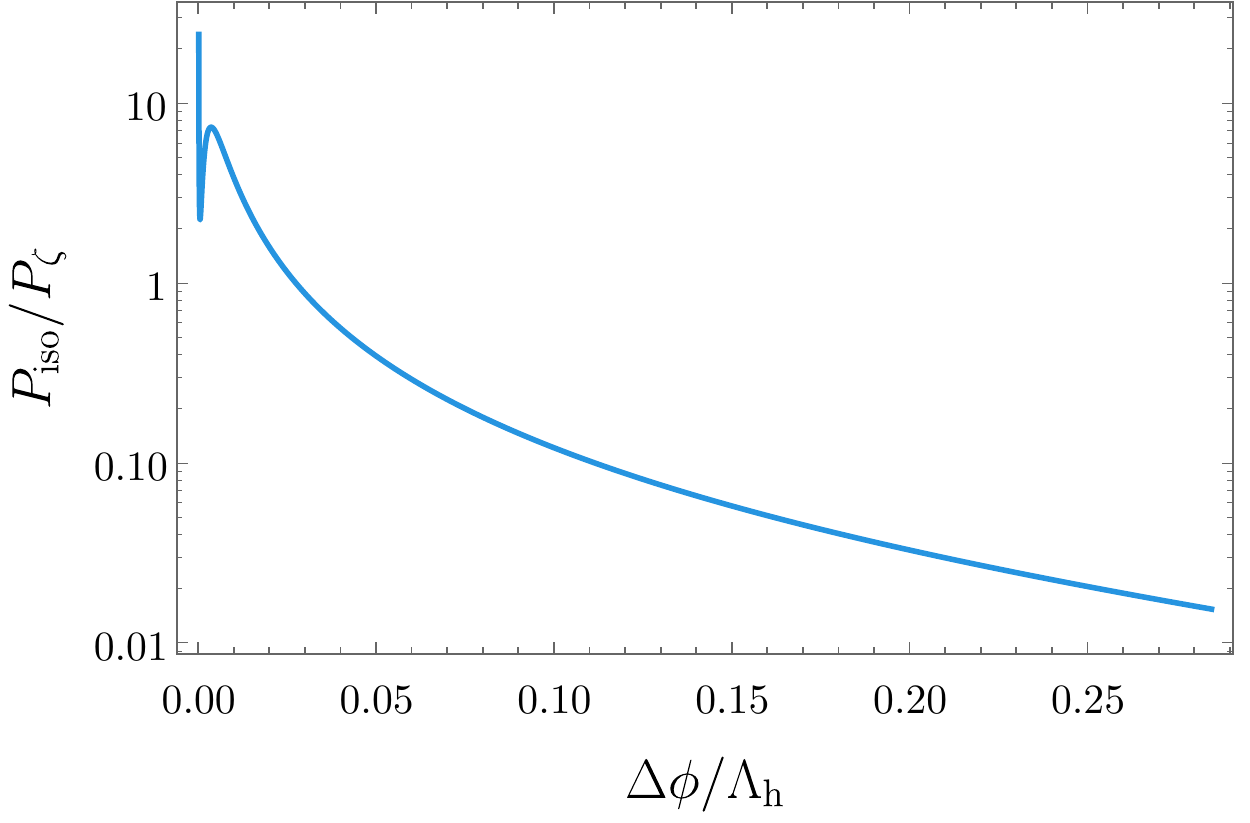}
   % \caption{Fraction of negative eigenvalues}

    \end{subfigure}
    \caption{Multifield effects in the 25-field model with large non-Gaussianity. }\label{fig:25fieldSHiso}
\end{figure}

In rare cases, 
we found 
 randomly generated manyfield models with 
 substantially non-Gaussian perturbations. 
 %large non-Gaussianities. 
 A particular example of this is a 25-field model with the uniform initial mass spectrum, $\Lh=0.4$, $\epsiloni=2\times10^{-9}$ and $\etai=-10^{-4}$. The final spectral index is $n_s=0.978$ with running  $\alpha_s=-0.0014$. The fractional increase of the power spectrum is 2.15, and the final ratio of the isocurvature and curvature power spectra is 0.015. The amplitude of local non-Gaussianities is given by $\fnl=1.42$, far above the typical values encountered.

%This model is that it was the only one with a uniform mass spectrum to give a large value for $\fnl$.
What sets this model apart from other models is that the two smallest eigenvalues of the Hessian remain close to each other throughout most of the trajectory, and even `bounce off' each other relatively early on during inflation. %This is shown, together with the rest of the graphs for this model, in
The evolution of the spectrum of the Hessian is given by the top panel of Figure \ref{fig:25fieldSHiso}. When the two eigenvalues are near each other there is significant power in the isocurvature modes and we see a drastic increase in $\fnl$. 
The generation of $\fnl$ in this model is consistent with equation \eqref{eq:PT}: as the two eigenvalues of the Hessian come very close to each other, slightly perturbed classical trajectories can become widely separated and experience drastically different levels of superhorizon evolution. 
%
%We note that $\eta_\perp$ is  not unusually large at any point in the trajectory, but %it appears that 
%there is significant power in isocurvature  to source the evolution of $\fnl$. We note moreover
We note 
 that $\eta_\perp$ grows during inflation and the isocurvature  decays rather slowly. The predictions of this model are likely to be sensitive to the physics of the post-inflationary era.

\subsection{Manyfield inflation in uplifted potentials }

For Gaussian random fields with zero mean, the minima of the potential typically appear at lower and lower values  as the number of fields is increased, and the radius of the Wigner semi-circle grows. As we live in a vacuum with a small, positive cosmological constant, it may therefore be interesting to consider GRF potentials which have been uplifted so that minima typically occur around $V=0$, cf. our discussion in section \ref{sec:differences}. From equation \eqref{eq:shiftedSemiCircle}, we see that the lower edge of the semi-circle will be at zero at $V=0$ if we lift the potential by $\sqrt{2}\Nf\Lv^4$.

\begin{figure}
    \centering
    \begin{subfigure}{0.48\textwidth}
    \centering
    \includegraphics[width=1\textwidth]{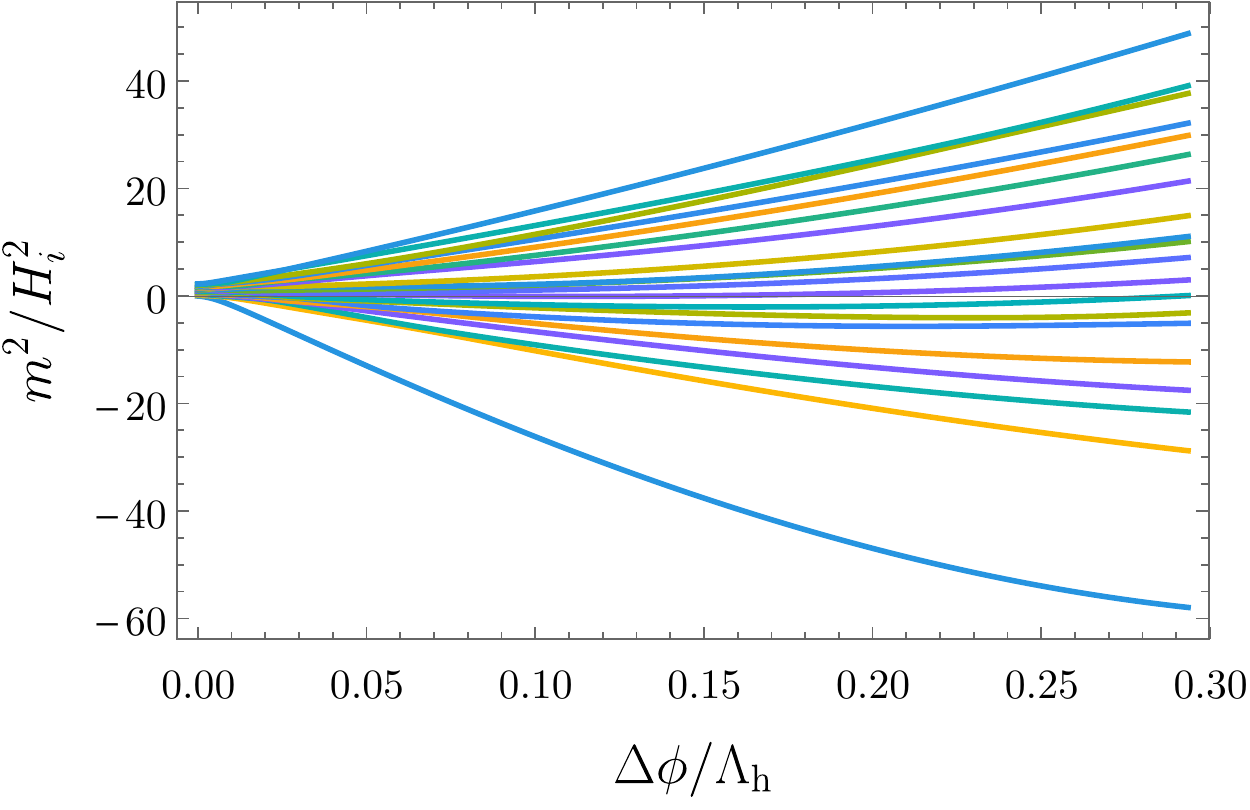}
    %\caption{$\fnl$ values from 1000 random 50-field potentials}
    \end{subfigure}
    ~ %add desired spacing between images, e. g. ~, \quad, \qquad, \hfill etc. 
    \begin{subfigure}{0.48\textwidth}
         \includegraphics[width=1\textwidth]{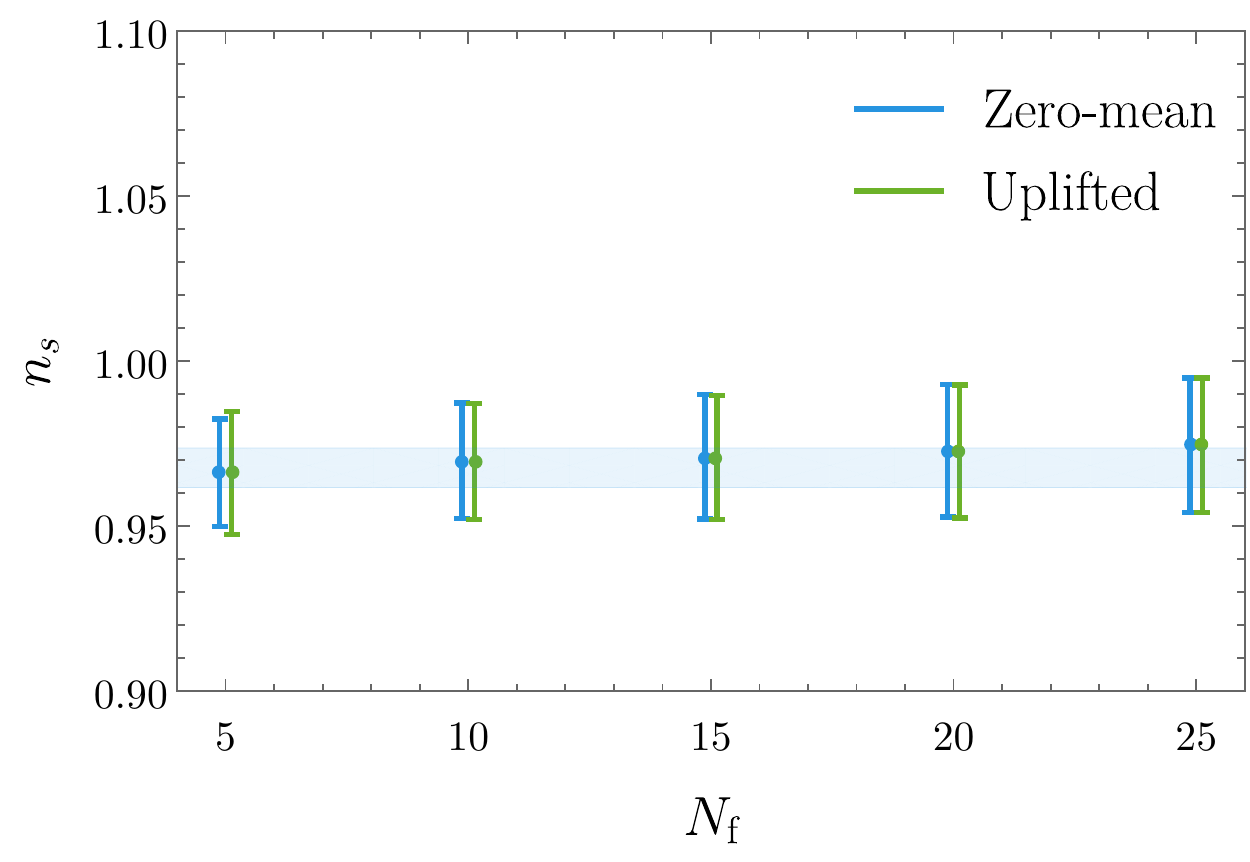}
   % \caption{Fraction of negative eigenvalues}

    \end{subfigure}
    \caption{Eigenvalue evolution for a 20-field model with uplifted potential (left) and comparison of predictions for spectral index computed from ensembles of 1000  models (right).}\label{fig:Lifted}
\end{figure}

The typical spectrum of these models is still too broad to make them useful tools to study multifield inflation, 
and we again consider the  uniform initial spectrum of equation \eqref{eq:spectra}.
%
%
%, we find that the squared masses of 
The evolution of the eigenvalues of the Hessian of
 these models behave somewhat differently from those of the simplest, mean-zero GRFs. 
%
%Since the masses are now at the bottom end of the Wigner semi-circle, the masses not only want to spread out, they also want to fluctuate upwards. 
An example of this is shown in Figure \ref{fig:Lifted}: while the dominant effect for small field values is still the eigenvalue repulsion and roughly conical spread of the eigenvalues,  for larger values of the field, the smallest eigenvalue tends upwards, towards the lower edge of the equilibrium spectrum.

The upturning behaviour is most clearly visible for small $\Nf$, which is as expected, as for large $\Nf$,  eigenvalue repulsion dominates. The predictions of these models for $n_s$ do not differ appreciably from the mean-zero GRF models with similar initial spectra, as shown in 
 %As an example, we compare the spectral indices between zero-mean and uplited GRFs in 
 Figure \ref{fig:Lifted}.  The uplifted models were generated with the same random seeds as the zero-mean ones, and  have identical odd-order Taylor coefficients (the even-order Taylor coefficients  differ however). The striking similarities in the predictions of these models can be understood as a consequence of the dominance of the cubic terms in the potential during a large fraction of the inflationary evolution. 

Clearly, there are many ways to use GRFs to construct inflationary models. We expect that the findings of this paper will extend also to other constructions of small-field, saddle-point inflation in which the spectrum relaxes during inflation (with some eigenvalues taking tachyonic values). However,  other, substantially different  constructions (e.g.~large-field models or other variants of uplifted models) can certainly lead to different predictions.

\bibliographystyle{JHEP}
\bibliography{references}

\providecommand{\href}[2]{#2}\begingroup\raggedright\begin{thebibliography}{100}

\bibitem{astro-ph/0210603}
J.~M. Maldacena, \emph{{Non-Gaussian features of primordial fluctuations in
  single field inflationary models}},
  \href{http://dx.doi.org/10.1088/1126-6708/2003/05/013}{\emph{JHEP} {\bf 05}
  (2003) 013}, [\href{https://arxiv.org/abs/astro-ph/0210603}{{\tt
  astro-ph/0210603}}].

\bibitem{Creminelli:2004yq}
P.~Creminelli and M.~Zaldarriaga, \emph{{Single field consistency relation for
  the 3-point function}},
  \href{http://dx.doi.org/10.1088/1475-7516/2004/10/006}{\emph{JCAP} {\bf 0410}
  (2004) 006}, [\href{https://arxiv.org/abs/astro-ph/0407059}{{\tt
  astro-ph/0407059}}].

\bibitem{Chen:2013aj}
X.~Chen, H.~Firouzjahi, M.~H. Namjoo and M.~Sasaki, \emph{{A Single Field
  Inflation Model with Large Local Non-Gaussianity}},
  \href{http://dx.doi.org/10.1209/0295-5075/102/59001}{\emph{EPL} {\bf 102}
  (2013) 59001}, [\href{https://arxiv.org/abs/1301.5699}{{\tt 1301.5699}}].

\bibitem{Mooij:2015yka}
S.~Mooij and G.~A. Palma, \emph{{Consistently violating the non-Gaussian
  consistency relation}},
  \href{http://dx.doi.org/10.1088/1475-7516/2015/11/025}{\emph{JCAP} {\bf 1511}
  (2015) 025}, [\href{https://arxiv.org/abs/1502.03458}{{\tt 1502.03458}}].

\bibitem{1502.02114}
{\scshape Planck} collaboration, P.~A.~R. Ade et~al., \emph{{Planck 2015
  results. XX. Constraints on inflation}},
  \href{http://dx.doi.org/10.1051/0004-6361/201525898}{\emph{Astron.
  Astrophys.} {\bf 594} (2016) A20},
  [\href{https://arxiv.org/abs/1502.02114}{{\tt 1502.02114}}].

\bibitem{Ade:2015ava}
{\scshape Planck} collaboration, P.~A.~R. Ade et~al., \emph{{Planck 2015
  results. XVII. Constraints on primordial non-Gaussianity}},
  \href{http://dx.doi.org/10.1051/0004-6361/201525836}{\emph{Astron.
  Astrophys.} {\bf 594} (2016) A17},
  [\href{https://arxiv.org/abs/1502.01592}{{\tt 1502.01592}}].

\bibitem{Dalal:2007cu}
N.~Dalal, O.~Dore, D.~Huterer and A.~Shirokov, \emph{{The imprints of
  primordial non-gaussianities on large-scale structure: scale dependent bias
  and abundance of virialized objects}},
  \href{http://dx.doi.org/10.1103/PhysRevD.77.123514}{\emph{Phys. Rev.} {\bf
  D77} (2008) 123514}, [\href{https://arxiv.org/abs/0710.4560}{{\tt
  0710.4560}}].

\bibitem{Matarrese:2008nc}
S.~Matarrese and L.~Verde, \emph{{The effect of primordial non-Gaussianity on
  halo bias}}, \href{http://dx.doi.org/10.1086/587840}{\emph{Astrophys. J.}
  {\bf 677} (2008) L77--L80}, [\href{https://arxiv.org/abs/0801.4826}{{\tt
  0801.4826}}].

\bibitem{Slosar:2008hx}
A.~Slosar, C.~Hirata, U.~Seljak, S.~Ho and N.~Padmanabhan, \emph{{Constraints
  on local primordial non-Gaussianity from large scale structure}},
  \href{http://dx.doi.org/10.1088/1475-7516/2008/08/031}{\emph{JCAP} {\bf 0808}
  (2008) 031}, [\href{https://arxiv.org/abs/0805.3580}{{\tt 0805.3580}}].

\bibitem{Desjacques:2010jw}
V.~Desjacques and U.~Seljak, \emph{{Primordial non-Gaussianity from the large
  scale structure}},
  \href{http://dx.doi.org/10.1088/0264-9381/27/12/124011}{\emph{Class. Quant.
  Grav.} {\bf 27} (2010) 124011}, [\href{https://arxiv.org/abs/1003.5020}{{\tt
  1003.5020}}].

\bibitem{Alvarez:2014vva}
M.~Alvarez et~al., \emph{{Testing Inflation with Large Scale Structure:
  Connecting Hopes with Reality}},  \href{https://arxiv.org/abs/1412.4671}{{\tt
  1412.4671}}.

\bibitem{Ferraro:2014jba}
S.~Ferraro and K.~M. Smith, \emph{{Using large scale structure to measure
  $f_{NL}, g_{NL}$ and $\tau_{NL}$}},
  \href{http://dx.doi.org/10.1103/PhysRevD.91.043506}{\emph{Phys. Rev.} {\bf
  D91} (2015) 043506}, [\href{https://arxiv.org/abs/1408.3126}{{\tt
  1408.3126}}].

\bibitem{Baldauf:2016sjb}
T.~Baldauf, M.~Mirbabayi, M.~Simonovic and M.~Zaldarriaga, \emph{{LSS
  constraints with controlled theoretical uncertainties}},
  \href{https://arxiv.org/abs/1602.00674}{{\tt 1602.00674}}.

\bibitem{1002.3110}
C.~T. Byrnes and K.-Y. Choi, \emph{{Review of local non-Gaussianity from
  multi-field inflation}},
  \href{http://dx.doi.org/10.1155/2010/724525}{\emph{Adv. Astron.} {\bf 2010}
  (2010) 724525}, [\href{https://arxiv.org/abs/1002.3110}{{\tt 1002.3110}}].

\bibitem{Vernizzi:2006ve}
F.~Vernizzi and D.~Wands, \emph{{Non-gaussianities in two-field inflation}},
  \href{http://dx.doi.org/10.1088/1475-7516/2006/05/019}{\emph{JCAP} {\bf 0605}
  (2006) 019}, [\href{https://arxiv.org/abs/astro-ph/0603799}{{\tt
  astro-ph/0603799}}].

\bibitem{Battefeld:2006sz}
T.~Battefeld and R.~Easther, \emph{{Non-Gaussianities in Multi-field
  Inflation}},
  \href{http://dx.doi.org/10.1088/1475-7516/2007/03/020}{\emph{JCAP} {\bf 0703}
  (2007) 020}, [\href{https://arxiv.org/abs/astro-ph/0610296}{{\tt
  astro-ph/0610296}}].

\bibitem{Senatore:2010wk}
L.~Senatore and M.~Zaldarriaga, \emph{{The Effective Field Theory of Multifield
  Inflation}}, \href{http://dx.doi.org/10.1007/JHEP04(2012)024}{\emph{JHEP}
  {\bf 04} (2012) 024}, [\href{https://arxiv.org/abs/1009.2093}{{\tt
  1009.2093}}].

\bibitem{Peterson:2010mv}
C.~M. Peterson and M.~Tegmark, \emph{{Non-Gaussianity in Two-Field Inflation}},
  \href{http://dx.doi.org/10.1103/PhysRevD.84.023520}{\emph{Phys. Rev.} {\bf
  D84} (2011) 023520}, [\href{https://arxiv.org/abs/1011.6675}{{\tt
  1011.6675}}].

\bibitem{Peterson:2011yt}
C.~M. Peterson and M.~Tegmark, \emph{{Testing multifield inflation: A geometric
  approach}}, \href{http://dx.doi.org/10.1103/PhysRevD.87.103507}{\emph{Phys.
  Rev.} {\bf D87} (2013) 103507}, [\href{https://arxiv.org/abs/1111.0927}{{\tt
  1111.0927}}].

\bibitem{1312.4035}
R.~Easther, J.~Frazer, H.~V. Peiris and L.~C. Price, \emph{{Simple predictions
  from multifield inflationary models}},
  \href{http://dx.doi.org/10.1103/PhysRevLett.112.161302}{\emph{Phys. Rev.
  Lett.} {\bf 112} (2014) 161302}, [\href{https://arxiv.org/abs/1312.4035}{{\tt
  1312.4035}}].

\bibitem{Price:2015qqb}
L.~C. Price, H.~V. Peiris, J.~Frazer and R.~Easther, \emph{{Designing and
  testing inflationary models with Bayesian networks}},
  \href{http://dx.doi.org/10.1088/1475-7516/2016/02/049}{\emph{JCAP} {\bf 1602}
  (2016) 049}, [\href{https://arxiv.org/abs/1511.00029}{{\tt 1511.00029}}].

\bibitem{wigner}
E.~P. Wigner, \emph{{On the statistical distribution of the widths and spacings
  of nuclear resonance levels}}, {\emph{Mathematical Proceedings of the
  Cambridge Philosophical Society} {\bf 47} (1951) 790 -- 798}.

\bibitem{Kuijlaars}
A.~B.~J. {Kuijlaars}, \emph{{Universality}}, {\emph{The Oxford Handbook on
  Random Matrix Theory, (G. Akemann, J. Baik, and P. Di Francesco, eds.),
  Oxford University Press} (2011) 103--134},
  [\href{https://arxiv.org/abs/1103.5922}{{\tt 1103.5922}}].

\bibitem{Deift}
P.~{Deift}, \emph{{Universality for mathematical and physical systems}},
  {\emph{ArXiv Mathematical Physics e-prints} (Mar., 2006) },
  [\href{https://arxiv.org/abs/math-ph/0603038}{{\tt math-ph/0603038}}].

\bibitem{Erdos}
L.~{Erdos}, \emph{{Universality for random matrices and log-gases}},
  {\emph{ArXiv e-prints} (Dec., 2012) },
  [\href{https://arxiv.org/abs/1212.0839}{{\tt 1212.0839}}].

\bibitem{DBM1}
M.~C.~D. Marsh, L.~McAllister, E.~Pajer and T.~Wrase, \emph{{Charting an
  Inflationary Landscape with Random Matrix Theory}},
  \href{http://dx.doi.org/10.1088/1475-7516/2013/11/040}{\emph{JCAP} {\bf 1311}
  (2013) 040}, [\href{https://arxiv.org/abs/1307.3559}{{\tt 1307.3559}}].

\bibitem{DBM2}
M.~Dias, J.~Frazer and M.~C.~D. Marsh, \emph{{Simple emergent power spectra
  from complex inflationary physics}},
  \href{http://dx.doi.org/10.1103/PhysRevLett.117.141303}{\emph{Phys. Rev.
  Lett.} {\bf 117} (2016) 141303},
  [\href{https://arxiv.org/abs/1604.05970}{{\tt 1604.05970}}].

\bibitem{DBM3}
M.~Dias, J.~Frazer and M.~D. Marsh, \emph{{Seven Lessons from Manyfield
  Inflation in Random Potentials}},
  \href{https://arxiv.org/abs/1706.03774}{{\tt 1706.03774}}.

\bibitem{1409.5135}
T.~Battefeld and C.~Modi, \emph{{Local random potentials of high
  differentiability to model the Landscape}},
  \href{http://dx.doi.org/10.1088/1475-7516/2015/03/010}{\emph{JCAP} {\bf 1503}
  (2015) 010}, [\href{https://arxiv.org/abs/1409.5135}{{\tt 1409.5135}}].

\bibitem{Freivogel}
B.~Freivogel, R.~Gobbetti, E.~Pajer and I.-S. Yang, \emph{{Inflation on a
  Slippery Slope}},  \href{https://arxiv.org/abs/1608.00041}{{\tt 1608.00041}}.

\bibitem{1611.07059}
F.~G. Pedro and A.~Westphal, \emph{{Inflation with a graceful exit in a random
  landscape}}, \href{http://dx.doi.org/10.1007/JHEP03(2017)163}{\emph{JHEP}
  {\bf 03} (2017) 163}, [\href{https://arxiv.org/abs/1611.07059}{{\tt
  1611.07059}}].

\bibitem{Wang:2016kzp}
G.~Wang and T.~Battefeld, \emph{{Random Functions via Dyson Brownian Motion:
  Progress and Problems}},
  \href{http://dx.doi.org/10.1088/1475-7516/2016/09/008}{\emph{JCAP} {\bf 1609}
  (2016) 008}, [\href{https://arxiv.org/abs/1607.02514}{{\tt 1607.02514}}].

\bibitem{Frazer:2011tg}
J.~Frazer and A.~R. Liddle, \emph{{Exploring a string-like landscape}},
  \href{http://dx.doi.org/10.1088/1475-7516/2011/02/026}{\emph{JCAP} {\bf 1102}
  (2011) 026}, [\href{https://arxiv.org/abs/1101.1619}{{\tt 1101.1619}}].

\bibitem{1111.6646}
J.~Frazer and A.~R. Liddle, \emph{{Multi-field inflation with random
  potentials: field dimension, feature scale and non-Gaussianity}},
  \href{http://dx.doi.org/10.1088/1475-7516/2012/02/039}{\emph{JCAP} {\bf 1202}
  (2012) 039}, [\href{https://arxiv.org/abs/1111.6646}{{\tt 1111.6646}}].

\bibitem{astro-ph/0410281}
M.~Tegmark, \emph{{What does inflation really predict?}},
  \href{http://dx.doi.org/10.1088/1475-7516/2005/04/001}{\emph{JCAP} {\bf 0504}
  (2005) 001}, [\href{https://arxiv.org/abs/astro-ph/0410281}{{\tt
  astro-ph/0410281}}].

\bibitem{Bachlechner:2014rqa}
T.~C. Bachlechner, \emph{{On Gaussian Random Supergravity}},
  \href{http://dx.doi.org/10.1007/JHEP04(2014)054}{\emph{JHEP} {\bf 04} (2014)
  054}, [\href{https://arxiv.org/abs/1401.6187}{{\tt 1401.6187}}].

\bibitem{1302.3842}
D.~J. Mulryne, \emph{{Transporting non-Gaussianity from sub to super-horizon
  scales}}, \href{http://dx.doi.org/10.1088/1475-7516/2013/09/010}{\emph{JCAP}
  {\bf 1309} (2013) 010}, [\href{https://arxiv.org/abs/1302.3842}{{\tt
  1302.3842}}].

\bibitem{1203.2635}
D.~Seery, D.~J. Mulryne, J.~Frazer and R.~H. Ribeiro, \emph{{Inflationary
  perturbation theory is geometrical optics in phase space}},
  \href{http://dx.doi.org/10.1088/1475-7516/2012/09/010}{\emph{JCAP} {\bf 1209}
  (2012) 010}, [\href{https://arxiv.org/abs/1203.2635}{{\tt 1203.2635}}].

\bibitem{1502.03125}
M.~Dias, J.~Frazer and D.~Seery, \emph{{Computing observables in curved
  multifield models of inflation --- A guide (with code) to the transport
  method}}, \href{http://dx.doi.org/10.1088/1475-7516/2015/12/030}{\emph{JCAP}
  {\bf 1512} (2015) 030}, [\href{https://arxiv.org/abs/1502.03125}{{\tt
  1502.03125}}].

\bibitem{Dias:2016rjq}
M.~Dias, J.~Frazer, D.~J. Mulryne and D.~Seery, \emph{{Numerical evaluation of
  the bispectrum in multiple field inflation -- the transport approach with
  code}}, \href{http://dx.doi.org/10.1088/1475-7516/2016/12/033}{\emph{JCAP}
  {\bf 1612} (2016) 033}, [\href{https://arxiv.org/abs/1609.00379}{{\tt
  1609.00379}}].

\bibitem{Berera:1996nv}
A.~Berera, \emph{{Thermal properties of an inflationary universe}},
  \href{http://dx.doi.org/10.1103/PhysRevD.54.2519}{\emph{Phys. Rev.} {\bf D54}
  (1996) 2519--2534}, [\href{https://arxiv.org/abs/hep-th/9601134}{{\tt
  hep-th/9601134}}].

\bibitem{Aazami:2005jf}
A.~Aazami and R.~Easther, \emph{{Cosmology from random multifield potentials}},
  \href{http://dx.doi.org/10.1088/1475-7516/2006/03/013}{\emph{JCAP} {\bf 0603}
  (2006) 013}, [\href{https://arxiv.org/abs/hep-th/0512050}{{\tt
  hep-th/0512050}}].

\bibitem{Easther:2005zr}
R.~Easther and L.~McAllister, \emph{{Random matrices and the spectrum of
  N-flation}},
  \href{http://dx.doi.org/10.1088/1475-7516/2006/05/018}{\emph{JCAP} {\bf 0605}
  (2006) 018}, [\href{https://arxiv.org/abs/hep-th/0512102}{{\tt
  hep-th/0512102}}].

\bibitem{MarchRussell:2006mj}
J.~March-Russell and F.~Riva, \emph{{Signals of Inflation in a Friendly String
  Landscape}},
  \href{http://dx.doi.org/10.1088/1126-6708/2006/07/033}{\emph{JHEP} {\bf 07}
  (2006) 033}, [\href{https://arxiv.org/abs/astro-ph/0604254}{{\tt
  astro-ph/0604254}}].

\bibitem{Tye:2008ef}
S.~H.~H. Tye, J.~Xu and Y.~Zhang, \emph{{Multi-field Inflation with a Random
  Potential}},
  \href{http://dx.doi.org/10.1088/1475-7516/2009/04/018}{\emph{JCAP} {\bf 0904}
  (2009) 018}, [\href{https://arxiv.org/abs/0812.1944}{{\tt 0812.1944}}].

\bibitem{Tye:2009ff}
S.~H.~H. Tye and J.~Xu, \emph{{A Meandering Inflaton}},
  \href{http://dx.doi.org/10.1016/j.physletb.2009.12.045}{\emph{Phys. Lett.}
  {\bf B683} (2010) 326--330}, [\href{https://arxiv.org/abs/0910.0849}{{\tt
  0910.0849}}].

\bibitem{Battefeld:2012qx}
D.~Battefeld, T.~Battefeld and S.~Schulz, \emph{{On the Unlikeliness of
  Multi-Field Inflation: Bounded Random Potentials and our Vacuum}},
  \href{http://dx.doi.org/10.1088/1475-7516/2012/06/034}{\emph{JCAP} {\bf 1206}
  (2012) 034}, [\href{https://arxiv.org/abs/1203.3941}{{\tt 1203.3941}}].

\bibitem{Battefeld:2013xwa}
D.~Battefeld and T.~Battefeld, \emph{{A Smooth Landscape: Ending Saddle Point
  Inflation Requires Features to be Shallow}},
  \href{http://dx.doi.org/10.1088/1475-7516/2013/07/038}{\emph{JCAP} {\bf 1307}
  (2013) 038}, [\href{https://arxiv.org/abs/1304.0461}{{\tt 1304.0461}}].

\bibitem{Pedro:2013nda}
F.~G. Pedro and A.~Westphal, \emph{{The Scale of Inflation in the Landscape}},
  \href{http://dx.doi.org/10.1016/j.physletb.2014.10.022}{\emph{Phys. Lett.}
  {\bf B739} (2014) 439--444}, [\href{https://arxiv.org/abs/1303.3224}{{\tt
  1303.3224}}].

\bibitem{Liu:2015dda}
J.~Liu, Y.~Wang and S.~Zhou, \emph{{Nonuniqueness of classical inflationary
  trajectories on a high-dimensional landscape}},
  \href{http://dx.doi.org/10.1103/PhysRevD.91.103525}{\emph{Phys. Rev.} {\bf
  D91} (2015) 103525}, [\href{https://arxiv.org/abs/1501.06785}{{\tt
  1501.06785}}].

\bibitem{Linde:2016uec}
A.~Linde, \emph{{Random Potentials and Cosmological Attractors}},
  \href{http://dx.doi.org/10.1088/1475-7516/2017/02/028}{\emph{JCAP} {\bf 1702}
  (2017) 028}, [\href{https://arxiv.org/abs/1612.04505}{{\tt 1612.04505}}].

\bibitem{Masoumi:2016eag}
A.~Masoumi, A.~Vilenkin and M.~Yamada, \emph{{Inflation in random Gaussian
  landscapes}},
  \href{http://dx.doi.org/10.1088/1475-7516/2017/05/053}{\emph{JCAP} {\bf 1705}
  (2017) 053}, [\href{https://arxiv.org/abs/1612.03960}{{\tt 1612.03960}}].

\bibitem{He:2017kqc}
Y.-H. He, V.~Jejjala, L.~Pontiggia, Y.~Xiao and D.~Zhou, \emph{{Flatness of
  Minima in Random Inflationary Landscapes}},
  \href{https://arxiv.org/abs/1704.08351}{{\tt 1704.08351}}.

\bibitem{Liu:2017dzi}
J.~Liu, \emph{{Artificial Neural Network in Cosmic Landscape}},
  \href{https://arxiv.org/abs/1707.02800}{{\tt 1707.02800}}.

\bibitem{Masoumi:2017xbe}
A.~Masoumi, A.~Vilenkin and M.~Yamada, \emph{{Inflation in multi-field random
  Gaussian landscapes}},  \href{https://arxiv.org/abs/1707.03520}{{\tt
  1707.03520}}.

\bibitem{Green:2014xqa}
D.~Green, \emph{{Disorder in the Early Universe}},
  \href{http://dx.doi.org/10.1088/1475-7516/2015/03/020}{\emph{JCAP} {\bf 1503}
  (2015) 020}, [\href{https://arxiv.org/abs/1409.6698}{{\tt 1409.6698}}].

\bibitem{Amin:2015ftc}
M.~A. Amin and D.~Baumann, \emph{{From Wires to Cosmology}},
  \href{http://dx.doi.org/10.1088/1475-7516/2016/02/045}{\emph{JCAP} {\bf 1602}
  (2016) 045}, [\href{https://arxiv.org/abs/1512.02637}{{\tt 1512.02637}}].

\bibitem{Amin:2017wvc}
M.~A. Amin, M.~A.~G. Garcia, H.-Y. Xie and O.~Wen, \emph{{Multifield Stochastic
  Particle Production: Beyond a Maximum Entropy Ansatz}},
  \href{https://arxiv.org/abs/1706.02319}{{\tt 1706.02319}}.

\bibitem{Agarwal:2011wm}
N.~Agarwal, R.~Bean, L.~McAllister and G.~Xu, \emph{{Universality in D-brane
  Inflation}},
  \href{http://dx.doi.org/10.1088/1475-7516/2011/09/002}{\emph{JCAP} {\bf 1109}
  (2011) 002}, [\href{https://arxiv.org/abs/1103.2775}{{\tt 1103.2775}}].

\bibitem{McAllister:2012am}
L.~McAllister, S.~Renaux-Petel and G.~Xu, \emph{{A Statistical Approach to
  Multifield Inflation: Many-field Perturbations Beyond Slow Roll}},
  \href{http://dx.doi.org/10.1088/1475-7516/2012/10/046}{\emph{JCAP} {\bf 1210}
  (2012) 046}, [\href{https://arxiv.org/abs/1207.0317}{{\tt 1207.0317}}].

\bibitem{Dias:2012nf}
M.~Dias, J.~Frazer and A.~R. Liddle, \emph{{Multifield consequences for D-brane
  inflation}}, \href{http://dx.doi.org/10.1088/1475-7516/2013/03/E01,
  10.1088/1475-7516/2012/06/020}{\emph{JCAP} {\bf 1206} (2012) 020},
  [\href{https://arxiv.org/abs/1203.3792}{{\tt 1203.3792}}].

\bibitem{Starobinsky:1986fxa}
A.~A. Starobinsky, \emph{{Multicomponent de Sitter (Inflationary) Stages and
  the Generation of Perturbations}}, {\emph{JETP Lett.} {\bf 42} (1985)
  152--155}.

\bibitem{Sasaki:1995aw}
M.~Sasaki and E.~D. Stewart, \emph{{A General analytic formula for the spectral
  index of the density perturbations produced during inflation}},
  \href{http://dx.doi.org/10.1143/PTP.95.71}{\emph{Prog. Theor. Phys.} {\bf 95}
  (1996) 71--78}, [\href{https://arxiv.org/abs/astro-ph/9507001}{{\tt
  astro-ph/9507001}}].

\bibitem{Lyth:2005fi}
D.~H. Lyth and Y.~Rodriguez, \emph{{The Inflationary prediction for primordial
  non-Gaussianity}},
  \href{http://dx.doi.org/10.1103/PhysRevLett.95.121302}{\emph{Phys. Rev.
  Lett.} {\bf 95} (2005) 121302},
  [\href{https://arxiv.org/abs/astro-ph/0504045}{{\tt astro-ph/0504045}}].

\bibitem{GRF2}
T.~Bjorkmo and M.~D. Marsh, ``{To appear}.''

\bibitem{Bray:2007tf}
A.~J. Bray and D.~S. Dean, \emph{{Statistics of critical points of Gaussian
  fields on large-dimensional spaces}},
  \href{http://dx.doi.org/10.1103/PhysRevLett.98.150201}{\emph{Phys. Rev.
  Lett.} {\bf 98} (2007) 150201}.

\bibitem{cond-mat/0609651}
D.~S. Dean and S.~N. Majumdar, \emph{{Large deviations of extreme eigenvalues
  of random matrices}},
  \href{http://dx.doi.org/10.1103/PhysRevLett.97.160201}{\emph{Phys. Rev.
  Lett.} {\bf 97} (2006) 160201},
  [\href{https://arxiv.org/abs/cond-mat/0609651}{{\tt cond-mat/0609651}}].

\bibitem{1112.3034}
D.~Marsh, L.~McAllister and T.~Wrase, \emph{{The Wasteland of Random
  Supergravities}},
  \href{http://dx.doi.org/10.1007/JHEP03(2012)102}{\emph{JHEP} {\bf 03} (2012)
  102}, [\href{https://arxiv.org/abs/1112.3034}{{\tt 1112.3034}}].

\bibitem{Bachlechner:2012at}
T.~C. Bachlechner, D.~Marsh, L.~McAllister and T.~Wrase, \emph{{Supersymmetric
  Vacua in Random Supergravity}},
  \href{http://dx.doi.org/10.1007/JHEP01(2013)136}{\emph{JHEP} {\bf 01} (2013)
  136}, [\href{https://arxiv.org/abs/1207.2763}{{\tt 1207.2763}}].

\bibitem{Easther:2016ire}
R.~Easther, A.~H. Guth and A.~Masoumi, \emph{{Counting Vacua in Random
  Landscapes}},  \href{https://arxiv.org/abs/1612.05224}{{\tt 1612.05224}}.

\bibitem{Kachru:2003aw}
S.~Kachru, R.~Kallosh, A.~D. Linde and S.~P. Trivedi, \emph{{De Sitter vacua in
  string theory}},
  \href{http://dx.doi.org/10.1103/PhysRevD.68.046005}{\emph{Phys. Rev.} {\bf
  D68} (2003) 046005}, [\href{https://arxiv.org/abs/hep-th/0301240}{{\tt
  hep-th/0301240}}].

\bibitem{Ferrara:2014kva}
S.~Ferrara, R.~Kallosh and A.~Linde, \emph{{Cosmology with Nilpotent
  Superfields}}, \href{http://dx.doi.org/10.1007/JHEP10(2014)143}{\emph{JHEP}
  {\bf 10} (2014) 143}, [\href{https://arxiv.org/abs/1408.4096}{{\tt
  1408.4096}}].

\bibitem{Westphal:2006tn}
A.~Westphal, \emph{{de Sitter string vacua from Kahler uplifting}},
  \href{http://dx.doi.org/10.1088/1126-6708/2007/03/102}{\emph{JHEP} {\bf 03}
  (2007) 102}, [\href{https://arxiv.org/abs/hep-th/0611332}{{\tt
  hep-th/0611332}}].

\bibitem{Achucarro:2006zf}
A.~Achucarro, B.~de~Carlos, J.~A. Casas and L.~Doplicher, \emph{{De Sitter
  vacua from uplifting D-terms in effective supergravities from realistic
  strings}}, \href{http://dx.doi.org/10.1088/1126-6708/2006/06/014}{\emph{JHEP}
  {\bf 06} (2006) 014}, [\href{https://arxiv.org/abs/hep-th/0601190}{{\tt
  hep-th/0601190}}].

\bibitem{Dudas:2006vc}
E.~Dudas and Y.~Mambrini, \emph{{Moduli stabilization with positive vacuum
  energy}}, \href{http://dx.doi.org/10.1088/1126-6708/2006/10/044}{\emph{JHEP}
  {\bf 10} (2006) 044}, [\href{https://arxiv.org/abs/hep-th/0607077}{{\tt
  hep-th/0607077}}].

\bibitem{Rummel:2014raa}
M.~Rummel and Y.~Sumitomo, \emph{{De Sitter Vacua from a D-term Generated
  Racetrack Uplift}},
  \href{http://dx.doi.org/10.1007/JHEP01(2015)015}{\emph{JHEP} {\bf 01} (2015)
  015}, [\href{https://arxiv.org/abs/1407.7580}{{\tt 1407.7580}}].

\bibitem{Blaback:2015zra}
J.~Blaback, U.~H. Danielsson, G.~Dibitetto and S.~C. Vargas, \emph{{Universal
  dS vacua in STU-models}},
  \href{http://dx.doi.org/10.1007/JHEP10(2015)069}{\emph{JHEP} {\bf 10} (2015)
  069}, [\href{https://arxiv.org/abs/1505.04283}{{\tt 1505.04283}}].

\bibitem{Cicoli:2015ylx}
M.~Cicoli, F.~Quevedo and R.~Valandro, \emph{{De Sitter from T-branes}},
  \href{http://dx.doi.org/10.1007/JHEP03(2016)141}{\emph{JHEP} {\bf 03} (2016)
  141}, [\href{https://arxiv.org/abs/1512.04558}{{\tt 1512.04558}}].

\bibitem{Marsh:2014nla}
M.~C.~D. Marsh, B.~Vercnocke and T.~Wrase, \emph{{Decoupling and de Sitter
  Vacua in Approximate No-Scale Supergravities}},
  \href{http://dx.doi.org/10.1007/JHEP05(2015)081}{\emph{JHEP} {\bf 05} (2015)
  081}, [\href{https://arxiv.org/abs/1411.6625}{{\tt 1411.6625}}].

\bibitem{Gallego:2017dvd}
D.~Gallego, M.~C.~D. Marsh, B.~Vercnocke and T.~Wrase, \emph{{A New Class of de
  Sitter Vacua in Type IIB Large Volume Compactifications}},
  \href{https://arxiv.org/abs/1707.01095}{{\tt 1707.01095}}.

\bibitem{Bachlechner:2017hsj}
T.~C. Bachlechner, K.~Eckerle, O.~Janssen and M.~Kleban, \emph{{Systematics of
  Aligned Axions}},  \href{https://arxiv.org/abs/1709.01080}{{\tt 1709.01080}}.

\bibitem{Lyth:1998xn}
D.~H. Lyth and A.~Riotto, \emph{{Particle physics models of inflation and the
  cosmological density perturbation}},
  \href{http://dx.doi.org/10.1016/S0370-1573(98)00128-8}{\emph{Phys. Rept.}
  {\bf 314} (1999) 1--146}, [\href{https://arxiv.org/abs/hep-ph/9807278}{{\tt
  hep-ph/9807278}}].

\bibitem{Mulryne:2009ci}
D.~Mulryne, D.~Seery and D.~Wesley, \emph{{Non-Gaussianity constrains hybrid
  inflation}},  \href{https://arxiv.org/abs/0911.3550}{{\tt 0911.3550}}.

\bibitem{Chen:2009zp}
X.~Chen and Y.~Wang, \emph{{Quasi-Single Field Inflation and
  Non-Gaussianities}},
  \href{http://dx.doi.org/10.1088/1475-7516/2010/04/027}{\emph{JCAP} {\bf 1004}
  (2010) 027}, [\href{https://arxiv.org/abs/0911.3380}{{\tt 0911.3380}}].

\bibitem{Dias:2011xy}
M.~Dias and D.~Seery, \emph{{Transport equations for the inflationary spectral
  index}}, \href{http://dx.doi.org/10.1103/PhysRevD.85.043519}{\emph{Phys.
  Rev.} {\bf D85} (2012) 043519}, [\href{https://arxiv.org/abs/1111.6544}{{\tt
  1111.6544}}].

\bibitem{Anderson:2012em}
G.~J. Anderson, D.~J. Mulryne and D.~Seery, \emph{{Transport equations for the
  inflationary trispectrum}},
  \href{http://dx.doi.org/10.1088/1475-7516/2012/10/019}{\emph{JCAP} {\bf 1210}
  (2012) 019}, [\href{https://arxiv.org/abs/1205.0024}{{\tt 1205.0024}}].

\bibitem{Seery:2016lko}
D.~Seery, \emph{{CppTransport: a platform to automate calculation of
  inflationary correlation functions}},
  \href{https://arxiv.org/abs/1609.00380}{{\tt 1609.00380}}.

\bibitem{Mulryne:2016mzv}
D.~J. Mulryne and J.~W. Ronayne, \emph{{PyTransport: A Python package for the
  calculation of inflationary correlation functions}},
  \href{https://arxiv.org/abs/1609.00381}{{\tt 1609.00381}}.

\bibitem{Ronayne:2017qzn}
J.~W. Ronayne and D.~J. Mulryne, \emph{{Numerically evaluating the bispectrum
  in curved field-space - with PyTransport 2.0}},
  \href{https://arxiv.org/abs/1708.07130}{{\tt 1708.07130}}.

\bibitem{Lyth:1984gv}
D.~H. Lyth, \emph{{Large Scale Energy Density Perturbations and Inflation}},
  \href{http://dx.doi.org/10.1103/PhysRevD.31.1792}{\emph{Phys. Rev.} {\bf D31}
  (1985) 1792--1798}.

\bibitem{Wands:2000dp}
D.~Wands, K.~A. Malik, D.~H. Lyth and A.~R. Liddle, \emph{{A New approach to
  the evolution of cosmological perturbations on large scales}},
  \href{http://dx.doi.org/10.1103/PhysRevD.62.043527}{\emph{Phys. Rev.} {\bf
  D62} (2000) 043527}, [\href{https://arxiv.org/abs/astro-ph/0003278}{{\tt
  astro-ph/0003278}}].

\bibitem{GrootNibbelink:2001qt}
S.~Groot~Nibbelink and B.~J.~W. van Tent, \emph{{Scalar perturbations during
  multiple field slow-roll inflation}},
  \href{http://dx.doi.org/10.1088/0264-9381/19/4/302}{\emph{Class. Quant.
  Grav.} {\bf 19} (2002) 613--640},
  [\href{https://arxiv.org/abs/hep-ph/0107272}{{\tt hep-ph/0107272}}].

\bibitem{GarciaBellido:1995qq}
J.~Garcia-Bellido and D.~Wands, \emph{{Metric perturbations in two field
  inflation}}, \href{http://dx.doi.org/10.1103/PhysRevD.53.5437}{\emph{Phys.
  Rev.} {\bf D53} (1996) 5437--5445},
  [\href{https://arxiv.org/abs/astro-ph/9511029}{{\tt astro-ph/9511029}}].

\bibitem{Gordon:2000hv}
C.~Gordon, D.~Wands, B.~A. Bassett and R.~Maartens, \emph{{Adiabatic and
  entropy perturbations from inflation}},
  \href{http://dx.doi.org/10.1103/PhysRevD.63.023506}{\emph{Phys. Rev.} {\bf
  D63} (2001) 023506}, [\href{https://arxiv.org/abs/astro-ph/0009131}{{\tt
  astro-ph/0009131}}].

\bibitem{Lyth:1996im}
D.~H. Lyth, \emph{{What would we learn by detecting a gravitational wave signal
  in the cosmic microwave background anisotropy?}},
  \href{http://dx.doi.org/10.1103/PhysRevLett.78.1861}{\emph{Phys. Rev. Lett.}
  {\bf 78} (1997) 1861--1863},
  [\href{https://arxiv.org/abs/hep-ph/9606387}{{\tt hep-ph/9606387}}].

\bibitem{1612.05138}
A.~Pourtsidou, \emph{{Synergistic tests of inflation}},
  \href{https://arxiv.org/abs/1612.05138}{{\tt 1612.05138}}.

\bibitem{Amendola:2001ni}
L.~Amendola, C.~Gordon, D.~Wands and M.~Sasaki, \emph{{Correlated perturbations
  from inflation and the cosmic microwave background}},
  \href{http://dx.doi.org/10.1103/PhysRevLett.88.211302}{\emph{Phys. Rev.
  Lett.} {\bf 88} (2002) 211302},
  [\href{https://arxiv.org/abs/astro-ph/0107089}{{\tt astro-ph/0107089}}].

\bibitem{Wands:2002bn}
D.~Wands, N.~Bartolo, S.~Matarrese and A.~Riotto, \emph{{An Observational test
  of two-field inflation}},
  \href{http://dx.doi.org/10.1103/PhysRevD.66.043520}{\emph{Phys. Rev.} {\bf
  D66} (2002) 043520}, [\href{https://arxiv.org/abs/astro-ph/0205253}{{\tt
  astro-ph/0205253}}].

\bibitem{Byrnes:2006fr}
C.~T. Byrnes and D.~Wands, \emph{{Curvature and isocurvature perturbations from
  two-field inflation in a slow-roll expansion}},
  \href{http://dx.doi.org/10.1103/PhysRevD.74.043529}{\emph{Phys. Rev.} {\bf
  D74} (2006) 043529}, [\href{https://arxiv.org/abs/astro-ph/0605679}{{\tt
  astro-ph/0605679}}].

\bibitem{Bernardeau:2002jy}
F.~Bernardeau and J.-P. Uzan, \emph{{NonGaussianity in multifield inflation}},
  \href{http://dx.doi.org/10.1103/PhysRevD.66.103506}{\emph{Phys. Rev.} {\bf
  D66} (2002) 103506}, [\href{https://arxiv.org/abs/hep-ph/0207295}{{\tt
  hep-ph/0207295}}].

\bibitem{Bernardeau:2002jf}
F.~Bernardeau and J.-P. Uzan, \emph{{Inflationary models inducing non-Gaussian
  metric fluctuations}},
  \href{http://dx.doi.org/10.1103/PhysRevD.67.121301}{\emph{Phys. Rev.} {\bf
  D67} (2003) 121301}, [\href{https://arxiv.org/abs/astro-ph/0209330}{{\tt
  astro-ph/0209330}}].

\bibitem{Rigopoulos:2005us}
G.~I. Rigopoulos, E.~P.~S. Shellard and B.~J.~W. van Tent, \emph{{Quantitative
  bispectra from multifield inflation}},
  \href{http://dx.doi.org/10.1103/PhysRevD.76.083512}{\emph{Phys. Rev.} {\bf
  D76} (2007) 083512}, [\href{https://arxiv.org/abs/astro-ph/0511041}{{\tt
  astro-ph/0511041}}].

\bibitem{Rigopoulos:2005ae}
G.~I. Rigopoulos, E.~P.~S. Shellard and B.~J.~W. van Tent, \emph{{Large
  non-Gaussianity in multiple-field inflation}},
  \href{http://dx.doi.org/10.1103/PhysRevD.73.083522}{\emph{Phys. Rev.} {\bf
  D73} (2006) 083522}, [\href{https://arxiv.org/abs/astro-ph/0506704}{{\tt
  astro-ph/0506704}}].

\bibitem{Bartolo:2001cw}
N.~Bartolo, S.~Matarrese and A.~Riotto, \emph{{Nongaussianity from inflation}},
  \href{http://dx.doi.org/10.1103/PhysRevD.65.103505}{\emph{Phys. Rev.} {\bf
  D65} (2002) 103505}, [\href{https://arxiv.org/abs/hep-ph/0112261}{{\tt
  hep-ph/0112261}}].

\bibitem{Wang:2010si}
T.~Wang, \emph{{Note on Non-Gaussianities in Two-field Inflation}},
  \href{http://dx.doi.org/10.1103/PhysRevD.82.123515}{\emph{Phys. Rev.} {\bf
  D82} (2010) 123515}, [\href{https://arxiv.org/abs/1008.3198}{{\tt
  1008.3198}}].

\bibitem{Byrnes:2008wi}
C.~T. Byrnes, K.-Y. Choi and L.~M.~H. Hall, \emph{{Conditions for large
  non-Gaussianity in two-field slow-roll inflation}},
  \href{http://dx.doi.org/10.1088/1475-7516/2008/10/008}{\emph{JCAP} {\bf 0810}
  (2008) 008}, [\href{https://arxiv.org/abs/0807.1101}{{\tt 0807.1101}}].

\bibitem{Byrnes:2008zy}
C.~T. Byrnes, K.-Y. Choi and L.~M.~H. Hall, \emph{{Large non-Gaussianity from
  two-component hybrid inflation}},
  \href{http://dx.doi.org/10.1088/1475-7516/2009/02/017}{\emph{JCAP} {\bf 0902}
  (2009) 017}, [\href{https://arxiv.org/abs/0812.0807}{{\tt 0812.0807}}].

\bibitem{Tanaka:2010km}
T.~Tanaka, T.~Suyama and S.~Yokoyama, \emph{{Use of delta N formalism -
  Difficulties in generating large local-type non-Gaussianity during inflation
  -}}, \href{http://dx.doi.org/10.1088/0264-9381/27/12/124003}{\emph{Class.
  Quant. Grav.} {\bf 27} (2010) 124003},
  [\href{https://arxiv.org/abs/1003.5057}{{\tt 1003.5057}}].

\bibitem{Chen:2011zf}
X.~Chen, \emph{{Primordial Features as Evidence for Inflation}},
  \href{http://dx.doi.org/10.1088/1475-7516/2012/01/038}{\emph{JCAP} {\bf 1201}
  (2012) 038}, [\href{https://arxiv.org/abs/1104.1323}{{\tt 1104.1323}}].

\bibitem{Achucarro:2010jv}
A.~Achucarro, J.-O. Gong, S.~Hardeman, G.~A. Palma and S.~P. Patil, \emph{{Mass
  hierarchies and non-decoupling in multi-scalar field dynamics}},
  \href{http://dx.doi.org/10.1103/PhysRevD.84.043502}{\emph{Phys. Rev.} {\bf
  D84} (2011) 043502}, [\href{https://arxiv.org/abs/1005.3848}{{\tt
  1005.3848}}].

\bibitem{Linde:1996gt}
A.~D. Linde and V.~F. Mukhanov, \emph{{Nongaussian isocurvature perturbations
  from inflation}},
  \href{http://dx.doi.org/10.1103/PhysRevD.56.R535}{\emph{Phys. Rev.} {\bf D56}
  (1997) R535--R539}, [\href{https://arxiv.org/abs/astro-ph/9610219}{{\tt
  astro-ph/9610219}}].

\bibitem{Lyth:2001nq}
D.~H. Lyth and D.~Wands, \emph{{Generating the curvature perturbation without
  an inflaton}},
  \href{http://dx.doi.org/10.1016/S0370-2693(01)01366-1}{\emph{Phys. Lett.}
  {\bf B524} (2002) 5--14}, [\href{https://arxiv.org/abs/hep-ph/0110002}{{\tt
  hep-ph/0110002}}].

\bibitem{Moroi:2002rd}
T.~Moroi and T.~Takahashi, \emph{{Cosmic density perturbations from late
  decaying scalar condensations}},
  \href{http://dx.doi.org/10.1103/PhysRevD.66.063501}{\emph{Phys. Rev.} {\bf
  D66} (2002) 063501}, [\href{https://arxiv.org/abs/hep-ph/0206026}{{\tt
  hep-ph/0206026}}].

\bibitem{Lyth:2003}
D.~H. Lyth, C.~Ungarelli and D.~Wands, \emph{Primordial density perturbation in
  the curvaton scenario},
  \href{http://dx.doi.org/10.1103/PhysRevD.67.023503}{\emph{Phys. Rev. D} {\bf
  67} (Jan, 2003) 023503}.

\bibitem{1612.05248}
R.~de~Putter, J.~Gleyzes and O.~Dor\'e, \emph{{The next non-Gaussianity
  frontier: what can a measurement with $\sigma(f_{\rm NL}) \lesssim 1$ tell us
  about multifield inflation?}},  \href{https://arxiv.org/abs/1612.05248}{{\tt
  1612.05248}}.

\bibitem{Leung:2012ve}
G.~Leung, E.~R.~M. Tarrant, C.~T. Byrnes and E.~J. Copeland, \emph{{Reheating,
  Multifield Inflation and the Fate of the Primordial Observables}},
  \href{http://dx.doi.org/10.1088/1475-7516/2012/09/008}{\emph{JCAP} {\bf 1209}
  (2012) 008}, [\href{https://arxiv.org/abs/1206.5196}{{\tt 1206.5196}}].

\bibitem{Meyers:2013gua}
J.~Meyers and E.~R.~M. Tarrant, \emph{{Perturbative Reheating After
  Multiple-Field Inflation: The Impact on Primordial Observables}},
  \href{http://dx.doi.org/10.1103/PhysRevD.89.063535}{\emph{Phys. Rev.} {\bf
  D89} (2014) 063535}, [\href{https://arxiv.org/abs/1311.3972}{{\tt
  1311.3972}}].

\bibitem{Hardwick:2016whe}
R.~J. Hardwick, V.~Vennin, K.~Koyama and D.~Wands, \emph{{Constraining
  Curvatonic Reheating}},
  \href{http://dx.doi.org/10.1088/1475-7516/2016/08/042}{\emph{JCAP} {\bf 1608}
  (2016) 042}, [\href{https://arxiv.org/abs/1606.01223}{{\tt 1606.01223}}].

\bibitem{Hotinli:2017vhx}
S.~C. Hotinli, J.~Frazer, A.~H. Jaffe, J.~Meyers, L.~C. Price and E.~R.~M.
  Tarrant, \emph{{Effect of reheating on predictions following multiple-field
  inflation}}, \href{http://dx.doi.org/10.1103/PhysRevD.97.023511}{\emph{Phys.
  Rev.} {\bf D97} (2018) 023511}, [\href{https://arxiv.org/abs/1710.08913}{{\tt
  1710.08913}}].

\bibitem{Silverstein:2003hf}
E.~Silverstein and D.~Tong, \emph{{Scalar speed limits and cosmology:
  Acceleration from D-cceleration}},
  \href{http://dx.doi.org/10.1103/PhysRevD.70.103505}{\emph{Phys. Rev.} {\bf
  D70} (2004) 103505}, [\href{https://arxiv.org/abs/hep-th/0310221}{{\tt
  hep-th/0310221}}].

\bibitem{Renaux-Petel:2015mga}
S.~Renaux-Petel and K.~Turzynski, \emph{{Geometrical Destabilization of
  Inflation}},
  \href{http://dx.doi.org/10.1103/PhysRevLett.117.141301}{\emph{Phys. Rev.
  Lett.} {\bf 117} (2016) 141301},
  [\href{https://arxiv.org/abs/1510.01281}{{\tt 1510.01281}}].

\bibitem{Renaux-Petel:2017dia}
S.~Renaux-Petel, K.~Turzynski and V.~Vennin, \emph{{Geometrical
  destabilization, premature end of inflation and Bayesian model selection}},
  \href{https://arxiv.org/abs/1706.01835}{{\tt 1706.01835}}.

\bibitem{1612.04505}
A.~Linde, \emph{{Random Potentials and Cosmological Attractors}},
  \href{http://dx.doi.org/10.1088/1475-7516/2017/02/028}{\emph{JCAP} {\bf 1702}
  (2017) 028}, [\href{https://arxiv.org/abs/1612.04505}{{\tt 1612.04505}}].

\bibitem{Brown:2017osf}
A.~R. Brown, \emph{{Hyperinflation}},
  \href{https://arxiv.org/abs/1705.03023}{{\tt 1705.03023}}.

\bibitem{Green:2013rd}
D.~Green, M.~Lewandowski, L.~Senatore, E.~Silverstein and M.~Zaldarriaga,
  \emph{{Anomalous Dimensions and Non-Gaussianity}},
  \href{http://dx.doi.org/10.1007/JHEP10(2013)171}{\emph{JHEP} {\bf 10} (2013)
  171}, [\href{https://arxiv.org/abs/1301.2630}{{\tt 1301.2630}}].

\bibitem{Assassi:2013gxa}
V.~Assassi, D.~Baumann, D.~Green and L.~McAllister, \emph{{Planck-Suppressed
  Operators}},
  \href{http://dx.doi.org/10.1088/1475-7516/2014/01/033}{\emph{JCAP} {\bf 1401}
  (2014) 033}, [\href{https://arxiv.org/abs/1304.5226}{{\tt 1304.5226}}].

\bibitem{Kaiser:2013sna}
D.~I. Kaiser and E.~I. Sfakianakis, \emph{{Multifield Inflation after Planck:
  The Case for Nonminimal Couplings}},
  \href{http://dx.doi.org/10.1103/PhysRevLett.112.011302}{\emph{Phys. Rev.
  Lett.} {\bf 112} (2014) 011302}, [\href{https://arxiv.org/abs/1304.0363}{{\tt
  1304.0363}}].

\bibitem{Schutz:2013fua}
K.~Schutz, E.~I. Sfakianakis and D.~I. Kaiser, \emph{{Multifield Inflation
  after Planck: Isocurvature Modes from Nonminimal Couplings}},
  \href{http://dx.doi.org/10.1103/PhysRevD.89.064044}{\emph{Phys. Rev.} {\bf
  D89} (2014) 064044}, [\href{https://arxiv.org/abs/1310.8285}{{\tt
  1310.8285}}].

\bibitem{DeCross:2015uza}
M.~P. DeCross, D.~I. Kaiser, A.~Prabhu, C.~Prescod-Weinstein and E.~I.
  Sfakianakis, \emph{{Preheating after Multifield Inflation with Nonminimal
  Couplings, I: Covariant Formalism and Attractor Behavior}},
  \href{https://arxiv.org/abs/1510.08553}{{\tt 1510.08553}}.

\bibitem{DeCross:2016fdz}
M.~P. DeCross, D.~I. Kaiser, A.~Prabhu, C.~Prescod-Weinstein and E.~I.
  Sfakianakis, \emph{{Preheating after multifield inflation with nonminimal
  couplings, II: Resonance Structure}},
  \href{https://arxiv.org/abs/1610.08868}{{\tt 1610.08868}}.

\bibitem{DeCross:2016cbs}
M.~P. DeCross, D.~I. Kaiser, A.~Prabhu, C.~Prescod-Weinstein and E.~I.
  Sfakianakis, \emph{{Preheating after multifield inflation with nonminimal
  couplings, III: Dynamical spacetime results}},
  \href{https://arxiv.org/abs/1610.08916}{{\tt 1610.08916}}.

\bibitem{Dyson}
F.~J. Dyson, \emph{{A Brownian-Motion Model for the Eigenvalues of a Random
  Matrix}}, {\emph{Journal of Mathematical Physics} {\bf 3} (1962) 1191 --
  1198}.

\bibitem{Uhlenbeck}
G.~Uhlenbeck and L.~Ornstein, \emph{{On the Theory of the Brownian Motion}},
  {\emph{Phys.Rev.} {\bf 36} (1930) 823 -- 841}.

\end{thebibliography}\endgroup

%\begin{thebibliography}{99}

%\bibitem{a}
%Author, \emph{Title}, \emph{J. Abbrev.} {\bf vol} (year) pg.
%
%\bibitem{b}
%Author, \emph{Title},
%arxiv:1234.5678.
%
%\bibitem{c}
%Author, \emph{Title},
%Publisher (year).

% Please avoid comments such as "For a review'', "For some examples",
% "and references therein" or move them in the text. In general,
% please leave only references in the bibliography and move all
% accessory text in footnotes.

% Also, please have only one work for each \bibitem.

%\end{thebibliography}
\end{document}